\documentclass[12pt]{article}

\usepackage{multirow}
\usepackage{calc}
\usepackage{amsmath,amssymb,amsthm,amscd}
\numberwithin{equation}{section}
\usepackage[bf]{caption}
\usepackage{longtable}
\usepackage{array}
\usepackage{enumerate}
\usepackage{float}
\usepackage{subfig}
\usepackage{multicol}
\usepackage{multirow}
\usepackage{epsfig}
\usepackage{graphicx}
\usepackage{pict2e}
\usepackage{color}
\usepackage{authblk}
\usepackage{cite}
\usepackage[all]{xy}
\usepackage[width=1.09\textwidth]{caption}
\usepackage{bbm}
\usepackage{hyperref}

\usepackage[utf8]{inputenc}
\usepackage[T1]{fontenc}
\usepackage[english]{babel}
\usepackage{xspace}
\usepackage{pifont}

\newcommand{\U}[1]{\text{U(#1)}\xspace}
\newcommand{\SU}[1]{\text{SU(#1)}\xspace}

\def\cS{{\mathcal S}}

\newcommand{\one}[0]{\ensuremath{\mathbf{1} }\xspace}
\newcommand{\two}[0]{\ensuremath{\mathbf{2} }\xspace}
\newcommand{\eight}[0]{\ensuremath{\mathbf{8} }\xspace}
\newcommand{\fiveteen}[0]{\ensuremath{\mathbf{15} }\xspace}
\newcommand{\three}[0]{\ensuremath{\mathbf{3} }\xspace}

\newcommand{\four}[0]{\ensuremath{\mathbf{4} }\xspace}
\newcommand{\six}[0]{\ensuremath{\mathbf{6} }\xspace}

\newcommand{\beq}{\begin{equation}}
\newcommand{\eeq}{\end{equation}}
\newcommand{\bea}{\begin{eqnarray}}
\newcommand{\eea}{\end{eqnarray}}

\newcommand{\nn}{\nonumber}

\begin{document}

\baselineskip=15pt

\begin{titlepage}
\begin{flushright}
\parbox[t]{.91in}{UPR-1264-T}\\
\parbox[t]{1.73in}{CERN-PH-TH-2014-171}\\
\parbox[t]{1.27in}{Bonn-TH-2014-13}
\end{flushright}

\begin{center}

\vspace*{ 0.0cm}
{\Large {\bf F-Theory on all Toric Hypersurface Fibrations\\[0.2cm]
 and its Higgs Branches}
}
\\[0pt]
\vspace{-0.1cm}
\bigskip
\bigskip {
{{\bf  Denis Klevers}$^{\,\text{a},\,\text{b}}$},
{{\bf  Dami\'an Kaloni Mayorga Pe\~na}$^{\,\text{c}}$},\\
{{\bf Paul-Konstantin~Oehlmann}$^{\,\text{c}}$},
{{\bf Hernan~Piragua}$^{\,\text{a}}$},
{{\bf Jonas~Reuter}$^{\,\text{c}}$}
\bigskip }\\[3pt]
\vspace{0.cm}
{\it \small 
${}^{\text{a}}$ Department of Physics and Astronomy,
University of Pennsylvania,\\ Philadelphia, PA 19104-6396, USA\\[8pt]
${}^{\text{b}}$Theory Group, Physics Department, CERN, CH-1211, Geneva 23, Switzerland\\[8pt]
${}^{\text{c}}$ Bethe~Center~for~Theoretical~Physics, Physikalisches~Institut~der~Universit\"at~Bonn,\\ Nussallee~12,~53115~Bonn,~Germany\\[0.5cm]}

{\small klevers@sas.upenn.edu, damian@th.physik.uni-bonn.de,
oehlmann@th.physik.uni-bonn.de, hpiragua@sas.upenn.edu, jreuter@th.physik.uni-bonn.de}
\\[0.4cm]
\end{center}

\begin{center} {\bf Abstract } \end{center}
\vspace{0.4Em}

We consider F-theory compactifications on genus-one fibered
Calabi-Yau manifolds with their fibers realized 
as hypersurfaces in the toric varieties associated to the 16 
reflexive 2D polyhedra. We present a base-independent 
analysis of the codimension one, two and three singularities of these 
fibrations.  We use these geometric results to determine the gauge groups, 
matter representations, 6D matter 
multiplicities and 4D Yukawa couplings of the corresponding effective 
theories. All these theories have a non-trivial gauge 
group and matter content.  
We explore the network of Higgsings relating these theories.
Such Higgsings geometrically correspond to extremal transitions 
induced by blow-ups in the 2D toric varieties.
We recover the 6D effective theories of all 16 toric 
hypersurface fibrations by repeatedly Higgsing
the theories that exhibit Mordell-Weil torsion. 
We find that the three Calabi-Yau manifolds without section, whose 
fibers are given by the toric hypersurfaces in $\mathbb{P}^2$, 
$\mathbb{P}^1\times \mathbb{P}^1$ and  the recently studied $\mathbb{P}^2(1,1,2)$, yield F-theory realizations  of SUGRA 
theories with discrete gauge groups $\mathbb{Z}_3$, $\mathbb{Z}_2$ and $\mathbb{Z}_4$. This opens up a whole new 
arena for model building with discrete global symmetries in F-theory. In these three 
manifolds, we also find codimension two $I_2$-fibers supporting matter 
charged only under these discrete gauge groups. Their 6D matter 
multiplicities are computed employing ideal techniques and the 
associated Jacobian fibrations.  We also show that the Jacobian of the 
biquadric fibration has one rational section, yielding one U(1)-gauge 
field  in F-theory. Furthermore, the elliptically fibered Calabi-Yau 
manifold based on $dP_1$ has a U(1)-gauge field induced by a non-toric 
rational section. In this model, we find the first F-theory realization 
of matter with U(1)-charge $q=3$.

\begin{flushright}
\parbox[t]{1.2in}{August, 2014}
\end{flushright}
\end{titlepage}
\clearpage
\setcounter{footnote}{0}
\setcounter{tocdepth}{2}
\tableofcontents
\clearpage
%%%%%%%%%%%%%%%%%%%%%%%%%%%%%%%%%%%%%%
%UNTIL HERE!!
%%%%%%%%%%%%%%%%%%%%%%%%%%%%%%%%%%%%%%

% #################################
% #            Main part          #
% #################################
%\setcounter{page}{1}
\section{Introduction \& Summary of Results}
\label{sec:intro}

F-theory \cite{Vafa:1996xn,Morrison:1996na,Morrison:1996pp} is a
non-perturbative formulation of Type IIB string
theory with backreacted 7-branes, that is manifestly invariant
under the SL$(2,\mathbb{Z})$-duality symmetry of the theory.
String backgrounds constructed via F-theory are not only located in
the heart of the web of string dualities, but also allow
for the construction of phenomenologically appealing local GUT-models
\cite{Donagi:2008ca,Beasley:2008dc,Hayashi:2008ba,Beasley:2008kw},
which has recently rekindled a lot of interest into the subject. The
basic idea of F-theory is to replace the axio-dilaton
$\tau=C_0+ie^{-\phi}$ , that is only defined up to
SL$(2,\mathbb{Z})$-transformations, by a quantity, that only depends
on the SL$(2,\mathbb{Z})$-equivalence class of $\tau$. The canonical
geometrical object with this property is a two-torus $T^2(\tau)$, whose
complex structure is identified with $\tau$.
Thus, replacing $\tau$ by this auxiliary $T^2(\tau)$ provides
an SL$(2,\mathbb{Z})$-invariant formulation of Type IIB.
Non-trivial  backgrounds of $\tau$,  which are
sourced by 7-branes, on manifolds $B$  are
mapped under this replacement to torus-fibrations over $B$.
In particular, for a supersymmetric  and tadpole-canceling
setup of 7-branes on a complex K\"ahler manifold $B$ the total space of
this $T^2(\tau)$-fibration is a Calabi-Yau manifold $\pi:X\rightarrow B$.

Most of the torus-fibered Calabi-Yau manifolds $X$ that
have been studied 
are algebraic, that is they are realized as complete intersections in
some ambient space.\footnote{For recent advances on
Calabi-Yau manifolds constructed as determinantal and Pfaffian
varieties, see \cite{Hori:2006dk,Jockers:2012zr}.}
In these constructions, the torus fiber over $B$ is realized as an
algebraic curve $\mathcal{C}$ of genus one.
In addition, many examples considered in the literature are elliptically
fibered, meaning that 
$X$ has a section $B\rightarrow X$,
which was traditionally assumed to be holomorphic.
These elliptically fibered Calabi-Yau manifolds have fruitful
applications, e.g.~for the construction of semi-realistic GUTs in global
F-theory compactifications starting with \cite{Marsano:2009gv,Blumenhagen:2009yv} or the
classification and study of 6D SCFTs
\cite{Heckman:2013pva,DelZotto:2014hpa,Heckman:2014qba}.

Despite these successes, addressing open conceptual 
questions e.g.~regarding the finiteness of the
F-theory landscape\footnote{See
\cite{Douglas:2006xy} and the recent \cite{Cvetic:2014gia} for a finiteness proof in related Type I compactifications.} or 
which consistent 6D and 4D supergravity (SUGRA) theories can be realized in F-theory,\footnote{F-theory compactifications
to 8D are well-studied and classified, see e.g.~the recent
toric analysis of \cite{Grassi:2012qw} and the classification of
elliptic fibrations on K3-surfaces in \cite{Braun:2013yya}. Last
subtleties in the understanding of the gauge group of a generic K3
have been understood in \cite{Douglas:2014ywa}.}
as well as the understanding of the geometric origins of discrete 
symmetries or analogous field theoretic mechanisms, crucial for the phenomenology of F-theory models, 
requires to broaden the class of 
Calabi-Yau manifolds $X$ used for F-theory
compactifications.
In fact, using the well-developed  map between the geometry of
F-theory and SUGRA theories, see
\cite{Kumar:2009ac,Kumar:2010ru} for
the complete map in 6D and \cite{Grimm:2012yq} for results about certain
topological terms in 4D\footnote{SUGRA theories from string theory can also be constrained using tools from
heterotic/F-theory duality \cite{Anderson:2014gla}.},
one finds that the Calabi-Yau manifolds realizing many known
consistent SUGRA theories, in particular those with U(1) symmetries 
\cite{Park:2011wv} or discrete gauge groups, are still 
unknown \cite{Kumar:2010am}.\footnote{Of course
it is a logical possibility that some of the SUGRA theories without a
known F-theory realization are not consistent effective theories due to
a violation of consistency constraints that are unknown at this point.} 
For the search of an F-theory realization of these theories it is
crucial to construct new classes of Calabi-Yau manifolds $X$ admitting
new geometric features and to deduce the general SUGRA
theories that arise in F-theory compactifications on these $X$.\footnote{Compactifications of F-theory on Spin $7$ manifolds,
considered recently \cite{Bonetti:2013fma,Bonetti:2013nka}, have not
yet produced SUGRA theories that cannot be obtained by a Calabi-Yau compactification.}

There has been a lot of recent progress in systematically extending
the set of Calabi-Yau manifolds $X$ that can be used for
F-theory compactifications.
The different approaches can be roughly sorted into two groups. The
first group of approaches focuses on the classification and construction
of all bases $B$ that are admissible for F-theory
\cite{Morrison:2012np,Morrison:2012js,Martini:2014iza}. The second
group, to which this work belongs,
focuses on generalizing the type
of fiber $\mathcal{C}$ and the ways in which it can be fibered in a Calabi-Yau manifold $X$. There are three major extensions in this direction:
\begin{itemize}
\item Elliptic fibers with an increasing number of rational points and their corresponding elliptically fibered Calabi-Yau manifolds $X$ 
with a \textit{Mordell-Weil group} (MW-group) of rational sections
of increasing rank have been systematically constructed and studied 
\cite{Grimm:2010ez,Krause:2011xj,Grimm:2011fx,Park:2011ji,Cvetic:2012xn,Mayrhofer:2012zy,Braun:2013yti,Borchmann:2013jwa,Cvetic:2013nia,Grimm:2013oga,Braun:2013nqa,Cvetic:2013uta,Borchmann:2013hta,Cvetic:2013jta,Cvetic:2013qsa,Mayrhofer:2014opa}.\footnote{Certain aspects of models with a higher rank MW-group are studied already in
\cite{Aldazabal:1996du,Klemm:1996hh,Klemm:2004km}, see also
\cite{Esole:2011cn} for an analysis of models with $D_5$-fiber.}
The free part of the MW-group leads to U(1)-gauge fields in
F-theory\footnote{See \cite{Grimm:2011tb,Cvetic:2012xn,Braun:2014nva}
for a discussion of (geometrically) massive U(1)'s.}
\cite{Morrison:1996na} and the  torsion part yields non-simply
connected gauge groups \cite{Aspinwall:1998xj}.
\item Elliptic fibrations $X$ with a \textit{non-holomorphic zero section}
have been considered recently, starting with
\cite{Braun:2013yti,Cvetic:2013nia,Borchmann:2013hta}.
This permits  
the introduction of discrete degrees of freedom in the
construction of the fibration of the elliptic curve over the
base $B$ yielding a finite number of strata in the moduli space of $X$. 
\item Algebraic curves $\mathcal{C}$ of 
 one without 
any (rational) point have been used to construct 
\textit{genus-one fibrations}
\cite{Braun:2014oya,Morrison:2014era,Anderson:2014yva}. These are 
Calabi-Yau manifolds $X$ which do not have 
a section, but only multi-sections. These models can be analyzed 
employing their associated Jacobian fibration $J(X)$, which does exhibit 
a zero section, and its Weierstrass form.
\end{itemize}

The Calabi-Yau manifolds $X$ we consider
in this work invoke all three of these extensions.
We study all F-theory compactifications on Calabi-Yau manifolds 
$X_{F_i}$
with genus-one fiber $\mathcal{C}_{F_i}$
given as a
hypersurface in the 
toric varieties associated to the 16 2D reflexive
polyhedra, denoted by $F_i$, $i=1,\ldots, 16$.\footnote{These
genus-one curves have also been used in 
\cite{Huang:2013yta,Huang:2014nwa} 
as mirror curves for the computation of refined stable pair 
invariants in the refined topological string.} 
We refer to these
Calabi-Yau
manifolds as \emph{toric hypersurface fibrations}. We determine the generic
and intrinsic features of these $X_{F_i}$ that are relevant to F-theory:
the generic 
gauge group, the corresponding matter spectrum and the 4D
Yukawa couplings corresponding to the codimension one, two and three
singularities of $X_{F_i}$. These geometric results completely
determine the 6D and non-chiral 4D SUGRA theories obtained by
compactifying F-theory on Calabi-Yau threefolds and fourfolds
without $G_4$-flux. We prove completeness of
our  analysis of codimension one and two singularities by
checking cancellation of all 6D anomalies.
All these results are base-independent
in the sense that they
follow directly from the geometry of the fiber $\mathcal{C}_{F_i}$.
The only dependence on the base $B$ enters through the choice of two
divisors on $B$ that label the possible Calabi-Yau fibrations of
$\mathcal{C}_{F_i}$ \cite{Cvetic:2013nia}.

We highlight the following interesting geometrical findings of our
analysis of F-theory on the Calabi-Yau manifolds $X_{F_i}$:
\begin{itemize}
\item Every $X_{F_i}$ has an associated
minimal gauge group $G_{F_i}$ that is completely determined by the 
polyhedron $F_i$. In other words this gauge group is present without
tuning the complex structure of $X_{F_i}$ by means of Tate's algorithm
\cite{kodaira1963compact,tate1975algorithm,Bershadsky:1996nh} (see 
\cite{Katz:2011qp,Lawrie:2012gg} for recent refinements) 
or upon addition of tops.
The gauge groups $G_{F_i}$ and $G_{F_i^*}$
associated to $F_i$ and its dual polyhedron $F_i^*$ obey the rank
relation
\beq \label{eq:SumOfRanks}
\text{rk}(G_{F_i})+\text{rk}(G_{F_i^*})=6\,.
\eeq
\item We consider three
Calabi-Yau manifolds $X_{F_i}$, $i=1,2,4$, without section. Their fibers
are the general cubic in $\mathbb{P}^2$, the general biquadric in
$\mathbb{P}^1\times \mathbb{P}^1$ and the general quartic in
$\mathbb{P}^2(1,1,2)$,
respectively, where the latter is also studied
in \cite{Braun:2014oya,Morrison:2014era,Anderson:2014yva}.
The fibrations $X_{F_i}$, $i=1,2,4$, only have a genus-one fibration
with a three-, a two- and a two-section, respectively. As a direct
consequence of this absence of sections,
F-theory has discrete gauge group factors given by
$\mathbb{Z}_3$, $\mathbb{Z}_2$ and $\mathbb{Z}_4$, respectively.
We show that these Calabi-Yau manifolds, most notably the fibration of
the cubic, $X_{F_1}$,
have $I_2$-fibers at codimension two that support singlet matter with
charge $1$ under the respective discrete gauge groups.
We explain how the charge of all matter fields under these discrete 
groups  are computed from the intersections of the 
multi-sections with the relevant codimension two fibers.
\item We show that both $X_{F_2}$ and $X_{F_3}$ give rise to
one U(1)-factor, namely
$G_{F_2}=\text{U(1)}\times \mathbb{Z}_2$ and
$G_{F_3}=\text{U(1)}$. To this end, we show that unlike $X_{F_2}$, the
Weierstrass form of the Jacobian $J(X_{F_2})$ has one rational section,
whereas already $X_{F_3}$ has two sections: a toric and a non-toric one. 
In both cases, we determine
the coordinates of all sections explicitly.
\item For the first time, we find F-theory compactifications with matter
of U(1)-charge three. This matter is supported at a codimension two locus of $X_{F_3}$ with an $I_2$-fiber where
both the zero section and the non-toric rational section are ill-behaved
and each ``wrap'' one irreducible fiber component.
\end{itemize}

We note that the 16 toric hypersurface fibrations $X_{F_i}$  were
considered in \cite{Braun:2013nqa}, were a thorough classification of 
their toric MW-groups was carried out. 
Further specializations of the $X_{F_i}$ corresponding to toric tops 
\cite{Candelas:1996su,Bouchard:2003bu} (see \cite{Kuntzler:2014ila} for 
a systematic approach based on Tate's algorithm for elliptic fibers
in $\text{Bl}_1\mathbb{P}^2(1,1,2)$) permitted the engineering of 
toric F-theory models with certain gauge groups, in particular 
with an SU(5) GUT-group.
Some 4D examples of chiral $\SU5$ GUTs were constructed in this manner 
\cite{Cvetic:2013uta,Borchmann:2013hta}. Since we determine here the 
intrinsic gauge groups and the non-toric MW-groups, as well as the full 
matter spectrum and the Yukawa couplings of $X_{F_i}$, our
approach is complementary to these previous works.   

We have to remark that none of the fibrations $X_{F_i}$ yield an $\SU5$ 
gauge factor in their low-energy effective theories. Hence, strictly 
speaking, the intrinsic gauge symmetries associated to the toric 
hypersurface fibrations do not suffice to engineer $\SU5$ F-theory GUTs. 
There are, however, some arguments that challenge the simplest GUT 
picture in F-theory, and therefore, draw our attention towards 
alternative schemes which may be promising for particle physics models. 
In this spirit, we would like to briefly highlight some of the effective 
theories we obtain, which can potentially be
used to construct promising particle physics models in
F-theory. We find models with
discrete symmetries and up to three $\U1$ factors. These additional
symmetries can be used in order to forbid dangerous operators which
would render the theory incompatible with observations, e.g.~by 
mediating fast proton decay. In addition, we observe theories with
interesting gauge groups and spectra. In fact, $X_{F_{11}}$ precisely
leads to an effective theory with the Standard Model gauge group and the 
usual representations\footnote{See \cite{Lin:2014qga} for a different 
realization of a standard model like theory based on tops of 
$X_{F_5}$.}, and we further identify the trinification
group for $X_{F_{16}}$ as well as the Pati-Salam group for $X_{F_{13}}$. 
As we demonstrate explicitly, the matter spectra we obtain are very 
close to those one usually has in both of these theories.

\begin{figure}[ht!]
\center
\includegraphics[scale=0.4]{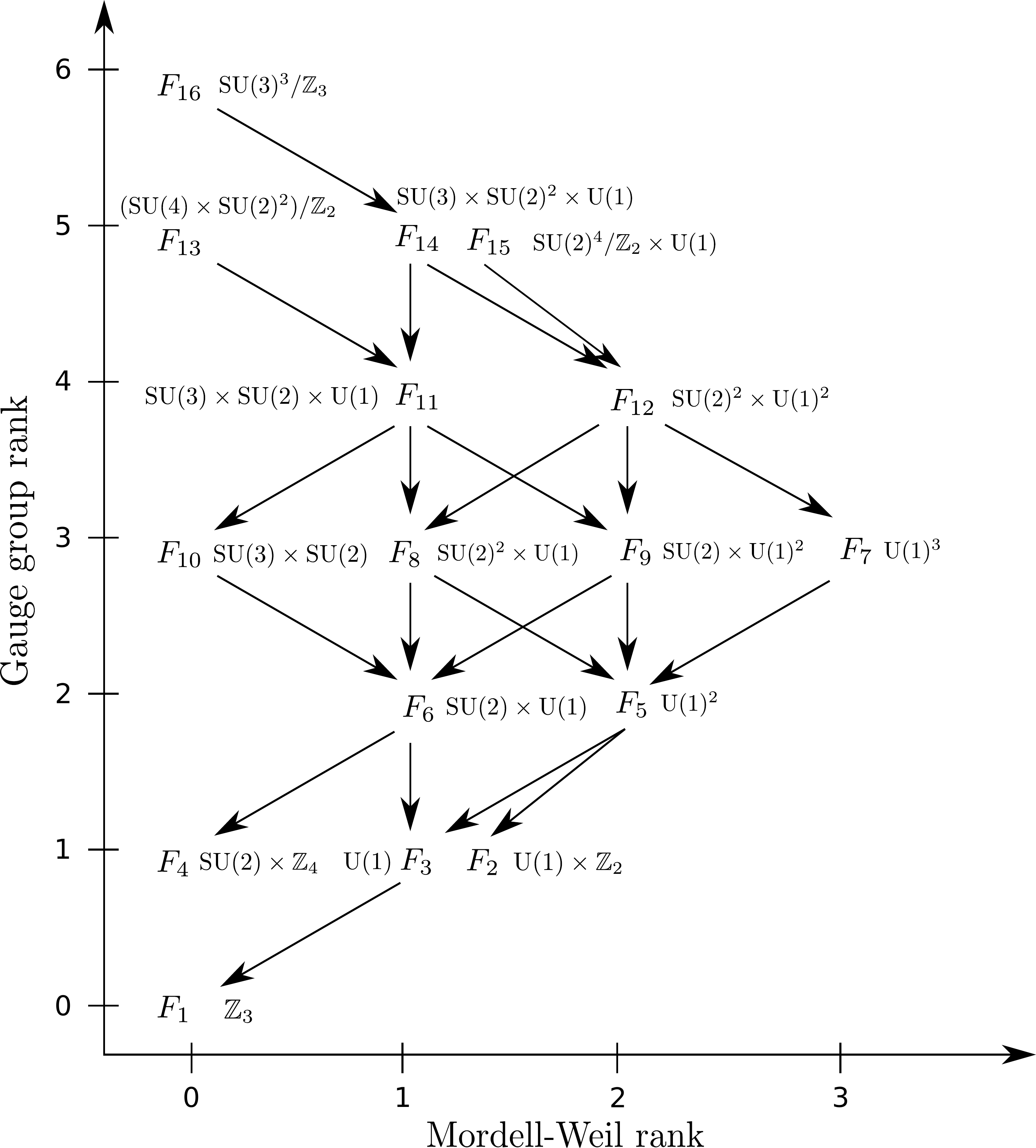}
\caption{\label{fig:network} The network of
Higgsings between all
F-theory compactifications on toric hypersurface fibrations
$X_{F_i}$.
The axes show the rank of the MW-group and the total rank of the gauge group of $X_{F_i}$. 
Each Calabi-Yau $X_{F_i}$ is abbreviated by $F_i$ and its
corresponding gauge group is shown. The arrows indicate
the existence of a Higgsing between two Calabi-Yau manifolds.}
\end{figure}

In this paper we also work out the entire network of Higgsings relating 
the effective theories of F-theory on the toric hypersurface fibrations
$X_{F_i}$. It is well-known that the toric varieties corresponding to 
the $16$ 2D reflexive polyhedra $F_i$ are related by blow-downs.
Consequently, the Calabi-Yau manifolds $X_{F_i}$ are related by the 
extremal transitions induced by these birational maps and a subsequent 
toric complex structure deformation. 
These transitions can be understood as 
Higgsings in the effective SUGRA theories arising from F-theory on
the $X_{F_i}$: given two polyhedra $F_i$ and $F_{i'}$ related by
a blow-down as $F_i\rightarrow F_{i'}$, we explicitly determine the
Higgsing that relates the effective theory of F-theory on $X_{F_i}$ to
that on $X_{F_{i'}}$. The resulting diagram
of all those Higgsings is given in Figure~\ref{fig:network}.
Since this chain of Higgsings is only a sub-branch of the full Higgs 
branch of the effective SUGRA theories of F-theory 
on $X_{F_i}$, we refer to it as the \emph{toric Higgs branch}. We  check 
that both the charged and the neutral spectrum of the SUGRA theories 
in 6D match. This involves the computation of the number of neutral 
hyper multiplets, that in turn can
be obtained from the Euler numbers of all $X_{F_i}$, which we also 
compute explicitly. 

We point out some interesting observations about the Higgsing
diagram in Figure~\ref{fig:network}:
\begin{itemize}
\item
All effective theories can be obtained by appropriately Higgsing the
theories with maximal gauge groups and matter spectra obtained from
F-theory on $X_{F_i}$, $i=13,15,16$.
We note that these are precisely the theories with non-trivial
Mordell-Weil torsion.
\item The network of Higgsings is symmetric around the horizontal line
where the total rank of the gauge group is $3$.
Reflection along this line exchanges the polyhedron with its dual.
This symmetry of Figure~\ref{fig:network} reflects the rank condition
\eqref{eq:SumOfRanks}. We emphasize that this symmetry maps
theories with discrete gauge groups to theories with non-simply
connected group, suggesting that multi-sections are somehow ``dual'' to MW-torsion.
\item  The three theories with discrete gauge groups 
arise by Higgsing theories with $\U1$'s. It is also remarkable that all 
discrete symmetries found are surviving remnants of $\U1$ 
symmetries. It seems that discrete symmetries in F-theory are 
automatically in agreement with the early observation 
\cite{Krauss:1988zc,Banks:1989ag,Ibanez:1991wt} that in a consistent 
theory of gravity, discrete global symmetries must be always embeddable 
into a local continuous symmetry. 
\item The toric Higgsings cannot change the rank of the F-theory gauge
group by more than $1$. This
explains why there are no arrows with slope below 45 degrees.
\end{itemize}

This paper is organized as follows. Section~\ref{sec:Genus1Fibrations} contains a summary of the geometry
of Calabi-Yau manifolds constructed as genus-one fibrations and
the physics of F-theory compactified on them.
We also present a basic account on toric geometry.
In Section~\ref{sec:AnalysisOfPolytopes} we discuss the construction and the different types of toric
hypersurface fibrations $X_{F_i}$. Their
codimension one, two and three singularities are analyzed, and 
the number of their complex structure moduli is computed. The F-theory 
gauge group, matter spectrum and Yukawa couplings are extracted from these results. Section~\ref{sec:Higgsings} is devoted to the
study of the toric Higgs branch of F-theory compactified on
the $X_{F_i}$. One particular Higgsing is
discussed in detail in order to illustrate the relevant techniques. Here 
we also present the Higgsings leading to the effective 
theories with discrete gauge groups.
We further elaborate on the details of the entire Higgsing chain in Appendices~\ref{app:higgschain} and \ref{app:decomposition}. 
Our conclusions can be found in Section~\ref{sec:conclusion}. This work
contains additional Appendices on
6D anomalies (Appendix~\ref{app:Anomalies}), 
additional geometrical data  of the $X_{F_i}$ (Appendix~\ref{app:AppPoly}) and 
the explicit Euler numbers of all Calabi-Yau threefolds
$X_{F_i}$ (Appendix~\ref{app:Eulernumbers}).

\section{Geometry \& Physics of F-theory Backgrounds}
\label{sec:Genus1Fibrations}

In this section, we summarize the key geometrical properties of 
Calabi-Yau manifolds $X$ that are genus-one fibrations over a base $B$ 
which are relevant for the study of  F-theory compactifications, 
see Sections \ref{sec:ellipticCurvesWithRP} and \ref{sec:divsonGenus1}. 
The structure of the 6D effective SUGRA theories obtained by 
compactifying F-theory on these manifolds is discussed in Section~\ref{sec:6DSUGRA}.
 Since we study in this work Calabi-Yau manifolds $X$ 
with their genus-one fibers realized as toric hypersurfaces, we 
introduce the necessary elements of toric geometry in Section~\ref{sec:CYin2DToric}. 

Readers familiar with  the tools and definitions presented here can safely skip this 
section and continue directly with Section~\ref{sec:AnalysisOfPolytopes}.
 
%%%%%%%%%%%%%%%%%%%%%%%%%%%%%%%%%%%%%%%%%%%%%%%%%%%%%%%%%%%%%%%%%%%% 
\subsection{Genus-one, Jacobian and elliptic fibrations with 
Mordell-Weil groups}
\label{sec:ellipticCurvesWithRP}
%%%%%%%%%%%%%%%%%%%%%%%%%%%%%%%%%%%%%%%%%%%%%%%%%%%%%%%%%%%%%%%%%%%%

We consider a smooth Calabi-Yau manifold $\pi:\,X\rightarrow B$ with 
general fiber given by an algebraic curve  $\mathcal{C}$ of genus-one. 
 $\mathcal{C}$ is a non-singular 
curve defined over a field $K$ that is not necessarily algebraically closed. In 
particular, we can think of the fibration $X$ as an algebraic curve 
$\mathcal{C}$ defined over the field $K$ of meromorphic functions 
on  the base $B$, which is clearly in general not algebraically closed. 
Thus, there are two qualitatively different situations to consider.

\paragraph{Curves with points} 
~\\
First, if the curve $\mathcal{C}$ has points with coordinates in $K$, then it is called an 
elliptic curve, which we denote by $\mathcal{E}$, and $X$ is called an elliptic
fibration. The points on $\mathcal{E}$ form an Abelian group under 
addition: one point can be chosen 
as the zero point, denoted by $P_0$, and the additional points $P_m$, 
$m=1,\ldots, r$,  (more precisely 
the differences $P_m-P_0$) are the generators of the Mordell-Weil group  
of rational points of $\mathcal{E}$. The Mordell-Weil theorem states 
that this group is finitely  generated 
\cite{silverman2009arithmetic,lang1959rational}.
Thus, it splits into a free part isomorphic to $\mathbb{Z}^r$ and a 
torsion subgroup, where the latter has 
been fully classified for $K=\mathbb{Q}$ by Mazur 
\cite{mazur1977modular,mazur1978rational}, see  
\cite{silverman2009arithmetic} for a review.\footnote{For the field of 
meromorphic functions on $B$, there are more torsion subgroups possible 
than for $K=\mathbb{Q}$ 
\cite{Aspinwall:1998xj}.}
Every point on $\mathcal{E}$ gives rise to a section of the fibration 
$X$, i.e.~rational maps from the base $B$ into $X$. 
The section associated to $P_0$ is the zero section, denoted by 
$\hat{s}_0:\, B\rightarrow X$,  and the $r$ rational points $P_m$ induce
the rational sections $\hat{s}_m:\, B\rightarrow X$. The set 
$\{\hat{s}_m\}$ can be seen to form a group, the 
MW-group of rational sections of $X$, by defining the addition of 
rational sections by addition of their 
corresponding points on $\mathcal{E}$. 
The free part of the MW-group gives rise to Abelian gauge symmetry in 
F-theory \cite{Morrison:1996na} and its torsion part yields 
non-simply connected gauge groups \cite{Aspinwall:1998xj},
see also \cite{Mayrhofer:2014opa} for a recent discussion of MW-torsion.

Every elliptic fibration $X$ can be written in Weierstrass form (WSF) 
\cite{nakayama1985weierstrass}, i.e.~as a hypersurface in the weighted 
projective bundle 
$\mathbb{P}^{(1,2,3)}(\mathcal{O}_B\oplus\mathcal{L}^2\oplus\mathcal{L}^3)$ over $B$ of the 
form
\beq \label{eq:WSF}
		y^2=x^3+fxz^4+gz^6\,.
\eeq
Here, $\mathcal{O}_B$ is the trivial bundle on $B$ and the 
line-bundle $\mathcal{L}$ is fixed by the Calabi-Yau condition of $X$ 
as $\mathcal{L}=K_B^{-1}$,  with $K_B$ denoting the canonical bundle of 
the base $B$. Then, the coefficients $f$ and $g$ have to be sections of
$K_B^{-4}$ and $K_B^{-6}$, respectively. The map from the canonical 
presentation of $X$ inherited from the canonical presentation of the elliptic %algebraic 
curve $\mathcal{E}$ to the Weierstrass form \eqref{eq:WSF} is birational. The zero section $\hat{s}_0$ of $X$
maps to the holomorphic zero section $[z:x:y]=[0:\lambda^2:\lambda^3]$ in \eqref{eq:WSF} and the rational sections 
$\hat{s}_m$ map to rational sections in \eqref{eq:WSF} with certain coordinates $[z_{P_m}:x_{P_m}:y_{P_m}]$, that are rational 
expressions in $K$ (we can clear denominators to obtain holomorphic coordinates).

\paragraph{Curves without points} 
~\\
Second, if the genus-one curve $\mathcal{C}$ does not have a point, the fibration $\pi:\,X\rightarrow B$ is 
without section. Such a fibration is referred to as a genus-one fibration \cite{Braun:2014oya}. Given a
genus-one curve $\mathcal{C}$, one can construct an 
associated elliptic curve 
$\mathcal{E}=J(\mathcal{C})$, that is the Jacobian  of the curve 
$\mathcal{C}$, i.e.~the group of degree zero line bundles on 
$\mathcal{C}$. The zero point on $J(\mathcal{C})$ is the trivial line 
bundle. Thus, there exists an elliptic fibration 
$\pi:\,J(X)\rightarrow B$  
associated to $X$ with general fiber given by $J(C)$. This implies that 
$J(X)$ can be represented as a Weierstrass model \eqref{eq:WSF}. 
Furthermore, it is a key property for F-theory that the 
$\tau$-function and the discriminant of $X$ and $J(X)$ are identical 
\cite{Braun:2014oya}.

In this work, we consider concrete genus-one curves $\mathcal{C}$ with $K$-rational 
divisors of degree $n>1$, respectively.\footnote{We expect that there always exists a degree $n$ divisor on a given algebraic genus-one 
curve $\mathcal{C}$.}  The corresponding 
fibration $X$ does not have a section, but an $n$-section, that we denote by $\hat{s}^{(n)}:\,B\rightarrow X$. 
Locally, at a point $p$ on $B$, the function field $K$ reduces to $\mathbb{C}$ and the $n$-section 
$\hat{s}^{(n)}$ maps $p$ to $n$ points in the fiber $\mathcal{C}$. 
Globally, however, upon moving around branch 
loci in $B$ the individual points are exchanged by a monodromy 
action, so that only the collection 
of all $n$ points together induce a well-defined divisor in $X$. 

As we will see explicitly for concrete genus-one fibrations $X$, the map 
from $X$ to the Weierstrass form \eqref{eq:WSF} of $J(X)$ can be 
obtained by an algebraic field extension $L$ of $K$. This field 
extension is only necessary as an intermediate step, i.e.~the final WSF 
\eqref{eq:WSF} of $J(X)$ is again defined over $K$. In Sections 
\ref{sec:F1_poly}, \ref{sec:F2_poly} and \ref{sec:F4_poly} we  explicitly work out the maps from $X$ 
to the WSF of their 
Jacobian fibrations $J(X)$, namely for the fibration of the cubic in 
$\mathbb{P}^2$, that has a three-section,
and the two fibrations of the biquadric curve in 
$\mathbb{P}^{1}\times\mathbb{P}^1$ and the quartic curve in 
$\mathbb{P}^2(1,1,2)$, that both have a two-section.\footnote{These 
examples have been considered in the mathematics 
literature in \cite{an2001jacobians}.} We note that genus-one 
fibrations $X$ by the quartic in $\mathbb{P}^2(1,1,2)$  have been 
considered recently in an F-theory context in \cite{Braun:2014oya,Morrison:2014era,Anderson:2014yva}.

%%%%%%%%%%%%%%%%%%%%%%%%%%%%%%%%%%%%%%%%%%%%%%%%%%%%%%%%%%%%%%%%%%%%
\subsection{Divisors on genus-one fibrations and their intersections}
\label{sec:divsonGenus1}
%%%%%%%%%%%%%%%%%%%%%%%%%%%%%%%%%%%%%%%%%%%%%%%%%%%%%%%%%%%%%%%%%%%%

In F-theory we are particularly interested in Calabi-Yau manifolds $X$ 
that arise as a crepant resolution of singular genus-one or elliptic 
fibrations. These resolved manifolds exhibit three different classes of 
divisors, that we discuss in the following.

The first set of divisors on $X$ is formed by the vertical divisors, 
i.e.~divisors that arise as pullbacks of divisors on $B$ under the 
projection map $\pi:\, X\rightarrow B$. Hence, there are $h^{(1,1)}(B)$ 
such divisors on $X$. We denote the preimage 
under $\pi$ of a vertical divisor $D$ on $X$ by $D^b$ so that 
$D=\pi^*(D^b)$. Thus, $D$ is a fibration of the curve 
$\mathcal{C}$ (or its degenerations) over the base $D^b$.  

The second class of divisors are the exceptional divisors of $X$.
In more detail, if the discriminant $\Delta=-16(4f^3+27 g^2)$ of the 
WSF \eqref{eq:WSF} of $X$ or of its Jacobian 
$J(X)$ vanishes to order higher than $1$ at one of its irreducible 
components 
\beq \label{eq:defSb}
 \mathcal{S}^b_{G_I}:=\{\Delta_I=0\}\,,\qquad I=1,\ldots,N \, ,
\eeq 
in $B$, then the total space of the WSF is singular. These 
codimension one singularities are classified in 
\cite{kodaira1963compact,tate1975algorithm}. In the resolution $X$ 
the fiber  over each $\mathcal{S}^b_{G_I}$ splits into a number of 
rational curves whose intersection pattern often agrees with
the Dynkin diagram of a Lie group $G_{I}$.\footnote{The fibers 
that do not have an associated group $G_{I}$  are the unconventional 
fibers in Table 2 of \cite{Bershadsky:1996nh}.} The shrinkable 
irreducible components of the fiber at $\mathcal{S}^b_{G_I}$ represent 
the simple roots of $G_{I}$ and are denoted by $c_{-\alpha_{i}}^{G_I}$ 
for $i=1,\ldots,\text{rk}(G_{I})$. 
Thus, $X$  has a set of exceptional divisors $D_{i}^{G_{I}}$ given as
the fibration of $c_{-\alpha_{i}}^{G_I}$ over $S^b_{G_I}$ for 
every $I$, to which we refer to as Cartan divisors of $G_{I}$. 
The $D_{i}^{G_{I}}$ intersect the curves $c_{-\alpha_{i}}^{G_I}$ as 
\beq \label{eq:CartanMat}
	D^{G_I}_{i}\cdot c^{G_J}_{-\alpha_{j}}=-C_{ij}^{G_I}\delta_{IJ}\,,
\eeq
where $C^{G_{I}}_{ij}$ denotes the Cartan matrix of $G_{I}$.
The F-theory gauge group is then given by 
the product of all $G_{I}$, as discussed in Section~\ref{sec:6DSUGRA}.

Finally, the third class of divisors is induced by the independent
sections and $n$-sections of the fibration of $X$.
We denote the divisor classes of the zero section $\hat{s}_0$ and the 
generators of the MW-group of rational sections $\hat{s}_m$ by $S_0$ 
and $S_m$, respectively. The class of a multi-section $\hat{s}^{(n)}$ 
is denoted by $S^{(n)}$.  Then, the intersections of these divisors with 
the fiber $f\cong\mathcal{C},$ $\mathcal{E}$ read
\beq\label{eq:intsSectionF}
	S_0\cdot f=S_m\cdot f=\tfrac{1}{n}S^{(n)}\cdot f=1\,.
\eeq 
The divisor classes that support Abelian gauge fields in F-theory 
\cite{Park:2011ji,Morrison:2012ei,Cvetic:2013nia}, see also
\cite{Grimm:2010ez,Grimm:2010ks,Grimm:2012yq}, are obtained from the 
Shioda map $\sigma$ of the rational sections 
$\hat{s}_m$. To a given generator of the MW-group $\hat{s}_m$ the Shioda 
map assigns the divisor
\begin{align}\label{eq:ShiodaMap}	 
	\sigma(\hat{s}_m) &:= S_m-S_0+ [K_B]-\pi(S_m\cdot S_0)	+
	\sum_{I=1}^N(S_m\cdot c^{G_I}_{-\alpha_{i}})
	(C_{G_I}^{-1})^{ij}D_{j}^{G_I}\ .
\end{align}
Here $\pi(\cdot)$ denotes the projection of a codimension two variety in 
$X$ to a divisor in the base $B$ and $[K_B]$ is the canonical bundle of 
$B$. The last term encodes contributions from non-Abelian gauge groups  
$G_{I}$ in F-theory with
$(C_{G_I}^{-1})^{i_Ij_I}$ denoting the inverse of the 
Cartan matrix $C^{G_I}_{ij}$.

The Shioda map \eqref{eq:ShiodaMap} enables us to define the  
N\'{e}ron-Tate height pairing of two rational sections
$\hat{s}_m$, $\hat{s}_n$ as 
\beq\label{eq:anomalycoeff}
	\pi(\sigma(\hat{s}_m)\cdot\sigma(\hat{s}_n))=\pi(S_m\cdot S_n)+[K_B]-\pi(S_m\cdot S_0)-\pi(S_n\cdot S_0)+
	\sum_I(\mathcal{C}_{G_I}^{-1})^{ij}(S_m\cdot c_{-\alpha_{i}}^{G_I})(S_n\cdot c_{-\alpha_{j}}^{G_{I}})\mathcal{S}_{G_I}^{\rm b}\,,
\eeq
where $\mathcal{C}^{G_I}_{ij}$ is the coroot matrix of $G_{I}$.
We note that for evaluating this pairing in a concrete situation,
the universal intersection relations
\beq\label{eq:SP^2}
 \pi(S_P^2+[K_B^{-1}]\cdot S_P)=
  \pi(S_m^2+[K_B^{-1}]\cdot S_m)=0
\eeq
prove useful
(cf.~\cite{Morrison:2012ei,Cvetic:2012xn,Cvetic:2013nia,Cvetic:2013uta} 
for details), whereas $\pi(S_m\cdot S_n)$  and $\pi(S_m\cdot S_0)$
are model-dependent.

We note that in F-theory both the vertical divisor \eqref{eq:defSb} and 
the matrix \eqref{eq:anomalycoeff} of vertical divisors 
enter the coefficients of the Green-Schwarz terms 
\cite{Park:2011ji,Grimm:2012yq,Cvetic:2012xn} and are thus essential for 
anomaly cancellation, cf.~Appendix~\ref{app:Anomalies} for details.

%%%%%%%%%%%%%%%%%%%%%%%%%%%%%%%%%%%%%%%%%%%%%%%%%%%%%%%%%%%%%%%
\subsection{The spectrum of  F-theory on genus-one fibrations}
\label{sec:6DSUGRA}
\label{sec:SUGRA6D}
%%%%%%%%%%%%%%%%%%%%%%%%%%%%%%%%%%%%%%%%%%%%%%%%%%%%%%%%%%%%%%%

After the geometric preludes of sections \ref{sec:ellipticCurvesWithRP}
and \ref{sec:divsonGenus1}, we are prepared to extract the spectrum
of F-theory on a genus-one fibration $X$. The following discussion
applies most directly to F-theory compactifications to 6D with 
effective theory given by an $\mathcal{N}=(1,0)$ SUGRA theory. 
However certain 
statements directly generalize to 4D F-theory vacua without $G_4$-flux.

For a more detailed derivation of some of the following results, that 
oftentimes require M-/F-theory duality, we refer \cite{Denef:2008wq,Grimm:2010ks,Taylor:2011wt} and references therein.

\subsubsection*{Codimension one singularities}

All vector fields and certain hyper multiplets in F-theory arise from
the singularities of the WSF of $X$ that are induced 
by codimension one singularities of its fibration.
Over a given irreducible discriminant component $\mathcal{S}^b_{G_I}$ 
defined in \eqref{eq:defSb}, the fiber of $X$ is reducible. We assume 
that there is a Lie group $G_I$ associated to this codimension one fiber 
of $X$. Then, the  shrinkable holomorphic curves $c^{G_I}_{-\alpha}$ in the 
fiber over $\mathcal{S}^b_{G_I}$ represent  all the positive roots of 
$G_I$. By quantization of the moduli space of an M2-brane wrapping  
such a curve $c^{G_I}_{-\alpha}$  one finds BPS-states transforming in one charged vector multiplet 
and $2g_I$ 
charged half-hyper multiplets with charge-vector $-\alpha$ \cite{Witten:1996qb,Katz:1996ht}. Another 
vector multiplet and $2g_I$ 
half-hyper multiplets with  charges $+\alpha$ are contributed by an 
M2-brane wrapping $c^{G_I}_{-\alpha}$ with the opposite orientation. 
Here $g_I$ is the genus of the curve $\mathcal{S}^b_{G_I}$ in $B$, that is computed 
as
\beq \label{eq:genusformula}
 g_I=1+\tfrac{1}{2}\mathcal{S}^b_{G_I}\cdot(\mathcal{S}^b_{G_I}+[K_B])\,.
\eeq
All these charged states become massless in the F-theory limit, when
the volume of the class of the genus-one fiber of $X$ is taken to zero. 
In this limit, these BPS-states fall into representations of the group
$G_I$ as follows.

First, we focus on the vector multiplets.
All vector multiplets for every root $\alpha$ of $G_{I}$ are completed 
into one massless vector multiplet transforming in the adjoint 
representation $\text{\textbf{adj}}(G_I)$ of $G_I$. The additional 
vector multiplets are provided by the KK-reduction of the M-theory 
three-form $C_3$ along the harmonic $(1,1)$-forms in $X$ that are dual 
to the Cartan divisor $D_I^{G_I}$ of $G_I$. 
Thus, every irreducible component \eqref{eq:defSb} of the discriminant 
with respective codimension one fiber classified by a Lie group $G_I$, 
$I=1,\ldots,N$, gives rise to a $G_I$ gauge symmetry  in F-theory 
\cite{Morrison:1996na,Morrison:1996pp,Bershadsky:1996nh}. 
Furthermore, if $X$ has a MW-group of rank $r$, 
there are additional $(1,1)$-forms on $X$, which are the
duals of the divisors \eqref{eq:ShiodaMap}, that give rise to 
vector multiplets of Abelian gauge 
groups \cite{Morrison:1996pp}. Thus, the total 
gauge group $G_{X}$ of F-theory on $X$ is
\beq \label{eq:TotalGG}
	 G_{X}= \text{U}(1)^r \times\prod_{I=1}^N G_I\,.
\eeq 
This discussion and the results of Section~\ref{sec:divsonGenus1}
 imply further that the rank of $G_X$ can be directly 
computed  in terms of the Hodge 
numbers $h^{(1,1)}(X)$ and $h^{(1,1)}(B)$ of $X$ and $B$, respectively,  
as 
\beq \label{eq:rankGX}
	\text{rk}(G_X)=h^{(1,1)}(X)-h^{(1,1)}(B)-1\,.
\eeq
These results \eqref{eq:TotalGG} and \eqref{eq:rankGX} hold in 
compactifications to eight\footnote{In 8D vacua, 
no non-split fibers are possible, i.e.~all gauge groups are of 
$ADE$-type \cite{Bershadsky:1996nh}.}, six and four dimensions.

Second, we turn to the massless half-hyper multiplets over 
$\mathcal{S}^b_{G_I}$. In fact, also these fields  are completed into 
the adjoint representation $\text{\textbf{adj}}(G_I)$ of $G_I$.
In order to see this, we first note that there are neutral 
hyper multiplets induced by the complex structure moduli 
of $X$. Their total number, denoted by $H_{\text{neut}}$, is computed 
by the Hodge number $h^{(2,1)}(X)$ (or equivalently the Euler number $\chi(X)$) of $X$ as
\beq \label{eq:Hneutral}
	H_{\text{neut}}=h^{(2,1)}(X)+1=%\tfrac12 (h^{(1,1)}(B)+1+\text{rk}(G_X)-\chi(X))+1\,,
	  h^{(1,1)}(B)+2+\text{rk}(G_X)-\frac{1}{2}\chi(X)\,.
\eeq 
Then, for every group $G_I$, the $2g_I$ half-hyper multiplets with 
charges $-\alpha$ for all roots of $G_I$ combine with 
$g_I \cdot \text{rk}(G_I)$ neutral hyper multiplets from 
\eqref{eq:Hneutral} into $g_I$ hyper multiplets in the adjoint 
$\text{\textbf{adj}}(G_I)$ of $G_I$.
Thus, the number of hyper multiplets transforming in  
$\text{\textbf{adj}}(G_I)$ is given by 
\eqref{eq:genusformula} for every group $G_I$.

Let us emphasize that this discussion implies that $h^{(2,1)}(X)$ 
contains information about the gauge groups $G_I$ of $X$. Furthermore, 
also parts of the charged matter content from codimension two fibers
are counted by $h^{(2,1)}(X)$, however of \textit{another} theory 
related to the considered one by Higgsing. In Section~\ref{sec:Higgsings},
 for F-theory compactified on all toric hypersurface 
fibrations $X$, we  identify the part of $h^{(2,1)}(X)$ that comes from 
matter fields in all theories related by Higgsing.

\subsubsection*{Codimension two singularities}

The rest of the charged spectrum of F-theory on $X$ is encoded in
the codimension two singularities of the WSF of
$X$.

Non-Abelian charged matter is located at loci in
$\mathcal{S}^b_{G_I}$, where the vanishing order of the discriminant 
$\Delta$ of $X$ enhances. These loci are typically complete intersections of 
$\mathcal{S}^b_{G_I}$ with another divisor in $B$, that can be read
off from $\Delta$. 
The fiber of $X$ at these codimension two loci contains 
additional shrinkable rational curves $c$ that are not present in codimension 
one. These curves correspond to the weights of a representation 
$\mathbf{R}_{\underline{q}}$,  
under the gauge group $G_X$ in \eqref{eq:TotalGG}, where 
$\underline{q}=(q_1,\ldots,q_r)$ denotes the vector of U(1)-charges. 
The Dynkin labels $\lambda^{G_I}_i$  of $\mathbf{R}$ are computed according to 
\beq \label{eq:DynkinLabel}
	\lambda^{G_I}_i=D_i^{G_I}\cdot c\,,
\eeq
and the $m^{\text{th}}$ U(1)-charge $q_m$ is computed using \eqref{eq:ShiodaMap} as 
\cite{Park:2011ji,Morrison:2012ei}
\beq \label{eq:U1charge}
	q_m=\sigma(\hat{s}_m)\cdot c=(S_m\cdot c)
	-(S_0\cdot c)+\sum_I(S_m\cdot c_{-\alpha_{i_I}})
	(C_{(I)}^{-1})^{i_Ij_I}(D_{j_I}\cdot c)\,.
\eeq
We emphasize that these charges are automatically quantized, but not 
necessarily integers due to the usually fractional contribution form
the last term in \eqref{eq:U1charge}. 

We note that in the presence of U(1)'s, we automatically have additional 
matter that does not originate from intersections of codimension one 
discriminant components. In fact, the WSF of $X$ automatically has 
codimension two singularities for every rational section $\hat{s}_m$ 
with coordinates $[z_{P_m}:x_{P_m}:y_{P_m}]$  at the following locus
in $B$:
\beq \label{eq:charge1Matter}
	y_{P_m}=fz_{P_m}^4+x_{P_m}^2=0\,,\qquad m=1,\ldots, r\,.
\eeq
This can be seen by inserting $[z_{P_m}:x_{P_m}:y_{P_m}]$ into 
\eqref{eq:WSF}, which implies a relation between $f$ and $g$ which 
allows for a factorization of \eqref{eq:WSF} that reveals the presence
of conifold singularities in the WSF of $X$ precisely at 
\eqref{eq:charge1Matter}, see 
e.g.~\cite{Morrison:2012ei,Cvetic:2013nia} for details.
In the crepant resolution $X$, there is a reducible $I_2$-fiber
with one isolated rational curve at the codimension two loci  
\eqref{eq:charge1Matter}. The matter at the loci 
\eqref{eq:charge1Matter} are charged singlets 
$\mathbf{1}_{\underline{q}}$ with their U(1)-charges
computed according to \eqref{eq:U1charge}. This is clear as
generically \eqref{eq:charge1Matter} does not intersect any
discriminant component, which are the loci where the Cartan divisors 
$D_i^{G_I}$ are supported, so that  \eqref{eq:DynkinLabel} is trivial.

In concrete applications,  the complete intersection 
\eqref{eq:charge1Matter} describes a reducible variety in $B$ 
supporting multiple singlets with different charges. 
Matter at a generic point of \eqref{eq:charge1Matter} has U(1)-charge 
one, whereas matter at non-generic points, i.e.~points along which
other, oftentimes simpler constraints vanish, too, has
different U(1)-charges. From a technical point of view, we are
interested in all \emph{prime ideals}, denoted throughout the paper by
$I_{(k)}$, of the loci \eqref{eq:charge1Matter} for every $m$.
These are obtained by a primary decomposition of the 
complete intersection \eqref{eq:charge1Matter}, 
cf.~\cite{Cvetic:2013uta} for details.
The  codimension two variety in $B$ associated to $I_{(k)}$ is denoted 
by $V(I_{(k)})$, which is the standard notation in algebraic geometry 
for an algebraic set, i.e.~the set of points in $B$ so that all 
constraints in $I_{(k)}$ vanish. Then, we explicitly
analyze the $I_2$-fibers of $X$ over all these irreducible varieties 
$V(I_{(k)})$ in order to compute the respective U(1)-charges via 
\eqref{eq:U1charge}. 

In general, the multiplicity of matter in the representation 
$\mathbf{R}_{\underline{q}}$ is given by the homology class
of the corresponding codimension two locus in $B$. If the base $B$
is two-dimensional, which is the case in compactifications to 6D, this 
is just a set of points and the multiplicity is the number of these 
points. In F-theory compactifications to 4D, the  
homology class of a codimension two locus is the class of the
corresponding matter curve.

More specifically, the multiplicity of non-Abelian charged matter is 
computed easily as the homology class of the complete intersection with
$\mathcal{S}_{G_{I}}^b$. However, the determination of the multiplicity 
of singlets $\mathbf{1}_{\underline{q}}$ is more involved since they are located on the varieties $V(I_{(k)})$ associated to the 
usually very complex prime ideals $I_{(k)}$ of the 
complete intersection \eqref{eq:charge1Matter}.  The respective matter 
multiplicities are then  again given as the homology class of 
the variety $V(I_{(k)})$. It can be computed by the following 
procedure, see \cite{Cvetic:2013nia,Cvetic:2013uta,Cvetic:2013jta} for
more details: we first compute the homology class of the reducible 
complete intersection \eqref{eq:charge1Matter}. Given the list of its 
associated prime ideals $I_{(k)}$, we then subtract the 
multiplicities (homology classes) of those matter loci $V(I_{(k')})$, 
$\{k'\}\subset\{k\}$, we already know. Here we have to take into 
account the order $n_{k'}$ of the matter locus $V(I_{(k')})$ inside the 
complete intersection \eqref{eq:charge1Matter}. The order $n_{k'}$ is 
computed using the \emph{resultant technique} \cite{Cvetic:2013nia}. 
In all the cases considered  below in Section~\ref{sec:AnalysisOfPolytopes},
 this strategy yields the homology classes 
of all singlets $\mathbf{1}_{\underline{q}}$.

In summary, the 6D $\mathcal{N}=(1,0)$ SUGRA theory obtained by 
compactifying F-theory on a Calabi-Yau threefold $X$ has 
\begin{itemize}
\item a total number of vector multiplets $V$ reading
\beq \label{eq:Vformula} 
V=\text{\textbf{adj}}(G_X)=\sum_I \text{dim}(\text{\textbf{adj}}(G_I))+r\,,
\eeq
where $\text{\textbf{adj}}(G_X)$ and $\text{\textbf{adj}}(G_I)$ denote 
the adjoint representations of $G_I$ and $G_X$, respectively, and $r$ 
denotes the MW-rank of $X$,
\item a total number of hyper multiplets $H$ given by
\begin{align}
\begin{split} \label{eq:Hformula}	
	H&=H_{\text{codim}=2}+H_{\text{codim}=1}+H_{\text{mod}}\\
	  &=H_{\text{codim}=2}+\sum_{I=1}^n g_I(\text{dim}(\text{\textbf{adj}}(G_I))-\text{rk}(G_I))+h^{(2,1)}(X)+1\,,
\end{split}
\end{align}
where we split into contributions $H_{\text{codim}=2}$, 
$H_{\text{codim}=1}$ and $H_{\text{mod}}$ from codimension 
two fibers, from codimension one fibers over higher genus 
Riemann surfaces in $B$ and from complex structure moduli of $X$ (plus $1$), respectively, 
\item and a number of tensor multiplets $T$ counted by
\beq \label{eq:Tformula}
	T=h^{(1,1)}(B)-1=9-[K_B^{-1}]^2\,.
\eeq
\end{itemize}

For the second equality in \eqref{eq:Tformula} we have employed  the 
identity
\beq \label{eq:KB^2}
	[K_B^{-1}]\cdot [K_B^{-1}]=\int_B c_1(B)^2=10-h^{(1,1)}(B)\,.
\eeq
Here we used in the last equality the Euler number
$\chi(B)=(2+h^{(1,1)}(B))$ of a simply-connected base $B$ with 
$h^{(2,0)}=0$ and the index formula for the arithmetic genus 
$\chi_0(B)=1$,
\beq
	1=\chi_0(B)=\tfrac{1}{12}\int (c_2(B)+c_1(B)^2)=\tfrac{1}{12}(2+h^{(1,1)}(B)+\int_B c_1(B)^2)\,,
\eeq	
where $c_i(B)$, $i=1,2$, denote the Chern classes of $B$.

\subsubsection*{Codimension three singularities}
For completeness we note that codimension three singularities of the 
WSF of a Calabi-Yau fourfold $X$ support Yukawa points in
F-theory compactifications to 4D. The codimension three singularities 
are at the points in the threefold base $B$ of further enhancement of 
the vanishing order of the discriminant $\Delta$. All such enhancement points are given
as intersections of three matter curves in $B$, including 
self-intersections. Technically, given three matter curves
$V(I_{(1)})$, $V(I_{(2)})$ and $V(I_{(3)})$ we have to check
that the variety $V(I_{(1)})\cap V\big(I_{(2)})\cap V(I_{(3)}\big)$ 
contains a
codimension three component in $B$. This is achieved by
checking that the ideal  $I_{(1)}\cup I_{(2)}\cup I_{(3)}$ is 
codimension three in the ring of appropriate polynomials on $B$,  where 
we used the well-known equality 
$\bigcap_k V(I_{(k)})= V(\bigcup_k I_{(k)})$ for a family of algebraic sets $V(I_{(k)})$ \cite{hartshorne1977algebraic}.

As we see in Section~\ref{sec:AnalysisOfPolytopes}, 
all gauge-invariant Yukawa couplings are realized for the case of toric 
hypersurface fibrations $X_{F_i}$.

\subsection{Explicit examples: Calabi-Yau hypersurfaces in 2D toric 
varieties}
\label{sec:CYin2DToric}
 
All Calabi-Yau manifolds $X$ considered in this work are constructed
as fibrations of genus-one curves $\mathcal{C}$ that have a natural 
presentation as hypersurfaces in 2D toric varieties. These 
fibrations are automatically smooth, if the toric ambient spaces 
of the fiber $\mathcal{C}$ are fully resolved. In this section we 
present a very brief account on the construction of Calabi-Yau 
hypersurfaces in 2D toric varieties that are the basis for the
rest of this work. For a more complete account, we 
refer to  standard text books \cite{fulton1993introduction,david2011toric}.

\begin{figure}[h!]
\center
\includegraphics[scale=0.3]{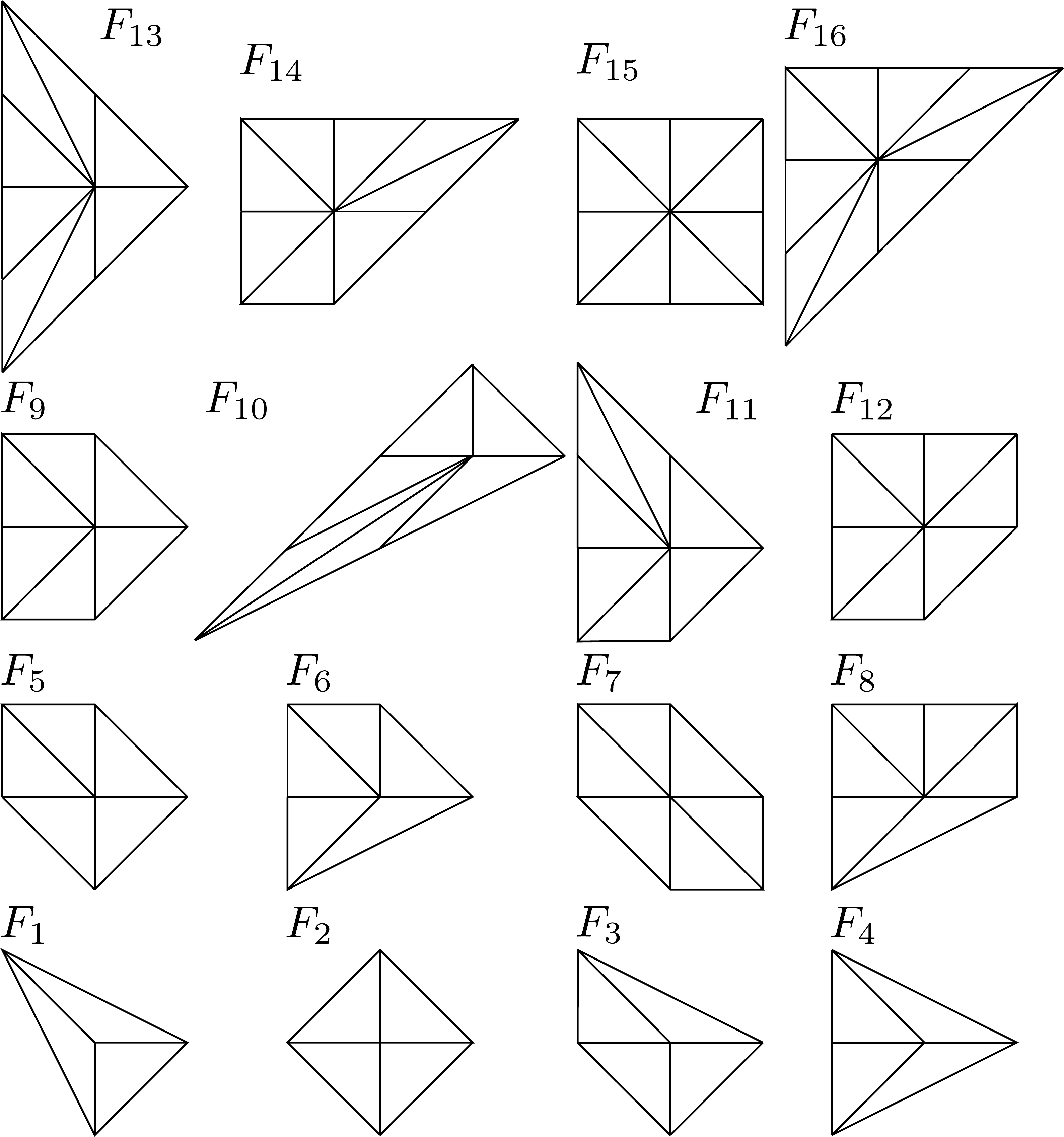}
\caption{\label{fig:16polytopes}The $16$ two dimensional reflexive polyhedra \cite{Grassi:2012qw}. The polyhedron $F_i$ and $F_{17-i}$ are dual for $i=1\dots 6$ and self-dual for $i=7\dots 10$.}
\end{figure}

A toric almost Fano surface is associated to each  of the 16 
two-dimensional reflexive polyhedra $F_i$, $i=1,\ldots, 16$, in a 
lattice $N=\mathbb{Z}^2$.\footnote{We refrain from
the common notation $\Delta$ for a polyhedron in order to avoid 
confusion with the discriminant.} 
These 16 reflexive polyhedra are given in a convenient presentation 
in Figure~\ref{fig:16polytopes} \cite{Grassi:2012qw}. As indicated 
there, the polyhedra
$F_i$ and $F_{17-i}$ for $i=1,\ldots,6$ are dual to each other, 
$F_i^*=F_{17-i}$, and the $F_i$ for $i=7,\ldots,10$ are self-dual, 
$F_i=F_i^*$, where the dual polyhedron $F_i^*$
is defined in the dual lattice $M=\mathbb{Z}^2$ of $N$ as
\beq \label{eq:dualpoly}
F^*_i=\{q\in M\otimes \mathbb{R}\vert\langle y,q\rangle\geq -1,\,\forall y\in F_i\}\,,
\eeq
where $\langle\cdot,\cdot\rangle$ is the pairing between $N$ and $M$.

For a given polyhedron $F_i$, we denote the associated toric variety by 
$\mathbb{P}_{F_i}$. Toric varieties are generalizations of weighted 
projective spaces \cite{Cox:1993fz}: to each integral point 
$v_k$, $k=1,\ldots,m+2$, except the origin of $F_i$,  we 
associate a coordinate $x_k$ in $\mathbb{C}$. Next, we introduce the 
lattice of relations between the $v_k$ 
with generators $\ell^{(a)}$ defined by
\beq \label{eq:ellvectors}
	\sum_{k=1}^{m+2} \ell^{(a)}_{k} v_k=0\,,\qquad a=1,\ldots ,m\,.
\eeq
Then, a smooth toric variety $\mathbb{P}_{F_i}$ is defined as the $(\mathbb{C}^*)^m$-quotient 
\beq \label{eq:ToricVariety}
\mathbb{P}_{F_i}=\frac{\mathbb{C}^{m+2}\backslash \text{SR}}{(\mathbb{C}^*)^m}=\{x_k\sim \prod_{a=1}^m \lambda_a^{\ell^{(a)}_k}x_k\vert\, \underline{x}\notin \text{SR}\,,\lambda_a\in\mathbb{C}^*\}\,,
\eeq
where the points $\underline{x}:=(x_1,\ldots, x_{m+2})$ are not allowed 
to lie in the Stanley-Reisner ideal SR. 

The construction \eqref{eq:ToricVariety} provides
a dictionary between the combinatorics of the polyhedron $F_i$
and the geometry of $\mathbb{P}_{F_i}$. For example, the toric divisor 
group on $\mathbb{P}_{F_i}$ is
generated by the divisors $D_k=\{x_k=0\}$ and the intersections of the 
$D_k$ are encoded in the $\text{SR}$
ideal. A full basis of the divisor group on 
$\mathbb{P}_{F_i}$ can be obtained 
using the linear equivalences between the $D_k$. Due to the relevance 
for the smoothness of a toric hypersurface fibration $X$,
we stress here that 
points that are not vertices in $F_i$ correspond to exceptional divisors 
resolving 
orbifold singularities in $\mathbb{P}_{F_i}$. 

The polyhedra $F_1$,
$F_3$, $F_5$, $F_7$ describe the generic del Pezzo surfaces 
$\mathbb{P}^2$ and $dP_i$, $i=1,2,3$, respectively, $F_2$ yields
 $\mathbb{P}^1\times \mathbb{P}^1$, $F_4$ describes 
$\mathbb{P}^2(1,1,2)$ and $F_{10}$ yields $\mathbb{P}^2(1,2,3)$.
In fact all other toric varieties $\mathbb{P}_{F_i}$ can be viewed as higher del
Pezzo surfaces at a special point in their respective complex structure moduli spaces.

Every toric variety $\mathbb{P}_{F_i}$ has an associated
Calabi-Yau hypersurface, i.e.~a genus-one curve $\mathcal{C}_{F_i}$.
It is defined as the generic section of its anti-canonical bundle
$K_{\mathbb{P}_{F_i}}^{-1}$.
The Calabi-Yau hypersurface in $\mathbb{P}_{F_i}$ is obtained by the Batyrev construction as the  following polynomial \cite{Batyrev:1994hm}
\begin{equation} \label{eq:BatyrevFormula}
 	p_{F_i}=\sum_{q\in F_i^*\cap M} a_q\prod_{k} x_k^{\langle v_k,q\rangle+1}\,,
\end{equation}
where $q$ denotes all integral points in $F_i^*$ and the $a_q$
are coefficients in the field $K$. 

We note that points $v_i$ 
interior to edges in $F_i$ are usually excluded in the product 
\eqref{eq:BatyrevFormula} because the 
corresponding divisors do not intersect the hypersurface  
$\mathcal{C}_{F_i}$. However, when considering 
Calabi-Yau fibrations $X_{F_i}$ of $\mathcal{C}_{F_i}$, as in Section~\ref{sec:AnalysisOfPolytopes},
these divisors intersect $X_{F_i}$ and resolve singularities of $X_{F_i}$
induced by singularities of its fibration, i.e.~these 
divisors are related to Cartan divisors $D_{i}^{G_I}$ discussed above in 
Section~\ref{sec:divsonGenus1}.

\section{Analysis of F-theory on Toric Hypersurface Fibrations}
\label{sec:AnalysisOfPolytopes}

In this section we analyze the geometry of the Calabi-Yau manifolds
$X_{F_i}$, that are constructed as fibration of the genus-one curves
$\mathcal{C}_{F_i}$ 
over a generic base $B$. For
each manifold we calculate the effective theory resulting from
compactifying F-theory on it. We calculate the gauge group, the charged
and neutral matter spectrum and the Yukawa couplings.

We start with a quick summary of some interesting results of this
study. There are three polyhedra leading to manifolds $X_{F_i}$ without
a section: $F_1$, $F_2$ and $F_4$, see sections \ref{sec:F1_poly},
\ref{sec:F2_poly} and \ref{sec:F4_poly}, respectively. They yield the discrete gauge groups
in F-theory. 
For three polyhedra we find associated gauge groups with Mordell-Weil torsion, giving
rise to non-simply connected gauge groups:
$F_{13}$, $F_{15}$ and $F_{16}$, see sections \ref{sec:polyF13},
\ref{sec:polyF15} and \ref{sec:polyF16}, respectively. The analysis of
the hypersurface $X_{F_3}$ and the corresponding effective theory of
F-theory, whose spectrum contains a charged singlet with U(1)-charge three, can be found in
section \ref{sec:F3_poly}.

We obtain the following list of gauge groups $G_{F_i}$ of
F-theory on the $X_{F_i}$:
\beq \nonumber
\text{
\begin{tabular}{|l|c||l|c||l|c|} \hline
$G_{F_{1}}$ & $\mathbb{Z}_3$ & $G_{F_{7}}$ & U(1)$^3$ & &  \rule{0pt}{13pt} \\
$G_{F_{2}}$ & U(1)$\times\mathbb{Z}_2$ & $G_{F_{8}}$ & SU(2)$^2\times$U(1) & $G_{F_{13}}$ & (SU(4)$\times$SU(2)$^2)/\mathbb{Z}_2$ \rule{0pt}{13pt} \\
$G_{F_{3}}$ & U(1) & $G_{F_{9}}$ & SU(2)$\times$U(1)$^2$ &  $G_{F_{14}}$ & SU(3)$\times$SU(2)$^2\times$U(1) \rule{0pt}{13pt} \\
$G_{F_{4}}$ & SU(2)$\times \mathbb{Z}_4$ & $G_{F_{10}}$ & SU(3)$\times$SU(2) & $G_{F_{15}}$ & SU(2)$^4/\mathbb{Z}_2\times$U(1) \rule{0pt}{13pt} \\
$G_{F_{5}}$ & U(1)$^2$ & $G_{F_{11}}$ & SU(3)$\times$SU(2)$\times$U(1)  & $G_{F_{16}}$ & SU(3)$^3/\mathbb{Z}_3$ \rule{0pt}{13pt}  \\
$G_{F_{6}}$ & SU(2)$\times$U(1) & $G_{F_{12}}$ & SU(2)$^2\times$U(1)$^2$ & & \rule{0pt}{13pt} \\ \hline
\end{tabular}
}
\eeq
From this and as a simple consequence of
\eqref{eq:rankGX}, we see that there is the following rule of thumb for computing
the rank of a gauge group $G_{F_i}$:
given a polyhedron $F_i$ with $3+n$
integral points without the origin, we have a gauge group with total rank $n$.

Let us outline the structure of this section. In the first subsection
\ref{sec:cubicquarticbiquadric} we briefly discuss the three different
representations of genus-one curves $\mathcal{C}_{F_i}$ realized as
toric hypersurfaces: the cubic, the biquadric and the quartic. There, we
define the line bundles of the base $B$ in which the
coefficients in these constraints have to take values in order to obtain
a genus-one fibered Calabi-Yau manifold. The functions $f$ and $g$
of the Weierstrass form \eqref{eq:WSF} for the cubic, the biquadric and
the quartic can be found in Appendix~\ref{app:AppPoly}. By appropriate
specializations of the coefficients, one can obtain $f$, $g$ and
$\Delta=4f^3+27g^2$ for all toric hypersurface fibration $X_{F_i}$.

In 
sections~\ref{sec:discSymmetry} to
\ref{sec:TorsionPolys}
we proceed to describe in detail each Calabi-Yau
manifold $X_{F_i}$. In each case we first discuss the genus-one curve
$\mathcal{C}_{F_i}$ realized as a toric hypersurface in
$\mathbb{P}_{F_i}$. We then construct the corresponding
toric hypersurface fibration $X_{F_i}$ and analyze its
codimension one and two singularities from which we extract the
non-Abelian gauge group and matter spectrum. If
$X_{F_i}$ has a non-trivial MW-group, we determine all its generators,
their Shioda maps and the height pairing. For genus-one fibrations, we
determine their discrete gauge groups. For completeness, we also 
determine the Yukawa couplings from codimension three singularities.
In each case we show as a
consistency check that the necessary 6D anomalies  (pure Abelian,
gravitational-Abelian, pure non-Abelian, non-Abelian gravitational,
non-Abelian-Abelian and purely gravitational) are canceled
implying consistency of the considered effective theories.

We organize the Calabi-Yau manifolds $X_{F_i}$ into five categories:
those with discrete gauge symmetries (Section~\ref{sec:discSymmetry}),
those with a gauge group of rank one and two but without discrete gauge
groups (Section~\ref{sec:Rk12Nodisc}), those with a gauge group of rank
three,
whose fiber polyhedra happen to be 
also self-dual (Section~\ref{sec:Rk3}), those with gauge groups of rank 
four and five without MW-torsion (Section~\ref{sec:Rk45NoTorsion}) and 
those $X_{F_i}$ with MW-torsion (Section~\ref{sec:TorsionPolys}).  This 
arrangement is almost in perfect agreement with the labeling of the 
polyhedra $F_i$ in Figure~\ref{fig:16polytopes} which facilitates the 
navigation through this section. We name the subsection containing the 
analysis of the specific manifold $X_{F_i}$ by its corresponding fiber 
polyhedron $F_{i}$.

%%%%%%%%%%%%%%%%%%%%%%%%%%%%%%%%%%%%%%%%%%%%%%%%%%%%%%%%%%%%%%%%%%%%%%%%%%%%%%%%%%%%%%%%%%%%%%%%%
\subsection{Three basic ingredients: the cubic, biquadric and
quartic}
\label{sec:cubicquarticbiquadric}
%%%%%%%%%%%%%%%%%%%%%%%%%%%%%%%%%%%%%%%%%%%%%%%%%%%%%%%%%%%%%%%%%%%%%%%%%%%%%%%%%%%%%%%%%%%%%%%%%

\subsubsection{Constructing toric hypersurface fibrations}
\label{sec:BasicIngredients}

In this section we explain the general construction of the Calabi-Yau
manifolds $X_{F_i}$ with toric hypersurface fiber $\mathcal{C}_{F_{i}}$
and base $B$. The following discussion applies to Calabi-Yau
$n$-folds $X_{F_i}$ with a general $(n-1)$-dimensional base $B$. The
cases of most relevance for F-theory and for this work are $n=3,4$. We
refer to \cite{Cvetic:2013nia,Cvetic:2013uta} for more details on the
following discussion.

The starting point of the construction of the genus-one
fibered Calabi-Yau manifold $X_{F_i}$ is the hypersurface equation
\eqref{eq:BatyrevFormula} of the curve $\mathcal{C}_{F_i}$.
In order to obtain the equation of $X_{F_i}$, the coefficients $a_q$ and
the variables $x_i$ of \eqref{eq:BatyrevFormula} have to be promoted to
sections of appropriate line bundles of the base $B$. We determine these
line bundles, by first constructing a fibration of the 2D toric variety
$\mathbb{P}_{F_i}$, which is the ambient space of $\mathcal{C}_{F_i}$,
over the same base $B$,
\beq
	\xymatrix{
	\mathbb{P}_{F_i} \ar[r] & 	\mathbb{P}^B_{F_i}(D,\tilde{D}) \ar[d]\\
	& B\,%_{n-1}\,
	}\,.
	\label{eq:PFfibration}
\eeq
Here $\mathbb{P}^B_{F_i}(D,\tilde{D})$ denotes the total space of this
fibration. The structure of its fibration is
parametrized by two divisors in $B$, denoted by $D$ and $\tilde{D}$.
This can be seen by noting that all  $m+2$ coordinates $x_k$ on the
fiber $\mathbb{P}_{F_i}$ are in general non-trivial sections of line
bundles on $B$. Then, we can use the $(\mathbb{C}^*)^m$-action of the
toric variety $\mathbb{P}_{F_i}$ to set $m$ variables to transform in
the trivial bundle of $B$. The divisors dual to the two remaining line
bundles are precisely $D,\, \tilde{D}$.

Next we impose equation \eqref{eq:BatyrevFormula} in
$\mathbb{P}_{F_i}(D,\tilde{D})$. Consistency fixes
the line bundles in which the coefficients $a_q$ have to take values
in terms of the two divisors $D$ and $\tilde{D}$. Then, we require
\eqref{eq:BatyrevFormula} to be a section of the anti-canonical bundle
$K^{-1}_{\mathbb{P}^B_{F_i}}$, which is the Calabi-Yau condition.
In addition, equation \eqref{eq:BatyrevFormula} imposed
in $\mathbb{P}^B_{F_i}(D,\tilde{D})$ clearly describes a genus-one
fibration over $B$, since for every generic point on $B$, the
hypersurface \eqref{eq:BatyrevFormula} describes exactly the curve
$\mathcal{C}_{F_i}$ in $\mathbb{P}_{F_i}$. The total Calabi-Yau
space resulting from the 
fibration of the toric hypersurface $\mathcal{C}_{F_i}$
is denoted by $X_{F_i}$ in the following. It enjoys the fibration structure
\beq \label{eq:Cfibration}
	\xymatrix{
	\mathcal{C}_{F_i} \ar[r] & 	X_{F_i} \ar[d]\\
	& B\,
	}\, .
\eeq

In principle, this procedure has to be carried out for all Calabi-Yau
manifolds $X_{F_{i}}$ associated to the 16 2D toric polyhedra $F_i$.
However, we observe that all the hypersurface constraints
of the $X_{F_{i}}$, except for $X_{F_2}$ and $X_{F_4}$, can be
obtained from the hypersurface constraint for  $X_{F_1}$, after setting appropriate coefficients to zero. This is possible because if $F_1$ is a
sub-polyhedron of $F_i$, then the corresponding toric variety
$\mathbb{P}_{F_i}$ is the blow-up of $\mathbb{P}_{F_1}= \mathbb{P}^2$ at
a given number of points, with the additional rays in $F_i$
corresponding to the blow-up divisors. However, adding rays to the
polyhedron $F_1$ removes rays from its dual polyhedron $F_1^*=F_{16}$.
By means of \eqref{eq:BatyrevFormula}, this removes coefficients from
hypersurface equation for $X_{F_1}$, i.e.~the hypersurface for $X_{F_i}$
is a certain specialization of the hypersurface of $X_{F_1}$ with some
$a_q\equiv 0$. We will be more explicit about this in the following
subsection (Section~\ref{sec:cubic}).

Thus, we only have to explicitly carry out the construction
of the toric hypersurfaces separately for the two Calabi-Yau manifolds 
$X_{F_2}$ and $X_{F_4}$.
The details of this are given in Sections \ref{sec:biquadric} and
\ref{sec:quartic}.

\subsubsection{Fibration by cubic curves: $X_{F_1}$ and its
specializations}
\label{sec:cubic}

We proceed to construct the Calabi-Yau manifold
$\mathcal{C}_{F_1}\rightarrow X_{F_1}\rightarrow B$ with fiber
given by the curve $\mathcal{C}_{F_1}$ in the toric variety
$\mathbb{P}_{F_1}$.  In addition, we argue how the Calabi-Yau manifolds
$X_{F_i}$, whose fiber polyhedron $F_i$ contains $F_1$,
can be obtained from $X_{F_1}$.

\begin{figure}[ht]
\centering
\includegraphics[scale=0.4]{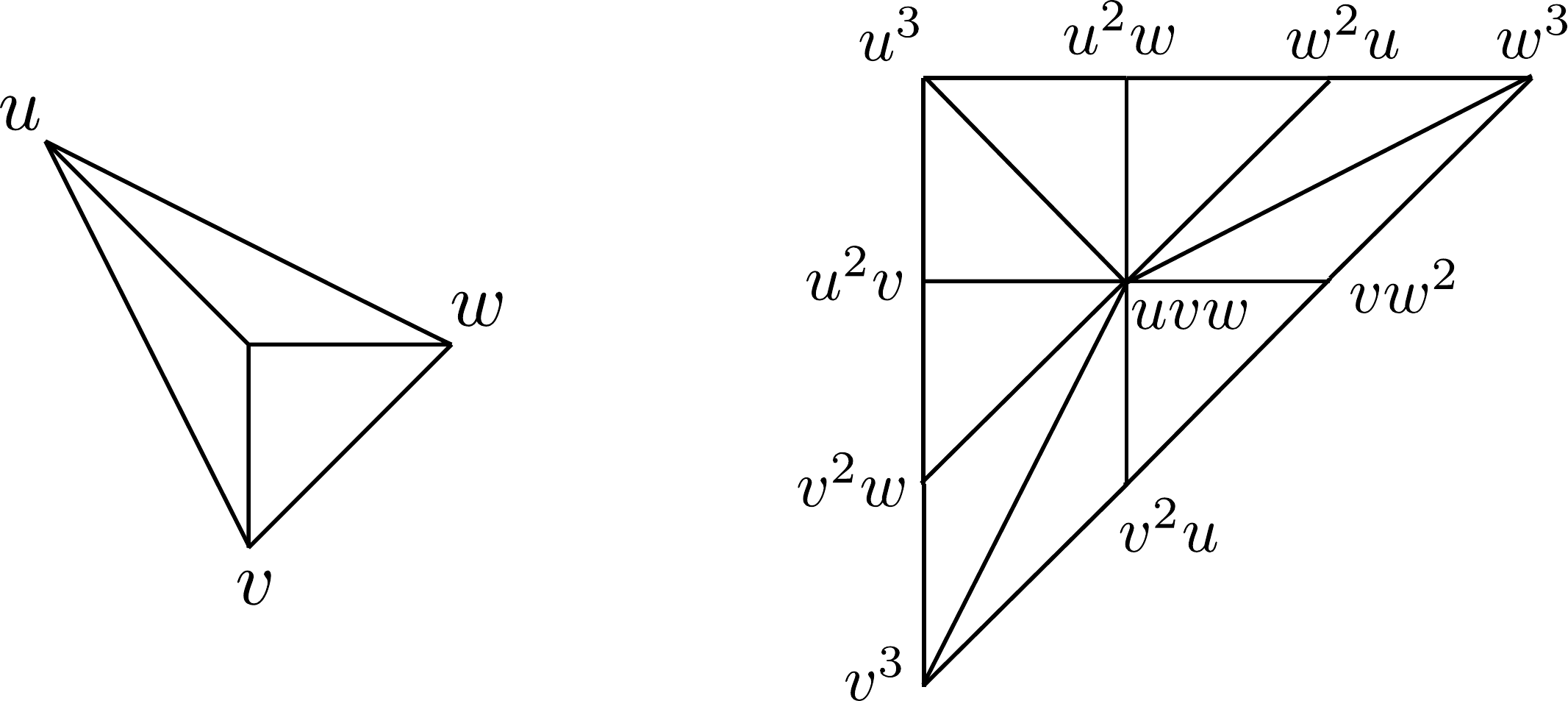}
\caption{\label{fig:poly1_toric} Polyhedron $F_1$ with choice of
projective coordinates and its dual with corresponding monomials.}
\end{figure}

The polyhedron $F_1$ and its dual are shown in
Figure~\ref{fig:poly1_toric}. The toric variety $\mathbb{P}_{F_1}$,
constructed using \eqref{eq:ToricVariety}, is the well-known projective
space $\mathbb{P}^2$. We introduce the projective coordinates $[u:v:w]$
on $\mathbb{P}^2$. In terms of these coordinates, we can read off the
SR-ideal from Figure~\ref{fig:poly1_toric} as
\beq
	SR=\{uvw\}\,.
\eeq
The divisor group of $\mathbb{P}^2$ is generated by the hyperplane
class $H$. The Calabi-Yau onefold in $\mathbb{P}^2$
is the degree three $\mathcal{C}_{F_1}$ in $3H$. Its defining equation,
constructed using \eqref{eq:BatyrevFormula} and Figure \ref{fig:poly1_toric}, is the most general cubic
\beq \label{eq:pF1}
p_{F_{1}} = s_1 u^3 + s_2 u^2 v+ s_3 u v^2 + s_4 v^3 + s_5 u^2 w + s_6 u v w + s_7 v^2 w + s_8 u w^2+s_9 v w^2 +s_{10} w^3\,,
\eeq
where the coefficients $s_i$ take values in the field $K$.

Next, we follow the discussion of Section \ref{sec:BasicIngredients}
to construct the toric hypersurface fibration $X_{F_1}$. We first
construct the ambient space \eqref{eq:PFfibration}, which in the case at
hand is a $\mathbb{P}^2$-fibration over the base $B$,
\beq \label{eq:P2fibration}
	\xymatrix{
	\mathbb{P}^2 \ar[r] & 	\mathbb{P}^2(\cS_7,\cS_9) \ar[d]\\
	& B\,
	}\,.
\eeq
The two divisors parametrizing this fibration are %two divisors
$\cS_7$
and $\cS_9$, cf.~\cite{Cvetic:2013nia,Cvetic:2013uta}. Upon
imposing  the constraint \eqref{eq:pF1} and requiring the Calabi-Yau
condition for $X_{F_1}$, we see that these two divisors are precisely
the classes of the coefficients $s_7$ and $s_9$, respectively. Indeed,
as mentioned above, we can use the $\mathbb{C}^{*}$-action on
$\mathbb{P}^2$ to turn e.g.~$w$ into a section of the trivial line
bundle of the base. Then, we choose the variables $u$ and $v$ as sections
of the bundles
\beq \label{eq:LBassignment}
	u\in \mathcal{O}_B(\cS_9+ [K_{B}])\,,\qquad v\in \mathcal{O}_B(\cS_9-\cS_7)\,.
\eeq
This allows us to compute the anti-canonical bundle of the
$\mathbb{P}^2$-fibration \eqref{eq:P2fibration} using adjunction as
\beq \label{eq:XF1-class}
 K^{-1}_{\mathbb{P}^2(\cS_7,\cS_9)} = \mathcal{O}(3H+2\cS_9-\cS_7)\;.
\eeq
Finally, we impose the Calabi-Yau condition on the constraint
\eqref{eq:pF1} for $X_{F_1}$ which fixes the divisor classes of the
coefficients $s_i$. We summarize the divisor classes of the homogeneous
coordinates $[u:v:w]$ and the coefficients $s_i$ in the following tables:
\beq \label{eq:cubicsections}
\text{
\begin{tabular}{c|c}
\text{section} & \text{Divisor Class}\\
\hline
	$u$&$H+\cS_9+[K_B]$\rule{0pt}{13pt} \\
	$v$&$H+\cS_9-\cS_7$\rule{0pt}{12pt} \\
	$w$&$H$\rule{0pt}{12pt} \vspace{2cm}\\
\end{tabular}
}\qquad \text{
\begin{tabular}{c|c}
\text{section} & \text{Divisor Class}\\
\hline
	$s_1$&$3[K_B^{-1}]-\cS_7-\cS_9$\rule{0pt}{13pt} \\
	$s_2$&$2[K_B^{-1}]-\cS_9$\rule{0pt}{12pt} \\
	$s_3$&$[K_B^{-1}]+\cS_7-\cS_9$\rule{0pt}{12pt} \\
	$s_4$&$2\cS_7-\cS_9$\rule{0pt}{12pt} \\
	$s_5$&$2[K_B^{-1}]-\cS_7$\rule{0pt}{12pt} \\
	$s_6$&$[K_B^{-1}]$\rule{0pt}{12pt} \\
	$s_7$&$\cS_7$\rule{0pt}{12pt} \\
	$s_8$&$[K_B^{-1}]+\cS_9-\cS_7$\rule{0pt}{12pt} \\
	$s_9$&$\cS_9$ \rule{0pt}{12pt} \\
	$s_{10}$&$2\cS_9-\cS_7$ \rule{0pt}{12pt}
\end{tabular}
}
\eeq
%
%\\*[3mm]
\subsubsection*{$X_{F_i}$ %\textit{
as specialized cubics}%}
%\\*[3mm]
%
As mentioned in the previous subsection, the equations of the Calabi-Yau
manifolds $X_{F_i}$  with $i\ne2,4$ can be expressed as specialized
versions of the cubic hypersurface equation \eqref{eq:pF1} of $X_{F_1}$.

In order to find the hypersurface equation for an $X_{F_i}$
we begin by calculating the anti-canonical class of
the fibration $\mathbb{P}_{F_i}^B(D,\tilde{D})$ defined in \eqref{eq:PFfibration}.
To this end, we first note that  toric ambient spaces $\mathbb{P}_{F_i}$
are obtained from $\mathbb{P}_{F_1}$ by a certain number of blow-ups at
points $P_j$. Assuming the number of blow-ups is $k$, we have
\beq
\mathbb{P}_{F_i}=\text{Bl }_{P_1,\cdots,P_k}\mathbb{P}_{F_1}\,.
\eeq
Each blow-up adds a $\mathbb{P}^1$ with an associated new
variable $e_j$ and divisor class $E_j$.
From the combinatorial point of view, this means
that there is an additional $\mathbb{C}^*$-action on
$\mathbb{P}_{F_i}$.

Next we note that the fibration $\mathbb{P}^B_{F_{i}}(D,\tilde{D})$ can 
be parametrized
by the same base divisors $\cS_7$ and $\cS_9$ as the fibration
\eqref{eq:P2fibration}, i.e.~we identify $D=\cS_7$ and
$\tilde{D}=\cS_9$.
Indeed, this is possible since we can use $(\mathbb{C}^*)$-actions,
including the new $(\mathbb{C}^*)$-actions from the $k$ blow-ups,
to make the variables $w$ and $e_j$ transform in the trivial bundle
on $B$
while maintaining the assignments \eqref{eq:LBassignment}
for $u$ and $v$.
Employing these results we calculate the anti-canonical
bundle of  $\mathbb{P}^B_{F_{i}}(\mathcal{S}_7,\mathcal{S}_9)$,
using the adjunction formula, yielding
\beq \label{eq:XFi-class}
 K^{-1}_{\mathbb{P}^B_{F_i}(\cS_7,\cS_9)} = \mathcal{O}(3\tilde{H}-E_1-E_2-\cdots -E_k+2\cS_9-\cS_7)\;.
\eeq
Here, $\tilde{H}$
denotes the pull-back $\tilde{H}=\tilde{\pi}^*(H)$ of the hyperplane
class $H$ on $\mathbb{P}^2$ under the blow-down map $\tilde{\pi}:\mathbb{P}_{F_i}\rightarrow \mathbb{P}^2$. By abuse of notation, we
will denote it throughout this work simply by $H$.
It is to be observed that if the coefficient $s_i$ is present in the constraint of $X_{F_i}$, i.e.~if it
is not removed by the $k$ blow-ups, 
its corresponding class $[s_i]$ remains unaltered from the
one given in Table~\eqref{eq:cubicsections}.

This relation of the hypersurface constraints of the $X_{F_i}$ for
$i\neq 2,4$ and all the bundles entering it to the hypersurface equation \eqref{eq:pF1} and
bundles \eqref{eq:cubicsections} of $X_{F_1}$ will facilitate our following
presentation. In particular, in the respective subsections on $X_{F_i}$
for  $i\ne 2,4$ only the classes for the variables $u$, $v$, $w$ and
$e_j$ have to be given explicitly.

\subsubsection{Fibration by the biquadric: $X_{F_2}$}
\label{sec:biquadric}

We construct the Calabi-Yau manifold $\mathcal{C}_{F_2}\rightarrow
X_{F_2}\rightarrow B$ as the fibration of the curve $\mathcal{C}_{F_2}$
in the toric variety $\mathbb{P}_{F_2}$ over $B$. As mentioned before,
its hypersurface equation cannot be described as a cubic. Thus,
$X_{F_2}$ has to be analyzed separately.

\begin{figure}[ht]
\centering
\includegraphics[scale=0.4]{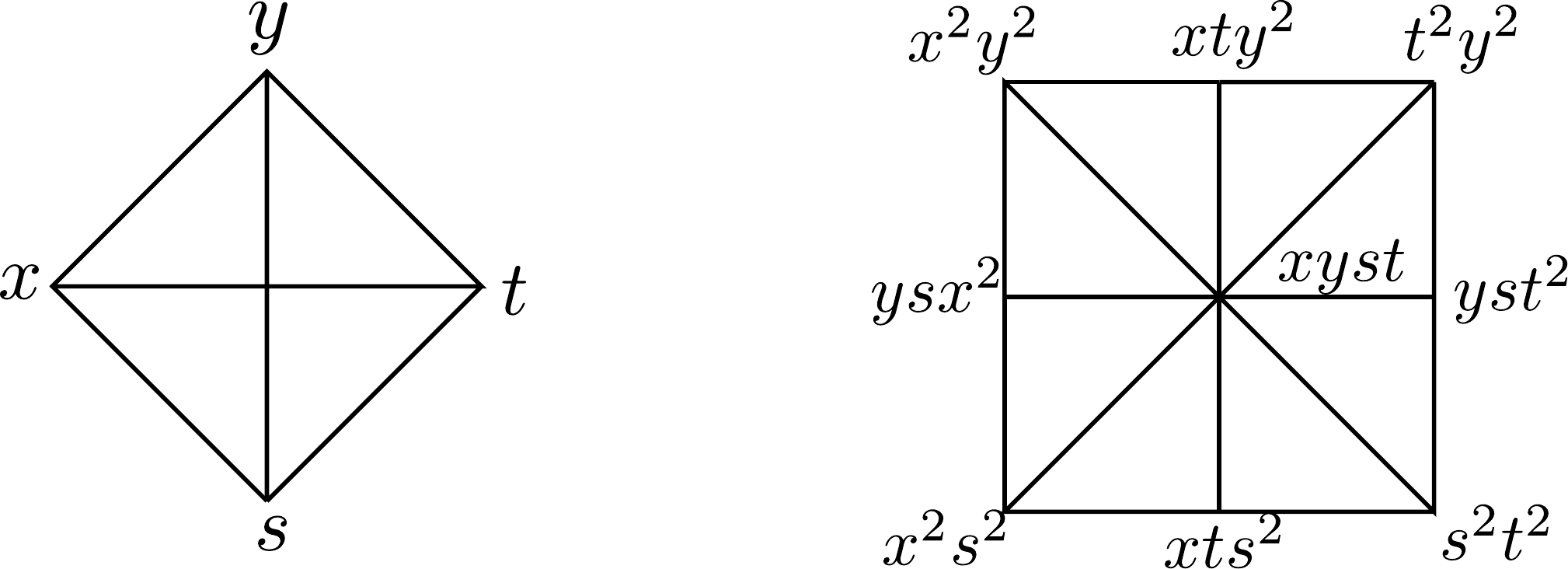}
\caption{\label{fig:poly2_toric} Polyhedron $F_2$ with choice of
projective coordinates and its dual with corresponding monomials.}
\end{figure}

The polyhedron $F_2$ and its dual are presented in
Figure~\ref{fig:poly2_toric}. The toric variety associated to it is
$\mathbb{P}_{F_2}=\mathbb{P}^1\times \mathbb{P}^1$ and we have
introduced the projective coordinates $[x:t]$ and $[y:s]$ on the two
$\mathbb{P}^1$'s, respectively.
The Stanley-Reisner ideal of $\mathbb{P}_{F_2}$ is given by
\beq \label{eq:SRF2}
SR = \{xt,ys \}\,.
\eeq
There are two divisor classes on
$\mathbb{P}_{F_2}$, that we denote by $H_1$ and $H_2$ with respective
representatives $\{x=0\}$ and $\{y=0\}$.
The Calabi-Yau onefold in $\mathbb{P}_{F_2}$ is the curve
$\mathcal{C}_{F_2}$ in the class $2H_1+2H_2$. It is a biquadric of the
form
\beq \label{eq:biquadric-F2}
p_{F_2} = (b_1 y^2+ b_2 s y + b_3 s^2) x^2 + (b_5 y^2+ b_6 s y + b_7 s^2) x t + (b_8 y^2+ b_9 s y + b_{10} s^2) t^2 \, ,
\eeq
as can be shown using \eqref{eq:BatyrevFormula} and
Figure~\ref{fig:poly2_toric}. Here the $b_i$ denote coefficients in
the field $K$.

In order to find $X_{F_2}$ we proceed to  construct
$\mathbb{P}_{F_2}^B(D,\tilde{D})$, the fibration of
$\mathbb{P}_{F_2}$ introduced in \eqref{eq:PFfibration}. It is possible
to consistently parametrize this fibration in terms of the same divisor 
classes $D=\cS_7$ and $\tilde{D}=\cS_9$ as in 
\eqref{eq:P2fibration}. In the hypersurface constraint \eqref{eq:biquadric-F2},
they correspond to the classes of the coefficients $b_7$ and $b_9$
respectively.\footnote{The
consistency of this assignment can be seen by noting that
$\mathbb{P}_{F_2}$ is related to $\mathbb{P}^2$ by the blow-up at
$x=y=0$ setting $b_{10}=0$  and the subsequent
blow-downs $x=y=1$. Then, \eqref{eq:biquadric-F2} precisely yields
\eqref{eq:pF1}.} This will  facilitate the matching of
the effective theories via Higgsings, as discussed in Section~\ref{sec:Higgsings}.
Next, we use the $(\mathbb{C}^*)^2$-actions on $\mathbb{P}_{F_2}$ to
achieve that the variables $x$ and $y$ transform in the trivial line
bundle on $B$. The other two variables $s$ and $t$ take values in the
following line bundles on $B$:
\beq
 \label{eq:LBQassignment}
	t\in \mathcal{O}_B([K^{-1}_{B}]-\cS_9 )\,,\qquad s\in \mathcal{O}_B([K^{-1}_{B}]-\cS_7 )\,.
\eeq
With this assignment of line bundles to the coordinates on
$\mathbb{P}_{F_2}$, the anti-canonical class of
$\mathbb{P}_{F_2}^B(\mathcal{S}_7,\mathcal{S}_9)$ is readily calculated as
\beq \label{eq:XF2-class}
 K^{-1}_{\mathbb{P}^B_{F_2}} = \mathcal{O}(2H_1+2H_2+3[K_B^{-1}]-\cS_7-\cS_9)\,.
\eeq
Finally, we require that the hypersurface \eqref{eq:biquadric-F2} is
Calabi-Yau, which fixes the divisor classes of the coefficients $b_i$
in terms of $\cS_7$, $\cS_9$ and $[K_B^{-1}]$.
In summary, we obtain that  the coordinates on $\mathbb{P}_{F_2}$
and the coefficients $b_i$ have the following divisor classes:
\beq \label{eq:biquadricsections}
\text{
\begin{tabular}{c|c}
\text{section} & \text{Divisor Class}\\
\hline
	$x$&$H_1$\rule{0pt}{13pt} \\
	$t$&$H_1+[K^{-1}_{B}]-\cS_9$\rule{0pt}{12pt} \\
	$y$&$H_2$\rule{0pt}{13pt} \\
	$s$&$H_2+[K^{-1}_{B}]-\cS_7$\rule{0pt}{12pt} \vspace{2cm}\\
\end{tabular}
}\qquad \text{
\begin{tabular}{c|c}
\text{Section} & \text{Divisor Class}\\
\hline
	$b_1$&$3[K_B^{-1}]-\cS_7-\cS_9$\rule{0pt}{13pt} \\
	$b_2$&$2[K_B^{-1}]-\cS_9$\rule{0pt}{12pt} \\
	$b_3$&$[K_B^{-1}]+\cS_7-\cS_9$\rule{0pt}{12pt} \\
	$b_5$&$2[K_B^{-1}]-\cS_7$\rule{0pt}{12pt} \\
	$b_6$&$[K_B^{-1}]$\rule{0pt}{12pt} \\
	$b_7$&$\cS_7$\rule{0pt}{12pt} \\
	$b_8$&$[K_B^{-1}]+\cS_9-\cS_7$\rule{0pt}{12pt} \\
	$b_9$&$\cS_9$ \rule{0pt}{12pt} \\
	$b_{10}$&$\cS_9+\cS_7-[K_B^{-1}]$ \rule{0pt}{12pt}
\end{tabular}
}
\eeq
We emphasize that the classes of the coefficients $b_i$, except for
$b_{10}$, agree with the classes of $s_i$ of the cubic $X_{F_1}$,
c.f.~\eqref{eq:cubicsections}, as expected.

\subsubsection{Fibration by the quartic: $X_{F_4}$ }
\label{sec:quartic}

We proceed to construct the Calabi-Yau manifold
$\mathcal{C}_{F_4}\rightarrow X_{F_4}\rightarrow B$ with general fiber
given by the curve $\mathcal{C}_{F_4}$ in the toric variety
$\mathbb{P}_{F_4}$. As we have mentioned above, its hypersurface
equation is not a special case of a cubic, which requires a separate
analysis of $X_{F_4}$.

\begin{figure}[H]
\center
\includegraphics[scale=0.4]{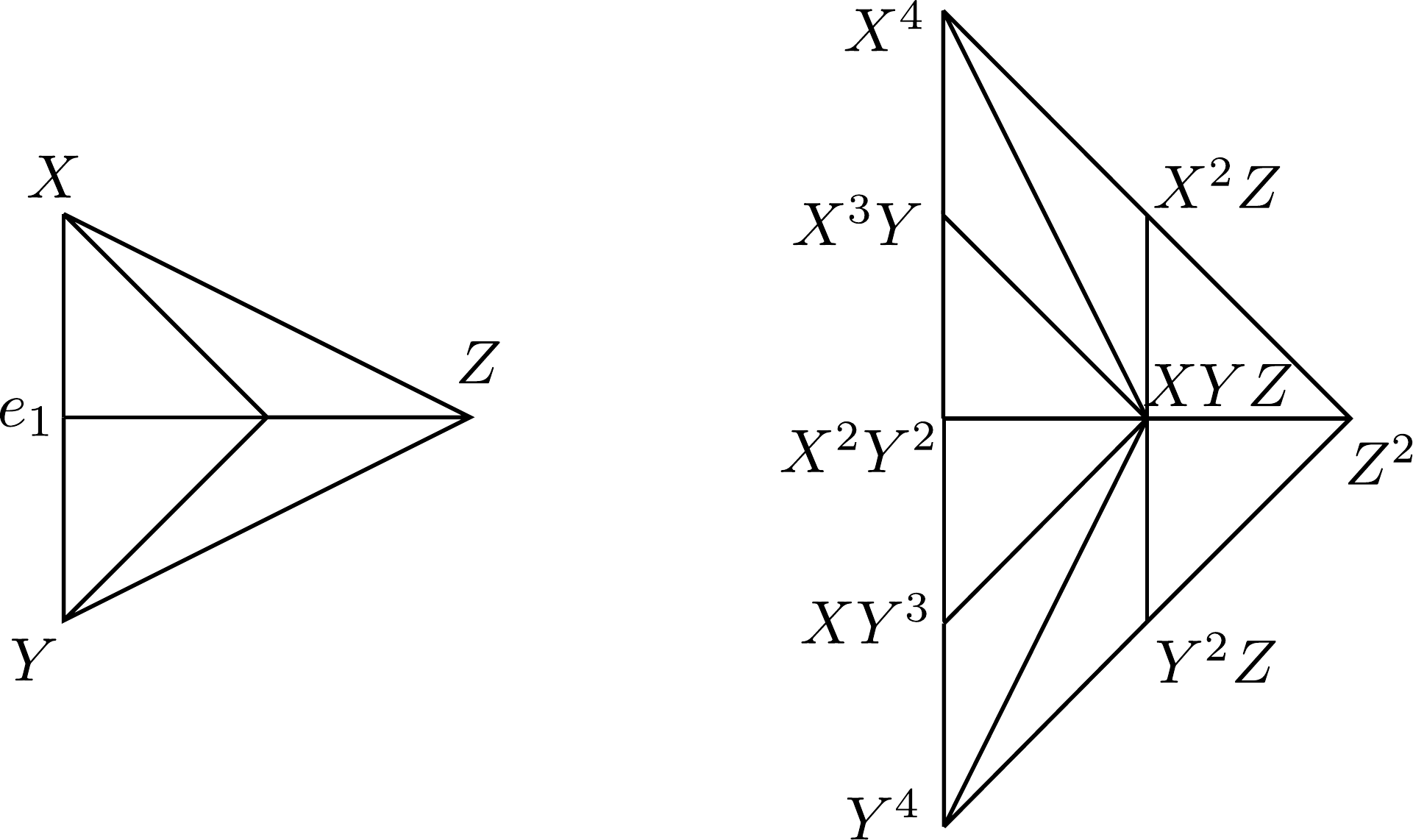}
\caption{\label{fig:poly4_toric}Polyhedron $F_4$ with choice of
projective coordinates and its dual with corresponding monomials where
the blow-up variable $e_1$ is suppressed.}
\end{figure}

The polyhedron $F_4$ and its dual polyhedron are shown in
Figure~\ref{fig:poly4_toric}. Its associated toric variety is
$\mathbb{P}_{F_4}=\mathbb{P}^{2}(1,1,2)$ and we introduce the homogeneous
coordinates $[X:Y:Z:e_1]$.   The Stanley-Reisner ideal of this toric
variety can be read off from Figure \ref{fig:poly4_toric} as
\beq \label{eq:SRF4}
SR = \{XY,Z e_1 \}\,.
\eeq
There are two divisor classes on $\mathbb{P}_{F_4}$, that are denoted
by $H$ and $E_1$ with representatives $\{X=0\}$ and $\{e_1=0\}$
respectively. We note that $\mathbb{P}_{F_4}$ automatically contains an
exceptional divisor $E_1$ corresponding to the point interior to
the edge of $F_4$.
The equation for the Calabi-Yau onefold $\mathcal{C}_{F_4}$ in
$\mathbb{P}_{F_4}$, which is a degree two curve, is the quartic equation
in the class $4H-2E_1$. It reads explicitly
\begin{align}
\begin{split}\label{eq:quartic_hypersurface}
p_{F_4}&= d_1 e_1^2 X^4 + d_2 e_1^2 X^3 Y + d_3 e_1^2 X^2 Y^2 + d_4 e_1^2 X Y^3 + d_5 e_1^2 Y^4 + d_6 e_1 X^2 Z \\
&\phantom{=} + d_7 e_1 X Y Z + d_8 e_1 Y^2 Z + d_9 Z^2\,,
\end{split}
\end{align}
as we infer from \eqref{eq:BatyrevFormula} and
Figure~\ref{fig:poly4_toric}. The coefficients $d_i$ take values in the
field $K$.

In order to find $X_{F_4}$ we construct the fibration
$\mathbb{P}_{F_4}^B(\cS_7,\cS_9)$, the fibration of $\mathbb{P}_{F_4}$
over $B$ introduced in \eqref{eq:PFfibration}. Again, it
is possible to parametrize the fibration by the identical divisors
$\cS_7$ and $\cS_9$ as in the fibration \eqref{eq:P2fibration} relevant
for the $X_{F_1}$.\footnote{We note that this assignment is a
consequence of the birational map between $\mathbb{P}_{F_4}$ and $\mathbb{P}_{F_1}$
induced by the blow-up at $X=Z=0$, setting $d_5=0$, and the blow-downs $X=1$ and $e_1=1$.}
As before, we can use the two $\mathbb{C}^*$-actions on $P_{F_4}$
to make two of its homogeneous coordinates transform in the trivial
bundles. However, it turns out that a convenient assignment of
sections, avoiding fractions, is given by
\beq \label{eq:coordsDivsF4}
     X\in\mathcal{O}_B(\mathcal{S}_9-[K_B^{-1}])\,,\quad Y\in\mathcal{O}_B(\cS_9-\cS_7)\,,\quad Z\in\mathcal{O}_B(\cS_9-[K_B^{-1}])\,.
\eeq
We use this to compute the anti-canonical bundle on
$\mathbb{P}_{F_4}^B(\mathcal{S}_7,\mathcal{S}_9)$ as
\begin{align}
K^{-1}_{\mathbb{P}^B_{F_4}} = \mathcal{O}(4H-2E_1+3\cS_9-\cS_7-[K_B^{-1}])\; .%3 [K_B^{-1}] - 2 E_1 + 4 H - \cS_7 + 3 \cS_9)\;.%2H-\cS_7+\cS_9)\; .
\end{align}
Imposing the Calabi-Yau condition on the constraint
\eqref{eq:quartic_hypersurface},
we fix the classes of all coefficients $d_i$.
The assignments of divisor classes to the coordinates on
$\mathbb{P}_{F_4}$ and the $d_i$ in summary read
\beq \label{eq:quarticsections}
\text{
\begin{tabular}{c|c}
\text{section} & \text{Divisor Class}\\
\hline
	$X$ & $H-E_1+\cS_9-[K_B^{-1}]$ \rule{0pt}{13pt} \\
	$Y$ & $H-E_1-\cS_7+\cS_9$ \rule{0pt}{12pt} \\ 
	$Z$ & $2H-E_1+\cS_9-[K_B^{-1}]$ \rule{0pt}{13pt} \\
	$e_1$ & $E_1$ \rule{0pt}{12pt} \vspace{2cm}\\
\end{tabular}
}\qquad \text{
\begin{tabular}{c|c}
\text{Section} & \text{Divisor Class}\\
\hline
	$d_1$ & $3[K_B^{-1}]-\cS_7-\cS_9$ \rule{0pt}{13pt} \\
	$d_2$ & $2[K_B^{-1}]-\cS_9$ \rule{0pt}{12pt} \\ 
	$d_3$ & $[K_B^{-1}]+\cS_7-\cS_9$ \rule{0pt}{12pt} \\ 
	$d_4$ & $2\cS_7-\cS_9$ \rule{0pt}{12pt} \\ 
	$d_5$ & $-[K_B^{-1}]+3\cS_7-\cS_9$ \rule{0pt}{12pt} \\
	$d_6$ & $2[K_B^{-1}]-\cS_7$ \rule{0pt}{12pt} \\ 
	$d_7$ & $[K_B^{-1}]$ \rule{0pt}{12pt} \\ 
	$d_8$ & $\cS_7$ \rule{0pt}{12pt} \\ 
	$d_9$ & $[K_B^{-1}]-\cS_7+\cS_9$ \rule{0pt}{12pt} 
\end{tabular}
}
\eeq
We emphasize that the classes of the coefficients $d_i$ slightly differ
from \eqref{eq:cubicsections} in the cubic $X_{F_1}$ due to the slightly
different assignments \eqref{eq:coordsDivsF4} of classes to the
coordinates.

%%%%%%%%%%%%%%%%%%%%%%%%%%%%%%%%%%%%%%%%%%%%%%%%%%%%%%%%%%%%%%%%%%%%%%%%%%%%%%%%%%%%%%%%%%%%%%%%%
\subsection{Fibration with discrete gauge symmetry}
%%%%%%%%%%%%%%%%%%%%%%%%%%%%%%%%%%%%%%%%%%%%%%%%%%%%%%%%%%%%%%%%%%%%%%%%%%%%%%%%%%%%%%%%%%%%%%%%%
\label{sec:discSymmetry}

In this section we analyze the toric hypersurface fibrations based on 
the fiber polyhedra $F_1$, $F_2$ and $F_4$. Since their fibrations 
do not have a section, but only multi-sections, they are genus-one 
fibrations. We analyze the codimension one, two and 
three singularities of these models, employing also their 
respective associated Jacobian fibrations. We show that 
the effective theories of F-theory on these Calabi-Yau 
manifolds exhibit discrete gauge groups and include matter that is 
charged only under the respective discrete group.

%%%%%%%%%%%%%%%%%%%%%%%%%%%%%%%%%%%%%%%%%%%%%%%%%%%%%%%%%%%%%%%%%%%%%%%%%%%%%%%%%%%%%%%%%%%%%%%%%
\subsubsection{Polyhedron $F_{1}$: $G_{F_1}=\mathbb{Z}_3$}
%%%%%%%%%%%%%%%%%%%%%%%%%%%%%%%%%%%%%%%%%%%%%%%%%%%%%%%%%%%%%%%%%%%%%%%%%%%%%%%%%%%%%%%%%%%%%%%%%
\label{sec:F1_poly}

We consider the genus-one fibration $X_{F_1}$ over an arbitrary base $B$
with genus-one fiber $\mathcal{C}_{F_1}$ realized as the Calabi-Yau
hypersurface in $\mathbb{P}_{F_1}=\mathbb{P}^2$. The toric data of
$\mathbb{P}_{F_1}=\mathbb{P}^2$ and the construction
of the Calabi-Yau manifolds $X_{F_1}$ have been discussed in Section
\ref{sec:cubic}.
The hypersurface equation for $X_{F_1}$ is given by \eqref{eq:pF1} with
the relevant divisor classes of the coordinates $[u:v:w]$ and
the coefficients $s_i$ summarized in \eqref{eq:cubicsections}.

The fibration  $\pi:X_{F_1}\rightarrow B$ does not have a section,
but only a three-section. Thus, $X_{F_1}$ is only a genus-one
fibration, cf.~the general discussion in Section
\ref{sec:ellipticCurvesWithRP}.
In order to obtain the WSF of $X_{F_1}$, given the absence of sections
of its fibration, we have to calculate the associated Jacobian fibration
$J(X_{F_1})$. The algorithm for computing $J(X_{F_1})$ is well known in
the mathematics literature, see for example \cite{an2001jacobians}, from
where we calculate $f$ and $g$, given explicitly in \eqref{eq:fcubic}
and \eqref{eq:gcubic}, and subsequently the  discriminant $\Delta$.
The discriminant does not factorize, which shows the absence of
codimension one singularities of $X_{F_1}$ and therefore, the absence of non-Abelian
gauge groups in the corresponding F-theory compactification.

The fibration $X_{F_1}$ has a three-section that is given by
\begin{align} \label{eq:secF1}
\hat{s}^{(3)}&= X_{F_1}\cap \{u=0\}:\,\,\, s_4 v^3 + s_7 v^2 w +s_9 v w^2 +s_{10} w^3=0\,,
\end{align}
as follows from the Calabi-Yau constraint \eqref{eq:pF1}.
We denote its divisor class, that agrees with $H+\mathcal{S}_9+[K_B]$,
by $S^{(3)}$.
Under the degree nine map from $X_1$ to its Jacobian this three-section
is mapped to the canonical zero section $z=0$ in the WSF of
$J(X_{F_1})$.
However, in $X_{F_1}$, the three-section $\hat{s}^{(3)}$ locally
maps a point on the base $B$ to three
points on the fiber $\mathcal{C}_{F_1}$. Globally, there exists a
monodromy group that interchanges
these three points, upon moving on the base $B$.
This fact, together with the existence of $I_2$-fibers in $X_{F_1}$ at
codimension two on which the mondromy group acts non-trivially,  
as we present next, and the results from Higgsing the U(1) gauge group in the
effective theory associated to  $X_{F_{3}}$, see
Section~\ref{S:discretesymm},
leads us to postulate the following \emph{discrete} gauge group of
$X_{F_1}$:
\beq \label{eq:GF1}
 G_{F_1}= \mathbb{Z}_{3}\,.
\eeq

In order to 
compute the charges of matter under this discrete group,
we have to associate a divisor class to the three-section. As certain
models with multi-section are related to models with multiple rational
sections by conifold transitions,
see~\cite{Morrison:2014era,Anderson:2014yva}, a
natural proposal for such a divisor class is an expression similar to
the Shioda map \eqref{eq:ShiodaMap}. We recall the three defining
properties of a Shioda map summarized on page 21 in \cite{Park:2011ji}.
Imposing these conditions on the divisor class associated to
\eqref{eq:secF1}, we obtain the following divisor class,
\beq \label{eq:ShiodathreesectionF1}
\sigma_{\mathbb{Z}_3} (\hat{s}^{(3)}) = S^{(3)}+[K_B] + \tfrac{4}{3} \mathcal{S}_9 - \tfrac{2}{3} \mathcal{S}_7 \, .
\eeq
We propose that matter charges under the discrete group $\mathbb{Z}_3$
should be computed using this class. In fact, we demonstrate next, that
the class \eqref{eq:ShiodathreesectionF1} allows us to compute
$\mathbb{Z}_3$-charges of matter-representations on $X_{F_1}$, that are
consistent with 6D anomaly cancellation and the Higgsing from
the model $X_{F_3}$, discussed in Section \ref{S:discretesymm}.

\subsubsection*{Charged and uncharged matter in $X_{F_1}$}

We proceed with determining the codimension two singularities of the
WSF of $J(X_{F_1})$. This analysis is most easily carried out directly
in the smooth fibration $X_{F_1}$.  The same techniques presented here
will also be used in a slightly modified form for the analysis of the
fibration $X_{F_2}$, $X_{F_3}$ and $X_{F_4}$. We note that the same
technique has been used recently in \cite{Braun:2014oya} and \cite{Anderson:2014yva}.
\
\subsubsection*{Finding the loci of $I_2$-fibers using elimination ideals}

We are looking for  loci of $B$ that support $I_2$-fibers in $X_{F_1}$.
At these loci, the genus-one fiber $\mathcal{C}_{F_1}$ of $X_{F_1}$ has
to degenerate into two $\mathbb{P}^1$'s, i.e.~the hypersurface equation \eqref{eq:pF1} has
to factor into two smooth polynomials. For a smooth cubic the only
factorization with this property is the one into a conic and a line,
i.e.~a factorization of \eqref{eq:pF1} of the form
\beq \label{eq:factF1}
 p_{F_1} \stackrel{!}{=} s_1(u+\alpha_1 v+\alpha_2 w)( u^2+ \beta_1 v^2 + \beta_2 w^2+ \beta_3 uv+ \beta_4 vw+\beta_5 uw)\,,
\eeq
where $\alpha_j$ and $\beta_k$ are seven unknown polynomials on $B$.
We note that we can assume $s_1\neq 0$, because otherwise we would
obtain a locus of codimension three or higher.
Making a comparison of coefficients on both sides of
\eqref{eq:factF1},
we obtain a set of constraints that defines an ideal in the ring
$K[s_i,\alpha_j,\beta_k]$, where $s_i$ are the coefficients in
\eqref{eq:pF1}. We denote this ideal by  $I_{(s_i,\alpha,\beta)}$.
We emphasize that there are two more constraints, namely nine,
in $I_{(s_i,\alpha,\beta,)}$ than unknowns $\alpha_j$,
$\beta_k$, i.e.~the system is over-determined.
Thus, there only exists a solution for $\alpha_j$,
$\beta_k$ satisfying  
\eqref{eq:factF1},
if two additional constraints on the $s_i$ are obeyed.
This implies that the ideal $I_{(s_i,\alpha,\beta,)}$ describes a
codimension two locus of the $s_i$.

In order to obtain the constraints that the $s_i$ have to obey for the
factorization \eqref{eq:factF1} to exist, we compute
the elimination ideal $I_{(s_i)}=I_{(s_i,\alpha,\beta)}\cap K[s_i]$,\footnote{Here we 
deviate from the notation in mathematics
literature, where the subscripts of the elimination ideal indicate the
eliminated variables.} where $K[s_i]$ is the polynomial ring only in the
variables $s_i$. We compute $I_{(s_i)}$, in the following abbreviated
as $I_{(1)}\equiv I_{(s_i)}$, explicitly using Singular \cite{Singular}
and obtain an ideal with $50$ generators. Furthermore, we calculate
its codimension in the ring $K[s_i]$ to be two.
Thus, its vanishing locus $V(I_{(1)})$ describes a codimension two
variety in $B$. In summary, we have shown that the factorization
\eqref{eq:factF1} corresponding to an $I_2$-fiber in $X_{F_1}$
happens at the codimension two locus $V(I_{(1)})$ in $B$.

We note that \eqref{eq:factF1} is
the only type of factorization that can occur. Thus, we do not expect
any further codimension two fibers and corresponding matter
representation in $X_{F_1}$. The spectrum of $X_{F_1}$ is summarized in
Table \ref{tab:poly1_matter}.
\begin{table}[t!]
\begin{center}
\footnotesize
\renewcommand{\arraystretch}{1.2}
\begin{tabular}{|c| c |p{3cm}@{} | c  |}\hline
Representation  & Multiplicity & \centering Fiber & Locus \\ \hline
$\one_{1}$ & $3\left( 6 [K_B^{-1}]^2 - \cS_7^2 + \cS_7 \cS_9 - \cS_9^2 + [K_B^{-1}](\cS_7 + \cS_9) \right) $ & \rule{0pt}{1.3cm}\parbox[c]{2.9cm}{\includegraphics[width=2.9cm]{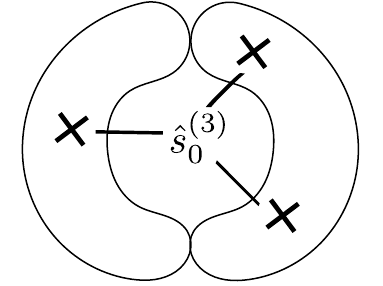}} & $V(I_{(1)})$  \\[0.9cm] \hline
\end{tabular}
\caption{\label{tab:poly1_matter}Charged matter representation under $\mathbb{Z}_3$ and corresponding codimension two fiber of $X_{F_1}$.}
\end{center}
\end{table}
Next we argue how to compute the charge of the matter located
at $V(I_{(1)})$ under the discrete gauge group $G_{F_1}=\mathbb{Z}_3$.
Due to the absence of a zero section on $X_{F_1}$, there is no
preferred curve in the $I_2$-fiber in Table~\ref{tab:poly1_matter}.
As can be observed from \eqref{eq:factF1},
the two $\mathbb{P}^1$'s in this $I_2$-fiber have intersection numbers
one and two with the three-section $\hat{s}_0^{(3)}$.
By naively applying \eqref{eq:U1charge} using
the divisor class \eqref{eq:ShiodathreesectionF1}, we compute the
charges $q=1$ and $q=2$ for the two rational curves, respectively. Thus,
it seems that there is no meaningful way to assign a discrete charge to
the matter located at $V(I_{(1)})$.
However, this seeming contradiction is resolved by noting that a 6D
hyper multiplet of charge
$q=1$ is the same as one with charge $q=-1$. In addition, employing the
discrete $\mathbb{Z}_3$ symmetry, we have $-1=2\,\text{ mod }3$, showing
that a 6D hyper multiplet of charge $q=1$ under a $\mathbb{Z}_3$
symmetry is physically equivalent to one with charge $q=2$. Thus, the
matter at $V(I_{(1)})$ has charge $q=1$ which is the same as $q=2$ under
the discrete gauge group $G_{F_1}=\mathbb{Z}_3$ .

We proceed to calculate the multiplicity of $V(I_{(1)})$. Unfortunately
given the size and number of polynomials in the ideal $I_{(1)}$, we are
unable to obtain its multiplicity geometrically with the available
computing power. Instead, we invoke the results
for its multiplicity that is obtained in Section~\ref{S:discretesymm}
using the Higgs transition $X_{F_3}\rightarrow X_{F_1}$. It is
shown in Table~\ref{tab:poly1_matter} for completeness.

We complete the discussion of the matter spectrum of $X_{F_1}$ by
calculating the number of neutral hyper multiplets. We use
\eqref{eq:Hneutral} and the explicit
expression for the Euler number $\chi(X_{F_1})$  of $X_{F_1}$ in
\eqref{eq:EulerNumbers} to obtain
\begin{align}
H_{\text{neut}} %&= h^{(2,1)}(X_{F_1})+1 \,, \nn\\
&= 12 + 11 [K_B^{-1}]^2 - 3 [K_B^{-1}] \cS_7 + 3 \cS_7^2 - 3 [K_B^{-1}] \cS_9 - 3 \cS_7 \cS_9 + 3 \cS_9^2\,.
\end{align}
Employing this together with the number of vector multiplets $V=0$ and
the charged spectrum in
Table~\ref{tab:poly1_matter}
we check cancellation of the 6D gravitational anomaly in \eqref{eq:6dAnomalies}.

\subsubsection*{Yukawa couplings in $X_{F_1}$}

We conclude this section by noting that there is only one
gauge-invariant Yukawa coupling possible:
\beq
\text{
\begin{tabular}{|c|c|}\hline
Yukawa & Locus \\ \hline
 $\one_1\cdot\one_1\cdot\one_1$ & $V(I_1)\cap V(I_1)\cap V(I_1)$  \\ \hline
\end{tabular}}
\eeq
Again, we cannot check for its presence explicitly due to the complexity of the ideal $I_{(1)}$.\footnote{The presence of this coupling can be deduced considering the Higgsing from $X_{F_3}$ 
to $X_{F_1}$ (see Section \ref{S:discretesymm}). Decomposing the states in $X_{F_3}$ in terms 
of states in $X_{F_1}$, we observe that after Higgsing, the Yukawa coupling 
$\one_1\cdot\one_1\overline{\one_2}$ in $X_{F_3}$ (see Table \ref{tab:poly3_yukawa}) gives 
rise to $\one_1\cdot\one_1\cdot\one_1$ in $X_{F_3}$.}
%
%
%
%

%%%%%%%%%%%%%%%%%%%%%%%%%%%%%%%%%%%%%%%%%%%%%%%%%%%%%%%%%%%%%%%%%%%%%%%%%%%%%%%%%%%%%%%%%%%%%%%%%
\subsubsection{Polyhedron $F_{2}$: $G_{F_2}=\text{U(1)}\times\mathbb{Z}_2$}
%%%%%%%%%%%%%%%%%%%%%%%%%%%%%%%%%%%%%%%%%%%%%%%%%%%%%%%%%%%%%%%%%%%%%%%%%%%%%%%%%%%%%%%%%%%%%%%%%
\label{sec:F2_poly}

Here, we analyze the genus-one fibration $X_{F_2}$ constructed as a
fibration of the Calabi-Yau onefold $\mathcal{C}_{F_2}$ in
$\mathbb{P}_{F_2}=\mathbb{P}^1\times\mathbb{P}^1$. The toric data
of $\mathbb{P}^1\times\mathbb{P}^1$ and the construction of
the toric hypersurface fibration $X_{F_2}$  have been presented in
Section~\ref{sec:biquadric}. The hypersurface constraint for $X_{F_2}$
is given in \eqref{eq:biquadric-F2} and the relevant divisor classes
are summarized in \eqref{eq:biquadricsections}.

First, we note that the fibration $\pi: X_{F_2}\rightarrow B$ 
does not have a section, i.e.~it is a genus-one fibration. We obtain its WSF
by computing its associated Jacobian fibration $J(X_{F_2})$, employing
again the straightforward algorithms from 
the mathematics literature \cite{an2001jacobians}. The results for the
functions $f$ and $g$ can be found in
\eqref{eq:f-F2} and \eqref{eq:g-F2}, from which the discriminant can be
readily computed. The discriminant does not factorize, which again shows
the absence of codimension one singularities. Thus, there is no
non-Abelian gauge symmetry for this 
F-theory compactifcation.

The fibration of $X_{F_2}$ has  two
independent two-sections, that are given by
\bea \label{eq:secF2}
\hat{s}^{(2)}_0= X_{F_2}\cap \{x=0\}:& b_8 y^2+b_9 s y+b_{10}s^2=0 \,,\nn\\
\hat{s}^{(2)}_1= X_{F_2} \cap \{y=0\}:& b_3 x^2+b_7 x t+b_{10}t^2=0\,,
\eea
where we used the hypersurface constraint \eqref{eq:biquadric-F2} and
the SR-ideal \eqref{eq:SRF2}.
We denote the two corresponding divisor classes, that agree with $H_1$
and $H_2$, by $S^{(2)}_0$ and $S^{(2)}_1$, respectively. Analogous to
the previous Section \ref{sec:F1_poly}, we expect a discrete
$\mathbb{Z}_2$ gauge group associated to the two-section $S^{(2)}_0$,
cf.~the similar discussion in \cite{Morrison:2014era}. We will provide independent evidence for this
by the analysis in Section \ref{S:discretesymm} of the Higgsing of the
effective theory of F-theory on $X_{F_5}$, that has a $\text{U}(1)^2$ gauge group, 
to the one arising from $X_{F_2}$.

The role of the other two-section
$\hat{s}^{(2)}_1$, however, is less clear in the biquadric
representation. Its meaning
for F-theory is unraveled by transforming the biquadric
\eqref{eq:biquadric-F2} defining $X_{F_2}$ into a cubic hypersurface
and then
by computing its Weierstrass form, which is precisely the WSF
of the Jacobian fibration of $X_{F_2}$, as we show.
This detour via the cubic yields a direct map to the Jacobian fibration
$J(X_{F_2})$, which allows us to follow the two-section
$\hat{s}^{(2)}_1$ in \eqref{eq:secF2}.

\subsubsection*{Map to the cubic in  $\mathbb{P}_{F_5}$  \& the
MW-group of $J(X_{F_2})$}

The curve $\mathcal{C}_{F_2}$ given as the biquadric
\eqref{eq:biquadric-F2} in $\mathbb{P}_{F_2}$ can be treated as the
cubic in $\mathbb{P}_{F_5}$ after an appropriate change of variables.
Indeed, by applying the transformation
$x \rightarrow x+\alpha t $ or  $y \rightarrow y +\beta s $, we can
set the coefficient of the monomial $s^2 t^2$ \eqref{eq:biquadric-F2} to
zero for an appropriate $\alpha$ or $\beta$. We note that both
$\alpha$ and $\beta$ have to involve square roots of the coefficients
$b_i$ in \eqref{eq:biquadric-F2}, i.e.~the two variable transformations
are only defined in a field extension. As we will see, 
this field extension will only be an intermediate step, since
all square roots will drop out in the final result of our computation. After
the change of variables, we obtain a polynomial of the following form:
\begin{equation} \label{eq:F2-as-cubic}
\tilde{p}=(\tilde{s}_1 y^2+ \tilde{s}_2 s y + \tilde{s}_3 s^2) x^2 + (\tilde{s}_5 y^2+ \tilde{s}_6 s y + \tilde{s}_7 s^2) x t + (\tilde{s}_8 y^2+ \tilde{s}_9 s y ) t^2\,,
\end{equation}
where the redefined coefficients $\tilde{s}_i$ depend on the variables
$b_i$ and are explicitly given in
\eqref{eq:trans-F5F2}. We note that $\tilde{p}$ is precisely
of the form of the cubic \eqref{eq:pF5} in $\mathbb{P}_{F_5}$ after
identifying
\beq \label{eq:varTrafo-F2-F5}
t\rightarrow w\,,\quad  s\rightarrow v\,,\quad x\rightarrow e_2\,,\quad
y\rightarrow e_1\,,\quad u=1\,.
\eeq
Since the curve $\tilde{p}=0$ is an elliptic curve,
we can compute its WSF, in particular the functions $f$ and $g$.
Inserting the explicit expressions
\eqref{eq:trans-F5F2} for the sections $\tilde{s}_i$ in
\eqref{eq:F2-as-cubic} in terms of the $b_i$ in \eqref{eq:biquadric-F2}
into the expressions for $f$ and $g$, we precisely recover
\eqref{eq:f-F2} and \eqref{eq:g-F2} obtained from the WSF of the Jacobian fibration $J(X_{F_2})$. Most notably, all square roots
in the coefficients $\tilde{s}_i$ have dropped out, as claimed.

Next, we note that the two-section $\hat{s}^{(2)}_1$ in \eqref{eq:secF2}
formally maps to the section
\beq
	s=1\,,\quad t=-\frac{\tilde{s}_3}{\tilde{s}_7}\,\quad y=0\,,
	\quad x=1\, ,
\eeq
in \eqref{eq:F2-as-cubic}. Under
the identification of coordinates \eqref{eq:varTrafo-F2-F5}, this is
precisely the section $\hat{s}_1$ of $X_{F_5}$ given in
\eqref{eq:sectionsF5}. Inserting the explicit expressions
\eqref{eq:trans-F5F2} for the $\tilde{s}_i$ into the WS-coordinates
\eqref{eq:secsF51} of $\hat{s}_1$, we obtain
\begin{align}
\begin{split}\label{eq:F2-section}
z_1=& 1\,, \\
x_1=& \tfrac{1}{12} (8 b_1 b_{10} + b_6^2 - 4 b_5 b_7 + 8 b_3 b_8 - 4 b_2 b_9)\, , \\
y_1=& \tfrac{1}{2} (b_{10} b_2 b_5 - b_1 b_{10} b_6 + b_3 b_6 b_8 - b_2 b_7 b_8 - b_3 b_5 b_9 +
   b_1 b_7 b_9)\,.
\end{split}
\end{align}
We emphasize that all square roots in the coefficients $b_i$ in
\eqref{eq:trans-F5F2} have dropped out and we obtain completely rational
WS-coordinates for the two-section $\hat{s}^{(2)}_1$. We double-check
that \eqref{eq:F2-section} solves the WSF of the Jacobian $J(X_{F_2})$.

In summary, we have shown for the first time
that the associated Jacobian fibration $J(X_{F_2})$ exhibits a
\emph{rank one} MW-group  of rational sections generated by the section
in \eqref{eq:F2-section}, which is precisely
the image of the two-section $\hat{s}^{(2)}_1$ in $X_{F_2}$ under the
map $X_{F_2}\rightarrow J(X_{F_2})$.
This means that there is an associated Abelian gauge field
in the F-theory compactified on $X_{F_2}$. We note that application
of the same logic to the two-section $\hat{s}_0^{(2)}$, which formally
maps to the section $\hat{s}_2$ defined in \eqref{eq:sectionsF5} in
$X_{F_{5}}$, does not lead to a rational section of the Jacobian
$J(X_{F_2})$ since its WS-coordinates \eqref{eq:secsF52} after
inserting \eqref{eq:trans-F5F2} still contain square roots. Hence,
$\hat{s}^{(2)}_0$ does not yield an additional U(1)-factor,
but corresponds to a discrete group $\mathbb{Z}_2$, as claimed.

Having proven the presence of a MW-group on $J(X_{F_2})$, we compute
the Shioda map of its generator. We note that the usual
expression \eqref{eq:ShiodaMap} has to be modified since
$\hat{s}_1^{(2)}$ is a two-section. It can be shown that the following
expression obeys all conditions listed in \cite{Park:2011ji} that have
to be obeyed by a Shioda map:
\beq
\label{eq:ShiodaF2}
\sigma (\hat{s}_1^{(2)}) = S_1^{(2)}-S_0^{(2)}+\tfrac{1}{2}([K_B]-\mathcal{S}_7+\mathcal{S}_9)\,
\eeq
Then we obtain the corresponding height pairing, using \eqref{eq:anomalycoeff}, as
\begin{align}\label{eq:heightpairingF2}
b_{11}=-\pi(\sigma(\hat{s}^{(2)}_1)\cdot\sigma(\hat{s}^{(2)}_1))= 2[K^{-1}_B]\,.
\end{align}
Here, we used the following intersections
\beq
\pi\big((S^{(2)}_0)^2)=-2([K_B^{-1}]-\mathcal{S}_9)\,,\qquad
\pi\big((S^{(2)}_1)^2)=-2([K_B^{-1}]-\mathcal{S}_7)\,,\quad
\pi\big(S^{(2)}_0\cdot S^{(2)}_1)= \mathcal{S}_7 +\mathcal{S}_9-[K_B^{-1}]\,.
\eeq
The first two equalities are just a translation of
the SR-ideal \eqref{eq:SRF2} into intersection relations of divisor
classes on $X_{F_2}$, employing \eqref{eq:biquadricsections} and
\eqref{eq:intsSectionF} for $n=2$. The third relation follows
by noting that according to \eqref{eq:secF2}  the two two-sections
$\hat{s}^{(2)}_0$ and $\hat{s}^{(2)}_1$ intersect precisely at
$b_{10}=0$, whose class is $[b_{10}]=\mathcal{S}_7 +\mathcal{S}_9-[K_B^{-1}]$, cf.~\eqref{eq:biquadricsections}.

We conclude by summarizing the full gauge group of the theory:
\beq
G_{F_2}=\text{U(1)}\times\mathbb{Z}_2 \,.
\eeq
We highlight again that the U(1) corresponds to a rational section in
the Jacobian fibration $J(X_{F_2})$, that is the image
of the two-section $\hat{s}^{(2)}_1$ under the degree four map $X_{F_2}\rightarrow J(X_{F_2})$.

As mentioned before, the discrete gauge $\mathbb{Z}_{2}$ %gauge 
symmetry
is induced by the two-section $\hat{s}_0^{(2)}$. For the computation of
charges of matter w.r.t.~the $\mathbb{Z}_2$, we have to associate a  
divisor class to it.  Imposing conditions on
this divisor class  similar  to the one that have lead to the Shioda map \eqref{eq:ShiodaMap} \cite{Park:2011ji}, we obtain
\beq \label{eq:ShiodatwoSecF2}
\sigma_{\mathbb{Z}_2} (\hat{s}_0^{(2)}) = S_0^{(2)}+[K_B^{-1}]-\mathcal{S}_9 \, .
\eeq
We use this divisor class to successfully compute the
$\mathbb{Z}_2$-charges of matter in the following.

\subsubsection*{Charged and uncharged matter in $X_{F_2}$}

Now that we know the gauge group of the theory, we proceed to derive
first the matter representation and then the corresponding
6D matter multiplicities. As in Section \ref{sec:F1_poly}, we use
the elimination ideal technique  to show directly  the presence
of three matter representations in $X_{F_2}$, namely $\one_{(1,+)}$, $\one_{(1,-)}$ and
$\one_{(0,-)}$, where $\pm$ denote the two possible
$\mathbb{Z}_2$-eigenvalues. Then, we compute their  multiplicities,
where we also invoke the equivalent presentation of $X_{F_2}$ as a
quartic.

In order to find the three $I_2$-fibers at codimension two in $X_{F_2}$, we first note that there
are three different possible ways to factorize the biquadric
\eqref{eq:biquadric-F2}, that correspond to the three inequivalent
ways to split its degree $(2,2)$ w.r.t.~the classes $H_1$ and $H_2$
in $\mathbb{P}^1\times\mathbb{P}^1$, namely as $(2,2)=(1,1)+(1,1)$,  $(2,2)=(1,0)+(1,2)$ and $(2,2)=(0,1)+(2,1)$ respectively.

The first type of factorization of \eqref{eq:biquadric-F2}
corresponding to $(2,2)=(1,1)+(1,1)$ is given by
\begin{equation} \label{eq:pF2-fact}
p_{F_2} \stackrel{!}{=} b_1 \big[ ( y + \alpha_1 s) x + (\alpha_2 y + \alpha_3 s) t \big ] \big[ ( y + \beta_1 s) x + (\beta_2 y + \beta_3 s) t \big]\,.
\end{equation}
Clearly, both factors are bilinear in $[x:t]$ and $[y:s]$, respectively,
as required. As before, we can factor out $b_1$ because it must not
vanish at a codimension two locus.
We note that there are six unknown polynomials $\alpha_j$ and $\beta_k$
and eight non-trivial constraints, as can be seen by a comparison of
coefficients on both sides. Thus, the ideal of constraints
is over-determined and imposes a codimension two condition 
on the coefficients $b_i$ for a solution to \eqref{eq:pF2-fact} to exist.
The elimination ideal, that we call $I_{(1)}$, obtained by eliminating
the unknowns $\alpha_j$ and $\beta_k$ from the ideal of constraints is
generated by $50$ polynomials. It is checked to be codimension two in
the ring, as expected, proving the existence of the factorization
\eqref{eq:pF2-fact} at codimension two. We denote
the zero set of $I_{(1)}$ by $V(I_{(1)})$, which is the geometric
codimension two locus in $B$.

Next, we note that each curve of the $I_2$-fiber described by
\eqref{eq:pF2-fact} has intersection one with both
two-sections $\hat{s}^{(2)}_0$ and $\hat{s}^{(2)}_1$. The U(1)-charge
computed using \eqref{eq:U1charge} and \eqref{eq:ShiodaF2} is zero. We
also note that the representation has charge $(-)$ under
the discrete symmetry because the two
intersection points of $\hat{s}^{(2)}_0$ with the fiber
are interchanged under a monodromy action. Formally, the charge under
$\mathbb{Z}_2$ is computed
using  \eqref{eq:U1charge} together with the divisor class
\eqref{eq:ShiodatwoSecF2}, showing that both curves in
the $I_2$-fiber have $\mathbb{Z}_2$-charge $(-)$. Thus, the
representation at the locus $V(I_{(1)})$ is $\one_{(0,-)}$ as shown in
Table~\ref{tab:poly2_matter}.
\begin{table}[t!]
\begin{center}
\renewcommand{\arraystretch}{1.2}
\begin{tabular}{|c| c |c| c  |}\hline
Representation  & Multiplicity & Fiber & Locus \\ \hline
$\one_{(0,-)}$ & $6[K^{-1}_B]^2+4[K_B^{-1}](\cS_7+\cS_9)-2\cS_7^2-2\cS_9^2$ & \rule{0pt}{1.35cm}\parbox[c]{2.9cm}{\includegraphics[width=2.9cm]{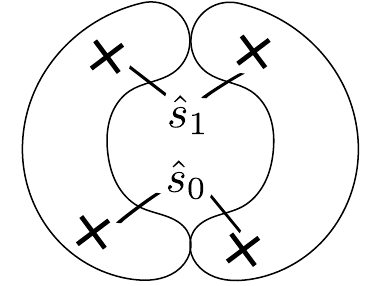}} & $V(I_{(1)})$  \\[0.95cm] \hline

$\one_{(1,-)}$ &  $6[K^{-1}_B]^2+4[K_B^{-1}](\cS_9-\cS_7)+2\cS_7^2-2\cS_9^2$ & \rule{0pt}{1.4cm}\parbox[c]{2.9cm}{\includegraphics[width=2.9cm]{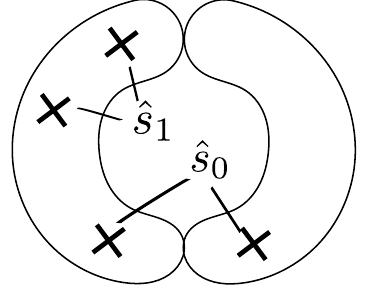}} & $V(I_{(2)})$  \\[0.95cm] \hline

$\one_{(1,+)}$ & $6[K^{-1}_B]^2+4[K_B^{-1}](\cS_7-\cS_9)-2\cS_7^2+2\cS_9^2$& \rule{0pt}{1.35cm}\parbox[c]{2.9cm}{\includegraphics[width=2.9cm]{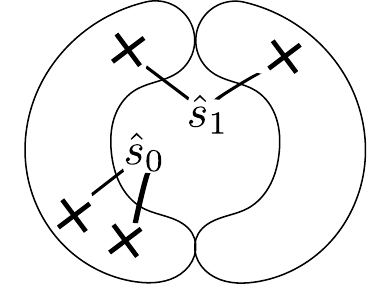}} & $V(I_{(3)})$  \\[0.95cm] \hline

\end{tabular}
\caption{\label{tab:poly2_matter}Charged matter representations under U(1)$\times\mathbb{Z}_2$ and corresponding codimension two fibers of $X_{F_2}$.}
\end{center}
\end{table}

The second type of factorization of \eqref{eq:biquadric-F2} into two
polynomials of degrees $(1,0)$ and $(1,2)$, respectively, takes the following explicit form 
\beq \label{eq:pF2-fact-2}
p_{F_2} \stackrel{!}{=} b_1 \big[ y + \alpha_1 s \big]  \big[ ( y + \beta_1 s) x^2 + (\beta_2 y + \beta_3 s) x t + (\beta_4 y + \beta_5 s) t^2 \big]\,,
\eeq
where $\alpha_1$ and the $\beta_k$ are six unknown polynomials.
We compute again the elimination ideal, denote $I_{(2)}$, that is
generated by eight polynomials in the $b_i$ and check that it is
codimension two in the ring. The corresponding codimension two locus
in $B$ supporting this type of $I_2$-fiber is denoted by $V(I_{(2)})$.
The intersection pattern of the two-sections with the $I_2$-fiber is
shown in the second entry of Table~\ref{tab:poly2_matter}. The U(1)- and
$\mathbb{Z}_2$-charges readily follow as discussed before and we find
the representation at this locus to be $\one_{(1,-)}$.

Finally, the last type of factorization corresponds to a split of
\eqref{eq:biquadric-F2} into two polynomials of degrees $(0,1)$ and
$(2,1)$. It can be written down explicitly and takes a similar form
as  
\eqref{eq:pF2-fact-2}.  The codimension two
elimination ideal corresponding to this factorization, denoted by
$I_{(3)}$, is generated by eight polynomials and its vanishing set is
denoted by $V(I_{(3)})$.  The intersection pattern of the two-sections
with this type of $I_2$-fiber is shown in the last entry of
Table~\ref{tab:poly2_matter}. Using the charge formula
\eqref{eq:U1charge} and the Shioda map \eqref{eq:ShiodaF2}, as well as 
\eqref{eq:ShiodatwoSecF2} we show that
the representation at $V(I_{(3)})$ is $\one_{(1,+)}$.

As a confirmation of the completeness of our analysis of codimension two
singularities of $X_{F_2}$ supporting U(1)-charged matter, we recall
that the codimension two locus supporting all $I_2$-singularities
associated to a U(1) is given by \eqref{eq:charge1Matter}. In the case
at hand we have to evaluate this constraint for the rational sections of
$J(X_2)$ with coordinates $[x_1:y_1:z_1]$ given in
\eqref{eq:F2-section}.
We calculate all associated prime ideals of the obtained
complete intersection using Singular \cite{Singular} and indeed
find  precisely the two prime ideals $I_{(2)}$ and $I_{(3)}$
corresponding to the two representations
$\one_{(1,-)}$ and $\one_{(1,+)}$ found previously using the elimination ideal technique.

As a next step, we calculate the homology classes in $B$ for the three
codimension two loci supporting the $I_2$-fibers, which determine,
according to Section \ref{sec:6DSUGRA}, the multiplicities of 6D charged
hyper multiplets in the corresponding  representations. We begin with
the variety $V(I_{(3)})$, whose multiplicity we denote by
$x_{\one_{(1,+)}}$, supporting the representation $\one_{(1,+)}$. Its
homology class is computed by taking two constraints of the ideal
$I_{(3)}$ and computing the homology class of the complete intersection
described by them. Then, we subtract (with their corresponding orders)
those components that are inside this complete intersection but do not
satisfy the other generators of the ideal $I_{(3)}$. We obtain:
\begin{align}
\begin{split}
 x_{\one_{(1,+)}}=&[b_2^2 b_{10}^2]\cdot[b_{10} b_2 b_5] - 2 ([b_2 b_{10}]\cdot[b_2 b_7]-[b_2]\cdot[b_3]) \,, \\
 =& 6[K^{-1}_B]^2+4[K_B^{-1}](\cS_7-\cS_9)-2\cS_7^2+2\cS_9^2 \,.
\end{split}
\end{align}
The multiplicity of $V(I_{(2)})$, denoted by $x_{\one_{(1,-)}}$, can be
calculated in a similar way. It
is given in the third row of Table~\ref{tab:poly2_matter}. As
a consistency check, we calculate the sum of both multiplicities and it
agrees with $[y_1]\cdot[f z_1^4]$ as it should, because the $I_{(2)}$
and $I_{(3)}$ are the two associated prime ideals of
\eqref{eq:charge1Matter} for the section \eqref{eq:F2-section}.

For the computation of the multiplicity of the variety $V(I_{(1)})$,
denoted by $x_{\one_{(0,-)}}$,
we cannot carry out the previously mentioned algorithm, due to
the size and complexity of the ideal $I_{(1)}$. Instead, $x_{\one_{(0,-)}}$ is obtained
by first calculating the multiplicity of all hyper multiplets charged
under the discrete symmetry, namely the %that are 
$\one_{(0,-)}$ and $\one_{(1,-)}$,
and then subtracting the number of hyper multiplets in the
representation $\one_{(1,-)}$, that we already know.

We  begin by noting that the total number of charge $(-)$ hyper
multiplets under the 
$\mathbb{Z}_2$ symmetry was calculated geometrically
in \cite{Morrison:2014era,Anderson:2014yva} for a genus-one fibration given by the  
quartic curve in $\mathbb{P}^{2}(1,1,2)$. Indeed, we can directly
use their results since we can transform the biquadric
\eqref{eq:biquadric-F2} to a quartic presentation. This quartic is to be
obtained by taking the discriminant of the biquadric
\eqref{eq:biquadric-F2} with respect to $y$.  To this end, we
rewrite the  biquadric in the suggestive form
\begin{equation}
 p = A(x,t) y^2 + B(x,t) y s + C(x,t) s^2 \,
\end{equation}
and then take the discriminant of this quadric in $y$ (we also set
$s=1$). We construct
a genus-one curve as the double cover over this discriminant, which is
then a quartic in $[x:t]$ and a new variable $w$ of weight two of the
form
\begin{equation} \label{eq:F2-quartic}
w^2 = B(x,t)^2-4 A(x,t) C(x,t) \equiv e_0 x^4 + e_1 x^3 t + e_2 x^2 t^2 + e_3 x t^3 + e_4 t^4 \, .
\end{equation}
Here we used the conventions of \cite{Morrison:2014era} in the last
equality. The coefficients $e_i$ can be expressed in terms of
the $b_i$ in \eqref{eq:biquadric-F2} by a comparison of coefficients.

In this form the reason for choosing the quadric w.r.t.~to $y$ in order to
construct the quartic \eqref{eq:F2-quartic} is
evident, because the two-section $\hat{s}^{(2)}_0=X_{F_2}\cap\{x=0\}$
is mapped to the two-section $x=0$, $w^2=e_4 t^4$ in
\eqref{eq:F2-quartic}. Using the results in
\cite{Morrison:2014era,Anderson:2014yva}, we calculate the
multiplicities of all charge $(-)$ hyper multiplets, both $\one_{(0,-)}$,
$\one_{(1,-)}$, using the one-to-one correspondence between
the loci of $I_2$-fibers in \eqref{eq:F2-quartic} with
the following complete intersection, c.f.~equation (2.22) in \cite{Morrison:2014era},
\begin{equation} \label{eq:MorrisonTayloreq}
\{e_1^4-8e_0e_1^2e_2+16 e_0^2 e_2^2-64 e_0^3 e_4 = 0\} \cap \{e_3=0\}\,.
\end{equation}
Its homology class is readily given as $[4e_1]\cdot[e_3]$, which
has to agree as mentioned before with the sum
$x_{\one_{(1,-)}}+x_{\one_{(0,-)}}$. Thus, the multiplicity
$x_{\one_{(0,-)}}$ follows by subtracting the multiplicity
$x_{\one_{(1,-)}}$  calculated previously from  $[4e_1]\cdot[e_3]$.
The result is given in the second row of Table~\ref{tab:poly2_matter}.

To complete the matter spectrum we calculate the number of neutral
hyper multiplets. Using \eqref{eq:Hneutral} and the explicit formula
for the Euler number of $X_{F_2}$ in \eqref{eq:EulerNumbers} of Appendix
\ref{app:Eulernumbers}, we obtain
\begin{align}
H_{\text{neut}} %&= h^{(2,1)}(X_{F_2})+1 \,, \nn\\
&= 13 + 11 [K_B^{-1}]^2 - 4 [K_B^{-1}] \cS_7 + 2 \cS_7^2 - 4 [K_B^{-1}] \cS_9 + 2 \cS_9^2\,.
\end{align}
Using this together with the charged matter spectrum in
Table~\ref{tab:poly2_matter}, 
the number of vector multiplets
$V=1$ and the height pairing
\eqref{eq:heightpairingF2} we confirm that all anomalies, including the purely gravitational one, are canceled.

\subsubsection*{Yukawa couplings in $X_{F_2}$}

We conclude this section by stating the geometrically realized Yukawa
couplings. We find the single Yukawa coupling in
Table~\ref{tab:poly2_yukawa}, by checking explicitly that the
corresponding varieties intersect at
codimension three, i.e.~that the ideal $I_{(1)}\cup I_{(2)}\cup I_{(3)}$
is codimension three in the ring generated by the coefficients $b_i$.
\begin{table}[H]
\begin{center}
\renewcommand{\arraystretch}{1.2}
\begin{tabular}{|c|c|}\hline
Yukawa & Locus \\ \hline
$\one_{(1,-)}\cdot\overline{\one_{(1,+)}}\cdot\one_{(0,-)}$\rule{0pt}{0.5cm} & $V(I_{(1)})\cap
 V(I_{(2)})\cap V(I_{(3)})$  \\[0.1cm] \hline
\end{tabular}
\caption{\label{tab:poly2_yukawa}Codimension three locus and corresponding Yukawa coupling for $X_{F_2}$. }
\end{center}
\end{table}

%%%%%%%%%%%%%%%%%%%%%%%%%%%%%%%%%%%%%%%%%%%%%%%%%%%%%%%%%%%%%%%%%%%%

%%%%%%%%%%%%%%%%%%%%%%%%%%%%%%%%%%%%%%%%%%%%%%%%%%%%%%%%%%%%%%%%%%%%
\subsubsection{Polyhedron $F_{4}$: $G_{F_4}=\text{SU(2)} \times \mathbb{Z}_4$}
%%%%%%%%%%%%%%%%%%%%%%%%%%%%%%%%%%%%%%%%%%%%%%%%%%%%%%%%%%%%%%%%%%%%
\label{sec:F4_poly}

In this section we study the genus one-fibration $X_{F_4}$ that
is constructed as the toric hypersurface fibration of
$\mathcal{C}_{F_4}$ in $\mathbb{P}_{F_4}=\mathbb{P}^{2}(1,1,2)$.
The toric data of $\mathbb{P}^{2}(1,1,2)$ as well as the construction
of the toric hypersurface fibration $X_{F_4}$ have been discussed in
Section \ref{sec:quartic}. The hypersurface constraint
of $X_{F_4}$ is shown in \eqref{eq:quartic_hypersurface} and the
relevant divisor classes can be found in  \eqref{eq:quarticsections}.
This model has recently received a lot of attention
\cite{Braun:2014oya,Morrison:2014era,Anderson:2014yva}. Here we provide
additional insights in
the nature of  the $\mathbb{Z}_4$
discrete gauge group of F-theory on $X_{F_4}$ as well as in the
computation of the charges of matter under this discrete group. We
also check 6D anomaly cancellation, which requires knowledge of all
multiplicities of 6D charged and uncharged hyper multiplets.

We begin by noting that the fibration $\pi:X_{F_4}\rightarrow B$
does not have a section, but only two- and four-sections, i.e.~$X_{F_4}$
is a genus-one fibration, see Section \ref{sec:ellipticCurvesWithRP}.
The three multi-sections induced by the ambient space $\mathbb{P}^{2}(1,1,2)$ of the fiber are
\begin{align}
\begin{split}\label{eq:secsF4}
\hat{s}_1^{(2)}=X_{F_4} \cap \{X=0\}&:\quad   d_5 e_1^2 + d_8 e_1  Z + d_9 Z^2=0\,,\\
\hat{s}_2^{(2)}=X_{F_4} \cap \{Y=0\}&:\quad d_1 e_1^2  + d_6 e_1  Z + d_9 Z^2=0 \,,\\
\hat{s}_3^{(4)}=X_{F_4} \cap \{Z=0\}&:\quad d_1  X^4 + d_2  X^3 Y + d_3  X^2 Y^2 + d_4  X Y^3 + d_5  Y^4 =0\,,
\end{split}
\end{align}
where we used the SR-ideal \eqref{eq:SRF4}. We denote the one
independent divisor classes of $\hat{s}_1^{(2)}$ by $S_1^{(2)}$. It
agrees with $H-E_1+\mathcal{S}_9-[K_B^{-1}]$ according to
\eqref{eq:quarticsections}, where $E_1$ is the class of the exceptional
divisor on $\mathbb{P}^2(1,1,2)$.

\begin{figure}[t!]
\center
\includegraphics[scale=0.6]{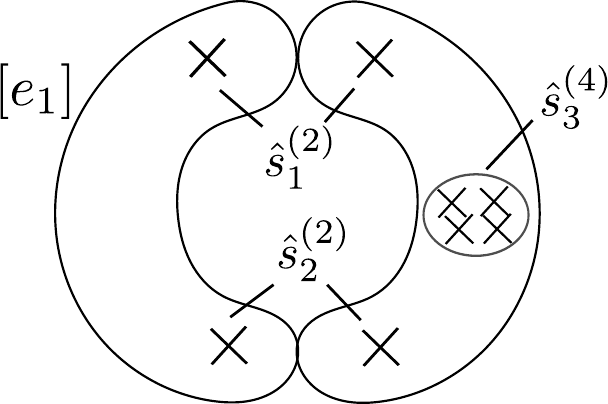}
\caption{\label{fig:poly4_codim1}Codimension one fibers of $X_{F_{4}}$. The crosses denote the intersections with the sections.}
\end{figure}

Since this fibration does not have a  section one has to utilize its
associated Jacobian fibration $J(X_{F_4})$ in order to find its WSF.
We readily compute the functions $f$ and $g$ in \eqref{eq:WSF}
using the algorithm in \cite{an2001jacobians}.
From the discriminant of this WSF, we find one $I_2$-singularity over
the divisors $\mathcal{S}^b_{\text{SU}(2)}=\{d_9=0\}\cap B$ in $B$.
Along this divisor, the constraint \eqref{eq:quartic_hypersurface}
factorizes as
\begin{align}
\SU2&:\quad p_{F_{4}}|_{d_9=0}= e_1 \cdot q_3 \, ,
\end{align}
where $q_3$ is the polynomial that remains after factoring out $e_1$.
The corresponding $I_2$-fiber is depicted in
Figure~\ref{fig:poly4_codim1}.
Due to the absence of a zero section, there is no preferred rational curve in this $I_2$-fiber. Thus, a possible choice for the
Cartan-divisor $D_1$  of the SU(2) is given by
\beq \label{eq:CartanSU2F4}
	D_1=[e_1]\,.
\eeq
In summary, the  gauge group of $X_{F_{4}}$ is given by
\begin{align} \label{eq:GF4}
	G_{F_{4}}=\text{SU}(2) \times \mathbb{Z}_4\,.
\end{align}

The discrete symmetry stems from the multi-sections. In order to
calculate the charges under the discrete symmetry we have to
orthogonalize the SU(2) such that it is not charged under the discrete
symmetry. This is done in a similar way as for U(1)'s via the Shioda map
\eqref{eq:ShiodaMap}. For the two two-sections $\hat{s}_m$ with $m=1,2$,
we propose
\beq \label{eq:ShiodatwoSecF4}
\sigma_{\mathbb{Z}_4} (\hat{s}_m^{(2)}) = S_m^{(2)} + \tfrac{1}{2} D_1+\tfrac{3}{4}[K_B^{-1}]-\tfrac{3}{4}\mathcal{S}_7 -
\tfrac{1}{4}\mathcal{S}_9 \, ,
\eeq
as the appropriate divisor class to compute charges under the discrete
gauge group $\mathbb{Z}_4$. Here we used that both two-sections intersect
each node in Figure \ref{fig:poly4_codim1} precisely once. For the
four-section $\hat{s}^{(4)}_3$ we note that the node corresponding to
the simple root of SU(2) in Figure \ref{fig:poly4_codim1} is not
intersected. Thus, the appropriate class for computing $\mathbb{Z}_4$
charges based on $\hat{s}^{(4)}_3$ is
\beq\label{eq:ShiodaFourSecF4}
	\sigma_{\mathbb{Z}_4}(\hat{s}_3^{(4)})=S_3^{(4)}+\tfrac{1}{2}[K_B^{-1}]-\tfrac32 \mathcal{S}_7+\tfrac12 \mathcal{S}_9\,.
\eeq
It is straightforward to check that the divisors
\eqref{eq:ShiodatwoSecF4} and \eqref{eq:ShiodaFourSecF4} obey all
properties of a Shioda map \cite{Park:2011ji}.

Next, we analyze the codimension two singularities of the WSF of
$J(X_{F_{4}})$ to determine the charged matter spectrum.
We find two codimension two
singularities leading to the matter
representations and the corresponding codimension two fibers in
$X_{F_{4}}$ that are given in the first and second entry
of Table~\ref{tab:poly4_matter}, respectively. We have also added the adjoint matter at the divisor
$\mathcal{S}_{\text{SU}(2)}^b=\mathcal{S}_9$ for completeness.
We have checked the representation content at the two codimension two
loci explicitly by computation of the Dynkin labels using
\eqref{eq:DynkinLabel} with $D_1$ given in \eqref{eq:CartanSU2F4}
and using the charge formula \eqref{eq:U1charge} for
\eqref{eq:ShiodatwoSecF4} or \eqref{eq:ShiodaFourSecF4}. We note that
the charges calculated from the two two-sections are half the  integral
charges computed from the four-section. However, both charges are
physically equivalent since in the case of the two-section we have to calculate
modulo two, whereas for the four-section, we calculate modulo four. In
other words, we obtain charges in two different conventions. Here we
choose the charge convention where all discrete charges are integer,
which agrees with the charges computed using  $\sigma_{\mathbb{Z}_4}(\hat{s}_3^{(4)})$.
\begin{table}[htb!]
\begin{center}
\renewcommand{\arraystretch}{1.2}
\begin{tabular}{|c|c|c|c|}\hline
Representation & Multiplicity & Fiber & Locus \\ \hline
$\two_1$ & $\begin{array}{c} (6 [K_B^{-1}] + 2\cS_7 - 2\cS_9) \\ \times([K_B^{-1}] - \cS_7 + \cS_9) \end{array}$ & \rule{0pt}{2.0cm}\parbox[c]{2.9cm}{\includegraphics[width=2.9cm]{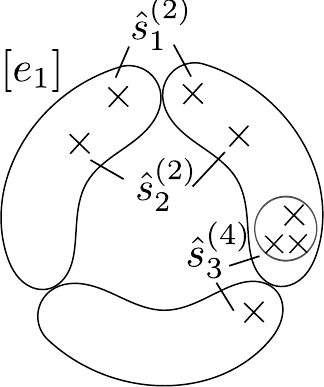}} & $V(I_{(1)})$, given by \eqref{eq:F4locus1} \\[1.6cm] \hline
$\one_2$ & $\begin{array}{c} 6 [K_B^{-1}]^2 + 13 [K_B^{-1}] \cS_7 - 3 \cS_7^2  \\ - 5 [K_B^{-1}] \cS_9 - 2 \cS_7 \cS_9 + \cS_9^2 \end{array}$ & \rule{0pt}{2cm}\parbox[c]{2.9cm}{\includegraphics[width=2.9cm]{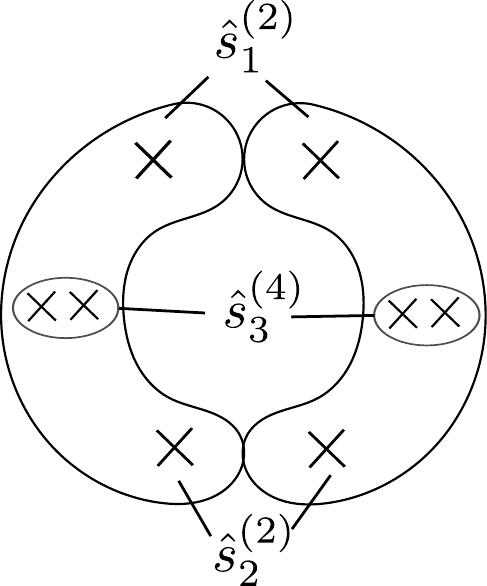}} & $V(I_{(2)})$, given by \eqref{eq:F4locus2} \\[1.6cm] \hline \hline
$\three$ & \rule{0pt}{0.45cm}$1+([K_B^{-1}]-\cS_7+\cS_9)\frac{\cS_9-\cS_7}{2}$ & Figure~\ref{fig:poly4_codim1} & $d_9=0$ \\ \hline
\end{tabular}
\caption{\label{tab:poly4_matter}Charged matter representations under SU$(2)\times \mathbb{Z}_4$ and corresponding codimension two fibers of $X_{F_{4}}$. The adjoint matter is included for completeness.}
\end{center}
\end{table}

The codimension two locus supporting the representation $\two_1$ is
given as the following complete intersection, that can
be read off directly from the discriminant of $J(X_{F_4})$:
\begin{align}
\begin{split}\label{eq:F4locus1}
I_{(1)}&:= \{ d_9, d_5^2 d_6^4 - d_4 d_5 d_6^3 d_7 + d_3 d_5 d_6^2 d_7^2 - d_2 d_5 d_6 d_7^3 + d_1 d_5 d_7^4 + d_4^2 d_6^3 d_8 - 2 d_3 d_5 d_6^3 d_8 \\
  &\phantom{=\{} - d_3 d_4 d_6^2 d_7 d_8 + 3 d_2 d_5 d_6^2 d_7 d_8 + d_2 d_4 d_6 d_7^2 d_8 - 4 d_1 d_5 d_6 d_7^2 d_8 - d_1 d_4 d_7^3 d_8 + d_3^2 d_6^2 d_8^2  \\
  &\phantom{=\{} - 2 d_2 d_4 d_6^2 d_8^2+ 2 d_1 d_5 d_6^2 d_8^2 - d_2 d_3 d_6 d_7 d_8^2 + 3 d_1 d_4 d_6 d_7 d_8^2 + d_1 d_3 d_7^2 d_8^2 + d_2^2 d_6 d_8^3 \\
  &\phantom{=\{} - 2 d_1 d_3 d_6 d_8^3- d_1 d_2 d_7 d_8^3 + d_1^2 d_8^4\}\,. \\
\end{split}
\end{align}
This ideal is easily checked to be prime. The discrete charge of the $\SU2$ doublet can be 
computed from the intersection of $\sigma_{\mathbb{Z}_4}(\hat{s}_3^{(4)})$ with the 
irreducible fiber components, which are depicted in Table \ref{tab:poly4_matter}. There one 
observes that the four-section intersects one node three times and another one just once. Recall 
again that 6D hyper multiplets with discrete charges $1$ and $3$ are physically
equivalent since a hyper multiplet contains two half-hypers with charges
$1$ and $-1$ and charge $-1$ is identified with charge $3$ under the
discrete $\mathbb{Z}_4$-symmetry.

The complete intersection
supporting the representation $\one_2$ is obtained by directly
searching for the loci of degeneration of the fiber of $X_{F_4}$ to
$I_2$. One makes a general factorization ansatz of
\eqref{eq:quartic_hypersurface} and determines the ideal of constraints
imposed by this factorization \cite{Morrison:2014era,Anderson:2014yva}.
Then one eliminates the unknown
variables introduced in this ansatz by computation of the elimination
ideal, see Sections \ref{sec:F1_poly} and \ref{sec:F2_poly} for an
explanation of this technique. The obtained elimination ideal,
denoted by $I_{(2)}$, is prime and codimension two in the ring of
coefficients $d_i$. It reads
\begin{align}
\begin{split}\label{eq:F4locus2}
I_{(2)}&:= \{ (d_8^2 - 4 d_5 d_9)^3 d_1 - (d_5 d_6 d_7^2 d_8^3 - d_5 d_6^2 d_8^4 - d_4 d_6 d_7 d_8^4 + d_3 d_6 d_8^5\! - d_5^2 d_7^4 d_9 \\
  &\phantom{=\{} - d_3^2 d_8^4 d_9 + 2 d_4 d_5 d_7^3 d_8 d_9 + 8 d_5^2 d_6^2 d_8^2 d_9 + 4 d_4 d_5 d_6 d_7 d_8^2 d_9 - d_4^2 d_7^2 d_8^2 d_9 - 2 d_3 d_5 d_7^2 d_8^2 d_9 \\
  &\phantom{=\{} + d_4^2 d_6 d_8^3 d_9 -d_4^4 d_9^3 -\!8 d_3 d_5 d_6 d_8^3 d_9 + 2 d_3 d_4 d_7 d_8^3 d_9  \!-16 d_5^3 d_6^2 d_9^2\! - 2 d_4^2 d_5 d_7^2 d_9^2 \!\\
  &\phantom{=\{} + 8 d_3 d_5^2 d_7^2 d_9^2 - 4 d_4^2 d_5 d_6 d_8 d_9^2+ 8 d_3 d_4^2 d_5 d_9^3 + 16 d_3 d_5^2 d_6 d_8 d_9^2 + 2 d_4^3 d_7 d_8 d_9^2 \\
  &\phantom{=\{} - 8 d_3 d_4 d_5 d_7 d_8 d_9^2 - 2 d_3 d_4^2 d_8^2 d_9^2 + 8 d_3^2 d_5 d_8^2 d_9^2- 16 d_3^2 d_5^2 d_9^3 - 4 d_5^2 d_6 d_7^2 d_8 d_9\!), \\ 
  &\phantom{=\}}\,\,\,\, (d_8^2 - 4 d_5 d_9)^2 d_2 - (d_5 d_7^3 d_8 - 2 d_5 d_6 d_7 d_8^2 - d_4 d_7^2 d_8^2 + d_4 d_6 d_8^3 + d_3 d_7 d_8^3 - 2 d_4^3 d_9^2 \\
  &\phantom{=\{} + 8 d_5^2 d_6 d_7 d_9- 2 d_4 d_5 d_7^2 d_9 - 4 d_4 d_5 d_6 d_8 d_9 + 3 d_4^2 d_7 d_8 d_9 - 4 d_3 d_5 d_7 d_8 d_9 \\
  &\phantom{=\{}- 2 d_3 d_4 d_8^2 d_9  +8 d_3 d_4 d_5 d_9^2)\}: I_\text{rest}^7 \, ,
\end{split}
\end{align}
where we have indicated the quotient by the ideal $I_\text{rest}:=\{ 2 d_4 d_9-d_7d_8,4 d_5 d_9 -d_8^2 \}$ by ``$\, :\,$''.

The multiplicity of 6D hyper multiplets
in the representation $\two_1$ is computed as the product of the
classes of the two constraints in the complete intersection $I_{(1)}$.
The  multiplicity of matter in the $\one_2$ representation is more
involved since the locus described by $I_{(2)}$, $V(I_{(2)})$,  is
one of three irreducible components of the complete intersection in the first ideal in
\eqref{eq:F4locus2}. Using the resultant technique \cite{Cvetic:2013nia} we decompose this 
complete intersection into $V(I_{(2)})+V(\{s_8,s_9\})+8\cdot V(\{ 2 d_4 d_9-d_7d_8,4 d_5 d_9 -d_8^2,d_4d_8-2d_5d_7 \}$.
This allows us to obtain the multiplicity shown in Table~\ref{tab:poly4_matter} that we double 
check following the arguments in \cite{Morrison:2014era} explained around
\eqref{eq:MorrisonTayloreq}.

We complete the matter spectrum of $X_{F_{4}}$ by the number of neutral
hyper multiplets, which is computed from \eqref{eq:Hneutral} using
the Euler number of $X_{F_4}$ in \eqref{eq:EulerNumbers}. It reads
\beq \label{eq:HneutF4}
H_{\text{neut}} = 13 + 11 [K_B^{-1}]^2 - 4 [K_B^{-1}] \cS_7 + 6 \cS_7^2 - 4 [K_B^{-1}] \cS_9 - 4 \cS_7 \cS_9 + 2 \cS_9^2\,.
\eeq
Finally, we use $S_{\text{SU}(2)}^b=\{d_9=0\}$,
the charged spectrum in Table~\ref{tab:poly4_matter} and
\eqref{eq:HneutF4} together with the number of vector multiplets
$V=1$ to check that all 6D
anomalies \eqref{eq:6dAnomalies} are canceled.

We conclude this section with the computation of the Yukawa
couplings. We find the single Yukawa coupling given in
Table~\ref{tab:poly4_yukawa}. In order to check that it is realized
at codimension three 
in $B$ we compute the associated prime ideals
of the ideal $I_{(1)}\cup I_{(2)}$. Indeed, we find that it is
codimension three in the ring,
as required for the existence of the Yukawa coupling.
\begin{table}[H]
\begin{center}
\renewcommand{\arraystretch}{1.2}
\begin{tabular}{|c|c|}\hline
Yukawa & Locus \\ \hline
$\two_1\cdot \two_1\cdot \overline{\one_{2}}$ & $V(I_{(1)})\cap V(I_{(1)})\cap V(I_{(2)}) $ \\ \hline
\end{tabular}
\caption{\label{tab:poly4_yukawa}Codimension three locus and corresponding Yukawa coupling for $X_{F_4}$. }
\end{center}
\end{table}

%%%%%%%%%%%%%%%%%%%%%%%%%%%%%%%%%%%%%%%%%%%%%%%%%%%%%%%%%%%%%%%%%%%%%%%%%%%%%%%%%%%%%%%%%%%%%%%%%
\subsection{Fibration with gauge groups of rank 1, 2 and no discrete gauge symmetry}
%%%%%%%%%%%%%%%%%%%%%%%%%%%%%%%%%%%%%%%%%%%%%%%%%%%%%%%%%%%%%%%%%%%%%%%%%%%%%%%%%%%%%%%%%%%%%%%%%
\label{sec:Rk12Nodisc}

In this section we analyze all toric hypersurface fibrations $X_{F_i}$ 
with gauge groups of rank one and two, but without discrete gauge 
symmetries. They are constructed using the fiber polyhedra $F_3$, $F_5$ 
and $F_6$. Apart from $X_{F_3}$, that possesses a non-toric section, all 
other $X_{F_i}$ considered here can be analyzed using techniques already 
developed e.g.~in \cite{Cvetic:2013nia,Cvetic:2013uta,Cvetic:2013jta}.
%%%%%%%%%%%%%%%%%%%%%%%%%%%%%%%%%%%%%%%%%%%%%%%%%%%%%%%%%%%%%%%%%%%%

%%%%%%%%%%%%%%%%%%%%%%%%%%%%%%%%%%%%%%%%%%%%%%%%%%%%%%%%%%%%%%%%%%%%
\subsubsection{Polyhedron $F_{3}$: $G_{F_3}=\text{U(1)}$}
%%%%%%%%%%%%%%%%%%%%%%%%%%%%%%%%%%%%%%%%%%%%%%%%%%%%%%%%%%%%%%%%%%%%
\label{sec:F3_poly}

\begin{figure}[ht]
\centering
\includegraphics[scale=0.4]{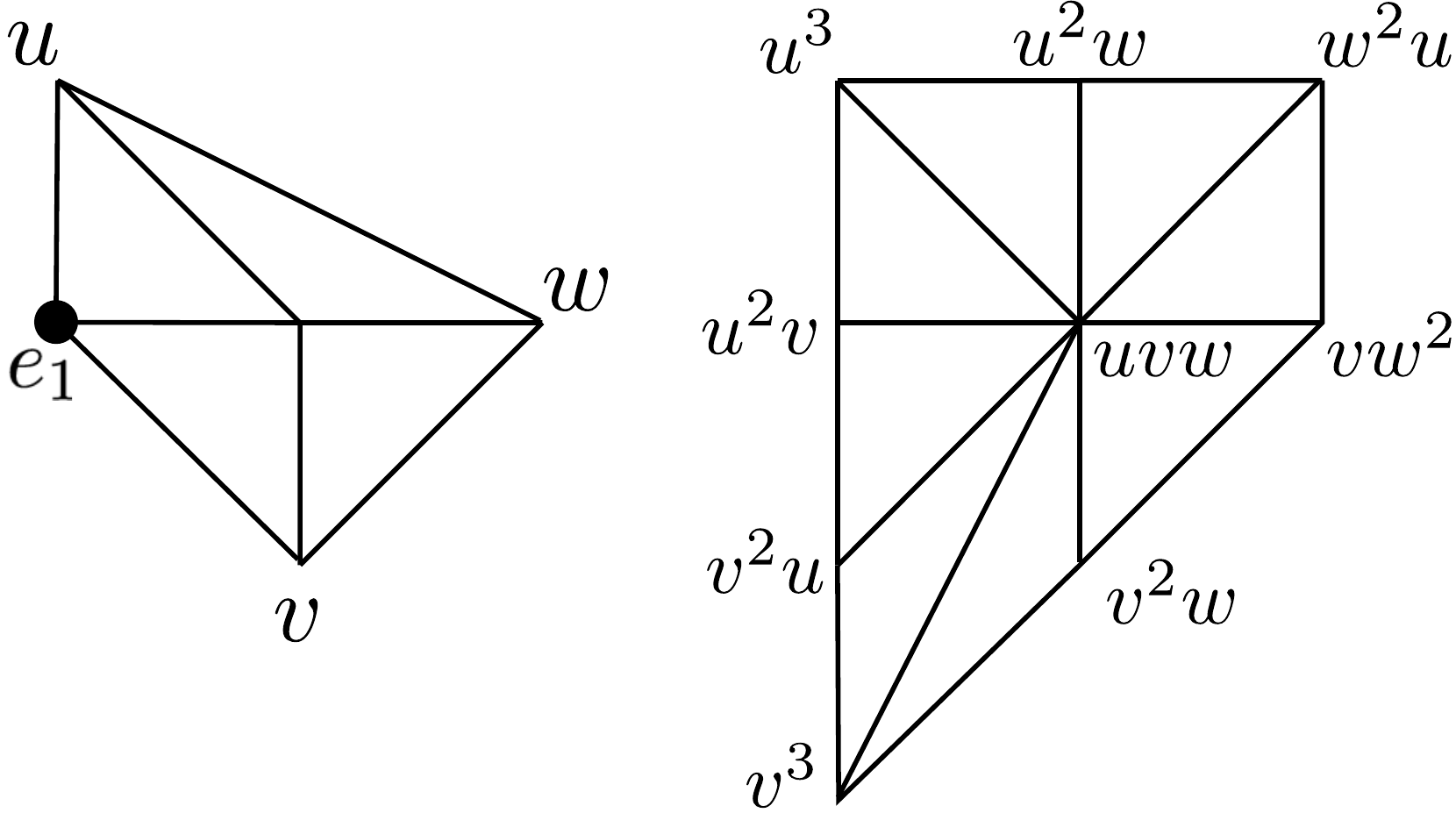}
\caption{Polyhedron $F_3$ with a choice of projective coordinates and its dual $F_{14}$ with the corresponding monomials. We have
set $e_1=1$ for brevity of our notation. The zero section is indicated by the dot.}
\label{fig:F3}
\end{figure}

We construct a Calabi-Yau manifold, denoted $X_{F_3}$, as a fibration of the toric
hypersurface in $\mathbb{P}_{F_3}=dP_1$ over a base $B$.
The polyhedron of $F_3$ along with a choice of projective coordinates
as well as its dual polyhedron 
are depicted in Figure~\ref{fig:F3}.
The coordinate $e_1$ vanishes on the exceptional divisor $E_1$
of $dP_1$ and $[u:v:w]$ are the pullback under the blow-down map $dP_1\rightarrow \mathbb{P}^2$ of the $\mathbb{P}^2$-coordinates.
The SR-ideal of $dP_1$ reads
\beq \label{eq:SRF3}
	SR_{F_3}=\{uv,w e_1 \}\,.
\eeq

Using \eqref{eq:BatyrevFormula} we construct the Calabi-Yau manifold $X_{F_3}$ as
the hypersurface
\beq \label{eq:PF3}
	p_{F_3}=s_1 u^3e_1^2 + s_2 u^2 ve_1^2  + s_3 u v^2 e_1^2  +  s_4 v^3e_1^2 +
  s_5 u^2 w e_1
+  s_6 u v w e_1 +  s_7 v^2 w e_1 + s_8 u w^2 + s_9 v w^2 \, ,
\eeq
in the ambient space \eqref{eq:PFfibration}, that in the case at hand is
a $dP_1$-fibration over $B$. The coordinates $[u:v:w:e_1]$ and
the coefficients $s_i$ take values in the line bundles in
\eqref{eq:cubicsections}.

The Calabi-Yau manifold $X_{F_3}$ is an elliptic fibration. This is
clear because
for a generic point on $B$ there is one marked point $P_0$ on its fiber,
which is the intersection of $e_1=0$ with  \eqref{eq:PF3}. This point
gives rise to a section of $X_{F_3}$, which we choose as the zero
section. Its generic coordinates read
\beq \label{eq:s0F3}
   \hat{s}_0= X_{F_3}\cap\{e_1=0\}:\,\,\, [s_9:-s_8:1:0]\,.
\eeq
There always exists a second section of $X_{F_3}$, which generates a rank one MW-group.\footnote{We note that this is not in contradiction with the results of \cite{Braun:2013nqa}. There
the toric Mordell-Weil group is computed, which is indeed trivial.} We emphasize that this second section is
not toric, i.e.~not given as the intersection of
a toric divisor in the fiber $\mathbb{P}_{F_3}$ with the
hypersurface \eqref{eq:PF3}, in contrast to the zero section \eqref{eq:s0F3}.

This can be seen as follows.
Without loss of generality, set $e_1=1$ in \eqref{eq:PF3} and consider
it as an elliptic curve $\mathcal{E}$ over a field $K$. Then construct
the tangent $t_P$ to the point $P_0$ which now is at $[u:v:w]=[0:0:1]$. It
is determined by requiring that along $t_P$  both $p_{F_3}$ and
its first derivative vanish at $P_0$, i.e.~that
$P_0$ is a point of intersection two of $\mathcal{E}$ and
$t_P$. It is described by
\beq \label{eq:tangentF3}
	t_P=s_8u+s_9v\,.
\eeq
Since \eqref{eq:PF3} is a curve of degree three, every line has to
intersect it at three points. Thus,  $t_P=0$
intersects $\mathcal{E}$ at a third point, denoted by $P_1$, which is
automatically rational. It gives rise
to a \emph{rational section} of $X_{F_3}$, with generic coordinates
\beq\label{eq:Q1F3}
	\hat{s}_1=X_{F_3}\cap\{t_P=0\} : \,\, [-s_9: s_8:s_1 s_9^3-s_4 s_8^3+s_3 s_9 s_8^2-s_2 s_9^2 s_8:s_7 s_8^2-s_6 s_9 s_8+s_5 s_9^2]\,.
\eeq
Thus, the elliptic fibration $X_{F_3}$ indeed has a rank one MW-group
with a non-toric generator, as claimed.
The Shioda map \eqref{eq:ShiodaMap} of the section $\hat{s}_1$ reads
\beq \label{eq:ShiodaF3}
	\sigma(\hat{s}_1)=S_1-S_0+3[K_B]+\cS_7-2\cS_9\,, 
\eeq
where $S_1$, $S_0$ are the divisor classes of the rational sections $\hat{s}_1$ and $\hat{s}_0$.

This result allows us to compute the height pairing of the section
$\hat{s}_1$. We obtain
\beq \label{eq:heightF3}
	 b_{11}=-2(3[K_B]+\cS_7-2\cS_9)\, ,
\eeq
where we employed \eqref{eq:anomalycoeff} along with the self-intersection
\eqref{eq:SP^2} for the section $\hat{s}_1$ as well as
\beq
	\pi(S_1\cdot S_0)=[z_1]=2[K_B^{-1}]+2\cS_9-\cS_7\,.
\eeq
This follows by noting that  $\pi(S_0\cdot S_1)$ is the  
locus in $B$ where the coordinates \eqref{eq:s0F3} and \eqref{eq:Q1F3}
of the two sections agree, which happens at
$z_1:=s_7 s_8^2 - s_6 s_8 s_9 + s_5 s_9^2=0$, that is precisely
the $z$-coordinate of $\hat{s}_1$ in the WSF, cf.~\eqref{eq:WSFQ1F3}. The
divisor class of $z_1$ is read off from \eqref{eq:cubicsections}.

\subsubsection*{Weierstrass form and gauge group}

We can apply Nagell's algorithm to the cubic \eqref{eq:PF3} with
respect to the point $P_0$ to obtain a birational map to its WSF.
We plug the  coordinates of the rational section \eqref{eq:Q1F3} into
this map to obtain its coordinates in  WSF,
\beq\label{eq:WSFQ1F3}
	z_{1}=s_7 s_8^2 - s_6 s_8 s_9 + s_5 s_9^2\,,\quad x_{1}= s_4^2 s_8^6+\ldots=p_8(\underline{s})\,,\quad y_{1}=- s_4^3 s_8^9+\ldots=p_{12}(\underline{s})\,.
\eeq
Here $p_8(\underline{s})$ and $p_{12}(\underline{s})$ are
two homogeneous polynomials in the coefficients $s_i$
of degree eight and twelve, respectively.  We have written out only
one monomial in $x_{Q_1}$ and $y_{Q_1}$, respectively, in order to
be able to determine their divisor classes. We refer the reader to
\eqref{eq:WScoordsSecF3} in Appendix~\ref{app:F3} for the explicit and
lengthy expressions for $p_8(\underline{s})$ and $p_{12}(\underline{s})$.

Furthermore, we determine the functions $f$, $g$ and the
discriminant $\Delta$ of the WSF for $X_{F_3}$. They are given by
specializing \eqref{eq:fcubic} as  $s_{10}=0$. We observe that
there is no factorization of $\Delta$ indicating the absence of
codimension one singularities and a non-Abelian gauge group.
Thus, the full gauge group on $X_{F_3}$ is given by the single
U(1) associated to its rank one MW-group,
\beq \label{eq:GF3}
	G_{F_3}=\text{U}(1)\,.
\eeq
We emphasize again that the generator \eqref{eq:Q1F3} of the
MW-group of $X_{F_3}$ is not toric.

\subsubsection*{Charged and uncharged matter}

Since the Calabi-Yau manifold $X_{F_3}$ has a non-trivial MW-group,
it automatically has $I_2$-fibers at codimension two in $B$, that
support U(1)-charged matter.

We first summarize the charged matter spectrum of $X_{F_3}$ before we
discuss its derivation in detail. The full charged matter spectrum is shown
in Table~\ref{tab:poly3_matter}, which includes the U(1)-charges and the 
multiplicities of 6D charged
hyper multiplets, 
as well as a schematic presentation of the corresponding reducible fibers and the
full expressions for the  codimension two loci. 
\begin{table}[ht!]
\begin{center}
\footnotesize
\renewcommand{\arraystretch}{1.2}
\begin{tabular}{|c|@{}c@{}|p{3cm}@{}|@{}c@{}|}
\hline
\!\! Representation \!\!\!& Multiplicity & \centering \hspace{-0.3cm}Fiber & Locus  \\ \hline
$\one_{3}$ \!\!\!& $\cS_9 ([K_B^{-1}]+\cS_9-\cS_7)$ \!& \rule{0pt}{1.2cm}\parbox[c]{1.8cm}{\includegraphics[width=2.8cm]{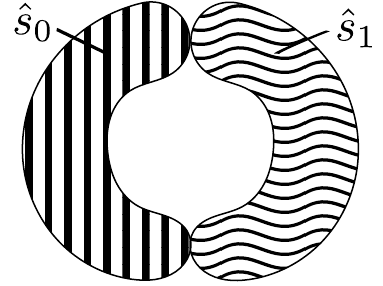}} & $V(I_{(3)}):=\{s_8 = s_9 = 0\}$   \\[0.9cm] \hline

$\one_{2}$ \!\!\!&\!\!  $\begin{array}{c}
 6 [K_B^{-1}]^2 \!+\![K_B^{-1}](4\cS_9\!-\!5  \cS_7 )\\
 + \cS_7^2  + 2 \cS_7 \cS_9 - 2 \cS_9^2
\end{array} \!\! $ & \rule{0pt}{1.2cm}\parbox[c]{1.8cm}{\includegraphics[width=2.8cm]{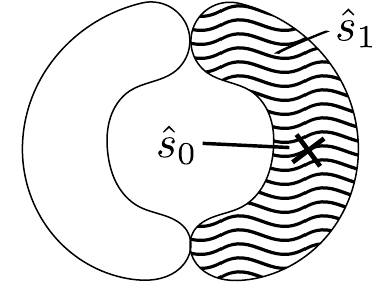}
}
 &\!\!  $\begin{array}{c}
 V(I_{(2)}):=\\
 \{s_4 s_8^3\!-\! s_3 s_8^2 s_9 \!+\! s_2 s_8 s_9^2\! -\! s_1 s_9^3\\
 =s_7 s_8^2  + s_5 s_9^2\!-\! s_6 s_8 s_9 =0\\
 \text{with   } (s_8,s_9)\neq (0,0)\}
\end{array}  $\!\!\!\!
 \\[0.9cm] \hline
$\one_{1}$  \!\!\!&\!\!\!\!
$\begin{array}{c}
 12 [K_B^{-1}]^2 +  [K_B^{-1}](8 \cS_7\!-\! \cS_9)\\
 - 4 \cS_7^2  + \cS_7 \cS_9 - \cS_9^2
\end{array}  $\!\! & \rule{0pt}{1.2cm}\parbox[c]{1.8cm}{\includegraphics[width=2.8cm]{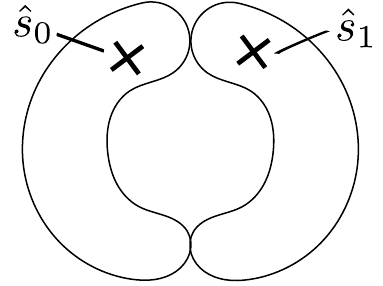}
} &\!\!\!\! $\begin{array}{c}
V(I_{(1)}):=\{\eqref{eq:F3charge1}\}\backslash\ (V(I_{(2)})\cup V(I_{(3)}))
\end{array}$\!\\[0.9cm] \hline
\end{tabular}
\caption{\label{tab:poly3_matter} Charged matter representation under U(1) and
codimension two fibers of $X_{F_3}$.}
\end{center}
\end{table}

The starting point for the derivation of the matter spectrum of
$X_{F_3}$ is, as discussed in
Section~\ref{sec:6DSUGRA}, the complete intersection
\eqref{eq:charge1Matter} in $B$ with the WS-coordinates
\eqref{eq:Q1F3} of the section $\hat{s}_1$ inserted:
\beq \label{eq:F3charge1}
	y_{1}=f z_1^4+3 x_1^2=0\,.
\eeq
We show that \eqref{eq:F3charge1} is a reducible variety with three
irreducible components supporting matter with charges one, two and
three. The corresponding codimension two loci are denoted
$V(I_{(1)})$, $V(I_{(2)})$ and $V(I_{(3)})$, respectively, with $I_{(1)}$, $I_{(2)}$ and $I_{(3)}$ denoting the corresponding
prime ideals, cf.~Table~\ref{tab:poly3_matter}.
In order to strictly prove that these three varieties  are all irreducible components of the complete intersection \eqref{eq:F3charge1}, we
have to compute all its associated prime ideals. Unfortunately, this is unfeasible  with the
available computing power and computer algebra programs, due to the high degree of the two polynomials in
\eqref{eq:F3charge1}. However, we explain that $X_{F_3}$ has three possible types of
$I_2$-fibers corresponding to the three possible factorizations of \eqref{eq:PF3} and that these factorization
happen precisely at the codimension two loci $V(I_{{(1)}})$, $V(I_{{(2)}})$ and $V(I_{{(3)}})$.
Thus, we claim that the corresponding ideals $I_{(1)}$, $I_{(2)}$ and $I_{(3)}$
are the only associated prime ideals of \eqref{eq:F3charge1}. We will
further substantiate this claim by checking 6D anomaly cancellation at the end of this section
as well as by reproducing the spectrum of $X_{F_3}$ by Higgsing
the effective theories of $X_{F_5}$ and $X_{F_6}$, see Section~\ref{sec:Higgsings}.

We begin by analyzing the fiber at the first two codimension two loci in Table~\ref{tab:poly3_matter}.
These are precisely the loci where the coordinates \eqref{eq:Q1F3} of the section $\hat{s}_1$
are ill-defined, since they are forbidden by the SR-ideal
\eqref{eq:SRF3}. This indicates, that the section $\hat{s}_{1}$
does not mark a point on the elliptic fiber of $X_{F_3}$, but
does wrap an entire $\mathbb{P}^1$. Since the rational section
is non-toric, determining the wrapped $\mathbb{P}^1$ is slightly more
involved than usual, as we demonstrate next.

First, we consider the locus $V(I_{(3)})=\{s_8=s_9=0\}$, which we
readily check to obey \eqref{eq:F3charge1}. At this locus the constraint \eqref{eq:PF3} factorizes as
\beq \label{eq:PF3ats8=s9=0}
p_{F_3}\big\vert_{s_8=s_9=0}=e_1 ( s_1 u^3e_1 +  s_2 u^2 v e_1 +  s_3 u v^2e_1 +  s_4 v^3e_1 + s_5 u^2 w +
   s_6 u v w + s_7 v^2 w)\,.
\eeq
Clearly, $V(I_{(3)})$ is the only codimension two locus where this
factorization can occur.
We immediately observe that the zero section $\hat{s}_0$ defined by
\eqref{eq:s0F3}
has wrapped 
the entire rational curve $e_1=0$ in
\eqref{eq:PF3ats8=s9=0}. The rational section $\hat{s}_1$ can be
identified at this locus by recalling the definition of the point
\eqref{eq:Q1F3} as the second
intersection point of the tangent to $P_0$ with $\mathcal{E}$. However, at
$s_8=s_9=0$ the curve \eqref{eq:PF3ats8=s9=0} is singular (after setting
$e_1=1$) precisely at $P_0$. Thus, \textit{every} line through $P_0$ is
automatically tangential at $P_0$. This simply
means that $P_1$ has become the \textit{entire} singular fiber at
$s_8=s_9=0$, since given any point on
\eqref{eq:PF3ats8=s9=0} (for $e_1=1$) we can construct a tangent at $P_0$
that passes through that point. Thus, at $s_8=s_9=0$ the section
$\hat{s}_1$ wraps the rational curve described by the parenthesis in
\eqref{eq:PF3ats8=s9=0}. The resulting fiber at $V(I_{(3)})$ is shown in
the second column of Table~\ref{tab:poly3_matter}.
We readily compute using the charge formula \eqref{eq:U1charge} that the
U(1) charge of the matter is indeed $q=3$ and its multiplicity is given
by $[s_8]\cdot[s_9]$, which after using
\eqref{eq:cubicsections}, yields the result shown in
Table~\ref{tab:poly3_matter}. We emphasize that this is the first
occurrence of matter with charge $q>2$ in models with
Abelian gauge symmetry in F-theory.

Second, we consider the locus $V(I_{(2)})$. The complete
intersection in $V(I_{(2)})$ shown in Table~\ref{tab:poly3_matter}
has two irreducible components, one of which given by $V(I_{(3)})$, that
we forbid by requiring $(s_8,s_9)\neq (0,0)$, and a second one
described by a prime ideal $I_{(2)}$ with
ten generators.\footnote{As all prime ideals in
this work, it is computed by the primary decomposition function in Singular
\cite{Singular}.}
The variety $V(I_{(2)})$ supports matter of charge two. We can check this locally by solving the complete
intersection inside $V(I_{(2)})$ e.g.~for $s_3$ and $s_6$ and by plugging this solution into \eqref{eq:PF3}.
Indeed, the fiber splits into a line and a non-singular quadric $q_2(e_1u,e_1v,w)$,
\beq \label{eq:pF3charge2}
	p_{F_3}\rightarrow(s_8u+s_9v)q_2(e_1u,e_1v,w)\,.
\eeq
Furthermore, we prove that $V(I_{(2)})$ is the only locus that can yield
an $I_2$-fiber of this type by computing the elimination ideal of the
ideal of constraints  necessary for the factorization
\eqref{eq:pF3charge2}.
We see that the zero section \eqref{eq:s0F3} is well-defined at $V(I_{(2)})$ and passes through the line. However,
the rational section \eqref{eq:Q1F3} is ill-defined. This is clear because the line in \eqref{eq:pF3charge2}
is precisely  the tangent $t_P$ at $P_0$ defined in
\eqref{eq:tangentF3}  and since the section
$\hat{s}_1$ is defined as the intersection of $t_P$ with $\mathcal{E}$.
Thus, the section $\hat{s}_1$ at $V(I_{(2)})$ wraps the entire rational curve given by the line in \eqref{eq:pF3charge2}.
Again we use \eqref{eq:U1charge} to show that the U(1)-charge is $q=2$,
as claimed in Table~\ref{tab:poly3_matter}.
The multiplicity of a 6D hyper multiplet in the representation
$\mathbf{1}_2$ is given by the homology class of $V(I_{(2)})$. It is
computed by first computing the homology class of the
complete intersection in $V(I_{(2)})$ in Table~\ref{tab:poly3_matter}
using \eqref{eq:cubicsections} and by
subtracting the class of the unwanted component $V(I_{(3)})$ with the
appropriate order. We determine it
to be six using the resultant technique of \cite{Cvetic:2013nia}, which precisely yields the
multiplicity in the third row of Table~\ref{tab:poly3_matter}.

Finally, we turn to the codimension two locus $V(I_{(1)})$ supporting matter
of charge one. In order for the charge formula \eqref{eq:U1charge} to
produce charge one for an $I_2$-fiber, both $\hat{s}_0$ and $\hat{s}_1$
have to be regular and pass through different rational curves in the
$I_2$-fiber. This can only happen for a factorization
of \eqref{eq:PF3} of the form (we can set $e_1=1$)
\beq \label{eq:factF3}
	p_{F_3}\rightarrow (d_1u+d_2v+d_3w)q_2(u,v,w)\,,
\eeq
with $q_2(u,v,w)$ denoting a quadric without the monomial $w^2$. We note
that all coefficients $d_i$,
$i=1,2,3$, have to be non-vanishing since $d_1=0$, $d_2=0$ or $d_3=0$
lead to a factorization in \eqref{eq:factF3}
that cannot happen at codimension two. We see that
$\hat{s}_0$ intersects the quadric $q_2=0$ and $\hat{s}_1$ intersects
the line, as required for matter with charge one. Furthermore, we
compute the elimination ideal, denoted by $I_{(1)}$, of the ideal of
constraints necessary for the factorization \eqref{eq:factF3}. It is
prime and of codimension two in the
ring, that means that the factorization \eqref{eq:factF3}, indeed, occurs
in codimension two in $B$. In addition, we check that the complete
intersection \eqref{eq:charge1Matter} is inside the ideal $I_{(1)}$ and
that $I_{(1)}$ is in turn not contained in $I_{(3)}$ or $I_{(2)}$, as
required. Thus, we identify $I_{(1)}$ as the third and last
associated prime ideal of \eqref{eq:F3charge1}.

Under the well-motivated assumption that $I_{(3)}$, $I_{(2)}$ and $I_{(1)}$ are the only
associated prime ideals of \eqref{eq:F3charge1}, we determine the multiplicity
of the $\one_1$-matter as follows. First, we determine the orders of the
loci $V(I_{(3)})$ and $V(I_{(2)})$ in
the complete intersection \eqref{eq:U1charge}.
Using the resultant technique of \cite{Cvetic:2013nia} and random
integers for some of the $s_i$ we find the
orders $81$ and $16$ for these loci, respectively. Then, we
subtract  their multiplicities with these orders from the class of the
complete intersection \eqref{eq:F3charge1} and obtain, using
\eqref{eq:cubicsections}, the multiplicity in the last row of
Table~\ref{tab:poly3_matter}.

The matter spectrum of $X_{F_3}$ is completed by the number of  neutral
hyper multiplets $H_{\text{neut}}$. Employing \eqref{eq:Hneutral} and
the Euler number $\chi(X_{F_3})$ of $X_{F_3}$ given in
\eqref{eq:EulerNumbers}, we obtain
\bea \label{eq:HneutF3}
	H_{\text{neutral}}&=& 13 + 11 [K^{-1}_B]^2 - 3 [K^{-1}_B] \cS_7 + 3 \cS_7^2 - 4 [K^{-1}_B] \cS_9 - 2 \cS_7 \cS_9 + 2 \cS_9^2\,.
\eea
Finally, we check anomaly-freedom of the full 6D SUGRA theory. To this
end we use \eqref{eq:heightF3}, the charged spectrum in
Table~\ref{tab:poly3_matter} and \eqref{eq:HneutF3}, together with
$V=1$, to show that  all relevant anomalies of the 6D SUGRA theory in
\eqref{eq:6dAnomalies} are canceled.

There is another quantum consistency condition the spectrum in Table~\ref{tab:poly3_matter} has
to pass. In order to have an effective theory that makes sense also in a quantum 
gravity model, it has been argued in \cite{Banks:2010zn} that all charges allowed by
Dirac quantization have to be present in the spectrum. Indeed, it is clear from the multiplicity 
formulas in Table~\ref{tab:poly3_matter} (e.g.~by evaluation for  a concrete base $B$, see 
Section \ref{sec:allowedRegionP2}) that if matter with a maximal charge $q$ is present
in the spectrum, also matter with all lower charges $q'<q$ is automatically there, as required.

For completeness, we include a discussion of the Yukawa
couplings. Forming the union of the ideals and computing their codimension to be three 
in the polynomial ring $K[s_i]$, we find the two Yukawa couplings given in 
Table~\ref{tab:poly3_yukawa}.
\begin{table}[H]
\begin{center}
\renewcommand{\arraystretch}{1.2}
\begin{tabular}{|c|c|}\hline
Yukawa & Locus \\ \hline
$\one_{1} \cdot \one_{1} \cdot \overline{\one_{2}} $ & $V(I_{(1)})\cap V(I_{(2)})$  \\ \hline
$\one_{1} \cdot \one_{2} \cdot \overline{\one_{3}} $ & $V(I_{(1)})\cap
 V(I_{(2)})\cap V(I_{(3)})$  \\ \hline
\end{tabular}
\caption{\label{tab:poly3_yukawa}Codimension three loci and corresponding Yukawa couplings for $X_{F_3}$. }
\end{center}
\end{table}

\subsubsection*{An alternative perspective: $X_{F_3}$ from $X_{F_5}$ by an extremal transition}

There is a second perspective on $X_{F_3}$ that provides an alternative
explanation for the presence of the rational point \eqref{eq:Q1F3} and
that will be useful for the understanding of the Higgs transition in
Section~\ref{sec:Higgsings}. The following can
be skipped on a first reading, as it is not important for the main
thread of this work.

We begin by noting that \eqref{eq:PF3} becomes singular if we tune the
complex structure so that $s_4\equiv 0$. The induced $I_2$-singularities
occur at codimension two and can be resolved by the blow-up
in the fiber at $u=w=0$. The Calabi-Yau manifold after this
extremal transition is precisely $X_{F_5}$, that we discuss below
in Section~\ref{sec:polyF5}. It has been shown that $X_{F_5}$ has a rank
two Mordell-Weil group \cite{Borchmann:2013jwa,Cvetic:2013nia}.

In the
singular fibration with all
exceptional divisors blown down, the three rational points on
the fiber $\mathcal{C}_{F_5}$ are the three intersection points with the
line $u=0$. One point agrees with the origin \eqref{eq:s0F3} of
$X_{F_3}$. We denote the other two points by $Q_1$, $Q_2$.
This implies
that the point $Q_1+Q_2$ is precisely given by
\eqref{eq:Q1F3}, in the limit $s_{4}\equiv 0$. Indeed, the
group law on a cubic curve is defined so that the point $Q_1+Q_2$
is found by first constructing the third intersection point of
the line through $Q_1$ and $Q_2$ and then by forming the line
through that point and the origin $P_0$. This line again has a third
intersection point with the curve, which is defined to be $Q_1+Q_2$
In our situation, the line through $Q_1$ and $Q_2$ is $u=0$. Thus,
the third intersection point of $u=0$ with $\mathcal{E}$
is the origin $P_0$. Consequently, the point $Q_1+Q_2$ is the
second intersection point of the \emph{tangent} through $P_0$ with the
elliptic curve. In fact, it can be checked by performing this
addition on the fiber of $X_{F_5}$ explicitly that the coordinates
of the point $Q_1+Q_2$ on the fiber of $X_{F_3}$ agree with the
coordinates \eqref{eq:Q1F3} after setting $s_4\equiv 0$.
Furthermore, we compute the Weierstrass coordinates of $Q_1+Q_2$
that also agree with \eqref{eq:WSFQ1F3} after setting $s_4\equiv 0$.\footnote{The coordinates of $Q_1+Q_2$ in WSF are obtained by
inserting its coordinates into the birational map from $X_{F_5}$ to
its WSF. We note that the result agrees with the
WS-coordinates of $Q_1+(-Q_2)$, \textit{not} $Q_1+Q_2$, where `$+$'
denotes here the addition in the WSF of $X_{F_5}$.}

This is not surprising since we recall that the $P_1$ in $X_{F_3}$
has been constructed as the second intersection of the tangent
\eqref{eq:tangentF3} to $P_0$. Thus, we see that the section
$\hat{s}_1$ can be understood as the sum of the sections
$\hat{s}_1+\hat{s}_2$ on $X_{F_5}$, which \textit{survives} the
extremal transition $X_{F_5}\leftrightarrow X_{F_3}$, i.e.~the
complex structure deformation associated to switching on $s_4$.
In contrast, the individual sections $\hat{s}_1$ and $\hat{s}_2$
on $X_{F_5}$ do not map to rational sections on $X_{F_3}$. As
consequence, cf.~Section~\ref{sec:Higgsings},
the U(1)-charges of matter in $X_{F_3}$ are given by the
sum of the U(1)-charges $q_1+q_2$ on $X_{F_5}$.

We can make these statements even more explicit by mapping $X_{F_3}$ to
$X_{F_5}$. The shift
\beq \label{eq:shiftF3F5}
	w\mapsto w-\frac{s_7 +  \sqrt{s_7^2 - 4 s_4 s_9}}{2 s_9}e_1 v\,
\eeq
precisely cancels the monomial proportional to $s_4$ in \eqref{eq:PF3}.
Clearly, this requires an extension of the field of meromorphic
functions on $B$ by the square root $\sqrt{s_7^2 - 4 s_4 s_9}$. Thus,
this map is certainly not birational. After this shift,
we precisely obtain the hypersurface of $X_{F_5}$,
cf.~\eqref{eq:pF5} for $e_2=1$. Due to the shift
\eqref{eq:shiftF3F5}, the coefficients $s_i$ in \eqref{eq:pF5}
have to be replaced by
\bea \label{eq:siF3toF5}
	s_2\!&\!\mapsto\! &\! s_2 - s_5\frac{ s_7+\sqrt{s_7^2 - 4 s_4 s_9}}{2 s_9} \,,\quad 
	s_3\mapsto s_3-\frac{s_4 s_8}{s_9} +\frac{(s_7 s_8-s_6s_9)(s_7+ \sqrt{s_7^2 - 4 s_4 s_9})}{s_9^2}\,,\nn\\
	 s_6\! &\!\mapsto\! &\! s_6 - s_8\frac{s_7 +\sqrt{s_7^2 - 4 s_4 s_9}}{s_9}\,, \quad s_7 \mapsto -\sqrt{s_7^2 - 4 s_4 s_9}\,,
\eea
with $s_1$, $s_5$, $s_8$ and $s_9$ unchanged. If we insert this
variable transformation into the expressions for $\hat{s}_1$ or
$\hat{s}_2$ in \eqref{eq:sectionsF5}, we introduce square roots,
i.e.~these sections do not map to rational sections on $X_{F_3}$.
However, if we insert \eqref{eq:siF3toF5} into the coordinates
for $\hat{s}_1+\hat{s}_2$ on $X_{F_5}$, we precisely reproduce
\eqref{eq:Q1F3}, i.e.~all square roots  cancel.

Furthermore, we can re-derive the Weierstrass coordinates
\eqref{eq:WSFQ1F3} of $\hat{s}_1$ on $X_{F_3}$ by first computing
the Weierstrass coordinates of $Q_1+Q_2$ and then inserting
\eqref{eq:siF3toF5}. In addition, $f$ and $g$ of the WSF of
$X_{F_3}$ can be obtained from the WSF for $X_{F_5}$ by insertion
of \eqref{eq:siF3toF5}.

%%%%%%%%%%%%%%%%%%%%%%%%%%%%%%%%%%%%%%%%%%%%%%%%%%%%%%%%%%%%%%%%%%%%%%%%%%%%%%%%%%%%%%%%%%%%%%%%%
\subsubsection{Polyhedron $F_5$: $G_{F_5}=\text{U(1)}^2$}
\label{sec:polyF5}
%%%%%%%%%%%%%%%%%%%%%%%%%%%%%%%%%%%%%%%%%%%%%%%%%%%%%%%%%%%%%%%%%%%%%%%%%%%%%%%%%%%%%%%%%%%%%%%%%

The toric hypersurface fibration $X_{F_5}$ is constructed 
as the fibration of the elliptic curve in $\mathbb{P}_{F_{5}}=dP_2$. 
As it is completely analyzed  in  
\cite{Cvetic:2013nia, Borchmann:2013hta}, we only state the results 
here for completeness.

\begin{figure}[h!]
\center
\includegraphics[scale=0.4]{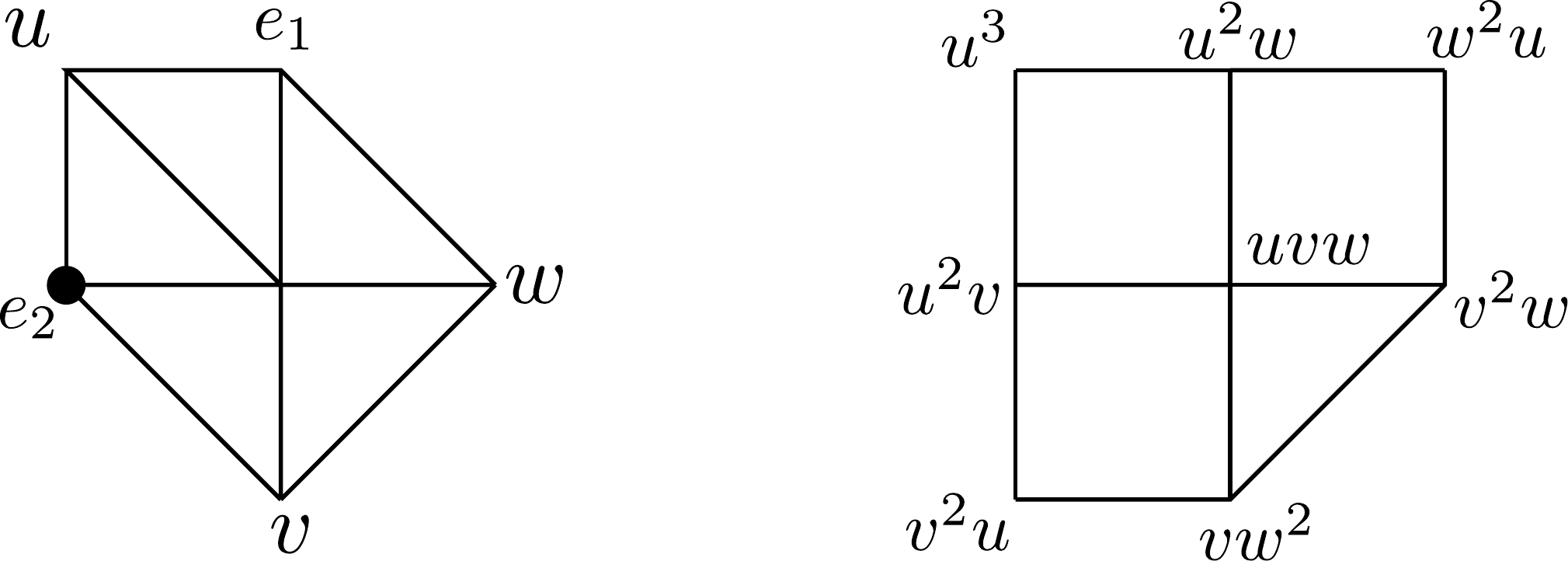}
\caption{\label{fig:poly5_toric}The toric diagram of polyhedron $F_5$ ($dP_2$) and its dual. The zero section is indicated by the dot.}
\end{figure}

The toric diagram of $F_5$ along with a choice of homogeneous coordinates as well as its dual polyhedron are depicted in Figure~\ref{fig:poly5_toric}. In the monomials corresponding to the integral points of $F_{12}$ by \eqref{eq:BatyrevFormula} we have set $e_i=1$, $\forall i$.
The toric variety $\mathbb{P}_{F_5}$ is the blow-up of $\mathbb{P}^2$, 
cf.~Section~\ref{sec:cubic}, at two points, i.e.~$dP_2$. The blow-up map
reads
\begin{align} \label{eq:F5blowup}
u \rightarrow e_1e_2 u\,, \quad v\rightarrow e_2 v \,, \quad w\rightarrow e_1 w\,.
\end{align}
The homogeneous coordinates on $dP_2$ after this blow-up are 
$[u:v:w:e_1:e_2]$ and are sections in line bundles associated to the 
divisors classes
\begin{align}
\label{tab:poly5_bundles}
\text{\begin{tabular}{|c|c|}\hline
Section & Divisor class \\ \hline
$u$ & $H-E_1-E_2+\cS_9+[K_B]$ \\ \hline
$v$ & $H-E_2+\cS_9-\cS_7$\\ \hline
$w$ & $H-E_1$\\ \hline
$e_1$ & $E_1$\\ \hline
$e_2$ & $E_2$\\ \hline
\end{tabular}
}
\end{align}
The Stanley-Reisner ideal of $\mathbb{P}_{F_5}$ is given by
\begin{align}
SR&=\{ w e_2,  w u, v e_1 , e_2 e_1, v u \}\, .
\end{align}
By use of \eqref{eq:BatyrevFormula} the hypersurface equation for
$X_{F_5}$ in the $dP_2$-fibration \eqref{eq:PFfibration} is given by
\beq
\label{eq:pF5}
p_{F_5} = s_1 e^2_2 e^2_1 u^3 + s_2 e^2_2 e_1 u^2 v + s_3 e^2_2 u v^2 + s_5 e_2 e^2_1 u^2 w + s_6 e_2 e_1 u v w 
 + s_7 e_2 v^2 w + s_8 e^2_1 u w^2 + s_9 e_1 v w^2 \, ,
\eeq
where the sections $s_i$ take values in the line bundles shown
in \eqref{eq:cubicsections}. We see that \eqref{eq:pF5} can also be
obtained from \eqref{eq:pF1} by the specialization $s_4=s_{10}=0$
and the map \eqref{eq:F5blowup}.

There are three rational sections of the fibration of $X_{F_5}$ with the coordinates
\begin{align}
\begin{split}\label{eq:sectionsF5}
\hat{s}_0=X_{F_5}\cap\{e_2=0\}&:\quad [s_9:-s_8:1:1:0]\,,\\
\hat{s}_1=X_{F_5}\cap\{e_1=0\}&:\quad [s_7:1:-s_3:0:1]\,,\\
\hat{s}_2=X_{F_5}\cap\{u=0\}&:\quad [0:1:1:s_7:-s_9]\,,
\end{split}
\end{align}
where we choose $\hat{s}_0$ as the zero section.

The Weierstrass form \eqref{eq:WSF} of \eqref{eq:pF5} can be computed using Nagell's algorithm.
The WS-coordinates of the sections $\hat{s}_1$ and $\hat{s}_2$ are given by \eqref{eq:secsF51} and \eqref{eq:secsF52}, respectively.
The functions $f$ and $g$ are given by \eqref{eq:fcubic} and \eqref{eq:gcubic}, respectively, after
setting $s_4=s_{10}=0$. After using this to calculate the discriminant we do not find any codimension one singularities.
Then the total gauge group of $X_{F_5}$ is
\beq \label{eq:GF5}
	G_{F_5}=\text{U}(1)^2 \,.
\eeq
Thus, the corresponding Shioda maps \eqref{eq:ShiodaMap} for $\hat{s}_1$ and $\hat{s}_2$ read
\begin{align}
\begin{split}\label{eq:ShiodaF5}
\sigma(\hat{s}_1) &= (S_1 - S_0 - [K_B^{-1}]) \, , \\
\sigma(\hat{s}_2) &= (S_2 - S_0 -[K_B^{-1}] - [s_9])\, ,
\end{split}
\end{align}
which allows us to compute the corresponding height pairing 
\eqref{eq:anomalycoeff} as
\begin{align}
\label{eq:bmnF5}
 b_{mn}= \begin{pmatrix} 2[K_B^{-1}] & [K_B^{-1}] + \cS_9-\cS_7\\
[K_B^{-1}] + \cS_9-\cS_7 & 2[K_B^{-1}] + 2\cS_9 \end{pmatrix}_{mn} \, .
\end{align}

To determine the 6D spectrum of charged hyper multiplets we analyze the 
codimension two singularities of the WSF of $X_{F_5}$. There are six 
singularities leading to the matter representations and the 
corresponding codimension two fibers in $X_{F_5}$ given in the first and 
third column of Table~\ref{tab:poly5_matter}. The detailed derivation
of these results can be found in  
\cite{Cvetic:2013nia,Cvetic:2013jta,Borchmann:2013jwa,Borchmann:2013hta}.

We complete the matter spectrum of $X_{F_5}$ by the number of neutral
hyper multiplets, which is computed from \eqref{eq:Hneutral} using
the Euler number \eqref{eq:EulerNumbers}. It reads
\beq \label{eq:HneutF5}
H_{\text{neut}} = 14 + 11 [K_B^{-1}]^2 - 4 [K_B^{-1}] \cS_7 + 2 \cS_7^2 - 4 [K_B^{-1}] \cS_9 - \cS_7 \cS_9 + 2 \cS_9^2 \,.
\eeq
\begin{table}[H]
\begin{center}
\footnotesize
\renewcommand{\arraystretch}{1.2}
\begin{tabular}{|c|@{}c@{}|c|@{}c@{}|}
\hline
Representation & Multiplicity & Fiber & Locus \\ \hline

$\one_{(1,-1)}$ & $\cS_7 ([K_B^{-1}]+\cS_7-\cS_9)$ & \rule{0pt}{1.4cm}\parbox[c]{2.9cm}{\includegraphics[width=2.9cm]{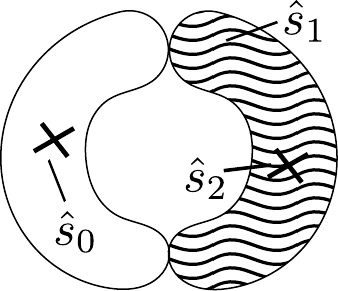}} & $V(I_{(1)}):=\{s_3 = s_7 = 0 \}$ \\[1.1cm] \hline

$\one_{(1,0)}$ & $\begin{array}{c} 6[K_B^{-1}]^2 + [K_B^{-1}](4\cS_7-5\cS_9) \\ - 2\cS_7^2 + \cS_7 \cS_9 +\cS_9^2 \end{array}$  & \rule{0pt}{1.4cm}\parbox[c]{2.9cm}{\includegraphics[width=2.9cm]{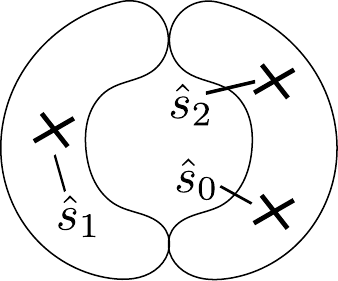}} & $\begin{array}{c} V(I_{(2)}):=\{s_2 s_7^2 + s_3^2 s_9 - s_3 s_6 s_7=0 \\ s_5 s_3 s_7 - s_3^2 s_8 - s_7^2 s_1 = 0 \}\backslash V(I_{(1)}) \end{array}$ \\[1.1cm] \hline

$\one_{(-1,-2)}$ & $\cS_9 ([K_B^{-1}]-\cS_7 +\cS_9)$ & \rule{0pt}{1.4cm}\parbox[c]{2.9cm}{\includegraphics[width=2.9cm]{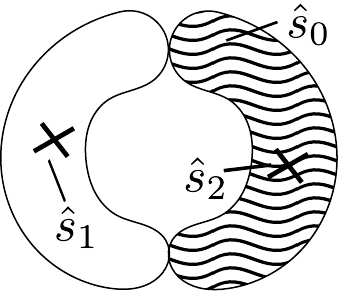}} & $V(I_{(3)}):=\{s_8 = s_9 = 0 \}$ \\[1.1cm] \hline

$\one_{(-1,-1)}$ & $\begin{array}{c} 6[K_B^{-1}]^2 + [K_B^{-1}](-5\cS_7+4\cS_9) \\ + \cS_7^2 + \cS_7 \cS_9 -2\cS_9^2 \end{array}$ & \rule{0pt}{1.4cm}\parbox[c]{2.9cm}{\includegraphics[width=2.9cm]{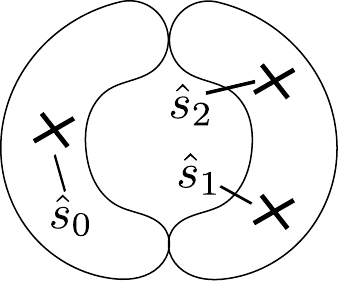}} & $\begin{array}{c} V(I_{(4)}):=\{ s_2 s_8 s_9 - s_3 s_8^2 - s_9^2 s_1=0 \\ s_5 s_9^2 - s_6 s_8 s_9 + s_8^2 s_7 = 0 \} \backslash(V(I_{(3)}) \end{array}$ \\[1.1cm] \hline

$\one_{(0,2)}$ & $\cS_7 \cS_9$ & \rule{0pt}{1.4cm}\parbox[c]{2.9cm}{\includegraphics[width=2.9cm]{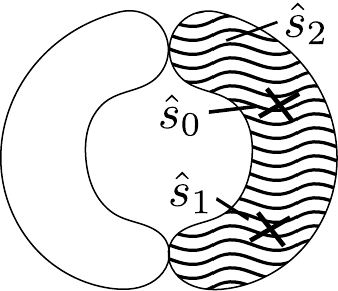}} & $V(I_{(5)}):=\{ s_9 = s_7 = 0 \}$ \\[1.1cm] \hline

$\one_{(0,1)}$ & $\begin{array}{c} 6[K_B^{-1}]^2 + [K_B^{-1}](4\cS_7+4\cS_9) \\ -2\cS_7^2 -2\cS_9^2 \end{array}$ & \rule{0pt}{1.4cm}\parbox[c]{2.9cm}{\includegraphics[width=2.9cm]{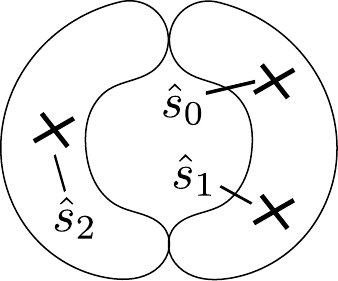}} & $\begin{array}{c} V(I_{(6)}):=\{ s_1 s_9^4 s_7^2 + (s_3 s_9^2+s_7 \\ \times (-s_6s_9+s_8s_7)) (s_3s_8s_9^2+s_7 \\ \times (-s_6s_8s_9+s_8^2s_7+s_9^2s_5))= 0 \\ s_2 s_9^3 s_7^2 + s_3^2s_9^4 - s_3s_6s_9^3s_7 \\ -s_7^3(-s_6s_8s_9+s_8^2s_7+s_9^2s_5)= 0 \} \\ \backslash(V(I_{(1)})\cup V(I_{(3)})\cup V(I_{(5)})) \end{array}$ \\[1.1cm] \hline
\end{tabular}
\caption{\label{tab:poly5_matter}Charged matter representations under U(1)$^2$ and corresponding codimension two fibers of $X_{F_5}$.}
\end{center}
\end{table}
The number $T$ of tensor multiplets is given by \eqref{eq:Tformula}
and the number of vector multiplets is $V=2$. 
Using the above results it can be checked that all 6D anomalies are canceled.
Finally we summarize the codimension three singularities
of the WSF of $X_{F_5}$. This leads to the Yukawa points given in 
Table~\ref{tab:poly5_yukawa}, see \cite{Cvetic:2013uta}.
\begin{table}[h!]
\begin{center}
\renewcommand{\arraystretch}{1.2}
\begin{tabular}{|c|c|}\hline
Yukawa & Locus\\ \hline
$\one_{(-1,-2)}\cdot\one_{(0,2)}\cdot\one_{(1,0)}$ & $s_7=s_8=s_9=0$\\ \hline
$\one_{(0,2)} \cdot \one_{(1,-1)} \cdot \one_{(-1,-1)}$ & $s_3=s_7=s_9=0$\\ \hline
$\one_{(-1,-1)} \cdot \one_{(1,0)} \cdot \one_{(0,1)}$ & $V(I_{(2)})\cup V(I_{(4)}) \cup V(I_{(6)})$\\ \hline
$\one_{(1,-1)} \cdot \overline{\one_{(1,0)}} \cdot \one_{(0,1)}$ & $V(I_{(1)}) \cup V(I_{(2)}) \cup V(I_{(6)})$\\ \hline
$\overline{\one_{(-1,-1)}} \cdot \one_{(-1,-2)} \cdot \one_{(0,1)}$ & $V(I_{(3)}) \cup V(I_{(4)}) \cup V(I_{(6)})$\\ \hline
\end{tabular}
\caption{\label{tab:poly5_yukawa}Codimension three loci and corresponding Yukawa couplings for polyhedron $F_5$. For the complicated loci we refer to the literature \cite{Cvetic:2013uta,Borchmann:2013hta}}
\end{center}
\end{table}

%%%%%%%%%%%%%%%%%%%%%%%%%%%%%%%%%%%%%%%%%%%%%%%%%%%%%%%%%%%%%%%%%%%%%%%%%%%%%%%%%%%%%%%%%%%%%%%%%
\subsubsection{Polyhedron $F_{6}$: $G_{F_6}=\text{SU}(2)\times\text{U}(1)$}
\label{sec:polyF6}
%%%%%%%%%%%%%%%%%%%%%%%%%%%%%%%%%%%%%%%%%%%%%%%%%%%%%%%%%%%%%%%%%%%%%%%%%%%%%%%%%%%%%%%%%%%%%%%%%

\begin{figure}[H]
\center
\includegraphics[scale=0.4]{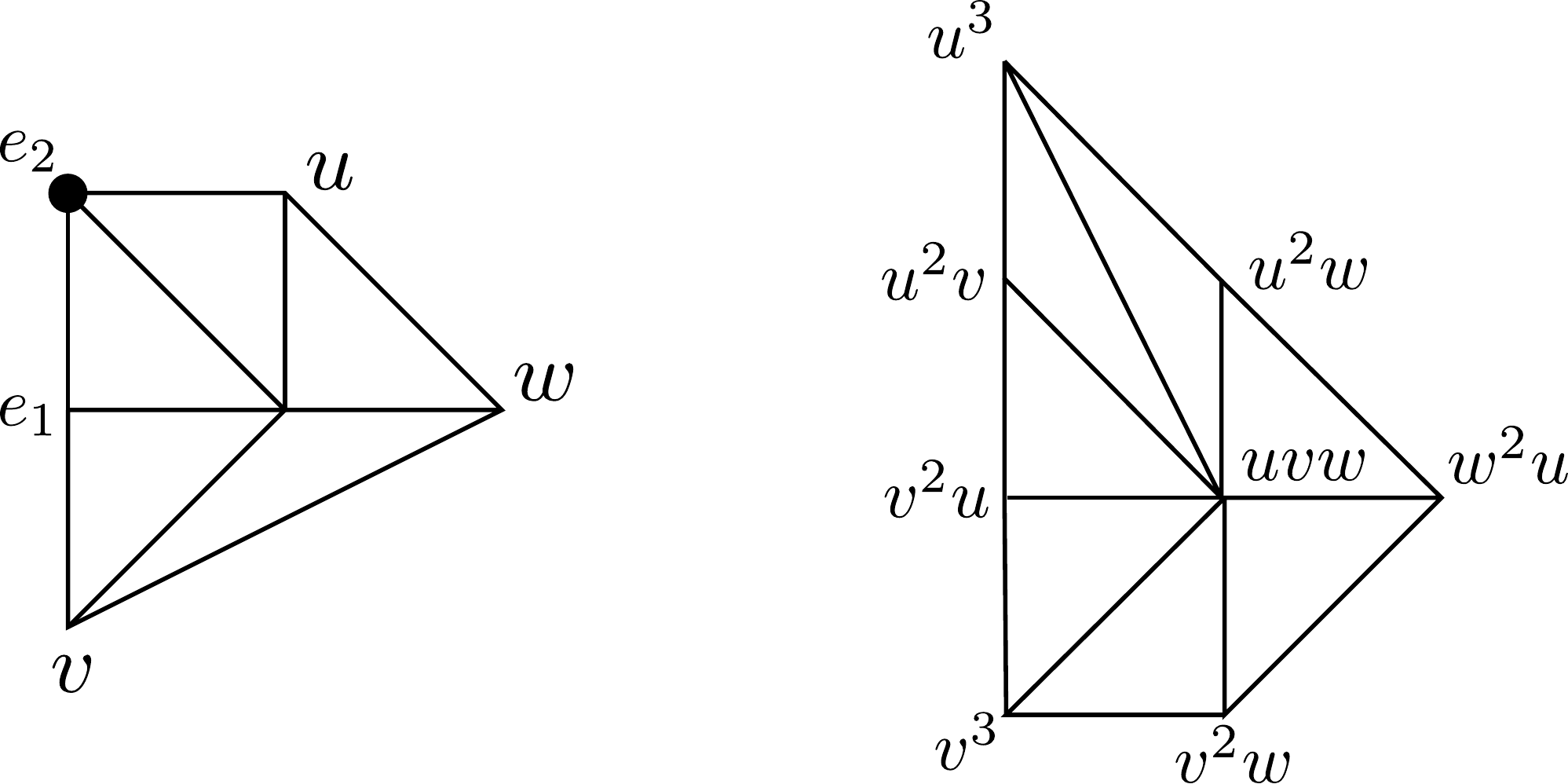}
\caption{\label{fig:poly6_toric}Polyhedron $F_6$ and its dual
$F_{11}$. The zero section is indicated by the dot.}
\end{figure}

We consider an elliptically fibered Calabi-Yau
manifold $X_{F_6}$ with an arbitrary base $B$ and general elliptic
fiber  given by the elliptic curve $\mathcal{E}$ in $\mathbb{P}_{F_6}$.
The toric data of $\mathbb{P}_{F_6}$ is summarized in
Figure~\ref{fig:poly6_toric}, where the polyhedron $F_6$ along with
a choice of homogeneous coordinates as well as its dual polyhedron
$F_{11}$ are shown. For brevity, we have set $e_i=1$, $\forall i$,
in the monomials that are associated to the integral points of $F_{11}$
by \eqref{eq:BatyrevFormula}.
We note that $\mathbb{P}_{F_6}$ is the blow-up of $\mathbb{P}^2$,
cf.~Section~\ref{sec:cubic}, at two points, defined by
\beq \label{eq:F6blowup}
u \rightarrow e_1e_2^2 u\,, \quad v\rightarrow e_1e_2 v\,.
\eeq
The homogeneous coordinates on the fiber after this blow-up are 
$[u:v:w:e_1:e_2]$ and take values in the line bundles associated to the 
following divisor classes:
\beq
\label{tab:poly6_bundles}
\text{
\begin{tabular}{|c|c|}\hline
Section & Divisor class\\ \hline
$u$ & $H-E_1-E_2+\cS_9+[K_B]$ \\ \hline
$v$ & $H-E_1-\cS_7+\cS_9$\\ \hline
$w$ & $H$\\ \hline
$e_1$ & $E_1-E_2$\\ \hline
$e_2$ & $E_2$\\ \hline
\end{tabular}}
\eeq
Here $H$ denotes the pullback of the hyperplane class on
$\mathbb{P}^2$ and the $E_i$ are the exceptional divisors of the
blow-up \eqref{eq:F6blowup}.
The Stanley-Reisner ideal of $\mathbb{P}_{F_6}$ then reads
\begin{align}
SR&=\{ uv, ue_1, we_1, we_2, ve_2 \} \, .
\end{align}
Employing \eqref{eq:BatyrevFormula} the hypersurface equation for
$X_{F_6}$ in the $\mathbb{P}_{F_6}$-fibration \eqref{eq:PFfibration}  is
\beq \label{eq:pF6}
p_{F_6}=s_1 e_1^2 e_2^4 u^3 + s_2e_1^2 e_2^3  u^2 v +s_3 e_1^2 e_2^2  u v^2 +s_4 e_1^2 e_2 v^3 + s_5e_1 e_2^2  u^2 w + s_6e_1 e_2  u v w  
+ s_7e_1  v^2 w + s_8 u w^2 \, ,
\eeq
where the sections $s_i$ take values in the line bundles shown
in \eqref{eq:cubicsections}. We note that \eqref{eq:pF6} is readily
obtained from \eqref{eq:pF1} by the specialization $s_9=s_{10}=0$
and the map \eqref{eq:F6blowup}.

There are two rational sections of the fibration of $X_{F_6}$.
Their coordinates are
\begin{align}
\begin{split}\label{eq:sectionsF6}
\hat{s}_0=X_{F_6}\cap\{e_2=0\}&:\quad [-s_7:1:s_8:1:0]\,,\\
\hat{s}_1=X_{F_6}\cap\{u=0\}&:\quad [0:1:s_4:1:-s_7]\,,
\end{split}
\end{align}
where we choose $\hat{s}_0$ as the zero section. The corresponding
points on $\mathcal{E}$ are denoted $P_0$ and $P_1$, respectively.

\begin{figure}[h!]
\center
\includegraphics[scale=0.6]{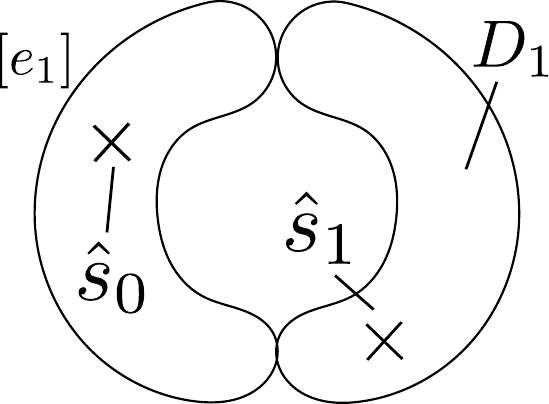}
\caption{\label{fig:poly6_codim1}Codimension one fiber
of $X_{F_6}$ at $s_8=0$ in $B$. The crosses denote the intersections
with the two sections.}
\end{figure}

We compute the Weierstrass form \eqref{eq:WSF} of \eqref{eq:pF6} using
Nagell's algorithm. The WS-coordinates of the section
$\hat{s}_1$ are given
by \eqref{eq:WScoordsSecF3} after setting $s_9=0$.
Furthermore, the functions $f$ and $g$ take the form of
\eqref{eq:fcubic} and \eqref{eq:gcubic}, respectively, after
setting $s_9=s_{10}=0$. From this the discriminant $\Delta$ is readily
computed. This allows us to find all codimension one singularities
of the WSF of $X_{F_6}$. We find one $I_2$-singularity over the divisor
$\mathcal{S}^b_{\text{SU}(2)}=\{s_8=0\}\cap B$ in $B$. Along this
divisor the constraint \eqref{eq:pF6} factorizes as
\beq \label{eq:SU2F6}
\SU2\,:\quad \left.p_{F_6}\right\vert_{s_8=0}= e_1 \cdot q_3\,,
\eeq
where $q_3$ is the polynomial that remains after factoring out
$e_1$. This is clearly an $I_2$-fiber, cf.~Figure~\ref{fig:poly6_codim1}, giving rise to an SU(2) gauge group.
In summary, the  gauge group of $X_{F_6}$ is
\begin{align} \label{eq:GF6}
	G_{F_6}=\text{SU}(2)\times\text{U}(1)\,.
\end{align}

The rational curve of the $I_2$-fiber in Figure~\ref{fig:poly6_codim1}
that is intersected by the
zero section $\hat{s}_0$ is the affine node and the other rational
curve, $c_{-\alpha_1}$, corresponds to the simple root of SU(2).
Thus, the class of the SU(2) Cartan divisor in $X_{F_6}$,
which is the fibration of  $c_{-\alpha_1}$ over $S^b_{\text{SU}(2)}$,
reads
\begin{align} \label{eq:CartansF6}
D_{1}&=[s_8]-[e_1]\, .
\end{align}
This can be seen by noting that $[s_8]$ is the class of the
complete $I_2$-fiber fibered over the base divisor
$S^b_{\text{SU}(2)}$, whereas $[e_1]$ is the class
of the affine node fibered over $S^b_{\text{SU}(2)}$.

With these results, we compute the Shioda map \eqref{eq:ShiodaMap}
of the section $\hat{s}_1$ as
\begin{align} \label{eq:ShiodaF6}
\sigma (\hat{s}_1)=S_1-S_0 +  [K_B]-\cS_7	+\frac{1}{2} D_{1}\, .
\end{align}
Here $S_0$, $S_1$ denote the divisor classes of the sections
$\hat{s}_0$, $\hat{s}_1$, respectively, and  we use
\beq
S_1 \cdot c_{-\alpha _1}= 1\,,
\eeq
which can be deduced from Figure~\ref{fig:poly6_codim1}.
Using \eqref{eq:ShiodaF6}, we  compute the height pairing
\eqref{eq:anomalycoeff},
\begin{align}
\label{eq:bmnF6}
 b_{11}=\frac{3}{2}[K_B^{-1}]+\frac{5}{2}\cS_7-\frac{1}{2}\cS_9 \,,
\end{align}
where we use \eqref{eq:SP^2} as well as
\beq
	\pi(S_1\cdot S_0)=\cS_7\,,
\eeq
that  follows since the
coordinates \eqref{eq:sectionsF6} of the two sections agree at $s_7=0$.

Next, we analyze the codimension two singularities of the WSF of
$X_{F_6}$ to determine the charged matter spectrum.
Here, the corresponding  representations under the gauge group
are determined following the general procedure outlined in 
Section~\ref{sec:ellipticCurvesWithRP} for the computation of Dynkin 
labels and U(1)-charges.
We find five codimension two singularities. Four of
these lead to the matter representations and the corresponding
codimension two fibers in $X_{F_6}$ given in the first and third column
of Table~\ref{tab:poly6_matter}, respectively.
\begin{table}[h!]
\begin{center}
\footnotesize
\renewcommand{\arraystretch}{1.2}
\begin{tabular}{|c|c|c|@{}c@{}|}\hline
Representation & Multiplicity & Fiber & Locus \\ \hline

$\two_{-3/2}$ & $\cS_7([K_B^{-1}]-\cS_7+\cS_9)$ & \rule{0pt}{1.6cm}\parbox[c]{2.9cm}{\includegraphics[width=2.9cm]{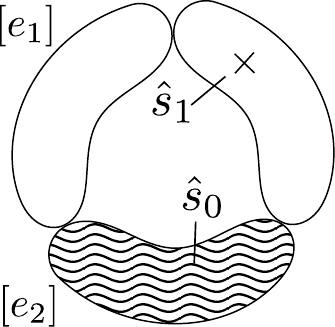}} & $V(I_{(1)}):=\{s_8 = s_7=0\}$ \\[1.3cm] \hline

$\two_{1/2}$ & $\begin{array}{c}
([K_B^{-1}]-\cS_7+\cS_9)\\
\times(6[K_B^{-1}]-2\cS_9+\cS_7)
\end{array}$
& \rule{0pt}{1.4cm}\parbox[c]{2.9cm}{\includegraphics[width=2.9cm]{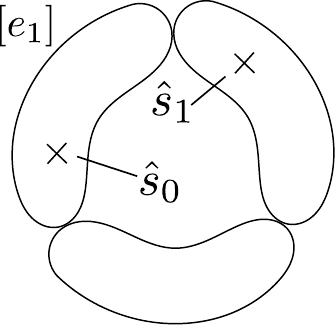}} & $\begin{array}{c}
 V(I_{(2)}):=\\
 \{s_8=s_4^2 s_5^3 - s_3 s_4 s_5^2 s_6+ s_2 s_4 s_5 s_6^2 \\
   - s_1 s_4 s_6^3+ s_3^2 s_5^2 s_7- 2 s_2 s_4 s_5^2 s_7 \\
  - s_2 s_3 s_5 s_6 s_7+3 s_1 s_4 s_5 s_6 s_7 \\
  + s_1 s_3 s_6^2 s_7+ s_2^2 s_5 s_7^2- 2 s_1 s_3 s_5 s_7^2 \\
  -s_1 s_2 s_6 s_7^2+ s_1^2 s_7^3=0\}
 \end{array}$  \\[1.1cm] \hline

$\one_{2}$ & $\cS_7 (-\cS_9+2\cS_7)$ & \rule{0pt}{1.4cm}\parbox[c]{2.9cm}{\includegraphics[width=2.9cm]{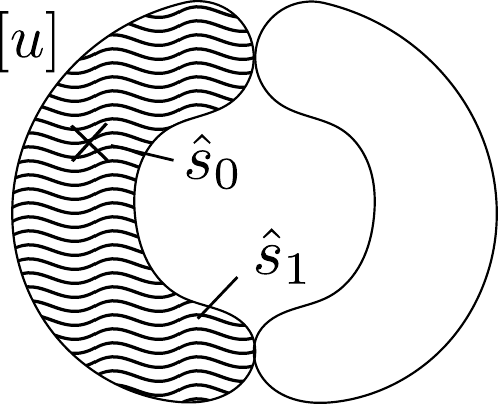}} & $V(I_{(3)}):=\{s_4=s_7=0\}$ \\[1.1cm] \hline %([K_B^{-1}]-\cS_9-2\cS_7)

$\one_{1}$ &
$\begin{array}{c}
6 [K_B^{-1}]^2 + 13 [K_B^{-1}] \mathcal{S}_7 - 3 \mathcal{S}_7^2 \\
- 5 [K_B^{-1}] \mathcal{S}_9 - 2 \mathcal{S}_7 \mathcal{S}_9 + \mathcal{S}_9^2
\end{array}$ & \rule{0pt}{1.4cm}\parbox[c]{2.9cm}{\includegraphics[width=2.9cm]{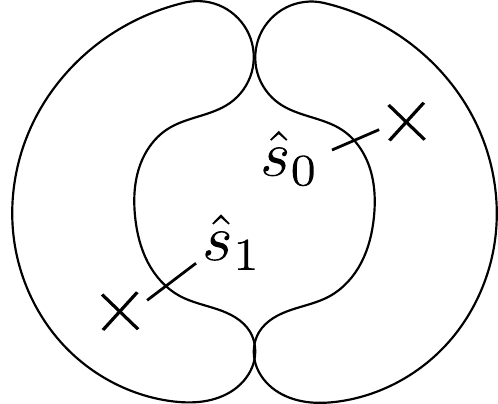}} &
 $\begin{array}{c}
 V(I_{(4)}):=\\
 \{y_Q=fz_Q^4+3x_Q^2=0\} \\
\backslash(V(I_{(1)})\cup V(I_{(3)}))\\
\text{with }x_Q, y_Q, f\text{ given in \eqref{eq:charge1matterF6}}
\end{array}$
\\[1.1cm] \hline \hline
$\three_{0}$ & \rule{0pt}{0.5cm}$1+([K_B^{-1}]-\cS_7+\cS_9)\frac{(-\cS_7+\cS_9)}{2}$ & Figure~\ref{fig:poly6_codim1}& $s_8=0$
 \\[0.1cm] \hline
\end{tabular}
\caption{\label{tab:poly6_matter}Charged matter representations under SU$(2)\times$U(1) and corresponding codimension two fibers of $X_{F_6}$. The adjoint matter is included for completeness.}
\end{center}
\end{table}
At the remaining locus
$s_8=s_6^2 - 4 s_5 s_7= 0$, the fiber is of Type $III$ , i.e.~it is a
degenerate version of the $I_2$-fiber in Figure~\ref{fig:poly6_codim1} with the two $\mathbb{P}^1$'s intersecting
in one point. Thus it does not support any additional matter.

The spectrum of charged singlets is determined starting from  
the complete intersection \eqref{eq:charge1Matter} for the section 
$\hat{s}_1$, see \eqref{eq:charge1matterF6} for its explicit expression. 
By computing its primary decomposition, we find  two 
associated prime ideals,
denoted $I_{(3)}$ and $I_{(4)}$, corresponding to  two 
different matter representations $\one_2$ and $\one_1$. We observe
that the ideal $I_{(3)}$ describes precisely the locus, where
the section \ref{eq:sectionsF6} is ill-defined and has to acquire 
a fiber component.

There is one subtlety since the constraint \eqref{eq:pF6} does not
factorize further at the locus $V(I_{(2)})$. In order to see the 
$I_3$-fiber of the representation $\two_{1/2}$ we have to compute the 
associated prime ideals of  \eqref{eq:pF6} at the locus $V(I_{(2)})$.  
Indeed, we find three prime ideals with the right intersections and, 
they thus,  correspond to the three 
irreducible components of an $I_3$-fiber. Two fiber components 
are described by prime ideals generated by more than three 
constraints.

The multiplicities of the charged hyper multiplets are presented in
Table~\ref{tab:poly6_matter}. These have been computed following
Section~\ref{sec:SUGRA6D}. For the case of the representations
$\two_{-3/2}$, $\two_{1/2}$ and $\one_{2}$, the multiplicities can be
directly computed from  \eqref{eq:cubicsections} and Table~\ref{tab:poly6_matter}, as the corresponding varieties $V(I_{(1)})$,
$V(I_{(2)})$ and $V(I_{(3)})$
are irreducible complete intersections. In contrast, $V(I_{(4)})$
supporting the $\one_1$-matter is not a complete intersection.
However, note
that  the varieties $V(I_{(4)})$,  $V(I_{(1)})$ and $V(I_{(3)})$
are the three irreducible components of the complete intersection
\eqref{eq:charge1Matter} for the section $\hat{s}_1$. Using its
WS-coordinates, given by \eqref{eq:WScoordsSecF3} for $s_9=0$, it reads
\bea\label{eq:charge1matterF6}
	y_1\!\!&\!\!=\!\!&\!\!s_4 s_6^2 s_7^2 - s_4 s_5 s_7^3 - s_3 s_6 s_7^3 + s_2 s_7^4 - 3 s_4^2 s_6 s_7 s_8 + 2 s_3 s_4 s_7^2 s_8 + 2 s_4^3 s_8^2=0\,,\nn \\
fz_1^4+3x_1^3\!\!&\!\!=\!\!&\!\!-12 s_4^3 s_6 s_7 s_8^2 + 6 s_4^4 s_8^3 + s_4^2 s_7^2 s_8 (7 s_6^2 - 4 s_5 s_7 + 8 s_3 s_8)+ s_7^5 (-s_2 s_6 + 2 s_1 s_7) \\ \!\!&\!\!\phantom{=}\!\!&\!\!  + s_7^4(s_3 (s_6^2 - 2 s_5 s_7) + 2 s_3^2 s_8)+ s_4 s_7^3 (-s_6^3 + 3 s_5 s_6 s_7 - 8 s_3 s_6 s_8 + 2 s_2 s_7 s_8)=0\,.\nn
\eea
Thus, the homology class of this complete intersection minus the
classes of $V(I_{(1)})$ and $V(I_{(3)})$ (with their respective
orders inside \eqref{eq:charge1matterF6}) yields the multiplicity
of the $\one_{1}$-matter.

The spectrum of charged matter is completed by the matter in the
adjoint representation $\mathbf{3}_0$ given in
the last row of Table~\ref{tab:poly6_matter}.
We recall that it  does not originate from codimension two fibers of
$X_{F_6}$, but is present if the divisor $S_{\text{SU}(2)}^b$ is a
higher genus curve in $B$, cf.~Section \ref{sec:6DSUGRA}. 
The multiplicity of charged hyper multiplets
in the adjoint is calculated using \eqref{eq:genusformula}.

We complete the matter spectrum of $X_{F_6}$ by the number of neutral
hyper multiplets, which is computed from \eqref{eq:Hneutral} using
the Euler number \eqref{eq:EulerNumbers} of $X_{F_6}$. It reads
\beq \label{eq:HneutF6}
H_{\text{neut}} = 14 + 11 [K_B^{-1}]^2 - 4 [K_B^{-1}] \cS_7 + 4 \cS_7^2 - 4 [K_B^{-1}] \cS_9 - 3 \cS_7 \cS_9 + 2 \cS_9^2 \,.
\eeq
The number $T$ of tensor multiplets is given by \eqref{eq:Tformula} and
we have $V=4$.
Finally, we use $S_{\text{SU}(2)}^b=\{s_8=0\}$, \eqref{eq:bmnF6},
the charged spectrum in Table~\ref{tab:poly6_matter} and
\eqref{eq:HneutF6} to check, following Appendix~\ref{app:Anomalies},
that all 6D anomalies are canceled.

We conclude this section by analyzing codimension three singularities
of the WSF of $X_{F_6}$. This determines the Yukawa points in a
compactification to 4D. All geometrically allowed Yukawa
couplings of the charged matter spectrum of $X_{F_6}$ are given
in Table~\ref{tab:poly6_yukawa}. In order to check the last Yukawa
coupling in Table~\ref{tab:poly6_yukawa}, we compute the minimal
associated primes of $I_{(2)}\cup I_{(4)}$. Indeed, it has a
codimension three associated prime, which confirms the presence of the
Yukawa coupling. We  emphasize that all Yukawa couplings allowed by gauge symmetry are indeed realized.
\begin{table}[htb!]
\begin{center}
\renewcommand{\arraystretch}{1.2}
\begin{tabular}{|c|c|}\hline
Yukawa & Locus \\ \hline
 \multirow{2}{*}{$\two_{-3/2}\cdot \two_{1/2}\cdot \one_{1} $}& $s_7 = s_8=0$ \\
&$s_4 s_5^3 - s_3 s_5^2 s_6 + s_2 s_5 s_6^2 - s_1 s_6^3=0$  \\ \hline
  $\two_{-3/2}\cdot \overline{\two_{1/2}} \cdot \one_{2} $ &$s_4 = s_7=s_8=0$\\ \hline
\multirow{2}{*}{$\overline{\one_{1}} \cdot \overline{\one_{1}} \cdot \one_{2} $}& $s_4 = s_7=0$  \\
&$s_3s_5^2-s_2s_5s_6+s_1s_6^2+s_2^2s_8-4s_1s_3s_8=0$  \\ \hline
 $\two_{1/2}\cdot \two_{1/2}\cdot \overline{\one_{1}}$ & $V(I_{(2)})\cap V(I_{(2)})\cap V(I_{(4)}) $ \\ \hline
\end{tabular}
\caption{\label{tab:poly6_yukawa}Codimension three loci and corresponding Yukawa couplings for $X_{F_6}$. }
\end{center}
\end{table}

%%%%%%%%%%%%%%%%%%%%%%%%%%%%%%%%%%%%%%%%%%%%%%%%%%%%%%%%%%%%%%%%%%%%%%%%%%%%%%%%%%%%%%%%%%%%%%%%%
\subsection{Fibrations with gauge groups of rank 3: selfdual polyhedra}
\label{sec:Rk3}
%%%%%%%%%%%%%%%%%%%%%%%%%%%%%%%%%%%%%%%%%%%%%%%%%%%%%%%%%%%%%%%%%%%%%%%%%%%%%%%%%%%%%%%%%%%%%%%%%

In the following section we analyze all toric hypersurface fibrations 
constructed from the four self-dual polyhedra $F_7$, $F_8$, $F_9$ and 
$F_{10}$. The rank of the gauge group of all these models is three and 
the rank of the MW-group assumes all values from zero to three. We 
encounter one novelty in the analysis of codimension two fibers in 
$X_{F_8}$ and $X_{F_{10}}$. There we find matter representations from 
non-split fibers at codimension two. The Calabi-Yau manifold
$X_{F_{10}}$ is also a generalization of the Tate form, allowing
for non-trivial coefficients of the monomials $x^3$ and $y^2$.
The vanishing loci of these coefficients support SU$(2)$ and SU$(3)$ 
gauge groups, respectively.

%%%%%%%%%%%%%%%%%%%%%%%%%%%%%%%%%%%%%%%%%%%%%%%%%%%%%%%%%%%%%%%%%%%%%%%%%%%%%%%%%%%%%%%%%%%%%%%%%
\subsubsection{Polyhedron $F_7$: $G_{F_7}=\text{U(1)}^3$}
\label{sec:polyF7}
%%%%%%%%%%%%%%%%%%%%%%%%%%%%%%%%%%%%%%%%%%%%%%%%%%%%%%%%%%%%%%%%%%%%%%%%%%%%%%%%%%%%%%%%%%%%%%%%%

\begin{figure}[H]
\center
\includegraphics[scale=0.4]{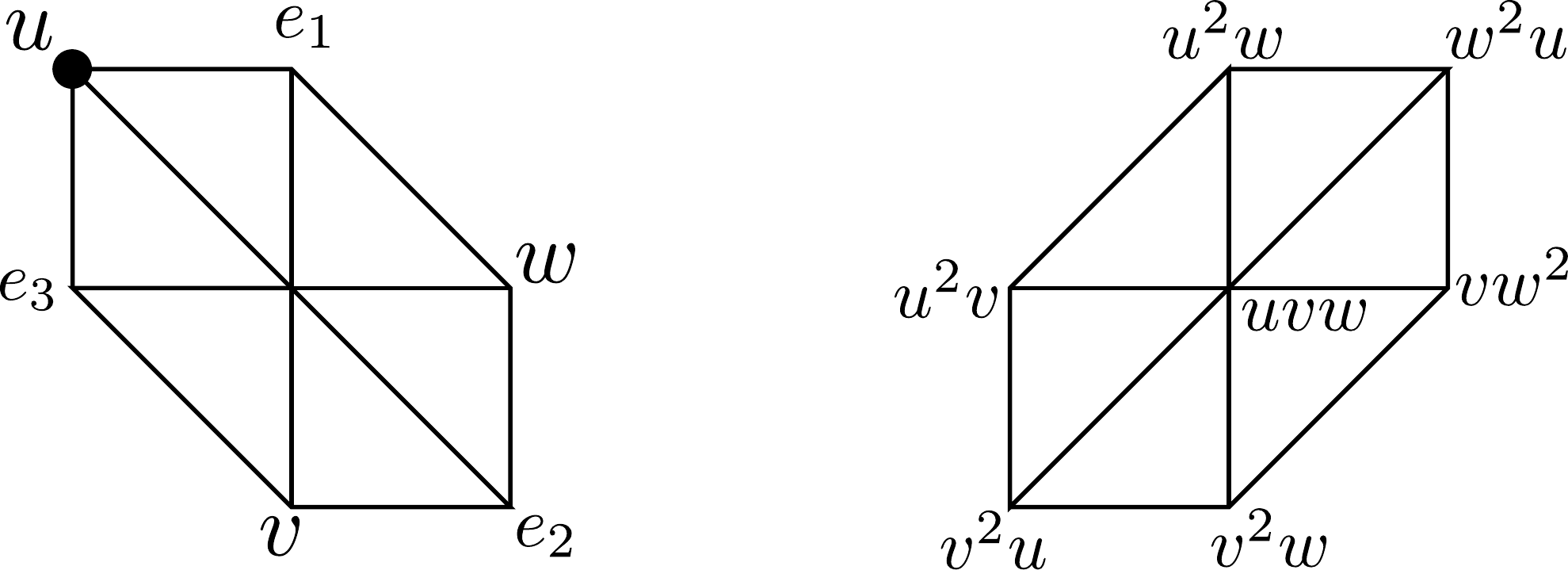}
\caption{\label{fig:poly7_toric}The toric diagram of polyhedron $F_7$ and its dual. The zero section is indicated by the dot.}
\end{figure}

We consider the elliptically fibered Calabi-Yau manifold $X_{F_7}$ with  
base $B$ and general elliptic fiber given by the elliptic curve $\mathcal{E}$ in $\mathbb{P}_{F_7}$.
The toric diagram of $\mathbb{P}_{F_7}=dP_3$ is depicted in
Figure~\ref{fig:poly7_toric}, where the polyhedron $F_7$ along with
a choice of homogeneous coordinates as well as its dual polyhedron
$F_{7}$ are shown. For brevity, we have set $e_i=1$, $\forall i$,
in the monomials  associated to the integral points in the dual polyhedron
by \eqref{eq:BatyrevFormula}.
The toric variety $\mathbb{P}_{F_7}$ is the del Pezzo surface $dP_3$, 
that is the blow-up of $\mathbb{P}^2$, cf.~Section \ref{sec:cubic}, at three points with blow-down map defined by
\beq \label{eq:F7blowup}
u \rightarrow e_1 e_3 u\,, \quad w\rightarrow e_1 e_2 w\,, \quad v\rightarrow e_2 e_3 v\,.
\eeq
The homogeneous coordinates on the fiber after this blow-up are 
$[u:v:w:e_1:e_2:e_3]$ and take values in the line bundles associated to 
the following divisor classes:
\begin{equation}
\label{tab:poly7_bundles}
\text{
\begin{tabular}{|c|c|}\hline
Section & Divisor class\\ \hline
$u$ & $H-E_1-E_3+\cS_9+[K_B]$ \\ \hline
$v$ & $H-E_2-E_3+\cS_9-\cS_7$\\ \hline
$w$ & $H-E_1-E_2$\\ \hline
$e_1$ & $E_1$\\ \hline
$e_2$ & $E_2$\\ \hline
$e_3$ &  $E_3$\\ \hline
\end{tabular}}
\end{equation}
The Stanley-Reisner ideal of $\mathbb{P}_{F_7}$ is given by
\begin{align}
SR=\{uw, ue_2, uv, e_1e_2, e_1v, e_1 e_3, wv,we_3,e_2 e_3\} \; .
\end{align}
Using \eqref{eq:BatyrevFormula} the hypersurface equation for
$X_{F_7}$ in the ambient space of the $\mathbb{P}_7$-fibration  
\eqref{eq:PFfibration} is obtained as
\beq\label{eq:pF7}
p_{F_7}= s_2 e_1e_3^2u^2v + s_3e_2e_3^2u v^2 + s_5 e_1^2 e_3 u^2 w + s_6 e_1 e_2 e_3 u v w + s_7 e_2^2 e_3 v^2 w  + s_8 e_1^2 e_2 u w^2 + s_9 e_1 e_2^2 w^2 v \, .
\eeq
Here the sections $s_i$ take values in the line bundles shown
in \eqref{eq:cubicsections}. We observe that \eqref{eq:pF7} can be
obtained from \eqref{eq:pF1} by the specialization $s_1=s_4=s_{10}=0$
and the map \eqref{eq:F7blowup}.

In total there are six rational sections of the elliptic 
fibration of $X_{F_7}$ with four of them being linearly independent 
\cite{Braun:2013nqa}. Their coordinates are
\begin{align}
\begin{split}\label{eq:sectionsF7}
\hat{s}_0=X_{F_7}\cap\{u=0\}:& [0:1:1:s_7:1:-s_9]\,,\\
\hat{s}_1=X_{F_7}\cap\{e_1=0\}:& [s_7:1:-s_3:0:1:1]\,,\\
\hat{s}_2=X_{F_7}\cap\{e_2=0\}:& [1:s_5:-s_2:1:0:1]\,,\\
\hat{s}_3=X_{F_7}\cap\{e_3=0\}:& [s_9:-s_8:1:1:1:0]\,,\\
X_{F_7}\cap\{v=0\}:& [1:0:1:1:s_5:-s_8]\,,\\
X_{F_7}\cap\{w=0\}:& [1:1:0:s_3:-s_2:1]\,,
\end{split}
\end{align}
where we choose $\hat{s}_0$ as the zero section and the sections 
$\hat{s}_m$, $m=1,2,3$, as the generators of the MW-group of $X_{F_7}$.

Using Nagell's algorithm we compute the Weierstrass form \eqref{eq:WSF} of \eqref{eq:pF7}.
The WS-coordinates of the section $\hat{s}_1$, $\hat{s}_2$ and $\hat{s}_3$ are given
by \eqref{eq:secsF71}, \eqref{eq:secsF72} and \eqref{eq:secsF73}, respectively. %, after setting the appropriate sections to zero.
Furthermore, we can obtain the functions $f$ and $g$ from \eqref{eq:fcubic} and \eqref{eq:gcubic}, respectively, by
setting $s_1=s_4=s_{10}=0$.
Since we do not find any codimension one singularity the total gauge group of $X_{F_7}$ is
\beq \label{eq:GF7}
	G_{F_7}=\text{U}(1)^3\,.
\eeq
Thus we compute the Shioda map \eqref{eq:ShiodaMap} of all rational
sections $\hat{s}_m$, $m=1,2,3$, as 
\begin{align} \label{eq:ShiodaF7}
\sigma (\hat{s}_m)=S_m-S_0 +[K_B]-\pi(S_m\cdot S_0)\,,
\end{align}
where we use the following intersection relations:
\begin{align}
\begin{split} \label{eq:intrelsF3}
\pi(S_1\cdot S_0)=\cS_7\,,&\,\,\,\,\,\pi(S_2\cdot S_0)=0\,,\qquad \pi(S_3\cdot S_0)=\cS_9\,,\\
 %\pi(S_1^2)=[K_B]\,,\quad  \,, \\
 %\pi(S_2^2)=[K_B]\,,\quad \pi(S_2\cdot S_0)=0 \,, \\
 %\pi(S_3^2)=[K_B]\,,\quad \pi(S_3\cdot S_0)=\cS_9 \,, \\
\pi(S_1 \cdot S_2)=0\,,&\,\, \pi(S_1\cdot S_3)=0 \,,\qquad \pi(S_2\cdot S_3)=0 \,.
\end{split}
\end{align}
Using \eqref{eq:ShiodaF7} and these intersection relations, together 
with \eqref{eq:SP^2}, we  compute the height pairing 
\eqref{eq:anomalycoeff} as
\begin{align}
\label{eq:bmnF7}
b_{mn}=- \pi(\sigma(\hat{s}_m)\sigma(\hat{s}_n))=\begin{pmatrix}
                                                  2[K_B^{-1}]+2\cS_7 & [K_B^{-1}]+\cS_7 & [K_B^{-1}]+\cS_7+\cS_9\\
                                                  [K_B^{-1}]+\cS_7 & 2[K_B^{-1}] & [K_B^{-1}]+\cS_9\\
						  [K_B^{-1}]+\cS_7+\cS_9 & [K_B^{-1}]+\cS_9 & 2[K_B^{-1}]+2\cS_9
                                                 \end{pmatrix}_{mn} \, .
\end{align}

Next, we analyze the codimension two singularities of the WSF of
$X_{F_7}$ to determine the charged matter spectrum.
We find ten codimension two singularities, which lead to the matter 
representations and the corresponding
codimension two fibers in $X_{F_7}$ given in the first and third column
of Table~\ref{tab:poly7_matter}, respectively. Given
all these codimension two fibers and the positions of the rational 
sections, we readily compute the U(1)-charges using \eqref{eq:U1charge}. 
Here, the starting point of the  analysis, that has led to the complete
charged matter spectrum in Table~\ref{tab:poly7_matter}, are the three 
complete intersections 
\eqref{eq:charge1Matter} evaluated for the three rational sections
$\hat{s}_1$, $\hat{s}_2$ and $\hat{s}_3$. Then, we determine all
their minimal associated prime ideals using Singular \cite{Singular},
which precisely produces all the ten ideals $I_{(k)}$, 
shown in Table~\ref{tab:poly7_matter}. 
\begin{table}[H]
\begin{center}
\renewcommand{\arraystretch}{1.2}
\begin{tabular}{|c|@{}c@{}|c|@{}c@{}|}\hline
Representation & Multiplicity & Fiber & Locus \\ \hline

$\one_{(1,1,0)}$ & $\begin{array}{c} (2 [K_B^{-1}]-\cS_9) \\ \times ([K_B^{-1}]+\cS_7-\cS_9) \end{array}$ & \rule{0pt}{1.15cm}\parbox[c]{2.3cm}{\includegraphics[width=2.3cm]{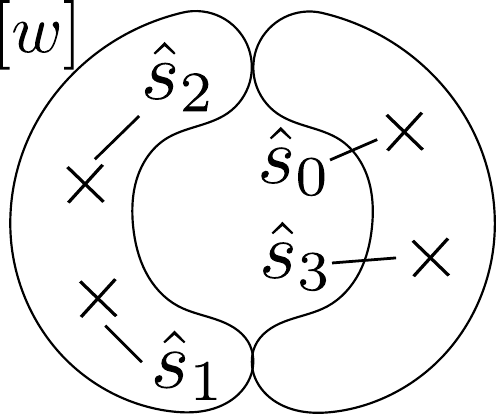}} & $V(I_{(1)}):=\{ s_2=s_3=0 \}$ \\[0.75cm] \hline

$\one_{(0,-1,0)}$ & $\begin{array}{c} (2 [K_B^{-1}]-\cS_9) \\ \times(2 [K_B^{-1}]-\cS_7) \end{array}$ & \rule{0pt}{1.15cm}\parbox[c]{2.3cm}{\includegraphics[width=2.3cm]{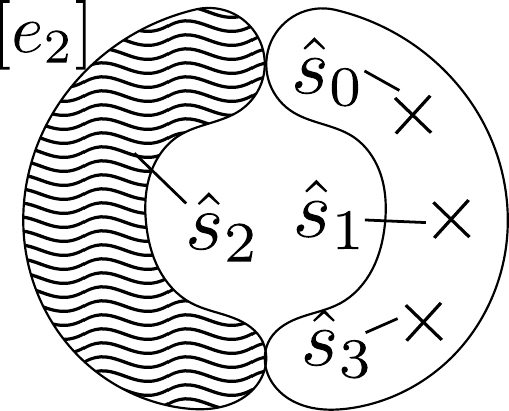}} & $V(I_{(2)}):=\{ s_2=s_5=0 \}$ \\[0.75cm] \hline

$\one_{(2,1,1)}$ & $\cS_7([K_B^{-1}]+\cS_7-\cS_9)$ & \rule{0pt}{1.15cm}\parbox[c]{2.3cm}{\includegraphics[width=2.3cm]{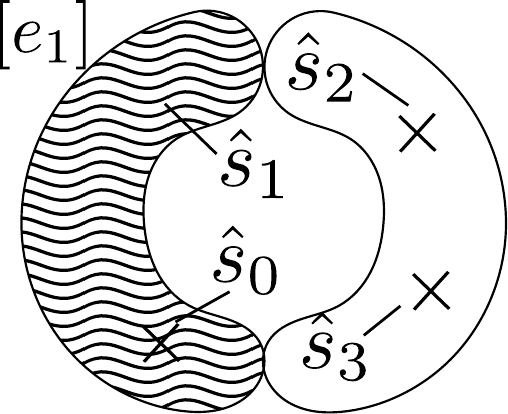}} & $V(I_{(3)}):=\{ s_3=s_7=0 \}$ \\[0.75cm] \hline

$\one_{(0,1,1)}$ & $\begin{array}{c} (2 [K_B^{-1}]-\cS_7) \\ \times([K_B^{-1}]-\cS_7+\cS_9) \end{array}$ & \rule{0pt}{1.15cm}\parbox[c]{2.3cm}{\includegraphics[width=2.3cm]{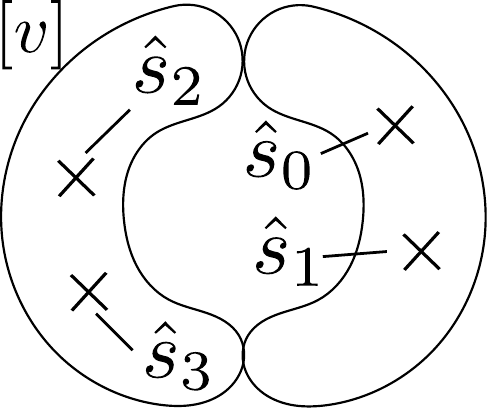}} & $V(I_{(4)}):=\{ s_5=s_8=0 \}$ \\[0.75cm] \hline

$\one_{(-2,-1,-2)}$ & $\cS_7 \cS_9$ & \rule{0pt}{1.15cm}\parbox[c]{2.3cm}{\includegraphics[width=2.3cm]{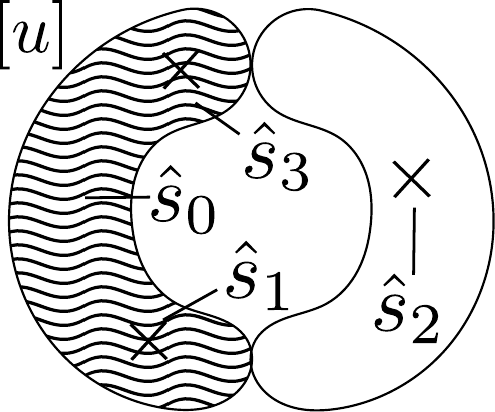}} & $V(I_{(5)}):=\{ s_7=s_9=0 \}$ \\[0.75cm] \hline

$\one_{(1,1,2)}$ & $\cS_9([K_B^{-1}]-\cS_7+\cS_9)$ & \rule{0pt}{1.15cm}\parbox[c]{2.3cm}{\includegraphics[width=2.3cm]{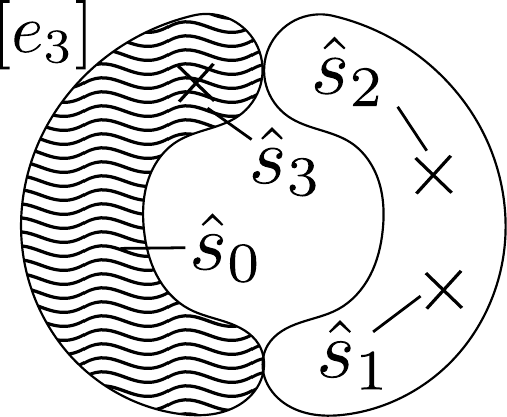}} & $V(I_{(6)}):=\{ s_8=s_9=0 \}$ \\[0.75cm] \hline

$\one_{(1,0,0)}$ & $\begin{array}{c} 2 [K_B^{-1}](4[K_B^{-1}]-2\cS_7+\cS_9) \\ -2 (2 [K_B^{-1}]-\cS_7) \\ \times([K_B^{-1}]-\cS_7+\cS_9) \end{array}$ & \rule{0pt}{1.15cm}\parbox[c]{2.3cm}{\includegraphics[width=2.3cm]{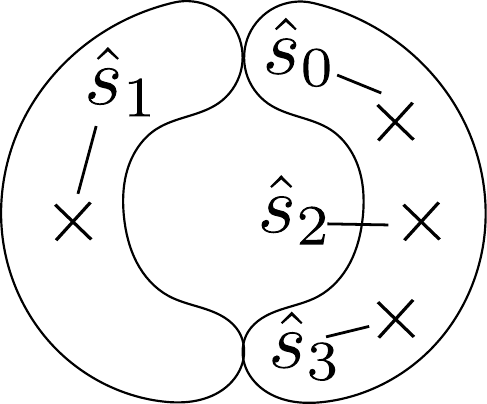}} & $\begin{array}{c} V(I_{(7)}):=\{ s_3s_8-s_5s_7=0 \\ s_2s_8^2-s_5s_6s_8+s_5^2s_9=0 \} \end{array}$ \\[0.75cm] \hline

$\one_{(0,0,1)}$ & $\begin{array}{c} 2 [K_B^{-1}](2[K_B^{-1}]-\cS_7+2\cS_9) \\ -2 \cS_9([K_B^{-1}]-\cS_7+\cS_9) \end{array}$ & \rule{0pt}{1.15cm}\parbox[c]{2.3cm}{\includegraphics[width=2.3cm]{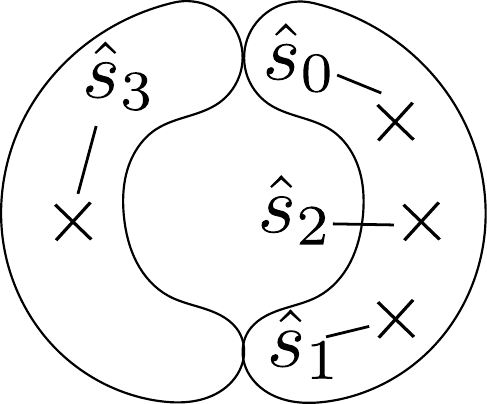}} & $\begin{array}{c} V(I_{(8)}):=\{ s_3s_8-s_2s_9=0 \\ s_5s_9^2-s_6s_8s_9+s_7s_8^2=0 \} \end{array}$  \\[0.75cm] \hline

$\one_{(1,0,1)}$ & $\begin{array}{c} 2 [K_B^{-1}]([K_B^{-1}]+\cS_7 \\ +\cS_9)-2\cS_7 \cS_9 \end{array}$ & \rule{0pt}{1.15cm}\parbox[c]{2.3cm}{\includegraphics[width=2.3cm]{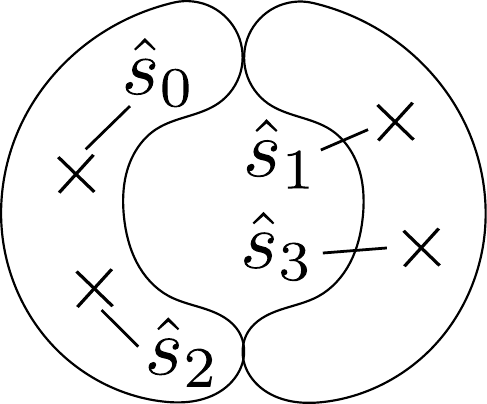}} & $\begin{array}{c} V(I_{(9)}):=\{ s_5s_7-s_2s_9=0 \\ s_3s_9^2-s_6s_7s_9+s_7^2s_8=0 \} \end{array}$ \\[0.75cm] \hline

$\one_{(1,1,1)}$ & $\begin{array}{c} 4[K_B^{-1}]^2 + 2[K_B^{-1}](\cS_7+\cS_9) \\ +2(-\cS_7^2+\cS_7\cS_9-\cS_9^2) \end{array}$
& \rule{0pt}{1.15cm}\parbox[c]{2.3cm}{\includegraphics[width=2.3cm]{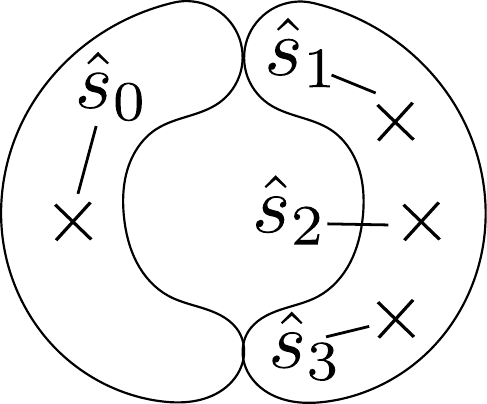}} & $\begin{array}{c} V(I_{(10)}):=\{ s_2^2s_5s_7+s_3^2s_5^2 \\ -s_2s_3s_5s_6+s_2^2s_3s_8=0 \\ s_2s_5^2s_9+s_3s_5^2s_8 \\ -s_2s_5s_6s_8+s_2^2s_8^2=0 \}\end{array} $ \\[0.75cm] \hline

\end{tabular}
\caption{\label{tab:poly7_matter}Charged matter representations under U(1)$^3$ and corresponding codimension two fibers of $X_{F_7}$.}
\end{center}
\end{table}
We note that
the varieties $V(I_{(k)})$, $k=1,\ldots,6$, are precisely the loci
where the six rational sections given in \eqref{eq:sectionsF7} are 
ill-defined and have to acquire a fiber component.

The multiplicities of the charged hyper multiplets are presented in
Table~\ref{tab:poly7_matter}. These have been computed following
Section~\ref{sec:SUGRA6D} as the homology classes of the respective 
codimension two varieties $V(I_{(k)})$, $k=1,\ldots,10$. In the cases 
where $V(I_{(k)})$ is not a complete intersection, the multiplicities 
are calculated using the resultant technique similarly 
as described in Section~\ref{sec:polyF6}.

To complete the matter spectrum of $X_{F_7}$ we compute the number of neutral
hyper multiplets from \eqref{eq:Hneutral} using
the Euler number $\chi(X_{F_6})$ given in \eqref{eq:EulerNumbers}. 
It reads
\beq \label{eq:HneutF7}
H_{\text{neut}} = 15 + 7 [K_B^{-1}]^2 - 2 [K_B^{-1}] \cS_7 + 2 \cS_7^2 - 2 [K_B^{-1}] \cS_9 - 2 \cS_7 \cS_9 + 2 \cS_9^2\,.
\eeq
We note that there are $T$ tensor multiplets and $V=3$ vector 
multiplets. These results together with the charged spectrum in 
Table~\ref{tab:poly7_matter} and \eqref{eq:HneutF7} as well as the 
height pairing  \eqref{eq:bmnF7}  allows us, following 
Appendix~\ref{app:Anomalies}, to check cancelation of all 6D anomalies.   

To conclude this section we analyze the codimension three singularities
of the WSF of $X_{F_7}$. This determines the Yukawa points in a
compactification to 4D. The geometrically allowed Yukawa couplings of 
$X_{F_7}$ are given in Table~\ref{tab:poly7_yukawa}. Here we confirm
the presence of each Yukawa coupling by checking that the intersection 
of the relevant varieties is codimension three in $B$.

\begin{table}[H]
\begin{center}
\renewcommand{\arraystretch}{1.2}
\begin{tabular}{|c|c|}\hline
Yukawa & Locus \\ \hline
$\one_{(1,1,0)} \cdot \one_{(0,-1,0)} \cdot \overline{\one_{(1,0,0)}}$ & $s_2=s_3=s_5=0$\\ \hline

$\one_{(1,1,0)} \cdot \overline{\one_{(2,1,1)}} \cdot \one_{(1,0,1)}$ & $s_2=s_3=s_7=0$ \\ \hline

$\one_{(1,1,0)} \cdot \one_{(0,0,1)} \cdot \overline{\one_{(1,1,1)}}$ & $s_2=s_3=s_5s_9^2-s_6s_8s_9+s_7s_8^2=0$ \\ \hline

$\one_{(0,-1,0)} \cdot \one_{(0,1,1)} \cdot \overline{\one_{(0,0,1)}}$ & $s_2=s_5=s_8=0$ \\ \hline

$\one_{(0,-1,0)} \cdot \overline{\one_{(1,0,1)}} \cdot \one_{(1,1,1)}$ & $s_2=s_5=s_3s_9^2-s_6s_7s_9+s_7^2s_8=0$ \\ \hline

$\one_{(2,1,1)} \cdot \one_{(-2,-1,-2)} \cdot \one_{(0,0,1)}$ & $s_3=s_7=s_9=0$ \\ \hline

$\overline{\one_{(2,1,1)}} \cdot \one_{(1,0,0)} \cdot \one_{(1,1,1)}$ & $s_3=s_7=s_2s_8^2-s_5s_6s_8+s_5^2s_9=0$ \\ \hline

$\one_{(0,1,1)} \cdot \overline{\one_{(1,1,2)}} \cdot \one_{(1,0,1)}$ & $s_5=s_8=s_9=0$\\ \hline

$\one_{(-2,-1,-2)} \cdot \one_{(1,1,2)} \cdot \one_{(1,0,0)}$ & $s_7=s_8=s_9=0$ \\ \hline

$\one_{(-2,-1,-2)} \cdot \one_{(1,0,1)} \cdot \one_{(1,1,1)}$ & $s_7=s_9=s_3s_5^2-s_2s_5s_6+s_2^2s_8=0$ \\ \hline

$\overline{\one_{(1,1,2)}} \cdot \one_{(0,0,1)} \cdot \one_{(1,1,1)}$ & $s_8=s_9=s_2^2s_7-s_2s_3s_6+s_3^2s_5=0$\\ \hline

\multirow{3}{*}{$\one_{(1,0,0)} \cdot \one_{(0,0,1)} \cdot \overline{\one_{(1,0,1)}}$} & $s_2s_9^3+s_7^2s_8^2-s_6s_7s_8s_9=0$ \\
 & $s_3s_9^2-s_6s_7s_9+s_7^2s_8=0$ \\
 & $s_5s_9^2-s_6s_8s_9+s_7s_8^2=0$ \\ \hline
\end{tabular}
\caption{\label{tab:poly7_yukawa}Codimension three loci and corresponding Yukawa couplings for $X_{F_7}$.}
\end{center}
\end{table}

%%%%%%%%%%%%%%%%%%%%%%%%%%%%%%%%%%%%%%%%%%%%%%%%%%%%%%%%%%%%%%%%%%%%

%%%%%%%%%%%%%%%%%%%%%%%%%%%%%%%%%%%%%%%%%%%%%%%%%%%%%%%%%%%%%%%%%%%%%%%%%%%%%%%%%%%%%%%%%%%%%%%%%
\subsubsection{Polyhedron $F_{8}$: $G_{F_8}=\text{SU(2)}^2\times\text{U(1)}$}
\label{sec:polyF8}
%%%%%%%%%%%%%%%%%%%%%%%%%%%%%%%%%%%%%%%%%%%%%%%%%%%%%%%%%%%%%%%%%%%%%%%%%%%%%%%%%%%%%%%%%%%%%%%%%

\begin{figure}[H]
\center
\includegraphics[scale=0.4]{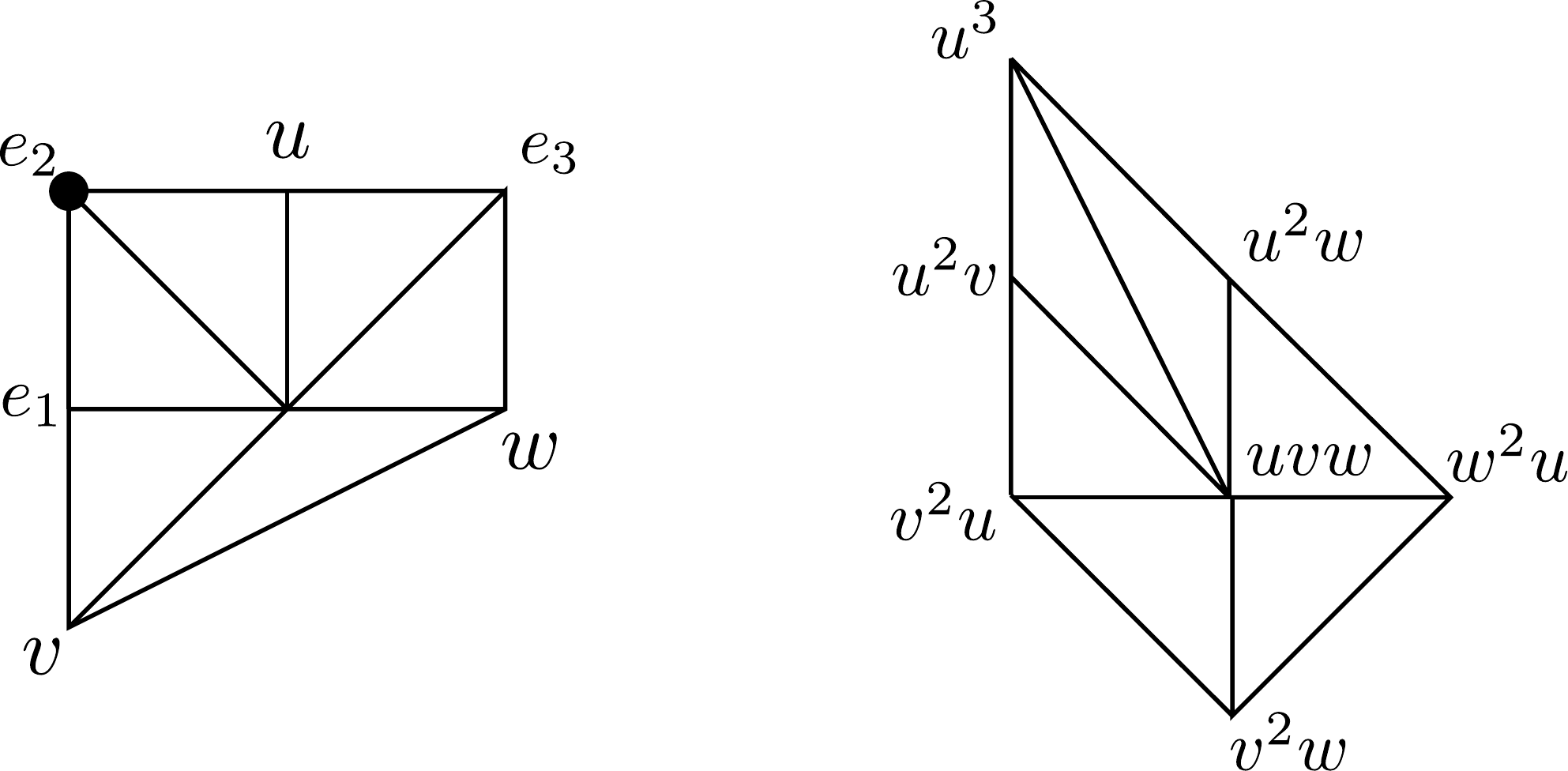}
\caption{\label{fig:poly8_toric}The toric diagram of polyhedron $F_8$ and its dual. The zero section is indicated by the dot.}
\end{figure}

In this section, we consider the elliptically fibered Calabi-Yau 
manifold $X_{F_8}$ over an arbitrary base $B$ and with general elliptic 
fiber  given by the elliptic curve $\mathcal{E}$ in $\mathbb{P}_{F_8}$.
In Figure~\ref{fig:poly8_toric} the toric diagram of $F_8$ and of its 
dual polyhedron are depicted.
For brevity, we have set $e_i=1$, $\forall i$, in the monomials that are 
associated by \eqref{eq:BatyrevFormula} 
to the integral points of the dual polyhedron.
The toric variety $\mathbb{P}_{F_8}$ is $\mathbb{P}^2$,
cf.~Section~\ref{sec:cubic}, blown-up at three non-generic points. The 
blow-down map reads
\beq \label{eq:F8blowup}
u \rightarrow e_1 e_2^2 e_3 u \,,\quad v\rightarrow e_1 v \,,\quad w\rightarrow e_3 w \, .
\eeq
The homogeneous coordinates on the fiber after this blow-up are 
$[u:v:w:e_1:e_2:e_3]$ and take values in the line bundles associated
to the divisors given by:
\beq
\label{tab:poly8_bundles}
\text{
\begin{tabular}{|c|c|}\hline
Section & Divisor class\\ \hline
$u$ & $H-E_1-E_2-E_3+\cS_9+[K_B]$ \\ \hline
$v$ & $H-E_1-\cS_7+\cS_9$\\ \hline
$w$ & $H-E_3$\\ \hline
$e_1$ & $E_1-E_2$\\ \hline
$e_2$ & $E_2$\\ \hline
$e_3$ &  $E_3$\\ \hline
\end{tabular}}
\eeq
The Stanley-Reisner ideal of $\mathbb{P}_{F_8}$ reads
\begin{align}
SR&=\{ u w, uv, ue_1,e_3v, e_3 e_1, e_3 e_2, w e_1, we_2, ve_2 \} \, .
\end{align}
Using \eqref{eq:BatyrevFormula}, the hypersurface equation for 
$X_{F_8}$ in the ambient space \eqref{eq:PFfibration} with $F_i=F_8$ is
\beq\label{eq:pF8}
p_{F_8}=s_1 e_1^2 e_2^4 e_3^2 u^3 + s_2 e_1^2 e_2^3 e_3 u^2 v + s_3 e_1^2 e_2^2 u v^2 +
 s_5 e_1 e_2^2 e_3^2 u^2 w + s_6 e_1 e_2 e_3 u v w +  s_7 e_1 v^2 w + s_8 e_3^2 u w^2\, ,
\eeq
where the classes of the sections $s_i$ are given 
in \eqref{eq:cubicsections}.
We note that restriction of \eqref{eq:pF1} as $s_4=s_9=s_{10}=0$
and application of the map \eqref{eq:F8blowup} also 
leads to \eqref{eq:pF8}.

There are two rational sections of the fibration of $X_{F_8}$.
Their coordinates are
\begin{align}
\begin{split}\label{eq:sectionsF8}
\hat{s}_0=X_{F_8}\cap\{e_2=0\}&:\quad [s_7:1:1:-s_8:0:1]\,,\\
\hat{s}_1=X_{F_8}\cap\{e_3=0\}&:\quad [s_7:1:-s_3:1:1:0]\,,
\end{split}
\end{align}
where we choose $\hat{s}_0$ as the zero section.

\begin{figure}[t!]
\center
\includegraphics[scale=0.8]{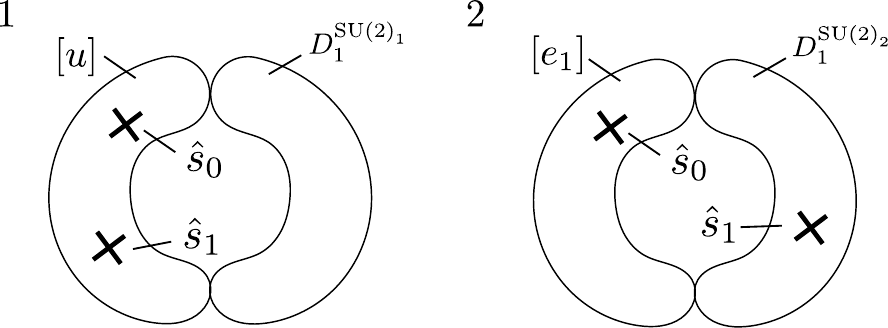}
\caption{\label{fig:poly8_codim1}Codimension one fibers of $X_{F_8}$ at $s_7=0$ and $s_8=0$ in $B$. The crosses denote the intersections with the two sections.}
\end{figure}

We compute the Weierstrass form \eqref{eq:WSF} of \eqref{eq:pF8} using
Nagell's algorithm. The WS-coordinates of the section
$\hat{s}_1$ are given
by \eqref{eq:WScoordsSecF3} after setting $s_4=s_9=0$.
Additionally, the functions $f$ and $g$ take the form of
\eqref{eq:fcubic} and \eqref{eq:gcubic}, respectively, after
setting $s_4=s_9=s_{10}=0$.
This allows us to find all codimension one singularities
of the WSF of $X_{F_8}$. We find two $I_2$-singularities over the divisors
$\mathcal{S}^b_{\text{SU}(2)_1}=\{s_7=0\}\cap B$ and 
$\mathcal{S}^b_{\text{SU}(2)_2}=\{s_8=0\}\cap B$ in $B$. Along these
divisors the constraint \eqref{eq:pF8} factorizes as
\begin{align}
\begin{split}\label{eq:SU2F8}
\SU2_1&:\quad p_{F_8}|_{s_7=0}= u \cdot q_2\, ,\\
\SU2_2&:\quad p_{F_8}\vert_{s_8=0}= e_1 \cdot q_3 \; .
\end{split}
\end{align}
Here $q_2$, $q_3$ are the polynomials of degree $n$ in $[u:v:w]$
that remain after factoring out $u$ and $e_1$, respectively. These are clearly $I_2$-fibers, cf.~Figure~\ref{fig:poly8_codim1}, giving rise to two SU(2) gauge groups.
In summary, the  gauge group of $X_{F_8}$ is
\beq \label{eq:GF8}
	G_{F_8}=\text{SU}(2)^2\times\text{U}(1)\,.
\eeq
Similar as in Section~\ref{sec:polyF6}, we obtain  the classes of the 
$\SU2$ Cartan divisors in $X_{F_8}$ given by
\begin{align}\label{eq:CartansF8}
D^{\SU2_1}_{1}=[s_7]-[u] \, , \quad
D^{\SU2_2}_{1}=[s_8]-[e_1]\, .
\end{align}
This allows us to compute the Shioda map \eqref{eq:ShiodaMap}
of the section $\hat{s}_1$ as
\begin{align} \label{eq:ShiodaF8}
\sigma (\hat{s}_1)&=S_1-S_0 +  [K_B]	+\frac{1}{2} D^{\SU2_2}_{1}\, ,
\end{align}
where we use that the two sections in \eqref{eq:sectionsF8} do not 
intersect, implying $\pi(S_1\cdot S_0)=0$, and
\begin{align}
S_1 \cdot C^{SU(2)_1}_{-\alpha _1}=0 \, ,\quad 
S_1 \cdot C^{SU(2)_2}_{-\alpha _1}=1\, ,
\end{align}
which follows from Figure~\ref{fig:poly8_codim1}.
We  compute the height pairing \eqref{eq:anomalycoeff} of 
$\hat{s}_1$ using \eqref{eq:SP^2} as
\begin{align}
\label{eq:bmnF8}
  b_{11}=\frac{3}{2}[K_B^{-1}]+\frac{1}{2}\cS_7-\frac{1}{2}\cS_9 \,.
\end{align}

In order to determine the charged matter spectrum we analyze the 
codimension two singularities of the WSF of $X_{F_8}$. For the singlets
we compute the associated prime ideals of 
the complete intersection \eqref{eq:charge1Matter} associated to 
$\hat{s}_1$.
We find seven codimension two singularities. \\\\Five of
these lead to the matter representations and the corresponding
codimension two fibers in $X_{F_8}$ given in the first and third column
of Table~\ref{tab:poly8_matter}, respectively.
The representations under $G_{F_8}$ have been determined 
following the general procedure outlined in Section \ref{sec:6DSUGRA}.
The remaining two loci, $s_7=s_6^2-4s_3s_8=0$ and $s_8=s_6^2-4s_5s_7=0$,
support Type $III$ singularities, which do not lead to additional matter.\\ \\

\begin{table}[H]
\begin{center}
\footnotesize
\renewcommand{\arraystretch}{1.2}
\begin{tabular}{|c|c|c|c|}\hline
Representation & Multiplicity & Fiber & Locus \\ \hline
$(\two,\two)_{-1/2}$ & $\cS_7([K_B^{-1}]-\cS_7+\cS_9)$ & \rule{0pt}{1.5cm}\parbox[c]{2.9cm}{\includegraphics[width=2.9cm]{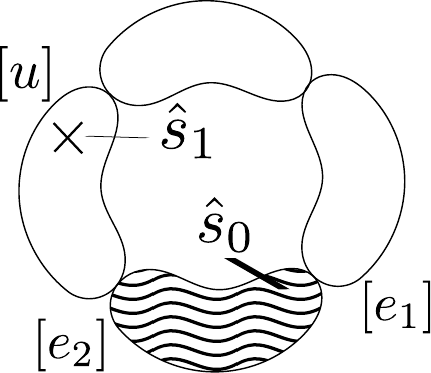}} & $V(I_{(1)}:=\{s_7=s_8=0\}$ \\[1.2cm] \hline

$(\one,\two)_{1/2}$ & $\begin{array}{c} 2([K_B^{-1}]-\cS_7+\cS_9) \\ (3[K_B^{-1}]-\cS_9) \end{array}$ & \rule{0pt}{1.5cm}\parbox[c]{2.9cm}{\includegraphics[width=2.9cm]{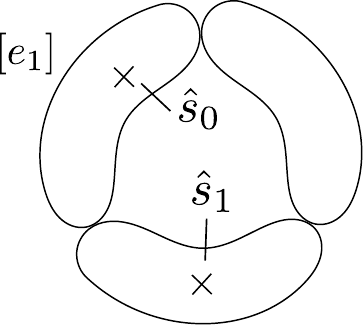}} & 
$\begin{array}{c} V(I_{(2)}):=\{s_8=0 \\ s_3^2 s_5^2 - s_2 s_3 s_5 s_6 + s_1 s_3 s_6^2 + s_2^2 s_5 s_7 \\- 2 s_1 s_3 s_5 s_7 - s_1 s_2 s_6 s_7 + s_1^2 s_7^2=0 \} \end{array}$ \\[1.2cm] \hline

$(\two,\one)_1$ & $\cS_7 ([K_B^{-1}]+\cS_7-\cS_9)$ & \rule{0pt}{1.5cm}\parbox[c]{2.9cm}{\includegraphics[width=2.9cm]{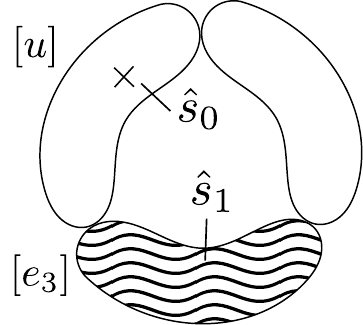}} & $V(I_{(3)}):=\{s_3=s_7=0\}$ \\[1.2cm] \hline

$(\two,\one)_0$ & $\cS_7 (5[K_B^{-1}]-\cS_7-\cS_9)$ & \rule{0pt}{1.5cm}\parbox[c]{2.9cm}{\includegraphics[width=2.9cm]{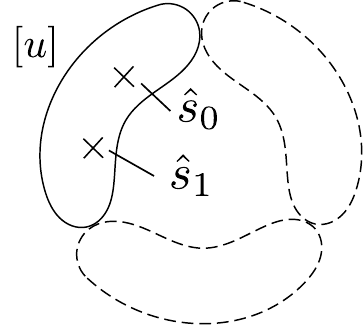}} & %\parbox[c]{\widthof{$s_3 s_5^2 - s_2 s_5 s_6 + s_1 s_6^2 +$}}{\center
$\begin{array}{c} V(I_{(4)}):=\{ s_7=0 \\ s_3 s_5^2 - s_2 s_5 s_6 + s_1 s_6^2 \\ + s_2^2 s_8 - 4 s_1 s_3 s_8=0 \} \end{array}$\\[1.2cm] \hline

$(\one,\one)_1$ & $\begin{array}{c} 6 [K_B^{-1}]^2 + 3 [K_B^{-1}] \cS_7 \\ - \cS_7^2 - 5 [K_B^{-1}] \cS_9 + \cS_9^2 \end{array}$ & \rule{0pt}{1.4cm}\parbox[c]{2.9cm}{\includegraphics[width=2.9cm]{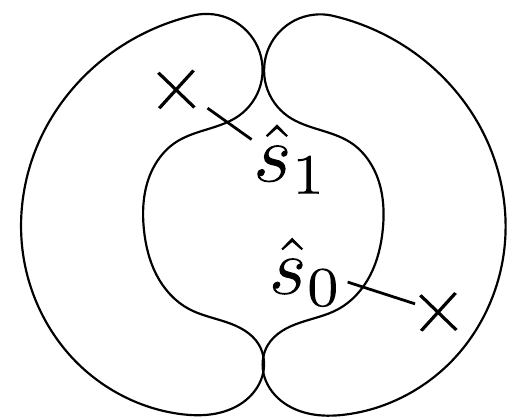}} &
$\begin{array}{c} V(I_{(5)}):=\{ s_3 s_6 - s_2 s_7=0 \\ s_7 (-s_2 s_6 + 2 s_1 s_7) + s_3 (s_6^2 \\ - 2 s_5 s_7) + 2 s_3^2 s_8=0 \} \backslash V(I_{(3)}) \end{array}$ \\[1.1cm] \hline \hline

$(\three,\one)_0$ & \rule{0pt}{0.5cm}$1+\cS_7 \frac{\cS_7-[K_B^{-1}]}{2}$ & Figure~\ref{fig:poly8_codim1} & $s_7=0$ \\[0.05cm] \hline

$(\one,\three)_0$ & \rule{0pt}{0.5cm}$\begin{array}{c} 1+\frac{\cS_9-\cS_7}{2} \\ \times ([K_B^{-1}]-\cS_7+\cS_9) \end{array}$ & Figure~\ref{fig:poly8_codim1} & $s_8=0$ \\[0.1cm] \hline
\end{tabular}
\caption{\label{tab:poly8_matter}Charged matter representations under SU$(2)^2\times$U(1) and corresponding codimension two fibers of $X_{F_8}$. The adjoint matter is included for completeness.}
\end{center}
\end{table}

We note the following subtlety. The fiber supporting the matter in the 
representation $(\two,\one)_0$ is non-split in the 
sense of \cite{Bershadsky:1996nh}.\footnote{We note that non-split 
fibers at codimension two are not classified.} This means that the 
constraint \eqref{eq:pF8} of the elliptic fiber at the codimension two 
locus $V(I_{(4)})$ does not fully factorize, as one expects, over the 
field $K$ of meromorphic functions on $B$. It only factorizes
in a field extension where certain square roots of the coefficients 
$s_i$ are allowed. In fact, the fiber we obtain at the locus 
$V(I_{(4)})$ allowing only for factorizations in $K$ is a line and a 
singular conic. However, a singular conic describes two 
lines, i.e.~the conic has to be factorized into two linear constraints 
describing two lines. This 
factorization requires introducing square roots of some combinations of 
the $s_i$. Geometrically, this means that these lines are interchanged 
by a codimension three monodromy  (that occur only on threefold bases  
$B$).\footnote{We thank Dave Morrison and Ron Donagi for explanations 
related to non-split codimension two fibers and singular conics. We also 
thank Sakura Sch\"afer-Nameki and Craig Lawrie for explaining to us the 
corresponding box graphs \cite{Hayashi:2014kca}. } The two lines of
the non-split fiber that are interchanged by this monodromy 
are the dashed $\mathbb{P}^1$'s in the fourth row of 
Table~\ref{tab:poly8_matter}. 

The multiplicities of the charged hyper multiplets are presented in
Table~\ref{tab:poly8_matter} in the second row. These have been computed 
following
Section \ref{sec:SUGRA6D}. Since the locus of the representation 
$(\one,\one)_1$ is not a complete intersection we compute its 
multiplicity as described in Section \ref{sec:polyF6} by subtraction
of the locus $V(I_{(3)})$ with its appropriate order.

The number of neutral hyper multiplets completes the matter spectrum of $X_{F_8}$. It is computed from \eqref{eq:Hneutral} using
the Euler number \eqref{eq:EulerNumbers} of $X_{F_8}$ and reads
\beq \label{eq:HneutF8}
H_{\text{neut}} = 15 + 11 [K_B^{-1}]^2 - 5 [K_B^{-1}] \cS_7 + 3 \cS_7^2 - 4 [K_B^{-1}] \cS_9 - 2 \cS_7 \cS_9 + 2 \cS_9^2\,.
\eeq
The base-dependent number $T$ of tensor multiplets is given by 
\eqref{eq:Tformula} and we have $V=7$.
Finally, we use this together with 
$\mathcal{S}^b_{\text{SU}(2)_1}=\mathcal{S}_7^b$,
$\mathcal{S}^b_{\text{SU}(2)_2}=\mathcal{S}_8^b$,
\eqref{eq:bmnF8}, the charged spectrum in Table 
\ref{tab:poly8_matter} and \eqref{eq:HneutF8} to check
cancelation of all 6D anomalies in \eqref{eq:6dAnomalies}.

To obtain the Yukawa points in a compactification to 4D we analyze codimension three singularities
of the WSF of $X_{F_8}$.
All Yukawa couplings of the charged matter spectrum of $X_{F_8}$ are 
given in Table \ref{tab:poly8_yukawa}. Clearly, all relevant loci here 
are codimension three.
\begin{table}[H]
\begin{center}
\renewcommand{\arraystretch}{1.2}
\begin{tabular}{|c|c|}\hline
Yukawa & Locus \\ \hline
$(\two,\two)_{-1/2} \cdot \overline{(\one,\two)_{1/2}} \cdot (\two,\one)_{1}$ & $s_8 = s_7=s_2 =0$ \\ \hline
$(\two,\two)_{-1/2} \cdot (\one, \two)_{1/2} \cdot (\two,\one)_{0} $ & $s_8 = s_7= s_2 s_5^2 - s_1 s_5 s_6 + s_0 s_6^2=0$ \\ \hline
$(\two,\one)_{1} \cdot (\two,\one)_{0} \cdot \overline{(\one,\one)_{1}} $ & $s_7 = s_2= -s_1 s_5 s_6 + s_0 s_6^2 + s_1^2 s_8 =0$ \\ \hline
\end{tabular}
\caption{\label{tab:poly8_yukawa}Codimension three loci and corresponding Yukawa couplings for $X_{F_8}$.}
\end{center}
\end{table}

%%%%%%%%%%%%%%%%%%%%%%%%%%%%%%%%%%%%%%%%%%%%%%%%%%%%%%%%%%%%%%%%%%%%%%%%%%%%%%%%%%%%%%%%%%%%%%%%%
\subsubsection{Polyhedron $F_9$: $G_{F_9}=\text{SU(2)}\times \text{U(1)}^2$}
\label{sec:polyF9}
%%%%%%%%%%%%%%%%%%%%%%%%%%%%%%%%%%%%%%%%%%%%%%%%%%%%%%%%%%%%%%%%%%%%%%%%%%%%%%%%%%%%%%%%%%%%%%%%%

\begin{figure}[H]
\center
\includegraphics[scale=0.4]{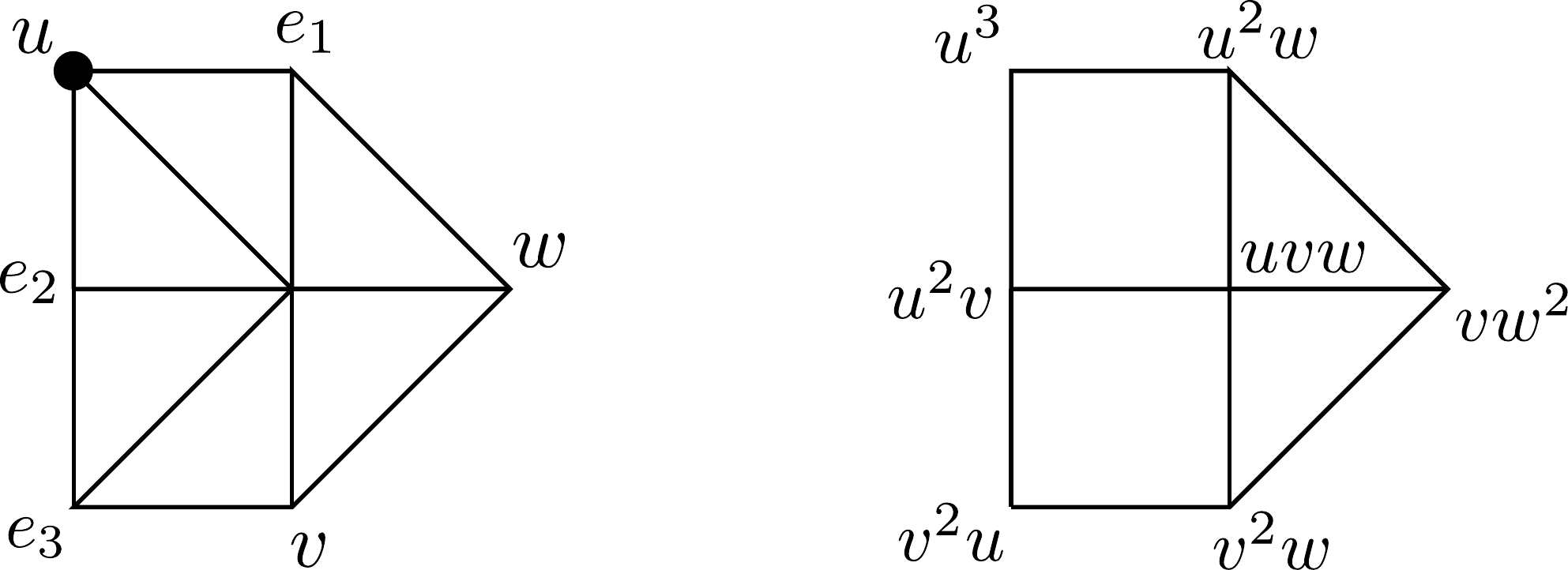}
\caption{\label{fig:poly9_toric}The toric diagram of polyhedron $F_9$ and its dual. The zero section is indicated by the dot.}
\end{figure}

Here, we consider the elliptically fibered Calabi-Yau manifold $X_{F_9}$ 
over a base $B$ and with general elliptic fiber  given by the toric 
hypersurface $\mathcal{C}_{F_9}$ in $\mathbb{P}_{F_9}$.
The toric data of $\mathbb{P}_{F_9}$ is depicted in
Figure \ref{fig:poly9_toric}. Here the polyhedron $F_9$ along with
a choice of homogeneous coordinates as well as its dual polyhedron is shown. For brevity, we have set $e_i=1$, $\forall i$,
in the monomials associated to the integral points of the dual
by \eqref{eq:BatyrevFormula}.
The toric variety $\mathbb{P}_{F_9}$ is obtained from $\mathbb{P}^2$,
cf.~Section \ref{sec:cubic}, by the three non-generic blow-ups
\begin{align}\label{eq:F9blowup}
u \rightarrow e_1 e_2 e_3 u \,,\quad w\rightarrow e_1 w \,,\quad v\rightarrow e_2 e_3^2 v \, .
%u \rightarrow e_1 u\; &, \; w\rightarrow e_1 w\\
%u \rightarrow e_2 u\; &, \; v\rightarrow e_2 v\\
%e_2 \rightarrow e_3 e_2\; &, \; v\rightarrow e_3 v
\end{align}
After these blow-ups the homogeneous coordinates on the fiber are 
$[u:v:w:e_1:e_2:e_3]$, which take values in the line bundles associated 
to the divisors:
\begin{equation}\label{tab:poly9_bundles}
\text{
\begin{tabular}{|c|c|}\hline
Section & Divisor class\\ \hline
$u$ & $H-E_1-E_2+\cS_9+[K_B]$ \\ \hline
$v$ & $H-E_2-E_3+\cS_9-\cS_7$\\ \hline
$w$ & $H-E_1$\\ \hline
$e_1$ & $E_1$\\ \hline
$e_2$ & $E_2-E_3$\\ \hline
$e_3$ &  $E_3$\\ \hline
\end{tabular}}
\end{equation}
The Stanley-Reisner ideal of $\mathbb{P}_{F_9}$ then reads
\begin{align}
SR=\{ uw ,u v, u e_3, e_1 v, e_1 e_3, e_1 e_2, w e_3, w e_2, v e_2\} \, .
\end{align}
Using \eqref{eq:BatyrevFormula} the hypersurface equation for $X_{F_9}$ in the ambient space \eqref{eq:PFfibration} with $F_i=F_9$ is
\beq
\label{eq:pF9}
p_{F_9}= s_1 e_1^2 e_2^2 e_3 u^3 + s_2 e_1 e_2^2 e_3^2 u^2 v + s_3 e_2^2e_3^3u v^2 + s_5 e_1^2 e_2 u^2 w + s_6 e_1 e_2 e_3 u v w + s_7 e_2 e_3^2 v^2 w + s_9 e_1 v w^2  ,
\eeq
where the divisor classes of the sections $s_i$ are given in 
\eqref{eq:cubicsections}. We see that \eqref{eq:pF9} can also be 
obtained from \eqref{eq:pF1} by the specialization $s_4=s_8=s_{10}=0$
and the map \eqref{eq:F9blowup}.

There are four rational sections on $X_{F_6}$ with one linear relation 
between them. Their coordinates are
\begin{align}
\begin{split}\label{eq:sectionsF9}
\hat{s}_0=X_{F_9}\cap\{u=0\}&:\quad [0:1:1:s_7:-s_9:1]\, ,\\
\hat{s}_1=X_{F_9}\cap\{e_3=0\}&:\quad [1:s_5:1:1:-s_9:0]\, ,\\
\hat{s}_2=X_{F_9}\cap\{e_1=0\}&:\quad [s_7:1:-s_3:0:1:1]\, , \\
X_{F_9}\cap\{v=0\}&:\quad [1:0:s_1:1:1:-s_5] \, .
\end{split}
\end{align}
We choose $\hat{s}_0$ as the zero section and $\hat{s}_m$, $m=1,2$,
as the generators of the MW-group of $X_{F_9}$.

\begin{figure}[t!]
\center
\includegraphics[scale=0.6]{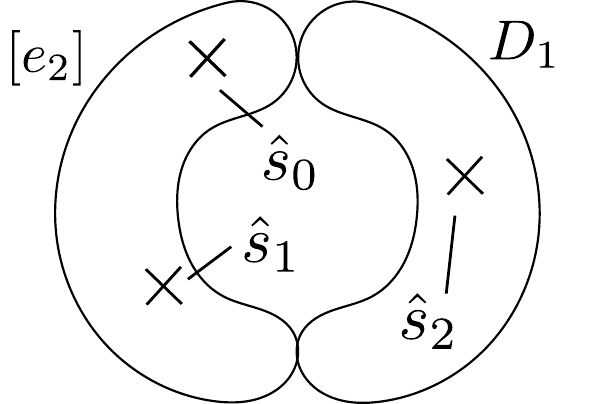}
\caption{\label{fig:poly9_codim1}Codimension one fiber
of $X_{F_9}$ at $s_9=0$ in $B$. The crosses denote the intersections
with the sections.}
\end{figure}

Employing Nagell's algorithm we compute the Weierstrass form \eqref{eq:WSF} of \eqref{eq:pF9}. The WS-coordinates of the sections
$\hat{s}_1$ and $\hat{s}_2$ are given
by \eqref{eq:secsF51} and \eqref{eq:secsF52}, respectively,
after setting the appropriate  
sections $s_i$ to zero.
To get the functions $f$ and $g$ we specialize \eqref{eq:fcubic} and \eqref{eq:gcubic} as $s_4=s_8=s_{10}=0$.
Using $f$ and $g$ we can compute the discriminant $\Delta$ to find all codimension one singularities
of the WSF of $X_{F_9}$.
We find one $I_2$-singularity over the divisor
$\mathcal{S}^b_{\text{SU}(2)}=\{s_9=0\}\cap B$ in $B$. Along this
divisor the constraint \eqref{eq:pF9} factorizes as
\begin{align}\label{eq:SU2F9}
\SU2&: \quad p_{F_9}|_{s_9=0}= e_2 \cdot q_3 \, ,
\end{align}
where $q_3$ is the polynomial that remains after factoring out
$e_2$. This is an $I_2$-fiber, cf.~Figure~\ref{fig:poly9_codim1}, giving rise to an $\SU2$ gauge group.
In summary, the  gauge group of $X_{F_9}$ is
\beq \label{eq:GF9}
	G_{F_9}=\text{SU}(2)\times\text{U}(1)^2\,.
\eeq
Similar as in Section \ref{sec:polyF6}, we compute the divisor 
class of the Cartan divisor as
\begin{align} \label{eq:CartansF9}
D_{1}=[s_9]-[e_2]=[s_9]-E_2+E_3 \, .
\end{align}
Employing this we obtain the Shioda map \eqref{eq:ShiodaMap}
of the sections as
\begin{align} \label{eq:ShiodaF9}
\sigma (\hat{s}_m)=S_m-S_0 +[K_B]-\delta_{m,2}\mathcal{S}_7+\frac{1}{2} \delta_{m,2} D_{1} \, .
\end{align}
Here $S_0$, $S_m$ denote the divisor classes of the sections
$\hat{s}_0$, $\hat{s}_m$, $m=1,2$, respectively, and  we used 
the relation
\begin{align}
S_m \cdot C_{-\alpha _1}&=\delta_{m,2}\,,\qquad 
\pi(S_0\cdot S_m)=\delta_{m,2}\mathcal{S}_7\,,\qquad \pi(S_1\cdot S_2)=0\,,
\end{align}
which can be deduced from Figure \ref{fig:poly9_codim1} and
the coordinates \eqref{eq:sectionsF9} of the sections, respectively.
Using these relations and \eqref{eq:SP^2}, we compute the height pairing 
\eqref{eq:anomalycoeff} of the $\hat{s}_m$ as
\begin{align}\label{eq:bmnF9}
 b_{mn}=\begin{pmatrix} 2[K_B^{-1}] & [K_B^{-1}]+\cS_7\\
 [K_B^{-1}]+\cS_7 & 2[K_B^{-1}]+2\cS_7-\frac12\cS_9 \end{pmatrix}_{mn} \, .
\end{align}
Turning to the charged matter spectrum we analyze the codimension two 
singularities of the WSF of $X_{F_9}$. The matter in non-trivial 
representations of the non-Abelian part of $G_{F_9}$ follows 
directly from the discriminant with $s_9=0$, whereas the charged singlets
are most easily seen from the primary decomposition of the complete 
intersections \eqref{eq:charge1Matter} corresponding to the 
two sections $\hat{s}_1$, $\hat{s}_2$, respectively.
We find eight codimension two singularities. Seven of
these lead to the matter representations and the corresponding
codimension two fibers in $X_{F_9}$ given in the first and third column
of Table~\ref{tab:poly9_matter}, respectively. Here, the corresponding  
representation under the gauge group are determined following the 
general procedure explained in Section \ref{sec:ellipticCurvesWithRP}.
\begin{table}[h!]
\begin{center}
\small
\renewcommand{\arraystretch}{1.2}
\begin{tabular}{|c|@{}c@{}|c|@{}c@{}|}\hline
Representation & Multiplicity & Fiber & Locus \\ \hline
$\one_{(1,2)}$ & $\cS_7 ([K_B^{-1}]+\cS_7-\cS_9)$ & \rule{0pt}{1.2cm}\parbox[c]{2.4cm}{\includegraphics[width=2.4cm]{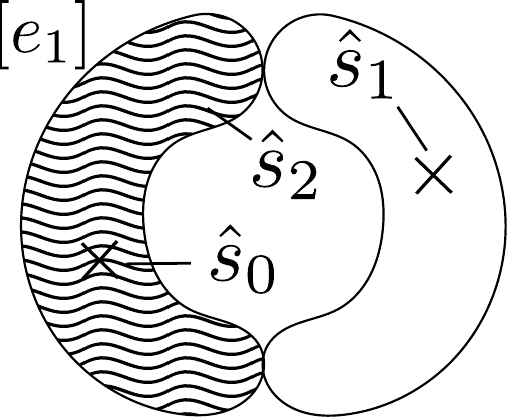}} & $V(I_{(1)}):=\{s_7=s_3=0 \}$ \\[0.9cm] \hline

$\one_{(1,0)}$ & $\begin{array}{c} (2[K_B^{-1}]-\cS_7) \\ (3[K_B^{-1}]-\cS_7-\cS_9) \end{array}$ & \rule{0pt}{1.2cm}\parbox[c]{2.4cm}{\includegraphics[width=2.4cm]{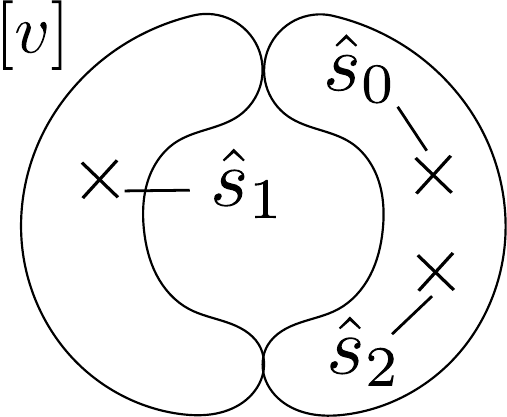}} & $V(I_{(2)}):=\{s_5=s_1=0 \}$ \\[0.9cm] \hline

$\one_{(0,1)}$ & $\begin{array}{c} (3[K_B^{-1}]-\cS_9) \\ \times(2[K_B^{-1}]+2\cS_7-\cS_9) \\ -2\cS_7([K_B^{-1}]+\cS_7-\cS_9) \end{array}$ & \rule{0pt}{1.2cm}\parbox[c]{2.4cm}{\includegraphics[width=2.4cm]{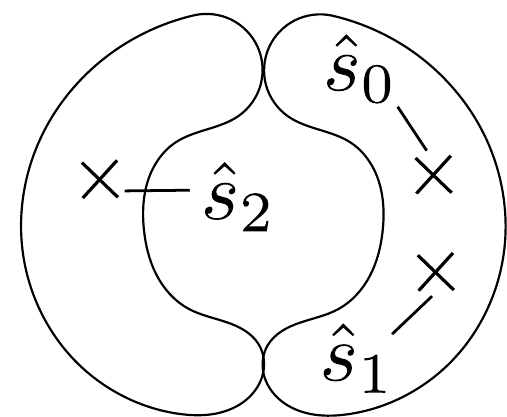}} & $\begin{array}{c} V(I_{(3)}):=\{ s_2 s_7^2+s_3^2s_9-s_3s_6s_7=0 \\ s_5 s_3-s_7s_1=0 \} \backslash V(I_{(1)}) \end{array}$ \\[0.9cm] \hline

$\one_{(1,1)}$ & $\begin{array}{c} 6[K_B^{-1}]^2+[K_B^{-1}] \\ \times(4\cS_7-2\cS_9) -2\cS_7^2 \end{array}$ & \rule{0pt}{1.2cm}\parbox[c]{2.4cm}{\includegraphics[width=2.4cm]{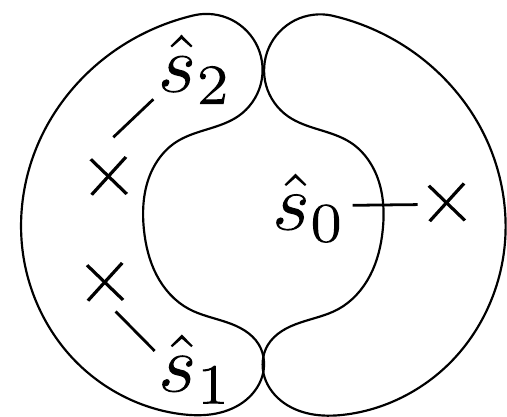}} & $\begin{array}{c} V(I_{(4)}):=\{s_2 s_9 s_7^2 +s_3^2 s_9^2 \\ -s_3s_6s_9s_7-s_7^3s_5=0 \\ s_1s_9s_7+s_5(s_3s_9-s_7s_6)=0 \} \\ \backslash (V(I_{(1)}) \cup V(I_{(5)}) \cup V(I_{(6)})) \end{array}$\\[0.9cm] \hline%$(2[K_B^{-1}]+2\cS_7)3[K_B^{-1}]-$\\$-2\cS_7([K_B^{-1}]+\cS_7-\cS_9)-$\\$-\cS_9(2[K_B^{-1}]-\cS_7)-3\cS_7 \cS_9$} \\ \hline

$\two_{(-1,-1/2)}$ & $\cS_9 (2[K_B^{-1}]-\cS_7)$ & \rule{0pt}{1.3cm}\parbox[c]{2.4cm}{\includegraphics[width=2.4cm]{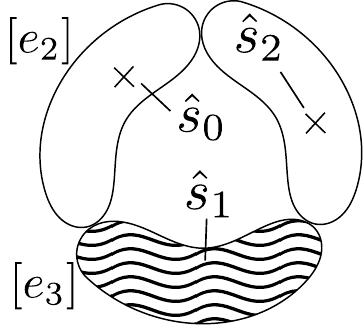}} & $V(I_{(5)}):=\{ s_9=s_5=0 \}$ \\[0.9cm] \hline

$\two_{(1,3/2)}$ & $\cS_7 \cS_9$ & \rule{0pt}{1.3cm}\parbox[c]{2.4cm}{\includegraphics[width=2.4cm]{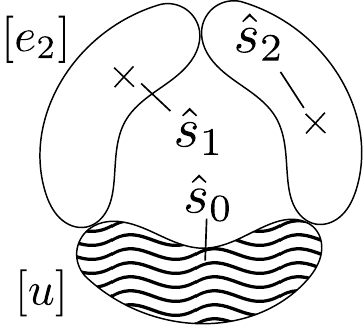}} & $V(I_{(6)}):=\{ s_9=s_7=0 \}$ \\[0.9cm] \hline

$\two_{(0,-1/2)}$ & $2\cS_9 (3[K_B^{-1}]-\cS_9)$ & \rule{0pt}{1.2cm}\parbox[c]{2.4cm}{\includegraphics[width=2.4cm]{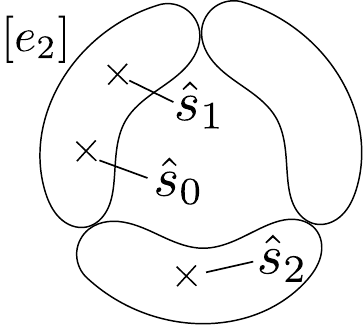}} & $\begin{array}{c} V(I_{(7)}):=\{ s_9=0 \\ s_3^2 s_5^2 + s_3 (-s_6 s_2 s_5 \\ + s_6^2 s_1 - 2 s_7 s_5 s_1) + s_7 (s_2^2 s_5 \\ - s_6 s_2 s_1 + s_7 s_1^2)=0 \} \end{array} $ \\[0.9cm] \hline \hline

$\three_{(0,0)}$ & \rule{0pt}{0.55cm}$1+\cS_9\frac{(\cS_9-[K_B^{-1}])}{2}$ & Figure \ref{fig:poly9_codim1} & $s_9=0$\\[0.05cm] \hline
\end{tabular}
\caption{\label{tab:poly9_matter}Charged matter representations under SU$(2)\times$U(1)$^2$ and corresponding codimension two fibers of $X_{F_9}$. The adjoint matter is included for completeness.}
\end{center}
\end{table}
At the remaining locus
$s_9=s_6^2 - 4 s_5 s_7= 0$, the fiber is of Type $III$, cf.~the 
discussion in Section \ref{sec:polyF6}. Thus it does not support any 
additional matter. 

We note that at the locus $V(I_{(2)})$ the linearly dependent section in 
\eqref{eq:sectionsF9} is singular.
At the locus $V(I_{(7)})$ corresponding to matter in the representation 
$\two_{(0,-1/2)}$ the elliptic curve \eqref{eq:pF9} does not naively 
factor into three rational curves. To correctly derive its splitting one 
needs to compute the associated prime ideals of the elliptic fiber at 
this locus. We find three ideals corresponding to
three rational curves, which indeed intersect as an $I_3$-fiber.

The multiplicities of the charged hyper multiplets are presented in
Table \ref{tab:poly9_matter}. These have been computed following
Section \ref{sec:SUGRA6D}.
The multiplicities of $V(I_{(3)})$ and $V(I_{(4)})$ must be calculated 
by appropriately subtracting  the  multiplicities of the loci 
$V(I_{(1)})$, $V(I_{(5)})$ and $V(I_{(6)})$, respectively, as described 
in Section \ref{sec:polyF6} and indicated in Table 
\ref{tab:poly9_matter}.

We complete the matter spectrum of $X_{F_9}$ by the number of neutral
hyper multiplets, which is computed from \eqref{eq:Hneutral} using
the Euler number $\chi(X_{F_9})$ given in\eqref{eq:EulerNumbers}. It 
reads
\beq \label{eq:HneutF9}
H_{\text{neut}} = 15 + 11 [K_B^{-1}]^2 - 4 [K_B^{-1}] \cS_7 + 2 \cS_7^2 - 6 [K_B^{-1}] \cS_9 + 2 \cS_9^2 \,.
\eeq
The number of tensor multiplets $T$ can be obtained by 
\eqref{eq:Tformula} and we have $V=5$ vector multiplets.
To check that the anomalies are canceled we use 
$S_{\text{SU}(2)}^b=\{s_9=0\}$, \eqref{eq:bmnF9},
the charged spectrum in Table \ref{tab:poly9_matter} and
\eqref{eq:HneutF9}, following the discussion in  Appendix \ref{app:Anomalies}.

We conclude this section by analyzing codimension three singularities
of the WSF of $X_{F_9}$ determining the  Yukawa points in a
compactification to 4D. In Table \ref{tab:poly9_yukawa} all 
geometrically allowed Yukawa couplings of the charged matter spectrum of 
$X_{F_9}$ are given.
\begin{table}[htb]
\begin{center}
\renewcommand{\arraystretch}{1.2}
\begin{tabular}{|c|c|}\hline
Yukawa & Locus \\ \hline
$\one_{(1,0)} \cdot \one_{(0,1)} \cdot \overline{\one_{(1,1)}}$ & $s_1=s_5=s_2s_7^2+s_3^2s_9-s_3s_6s_7$\\ \hline
$\overline{\one_{(0,1)}} \cdot \two_{(-1,-1/2)} \cdot \two_{(1,3/2)}$ & $s_5=s_7=s_9=0$ \\ \hline
$\one_{(1,1)} \cdot \two_{(-1,-1/2)} \cdot \two_{(0,-1/2)}$ & $s_5=s_9=0=s_1^3(s_3s_6^2+s_1s_7^2-s_2s_6s_7)$ \\ \hline
$\one_{(1,0)} \cdot \two_{(-1,-1/2)} \cdot \overline{\two_{(0,-1/2)}}$ & $s_5=s_9=s_1=0$ \\ \hline
$\one_{(1,2)} \cdot \overline{\two_{(1,3/2)}} \cdot \two_{(0,-1/2)}$ & $s_7=s_9=s_3=0$ \\ \hline
$\overline{\one_{(1,1)}} \cdot \two_{(1,3/2)} \cdot \two_{(0,-1/2)}$ & $s_7=s_9=0=s_1^2(s_3s_5^2+s_1s_6^2-s_2s_5s_6)$ \\ \hline
\end{tabular}
\caption{Codimension three loci and corresponding Yukawa couplings for $X_{F_9}$.}
\label{tab:poly9_yukawa}
\end{center}
\end{table}

%%%%%%%%%%%%%%%%%%%%%%%%%%%%%%%%%%%%%%%%%%%%%%%%%%%%%%%%%%%%%%%%%%%%%%%%%%%%%%%%%%%%%%%%%%%%%%%%%
\subsubsection{Polyhedron $F_{10}$ \& the generalized Tate form: $G_{F_{10}}=\text{SU(3)}\times\text{SU(2)}$}
\label{sec:polyF10}
%%%%%%%%%%%%%%%%%%%%%%%%%%%%%%%%%%%%%%%%%%%%%%%%%%%%%%%%%%%%%%%%%%%%%%%%%%%%%%%%%%%%%%%%%%%%%%%%%

The elliptically fibered Calabi-Yau manifold $X_{F_{10}}$ is constructed 
as the fibration of the elliptic curve $\mathcal{E}$ in 
$\mathbb{P}_{F_{10}}=\mathbb{P}^{2}(1,2,3)$ over an arbitrary base $B$.
Thus, the generic fiber in $\mathbb{P}_{F_{10}}$ is just the elliptic 
curve $\mathcal{E}$ in  the Tate form 
of the WS equation \eqref{eq:WSF}, however, with non-trivial 
coefficients in front of the terms $x^3$ and $y^2$, that are usually set 
to one.

\begin{figure}[t!]
\center
\includegraphics[scale=0.4]{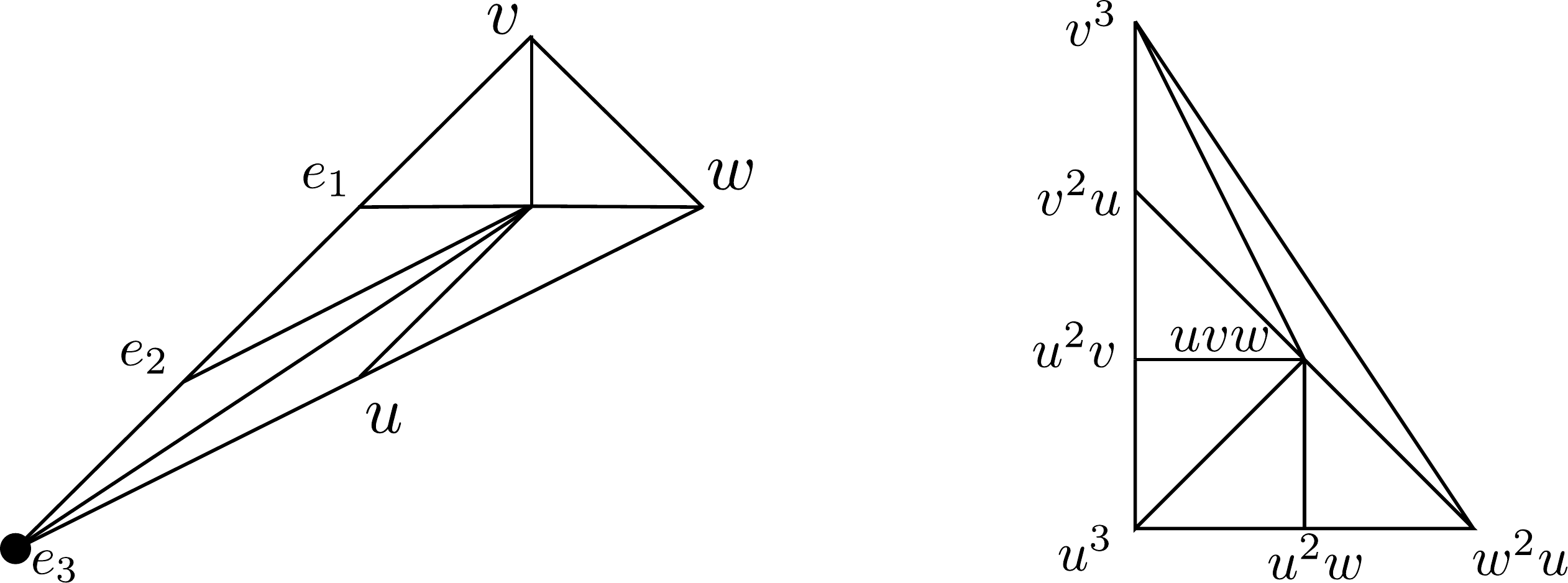}
\caption{\label{fig:poly10_toric}Polyhedron $F_{10}$ and its dual. The zero section is indicated by the dot.}
\end{figure}

The toric data of $\mathbb{P}_{F_{10}}$ is encoded in the polyhedron 
$F_{10}$, that is shown along with
a choice of homogeneous coordinates and together with its dual 
polyhedron in Figure~\ref{fig:poly10_toric}. In the dual polyhedron, 
we have set $e_i=1$, $\forall i$, in the monomials that are associated 
to its integral points by \eqref{eq:BatyrevFormula}. We obtain 
$\mathbb{P}_{F_{10}}=\mathbb{P}^2(1,2,3)$ by blowing-up $\mathbb{P}^2$, 
see Section \ref{sec:cubic}, in the following way:
\begin{align}\label{eq:F10blowup}
u \rightarrow e_1 e_2^2 e_3^3 u \,, \quad v\rightarrow e_1 e_2 e_3 v\, .
%u \rightarrow e_1 u\; &, \; v\rightarrow e_1 v\\
%u \rightarrow e_2 u\; &, \; e_1\rightarrow e_2 e_1\\
%u \rightarrow e_3 u\; &, \; e_2\rightarrow e_3 e_2
\end{align}
After these blow-ups, the homogeneous coordinates on the fiber  
are $[u:v:w:e_1:e_2:e_3]$ and take values in the line bundles associated 
to the divisor classes given by:
\begin{equation}\label{tab:poly10_bundles}
\text{
\begin{tabular}{|c|c|}\hline
Section & Divisor class\\ \hline
$u$ & $H-E_1-E_2-E_3+\cS_9+[K_B]$ \\ \hline
$v$ & $H-E_1+\cS_9-\cS_7$\\ \hline
$w$ & $H$\\ \hline
$e_1$ & $E_1-E_2$\\ \hline
$e_2$ & $E_2-E_3$\\ \hline
$e_3$ & $E_3$\\ \hline
\end{tabular}}
\end{equation}
The Stanley-Reisner ideal of $\mathbb{P}_{F_{10}}$ reads
\begin{align}
SR&=\{ u e_2, u e_1, uv, e_3 e_1, e_3 v, e_3 w, e_2 v, e_2 w, e_1 w\} \, .
\end{align}
Employing \eqref{eq:BatyrevFormula} we obtain the hypersurface equation 
for $X_{F_{10}}$ in the ambient space \eqref{eq:PF3} with general
fiber given by $\mathbb{P}^2(1,2,3)$ given by
\beq\label{eq:pF10}
p_{F_{10}}=s_1 e_1^2e_2^4e_3^6 u^3+s_2 e_1^2e_2^3e_3^4 u^2 v+s_3e_1^2e_2^2e_3^2 u v^2 + s_4 e_1^2 e_2 v^3 +s_5 e_1e_2^2e_3^3 u^2 w + s_6 e_1e_2e_3 u v w + s_8 u w^2 \, .
\eeq
Here the sections $s_i$ take values in the line bundles associated to 
the divisor classes shown in \eqref{eq:cubicsections}.
The hypersurface equation \eqref{eq:pF10} can also be obtained by the 
specialization $s_7=s_9=s_{10}=0$ and the map \eqref{eq:F10blowup} 
applied to \eqref{eq:pF1}.There is one rational section of the fibration 
of $X_{F_{10}}$. Its coordinate is
\begin{align}\label{eq:sectionsF10}
\hat{s}_0=X_{F_{10}}\cap \{e_3=0\}&:\quad [s_4:1:1:1:-s_8:0]\,.
\end{align}
Since this is the only section we naturally choose it as the zero section.

\subsubsection*{Comparison with the Tate form}

Before proceeding with the analysis of $X_{F_{10}}$, let us pause to 
compare with the standard elliptic fibration with fiber in 
$\mathbb{P}^{2}(1,2,3)$, that is the Tate form. We emphasize that 
\eqref{eq:pF10} describing $X_{F_{10}}$ can be viewed as a two-fold 
generalization of the standard Tate form of an elliptic fibration 
studied e.g.~in \cite{Morrison:1996na,Morrison:1996pp,Friedman:1997yq} , 
which is produced in the special case $s_8=1$ and $s_4=1$.

First, we identify the usual projective coordinates $[z:x:y]$ on 
$\mathbb{P}^{2}(1,2,3)$ and the Tate coefficients $a_i$. They read
\begin{align}
\begin{split}\label{eq:identTate}
	z\equiv e_3\,,\qquad x&\equiv v\,,\qquad y\equiv w\,,\\
	a_1\equiv s_6\,,\qquad a_2\equiv s_3\,,\qquad a_3&\equiv s_5\,,\qquad a_4\equiv s_2\,,\qquad a_6\equiv s_1\,.
\end{split}
\end{align}
Using this, we see that \eqref{eq:pF10} is indeed in Tate form.
 However, we note that there are two additional coefficients, namely 
$s_8$ and $s_4$, that do not have an analog in the standard Tate form, 
because they correspond to the  coefficients of $y^2$ and $x^3$, that 
are typically set to one.  As we see below, at the vanishing loci 
of these sections we find a SU$(3)$- and a SU$(2)$-singularity,
respectively.  Thus, allowing for non-trivial $s_4$, $s_8$,
is the first of the two aforementioned generalizations of $X_{F_{10}}$,
compared to the standard Tate form.

In addition, consistently imposing $s_4=s_8=1$
fixes the degrees of freedom in constructing the fibration of 
the elliptic curve $\mathcal{C}_{F_{10}}$ over the base $B$. 
Indeed, setting $s_4=s_8=1$ requires their divisor classes to be 
trivial, $[s_8]=0$, $[s_4]=0$. This fixes $\mathcal{S}_7$ and 
$\mathcal{S}_9$ according to \eqref{eq:cubicsections} as
\beq
	\mathcal{S}_7\stackrel{!}{=}[K_B]\,,\qquad \mathcal{S}_9\stackrel{!}{=}2\mathcal{S}_7=2[K_B]\,.
\eeq
Thus, the fibration $X_{F_{10}}$ is completely fixed in terms of the 
canonical bundle $K_B$ of the base $B$.
As we see from \eqref{tab:poly10_bundles}, the coordinates
$u$ and $v$ transform as a section of the line bundles $K_B^{3}$ and 
$K_B$, respectively. Using the $\mathbb{C}^*$-action on 
$\mathbb{P}_{F_{10}}$, this is equivalent, employing
\eqref{eq:identTate}, to
\beq
	x\in \mathcal{O}_B([K_B^{-2}])\,,\qquad 
 y\in\mathcal{O}_B([K_B^{-3}])\,.
\eeq
Thus, we see that by relaxing $s_4=s_8=1$, we also get more freedom,
parametrized in the divisors $\mathcal{S}_7$ and $\mathcal{S}_9$, 
in constructing the fibration of $\mathcal{C}_{F_{10}}$ over a given 
base $B$.  This is the second generalization of $X_{F_{10}}$ in contrast to the
standard Tate model.

\subsubsection*{Higher codimension singularities \& the spectrum of 
F-theory on $X_{F_{10}}$}

\begin{figure}[h!]
\center
\includegraphics[scale=0.8]{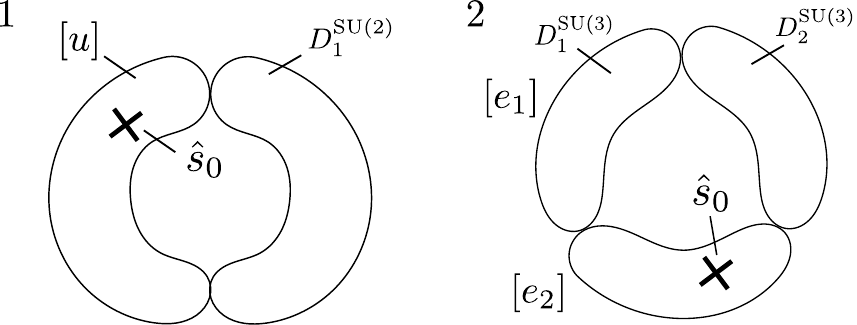}
\caption{\label{fig:poly10_codim1}Codimension one fibers of $X_{F_{10}}$. The crosses denote the intersections
with the zero section.}
\end{figure}

We consider in the following the most general elliptic 
fibration $X_{F_{10}}$ with general, non-trivial coefficients $s_4$ and 
$s_8$. In order to compute the Weierstrass form \eqref{eq:WSF} of the 
general hypersurface equation \eqref{eq:pF10} we apply Nagell's algorithm.
After setting $s_7=s_9=s_{10}=0$ in \eqref{eq:fcubic} and \eqref{eq:gcubic} we obtain the functions $f$ and $g$.
From this we compute the discriminant $\Delta$ to find all codimension one singularities
of the WSF of $X_{F_{10}}$. We find one $I_2$-singularity over the divisor
$\mathcal{S}^b_{\text{SU}(2)}=\{s_4=0\}\cap B$ in $B$ and one $I_3$-singularity over the divisor $\mathcal{S}^b_{\text{SU}(3)}=\{s_8=0\}\cap B$ in $B$. At the singularities the constraint \eqref{eq:pF10} factorizes as
\begin{align}
\begin{split}
\SU2&:\quad p_{F_{10}}|_{s_4=0}= u \cdot q_2 \, ,\\
\SU3&:\quad p_{F_{10}}|_{s_8=0}= e_1e_2 \cdot q_3 \, ,
\end{split}
\end{align}
where $q_2$ and $q_3$ are the remaining polynomials after factoring out
$u$ and $e_1e_2$, respectively. The reducible fibers at these loci are 
depicted in Figure \ref{fig:poly10_codim1}.
Thus, the total gauge group of $X_{F_{10}}$ is
\beq \label{eq:GF10}
	G_{F_{10}}=\text{SU}(3)\times\text{SU}(2)\,.
\eeq
The divisor classes of the corresponding
Cartan divisors can be calculated in a similar fashion as in Section \ref{sec:polyF6}. We obtain the classes
\begin{align}\label{eq:CartansF10}
D^{\SU2}_{1}=[s_4]-[u] \, ,\quad
D^{\SU3}_{1}=[e_1] \, ,\quad
D^{\SU3}_{2}=[s_8]-[e_1]-[e_2] \, .
\end{align}

Next, we turn to the charged matter spectrum, which is obtained by analyzing the codimension two singularities of the WSF of
$X_{F_{10}}$. All loci of codimension two singularities directly follow 
from the behavior of the discriminant and the representation content
under the gauge group $G_{F_{10}}$ is determined following the general 
procedure outlined in Section \ref{sec:ellipticCurvesWithRP}.
We find five codimension two singularities. Three of
these lead to the matter representations and the corresponding
codimension two fibers in $X_{F_{10}}$ in the first and third 
column of Table~\ref{tab:poly10_matter}, respectively.
We note that the fiber corresponding to matter in the representation
$(\two,\one)$ is non-split, cf.~the discussion in \ref{sec:polyF8}.
The two nodes that are identified by codimension three monodromies are 
drawn with dashed lines in Table~\ref{tab:poly10_matter}.
At the locus $s_4=s_6^2-4s_3s_8=0$ the fiber is of Type $III$ and at the 
locus $s_6=s_8=0$ it is of Type $IV$, i.e.~the fiber is a degeneration 
of the $I_3$-fiber at the locus $s_8=0$, where the three 
$\mathbb{P}^1$'s intersect in one point. Thus both loci
do not support any additional matter. The matter in the adjoint
representations has been added to Table~\ref{tab:poly10_matter} for
completeness.

In the second column of Table \ref{tab:poly10_matter} the multiplicities 
of the charged hyper multiplets are presented. They have been computed 
following Section \ref{sec:SUGRA6D}, directly from the classes
of all varieties $V(I_{k})$, $k=1,2,3$. This is straightforward, 
employing \eqref{eq:cubicsections}, as these varieties are irreducible 
complete intersections.

Finally, the number of neutral 
hyper multiplets is computed from \eqref{eq:Hneutral} using
the Euler number \eqref{eq:EulerNumbers} of $X_{F_{10}}$. It reads
\beq \label{eq:HneutF10}
H_{\text{neut}} = 15 + 11 [K_B^{-1}]^2 - 6 [K_B^{-1}] \cS_7 + 6 \cS_7^2 - 3 [K_B^{-1}] \cS_9 - 6 \cS_7 \cS_9 + 3 \cS_9^2 \,.
\eeq
The number $T$ of tensor multiplets is given by \eqref{eq:Tformula} and
we have $V=11$. 
We check that all 6D anomalies, cf.~Appendix \ref{app:Anomalies}, are 
canceled using $\mathcal{S}^b_{\text{SU}(2)}=\{s_4=0\}\cap B$,
$\mathcal{S}^b_{\text{SU}(3)}=\{s_8=0\}\cap B$, the charged spectrum in 
Table \ref{tab:poly10_matter} and \eqref{eq:HneutF10}.
\begin{table}[H]
\begin{center}
\renewcommand{\arraystretch}{1.2}
\begin{tabular}{|c|c|c|c|}\hline
Representation & Multiplicity & Fiber & Locus \\ \hline
$(\two,\three)$ & $\begin{array}{c} (2\cS_7-\cS_9) \\ ([K_B^{-1}]-\cS_7+\cS_9) \end{array}$ & \rule{0pt}{1.6cm}\parbox[c]{2.9cm}{\includegraphics[width=2.9cm]{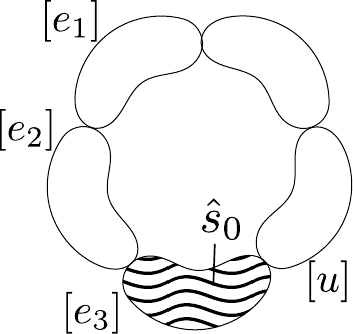}} & $\begin{array}{c} V(I_{(1)}):= \\ \{ s_4=s_8=0 \} \end{array}$ \\[1.2cm] \hline

$(\two, \one)$ & $\begin{array}{c} (2\cS_7-\cS_9) \\ (5[K_B^{-1}]-\cS_7-\cS_9) \end{array}$ & \rule{0pt}{1.55cm}\parbox[c]{2.9cm}{\includegraphics[width=2.9cm]{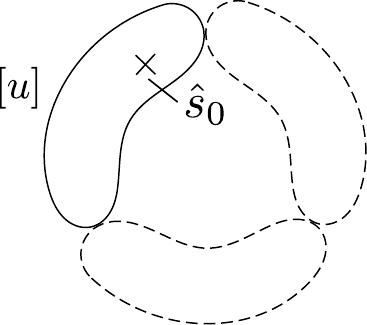}} & $\begin{array}{c} V(I_{(2)}):=\{ s_4=0= \\ -s_3 s_5^2 + s_2 s_5 s_6 - s_1 s_6^2 \\ - s_2^2 s_8 + 4 s_1 s_3 s_8 \}\end{array}$ \\[1.2cm] \hline

$(\one,\three)$ & $\begin{array}{c} ([K_B^{-1}]-\cS_7+\cS_9) \\ (6[K_B^{-1}]-\cS_7-\cS_9) \end{array}$ & \rule{0pt}{1.6cm}\parbox[c]{2.9cm}{\includegraphics[width=2.9cm]{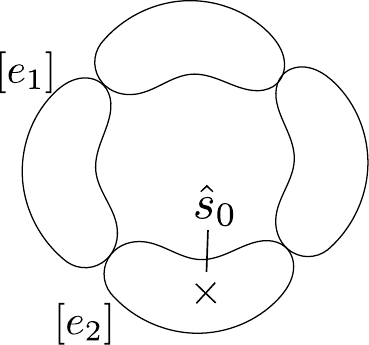}} & $\begin{array}{c} V(I_{(3)}):=\{ s_8=0 \\ s_4 s_5^3 - s_3 s_5^2 s_6 \\ + s_2 s_5 s_6^2 - s_1 s_6^3 = 0 \} \end{array}$ \\[1.2cm] \hline \hline

$(\mathbf{3},\one)_0$ & \rule{0pt}{0.85cm}$\begin{array}{c} 1+ \frac{2\cS_7-\cS_9 - [K_B^{-1}]}{2} \\ \times (2\cS_7-\cS_9) \end{array}$ & Figure \ref{fig:poly10_codim1} & $s_4=0$ \\[0.05cm] \hline

$(\one,\mathbf{8})_0$ & \rule{0pt}{0.8cm}$\begin{array}{c} 1+\frac{\cS_9-\cS_7}{2} \\ \times([K_B^{-1}]-\cS_7+\cS_9) \end{array}$ & Figure \ref{fig:poly10_codim1} & $s_8=0$ \\[0.05cm] \hline
\end{tabular}
\caption{\label{tab:poly10_matter}Charged matter representations under SU$(3)\times$SU(2) and corresponding codimension two fibers of $X_{F_{10}}$. The adjoint matter is included for completeness.}
\end{center}
\end{table}

We conclude this section by the analysis of codimension three 
singularities of the WSF of $X_{F_{10}}$ and the corresponding Yukawa 
points in compactifications to 4D. We find one possible Yukawa
coupling of the charged matter spectrum of $X_{F_{10}}$, which is given
in Table \ref{tab:poly10_yukawa}.
\begin{table}[htb]
\begin{center}
\renewcommand{\arraystretch}{1.2}
\begin{tabular}{|c|c|}\hline
Yukawa & Locus \\ \hline
$(\two,\three) \cdot (\two, \one) \cdot \overline{(\one,\three)}$ & $s_4=s_8=0=s_3 s_5^2 - s_2 s_5 s_6 + s_1 s_6^2$\\ \hline
\end{tabular}
\caption{\label{tab:poly10_yukawa}Codimension three loci and corresponding Yukawa couplings for $X_{F_{10}}$.}
\end{center}
\end{table}

%%%%%%%%%%%%%%%%%%%%%%%%%%%%%%%%%%%%%%%%%%%%%%%%%%%%%%%%%%%%%%%%%%%%
%%%%%%%%%%%%%%%%%%%%%%%%%%%%%%%%%%%%%%%%%%%%%%%%%%%%%%%%%%%%%%%%%%%%

%%%%%%%%%%%%%%%%%%%%%%%%%%%%%%%%%%%%%%%%%%%%%%%%%%%%%%%%%%%%%%%%%%%%%%%%%%%%%%%%%%%%%%%%%%%%%%%%%
\subsection{Fibrations with gauge groups of rank 4, 5 and no MW-torsion}
%%%%%%%%%%%%%%%%%%%%%%%%%%%%%%%%%%%%%%%%%%%%%%%%%%%%%%%%%%%%%%%%%%%%%%%%%%%%%%%%%%%%%%%%%%%%%%%%%
\label{sec:Rk45NoTorsion}

In this section we analyze toric hypersurface fibrations based on the 
fiber polyhedra $F_{11}$, $F_{12}$ and $F_{14}$. These are the 
fibrations that give rise to F-theory models 
with simply-connected gauge groups of maximal rank among all toric 
hypersurface fibrations, that is  four and five.
Most outstanding here is $X_{F_{11}}$ that exhibits the gauge group and 
the matter representations that coincide precisely with that of the 
\emph{Standard Model}.

%%%%%%%%%%%%%%%%%%%%%%%%%%%%%%%%%%%%%%%%%%%%%%%%%%%%%%%%%%%%%%%%%%%%%%%%%%%%%%%%%%%%%%%%%%%%%%%%%
\subsubsection{Polyhedron $F_{11}$: $G_{F_{11}}=\text{SU(3)}\times\text{SU(2)}\times\text{U(1)}$}
\label{sec:polyF11}
%%%%%%%%%%%%%%%%%%%%%%%%%%%%%%%%%%%%%%%%%%%%%%%%%%%%%%%%%%%%%%%%%%%%%%%%%%%%%%%%%%%%%%%%%%%%%%%%%

\begin{figure}[H]
\center
\includegraphics[scale=0.4]{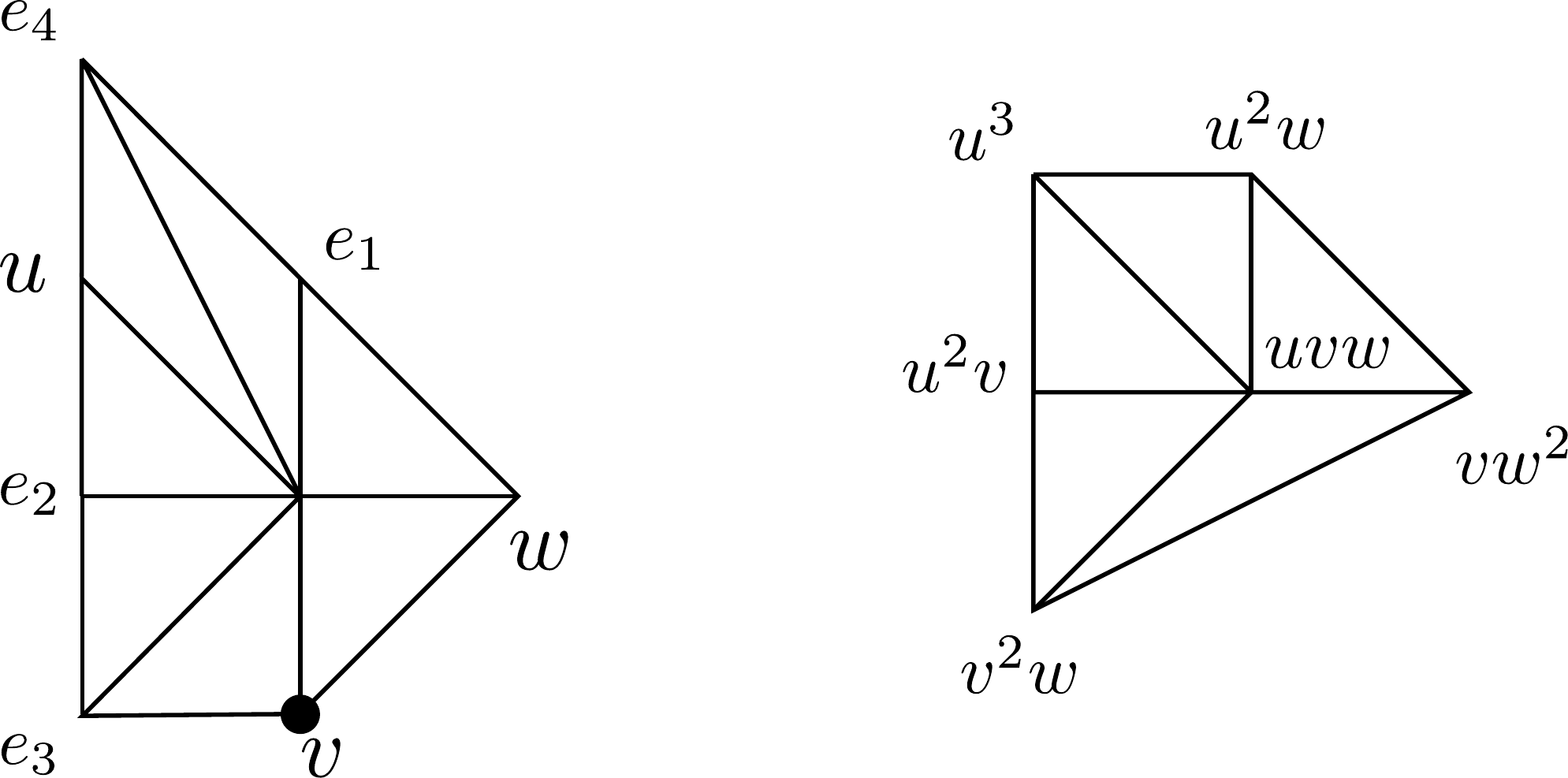}
\caption{\label{fig:poly11_toric}The toric diagram of polyhedron $F_{11}$ and its dual. The zero section is indicated by the dot.}
\end{figure}

We construct an elliptically fibered Calabi-Yau manifold $X_{F_{11}}$ 
with an arbitrary base $B$ and general elliptic fiber given by the 
elliptic curve $\mathcal{E}$ in $\mathbb{P}_{F_{11}}$.
The toric data of $\mathbb{P}_{F_{11}}$ is encoded in
Figure \ref{fig:poly11_toric}, where the corresponding 
polyhedron $F_{11}$, a choice of homogeneous coordinates as well as its 
dual polyhedron $F_{6}$ are shown. In the monomials that are associated 
to the integral points of $F_{6}$ according to 
\eqref{eq:BatyrevFormula}, we have set $e_i=1$, $\forall i$, for brevity 
of our notation.
Starting from $\mathbb{P}^2$, we obtain the toric variety 
$\mathbb{P}_{F_{11}}$ as a blow-up at four non-generic points. The 
blow-down map reads
\begin{align}\label{eq:F11blowup}
u \rightarrow e_1 e_2 e_3 e_4^2 u \,,\quad w\rightarrow e_1 e_4 w \,,\quad v\rightarrow e_2 e_3^2 v \, .
%u \rightarrow e_1 u\; &, \; w\rightarrow e_1 w\\
%u \rightarrow e_2 u\; &, \; v\rightarrow e_2 v\\
%e_2 \rightarrow e_3 e_2\; &, \; v\rightarrow e_3 v\\
%u \rightarrow e_4 u\; &, \; e_1\rightarrow e_4 e_1\\
\end{align}
After these blow-ups, the homogeneous coordinates on the fiber, given by 
$[u:v:w:e_1:e_2:e_3:e_4]$, take values in the line bundles associated to 
the following divisor classes:
\begin{equation}\label{tab:poly11_bundles}
\text{
\begin{tabular}{|c|c|}\hline
Section & Divisor class\\ \hline
$u$ & $H-E_1-E_2-E_4+\cS_9+[K_B]$ \\ \hline
$v$ & $H-E_2-E_3+\cS_9-\cS_7$\\ \hline
$w$ & $H-E_1$\\ \hline
$e_1$ & $E_1-E_4$\\ \hline
$e_2$ & $E_2-E_3$\\ \hline
$e_3$ & $E_3$\\ \hline
$e_4$ & $E_4$\\ \hline
\end{tabular}}
\end{equation}
The Stanley-Reisner ideal of $\mathbb{P}_{F_{11}}$ 
can be read off from Figure~\ref{fig:poly11_toric}. It is given by
\begin{align}
SR&=\{ u e_1, uw, uv, ue_3, e_4 w, e_4 v, e_4 e_3, e_4 e_2, e_1v, e_1 e_3, e_1e_2, w e_3, w e_2, ve_2\} \, .
\end{align}
We obtain the hypersurface equation of $X_{F_{11}}$ in the ambient space 
given by the $\mathbb{P}_{F_{11}}$-fibration \eqref{eq:PFfibration} 
either by applying \eqref{eq:BatyrevFormula} or by specializing
\eqref{eq:pF1} as $s_4=s_7=s_8=s_{10}=0$ and applying the map 
\eqref{eq:F11blowup}. It reads
\begin{align}
\begin{split}\label{eq:pF11}
p_{F_{11}}&=s_1 e_1^2 e_2^2 e_3 e_4^4 u^3 + s_2 e_1 e_2^2 e_3^2 e_4^2 u^2 v + s_3 e_2^2 e_3^2 u v^2 + s_5 e_1^2 e_2 e_4^3 u^2 w %\\ &\phantom{=}
+s_6 e_1 e_2 e_3 e_4 u v w + s_9 e_1 v w^2 \, ,
\end{split}
\end{align}
where the sections $s_i$ take values in the line bundles associated
to the divisor classes  in \eqref{eq:cubicsections}.

The elliptic fibration $X_{F_{11}}$ has three rational sections. Two of 
these are linear independent, that means the MW-group of $X_{F_{11}}$ 
has rank one. The coordinates of the sections read
\begin{align}
\begin{split}\label{eq:sectionsF11}
\hat{s}_0=X_{F_{11}}\cap\{v=0\}&:\quad [1:0:s_1:1:1:-s_5:1]\,,\\
X_{F_{11}}\cap\{e_3=0\}&:\quad[1:s_5:1:1:-s_9:0:1]\,,\\
\hat{s}_1=X_{F_{11}}\cap\{e_4=0\}&:\quad[s_9:1:1:-s_3:1:1:0]\,,
\end{split}
\end{align}
where we choose $\hat{s}_0$ as the zero section and $\hat{s}_1$
as the generator of the MW-group.

\begin{figure}[H]
\center
\includegraphics[scale=0.8]{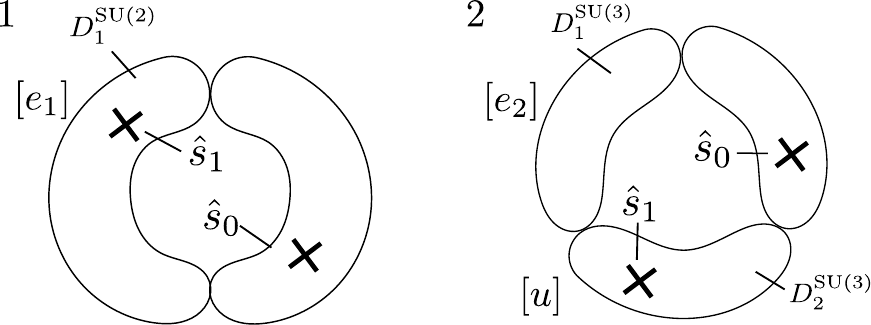}
\caption{\label{fig:poly11_codim1}Codimension one fibers of $X_{F_{11}}$. The crosses denote the intersections with the two sections.}
\end{figure}

The WSF \eqref{eq:WSF} of \eqref{eq:pF11} is computed using
Nagell's algorithm. The WS-coordinates of the section
$\hat{s}_1$ are given by restricting \eqref{eq:WScoordsSecF3} as 
$s_4=s_7=s_8=0$. The functions $f$ and $g$ of the WSF 
can be obtained by setting  $s_4=s_7=s_8=s_{10}=0$ in \eqref{eq:fcubic} 
and \eqref{eq:gcubic}, respectively. Using that we calculate the 
discriminant $\Delta$. This allows us to find all codimension one 
singularities of the WSF of $X_{F_{11}}$. We find one $I_2$-singularity 
over the divisor $\mathcal{S}^b_{\text{SU}(2)}=\{s_3=0\}\cap B$ and one 
$I_3$-singularity over the divisor 
$\mathcal{S}^b_{\text{SU}(3)}=\{s_9=0\}\cap B$ in $B$. Along these 
divisors the constraint \eqref{eq:pF11} factorizes as
\begin{align}
\begin{split}\label{eq:SU2F11}
\SU2&:\quad p_{F_{11}}|_{s_3=0}= e_1 \cdot q_3 \, , \\
\SU3&:\quad p_{F_{11}}|_{s_9=0}= e_2 u \cdot q_2 \, ,
\end{split}
\end{align}
where $q_2$, $q_3$ are the homogeneous polynomials in $[u:v:w]$
of degree two and three that remain after factoring out
$e_1$ and $e_2 u$. The corresponding $I_2$- and $I_3$-fibers are 
depicted in Figure \ref{fig:poly11_codim1}.
In summary, the total gauge group of $X_{F_{11}}$ is
\beq \label{eq:GF11}
	G_{F_{11}}=\SU3\times \text{SU}(2)\times\text{U}(1)\,.
\eeq
Following the path of Section \ref{sec:polyF6} we calculate the classes of the Cartan divisors as
\begin{align}\label{eq:CartansF11}
D^{\SU2}_{1}=[e_1] \, ,\quad
D^{\SU3}_{1}=[e_2] \, \quad
D^{\SU3}_{2}=[u] \, .
\end{align}
This enables the computation of the Shioda map \eqref{eq:ShiodaMap}
of the section $\hat{s}_1$. It reads
\begin{align} \label{eq:ShiodaF11}
\sigma (\hat{s}_1)=S_1-S_0 +  [K_B] + \frac{1}{2} D^{\SU2}_{1}	+ \frac{1}{3}\big(D^{\SU3}_{1}+2D^{\SU3}_{2}\big) \, ,
\end{align}
which follows since the section $\hat{s}_1$ does not 
intersect the zero section, see \eqref{eq:sectionsF11}, implying
\begin{align} \label{eq:S1S0F11}
	\pi(S_1\cdot S_0)=0\,,
\end{align}
and from the intersections of $\hat{s}_0$ and $\hat{s}_1$ 
with the codimension one fibers in Figure~\ref{fig:poly11_codim1}, 
yielding
\begin{align}
S_1 \cdot C^{\SU2}_{-\alpha _1}=1\,, \quad
S_1 \cdot C^{\SU3}_{-\alpha _1}=0\,,\quad
S_1 \cdot C^{\SU3}_{-\alpha _1}=1\,.
\end{align}
The data of the MW-group is completed by the height pairing 
\eqref{eq:anomalycoeff} of $\hat{s}_1$. It is computed as
\begin{align}
\label{eq:bmnF11}
 b_{11}= \frac{3}{2} [K_B^{-1}]-\frac{1}{2}\cS_7-\frac{1}{6}\cS_9 \, ,
\end{align}
where we use the universal intersection relation
\eqref{eq:SP^2} as well as \eqref{eq:S1S0F11}.

Next, we turn to the codimension two singularities of the WSF of
$X_{F_{11}}$ to calculate its charged matter spectrum.
Here, all  representations under the gauge group are determined using 
the methods outlined in Section \ref{sec:ellipticCurvesWithRP}.
The non-Abelian representations readily follow from the discriminant,
whereas the charged singlets are determined from the primary
decomposition of the complete intersection \eqref{eq:charge1Matter} for 
the section $\hat{s}_1$. Using this, we find seven singularities in 
codimension two. Five of these lead to the matter representations and 
the corresponding codimension two fibers in $X_{F_{11}}$ given in the 
first and third column of Table~\ref{tab:poly11_matter}, respectively.
The remaining singularities at $s_3=s_6^2-4s_2s_9=0$ and $s_6=s_9=0$
are of Type $III$ and $IV$, respectively. Since they are just 
degenerations of the codimension one fibers in 
Figure~\ref{fig:poly11_codim1} without 
additional $\mathbb{P}^1$'s, they do not yield further matter 
representations. The adjoint representations in 
the last two rows of Table~\ref{tab:poly11_matter} are shown 
for completeness.

The multiplicities of the charged hyper multiplets that are presented in
Table \ref{tab:poly11_matter} are straightforwardly computed  from
the homology class of all complete intersections $V(I_{(k)})$, 
$k=1,\ldots,5$.

We complete the matter spectrum of $X_{F_{11}}$ by the number of neutral
hyper multiplets, which is computed from \eqref{eq:Hneutral} using
the Euler number $\chi(X_{F_{11}})$ given in
\eqref{eq:EulerNumbers}. It reads
\begin{align} \label{eq:HneutF11}
H_{\text{neut}} = 16 + 11 [K_B^{-1}]^2 - 4 [K_B^{-1}] \cS_7 + 2 \cS_7^2 - 7 [K_B^{-1}] \cS_9 - \cS_7 \cS_9 + 3 \cS_9^2 \,.
\end{align}
There are $T$ tensors computed by \eqref{eq:Tformula} and
we have $V=12$ vector multiplets.
Using $S_{\text{SU}(2)}^b=\{s_3=0\}$, $S_{\text{SU}(3)}^b=\{s_9=0\}$, 
\eqref{eq:bmnF11},
the charged spectrum in Table \ref{tab:poly11_matter} and
\eqref{eq:HneutF11} we check cancelation of all 6D anomalies in 
\eqref{eq:6dAnomalies}, following the discussion of Appendix~\ref{app:Anomalies}.

We conclude our analysis with the Yukawa couplings of the charged 
matter spectrum of $X_{F_{11}}$, corresponding to the codimension three 
singularities of its WSF. All Yukawa points of $X_{F_{11}}$ are 
presented in Table \ref{tab:poly11_yukawa}.
\begin{table}[H]
\begin{center}
\renewcommand{\arraystretch}{1.2}
\begin{tabular}{|c|c|c|@{}c@{}|}\hline
Representation & Multiplicity & Splitting & Locus \\ \hline

$(\three,\two)_{-1/6}$ & $\cS_9([K_B^{-1}]+\cS_7-\cS_9)$ & \rule{0pt}{1.6cm}\parbox[c]{2.9cm}{\includegraphics[width=2.9cm]{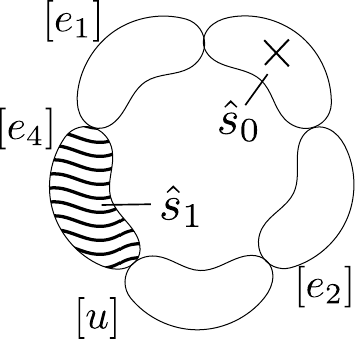}} & $V(I_{(1)}):=\{ s_3=s_9=0 \}$ \\[1.2cm] \hline

$(\one,\two)_{1/2}$ & $\begin{array}{c} ([K_B^{-1}]+\cS_7-\cS_9) \\ (6[K_B^{-1}]-2\cS_7-\cS_9) \end{array}$ & \rule{0pt}{1.6cm}\parbox[c]{2.9cm}{\includegraphics[width=2.9cm]{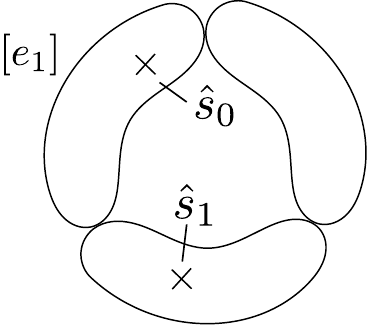}} & $\begin{array}{c} V(I_{(2)}):=\{ s_3=0 \\ s_2s_5^2 + s_1(s_1 s_9-s_5s_6)=0 \} \end{array}$ \\[1.2cm] \hline

$(\three,\one)_{-2/3}$ & $\cS_9(2 [K_B^{-1}]-\cS_7)$ & \rule{0pt}{1.6cm}\parbox[c]{2.9cm}{\includegraphics[width=2.9cm]{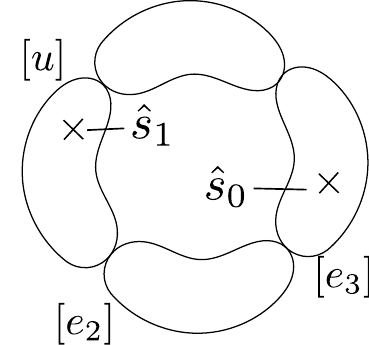}} & $V(I_{(3)}):=\{ s_5=s_9=0 \}$ \\[1.2cm] \hline

$(\three, \one)_{1/3}$ & $\cS_9 (5[K_B^{-1}]-\cS_7-\cS_9)$ & \rule{0pt}{1.6cm}\parbox[c]{2.9cm}{\includegraphics[width=2.9cm]{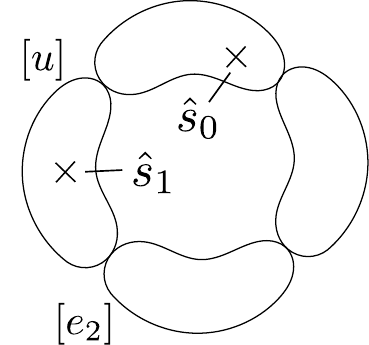}} & $\begin{array}{c} V(I_{(4)}):=\{ s_9=0 \\ s_3 s_5^2 + s_6(s_1 s_6-s_2 s_5)=0 \} \end{array}$ \\[1.2cm] \hline

$(\one,\one)_{-1}$ & $\begin{array}{c} (2[K_B^{-1}]-\cS_7) \\ (3[K_B^{-1}]-\cS_7-\cS_9) \end{array}$ & \rule{0pt}{1.3cm}\parbox[c]{2.9cm}{\includegraphics[width=2.9cm]{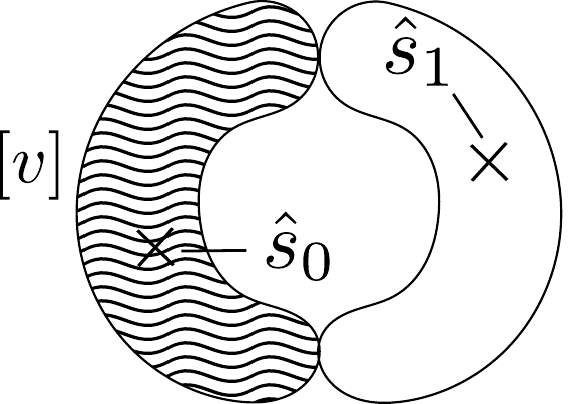}} & $V(I_{(5)}):=\{ s_1=s_5=0 \}$ \\[0.9cm] \hline \hline

$(\mathbf{8},\one)_0$ & \rule{0pt}{0.55cm}$1+\cS_9 \frac{\cS_9 - [K_B^{-1}]}{2}$ & Figure \ref{fig:poly11_codim1} & $s_9=0$ \\[0.05cm] \hline

$(\one,\mathbf{3})_0$ & \rule{0pt}{0.85cm}$\begin{array}{c}1+\frac{\cS_7-\cS_9}{2} \\ \times ([K_B^{-1}]+\cS_7-\cS_9) \end{array}$ & Figure \ref{fig:poly11_codim1} & $s_3=0$\\[0.05cm] \hline
\end{tabular}
\caption{\label{tab:poly11_matter}Charged matter representations under $\SU3 \times \SU2 \times \text{U}(1)$ and corresponding codimension two fibers of $X_{F_{11}}$. The adjoint matter is included for completeness.}
\end{center}
\end{table}
\begin{table}[htb!]
\begin{center}
\renewcommand{\arraystretch}{1.2}
\begin{tabular}{|c|c|}\hline
Yukawa & Locus \\ \hline
$\overline{(\three,\two)_{-1/6}}	\cdot (\three,\one)_{-2/3}	\cdot (\one,\two)_{1/2}$ & $s_3=s_5=s_9=0$ \\ \hline
$(\three,\two)_{-1/6}	\cdot \overline{(\three, \one)_{1/3}}	\cdot (\one,\two)_{1/2}$ & $s_3=s_9=0=s_1s_6-s_2s_5$ \\ \hline
$\overline{(\three,\one)_{-2/3}}	\cdot (\three, \one)_{1/3}	\cdot (\one,\one)_{-1}$ & $s_1=s_5=s_9=0$ \\ \hline
\end{tabular}
\caption{\label{tab:poly11_yukawa}Codimension three loci and corresponding Yukawa couplings for $X_{F_{11}}$.}
\end{center}
\end{table}

%%%%%%%%%%%%%%%%%%%%%%%%%%%%%%%%%%%%%%%%%%%%%%%%%%%%%%%%%%%%%%%%%%%%%%%%%%%%%%%%%%%%%%%%%%%%%%%%%
\subsubsection{Polyhedron $F_{12}$: $G_{F_{12}}=\text{SU(2)}^2\times\text{U(1)}^2$}
\label{sec:polyF12}
%%%%%%%%%%%%%%%%%%%%%%%%%%%%%%%%%%%%%%%%%%%%%%%%%%%%%%%%%%%%%%%%%%%%%%%%%%%%%%%%%%%%%%%%%%%%%%%%%

\begin{figure}[H]
\center
\includegraphics[scale=0.4]{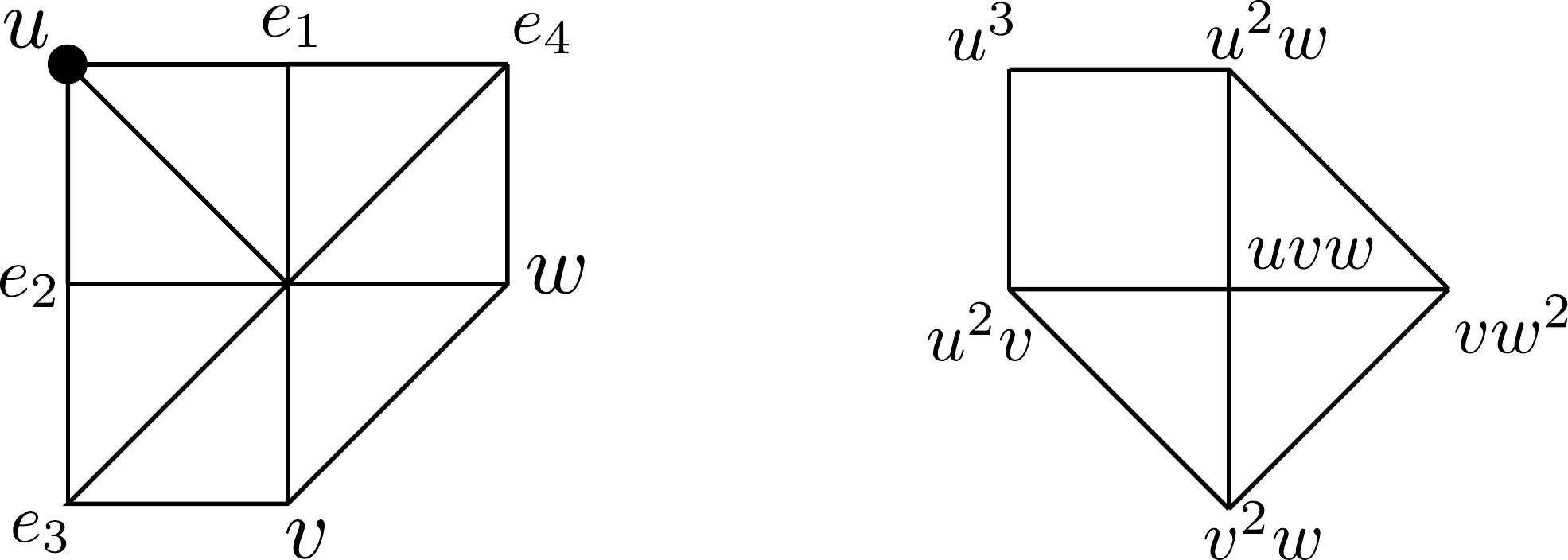}
\caption{\label{fig:poly12_toric}The toric diagram of polyhedron $F_{12}$ and its dual. The zero section is indicated by the dot.}
\end{figure}

In this section, we analyze the elliptically fibered Calabi-Yau manifold 
$X_{F_{12}}$ with base $B$ and general elliptic fiber  given by the 
elliptic curve $\mathcal{E}$ in $\mathbb{P}_{F_{12}}$.
The toric data of $\mathbb{P}_{F_{12}}$ can
be extracted from Figure \ref{fig:poly12_toric}, where the fiber 
polyhedron $F_{12}$ together with a choice of homogeneous coordinates as 
well as its dual polyhedron are shown.
As before, we have set $e_i=1$, $\forall i$, in the monomials associated 
to the integral points of $F_{5}$ by \eqref{eq:BatyrevFormula}.
The toric variety $\mathbb{P}_{F_{12}}$ is $\mathbb{P}^2$
blown-up at four non-generic points. In our conventions, the  blow-down 
map takes the form
\begin{align}\label{eq:F12blowup}
u \rightarrow e_1 e_2 e_3 e_4 u \,,\quad w\rightarrow e_1 e_4^2 w \,,\quad v\rightarrow e_2 e_3^2 v\, ,
%u \rightarrow e_1 u\; &, \; w\rightarrow e_1 w\\
%u \rightarrow e_2 u\; &, \; v\rightarrow e_2 v\\
%e_2 \rightarrow e_3 e_2\; &, \; v\rightarrow e_3 v\\
%e_1 \rightarrow e_4 e_1\; &, \; w\rightarrow e_4 w
\end{align}
so that the homogeneous coordinates on the fiber after this blow-up are 
$[u:v:w:e_1:e_2:e_3:e_4]$. In the total space of the 
$\mathbb{P}_{F_{12}}$-fibration constructed as in 
\eqref{eq:PFfibration}, these coordinates have the divisor classes given by:
\begin{equation}\label{tab:poly12_bundles}
\text{
\begin{tabular}{|c|c|}\hline
Section & Divisor class\\ \hline
$u$ & $H-E_1-E_2+\cS_9+[K_B]$ \\ \hline
$v$ & $H-E_2-E_3+\cS_9-\cS_7$\\ \hline
$w$ & $H-E_1-E_4$\\ \hline
$e_1$ & $E_1-E_4$\\ \hline
$e_2$ & $E_2-E_3$\\ \hline
$e_3$ &  $E_3$\\ \hline
$e_4$ &  $E_4$\\ \hline
\end{tabular}}
\end{equation}
The Stanley-Reisner ideal of $\mathbb{P}_{F_{12}}$ follows from
Figure~\ref{fig:poly12_toric} as
\begin{align}
SR=\{ u e_4, uw ,u v, u e_3, e_1 w, e_1 v, e_1 e_3, e_1 e_2, e_4 v, e_4 e_3, e_4 e_2, w e_3, w e_2,v e_2\} \, .
\end{align}
The hypersurface equation for $X_{F_{12}}$ can be obtained employing 
\eqref{eq:BatyrevFormula}. It reads
\begin{align}
\begin{split}\label{eq:pF12}
p_{F_{12}}&= s_1 e_1^2 e_2^2 e_3 e_4 u^3 + s_2e_1 e_2^2 e_3^2 u^2 v + s_5 e_1^2 e_2 e_4^2 u^2 w + s_6 e_1 e_2 e_3 e_4 u v w%\\ &\phantom{=} 
+ s_7 e_2 e_3^2 v^2 w + s_9 e_1 e_4^2 v w^2 \, ,
\end{split}
\end{align}
where the divisor classes of the sections $s_i$ are fixed by the 
Calabi-Yau condition  as shown in \eqref{eq:cubicsections}. We note that 
\eqref{eq:pF12} can also be obtained from \eqref{eq:pF1} by the 
specialization $s_3=s_4=s_8=s_{10}=0$ and the map \eqref{eq:F12blowup}.

There are five rational sections of the elliptic fibration of $X_{F_{12}}$. 
Their coordinates are
\begin{align}
\begin{split}\label{eq:sectionsF12}
\hat{s}_0=X_{F_{12}}\cap\{u=0\}&: \quad [0:1:1:s_7:-s_9:1:1]\, ,\\
\hat{s}_1=X_{F_{12}}\cap\{e_3=0\}&: \quad [1:s_5:1:1:-s_9:0:1]\, ,\\
\hat{s}_2=X_{F_{12}}\cap\{e_4=0\}&: \quad [1:1:s_2:-s_7:1:1:0]\, ,\\
X_{F_{12}}\cap \{v=0\}&: \quad [1:0:s_1:1:1:-s_5:1]\, ,\\
X_{F_{12}}\cap \{w=0\}&: \quad [1:s_1:0:1:1:1:-s_2]\, ,
\end{split}
\end{align}
where we choose $\hat{s}_0$ as the zero section. Clearly, only three of 
these sections are linearly independent. We choose $\hat{s}_1$
and $\hat{s}_2$ as the generators of the rank two MW-group of
$X_{F_{12}}$.

\begin{figure}[h!]
\center
\includegraphics[scale=0.8]{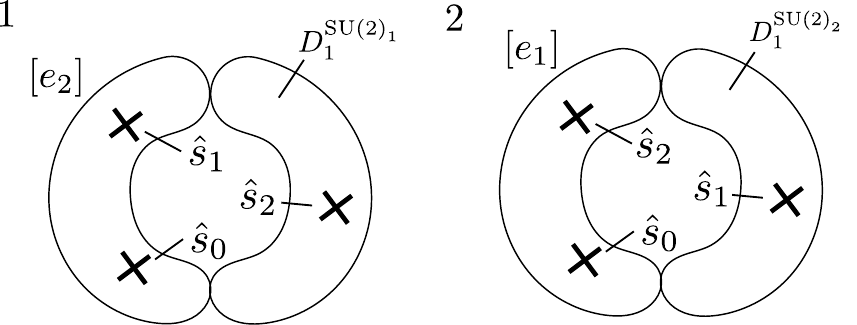}
\caption{\label{fig:poly12_codim1}Codimension one fibers of $X_{F_{12}}$. The crosses denote the intersections with the three sections.}
\end{figure}

We compute the Weierstrass form \eqref{eq:WSF} of \eqref{eq:pF12} using
Nagell's algorithm.
The WS-coordinates of the sections
$\hat{s}_1$ and $\hat{s}_2$ are given by \eqref{eq:secsF51} and 
\eqref{eq:secsF52}, respectively, in the limit  $s_3=s_4=s_8=s_{10}=0$.
Similarly, we obtain the functions $f$ and $g$ from \eqref{eq:fcubic} 
and \eqref{eq:gcubic} using this specialization.
From this the discriminant $\Delta$ is readily computed. The 
factorization of $\Delta$ shows the presence of two $I_2$-singularities
in $X_{F_{12}}$ over the divisors 
$\mathcal{S}^b_{\text{SU}(2)_1}=\{s_7=0\}\cap B$ and 
$\mathcal{S}^b_{\text{SU}(2)_2}=\{s_9=0\}\cap B$ in $B$. At these loci,
the constraint \eqref{eq:pF12} factorizes as
\begin{align}
\begin{split}\label{eq:SU2F12}
\SU2_1&:\quad p_{F_{12}}|_{s_7=0}= e_1 q_3 \, ,\\
\SU2_2&:\quad p_{F_{12}}|_{s_9=0}= e_2 q_3' \,,
\end{split}
\end{align}
where $q_3$, $q_3'$ are the remaining polynomials after factoring out
$e_1$ and $e_2$. The corresponding $I_2$-fibers are depicted in Figure 
\ref{fig:poly12_codim1}.
In summary, the total gauge group of $X_{F_{12}}$ is
\begin{align} \label{eq:GF12}
	G_{F_{12}}=\text{SU}(2)^2\times\text{U}(1)^2\,.
\end{align}
Analogous to Section \ref{sec:polyF6}, we obtain the divisor 
classes of the Cartan divisors of $X_{F_{12}}$ as
\begin{align}\label{eq:CartansF12}
D^{\SU2_1}_{1}=[s_7]-[e_1]\, ,\quad%=[s_7]-E_1+E_4 \, ,\\
D^{\SU2_2}_{1}=[s_9]-[e_2]\, .%=[s_9]-E_2+E_3 \, .
\end{align}
\begin{table}[H]
\begin{center}
\renewcommand{\arraystretch}{1.2}
\begin{tabular}{|c|c|c|@{}c@{}|}\hline
Representation & Multiplicity & Fiber & Locus \\ \hline

$(\two,\two)_{(1/2,1/2)}$ & $\cS_7\cdot\cS_9$ & \rule{0pt}{1.4cm}\parbox[c]{2.6cm}{\includegraphics[width=2.6cm]{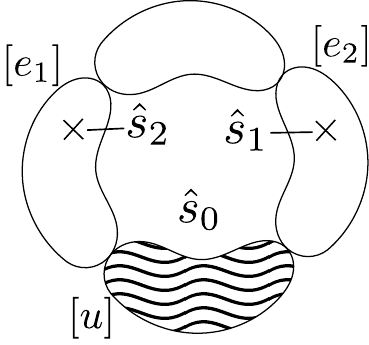}} & $V(I_{(1)}):=\{ s_9=s_7=0 \}$ \\[1cm] \hline

$(\one,\two)_{(-1,-1/2)}$ & $\cS_9\cdot(2[K_B^{-1}]-\cS_7)$ & \rule{0pt}{1.35cm}\parbox[c]{2.6cm}{\includegraphics[width=2.6cm]{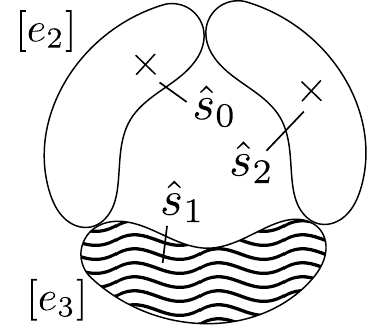}} & $V(I_{(2)}):=\{ s_9=s_5=0 \}$ \\[0.95cm] \hline

$(\two,\one)_{(-1/2,-1)}$ & $\cS_7\cdot(2[K_B^{-1}]-\cS_9)$ & \rule{0pt}{1.35cm}\parbox[c]{2.6cm}{\includegraphics[width=2.6cm]{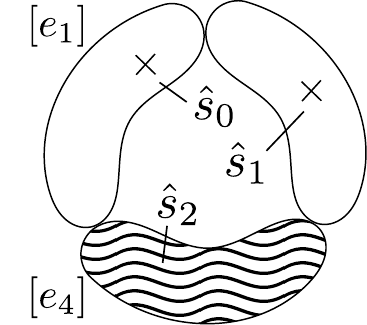}} & $V(I_{(3)}):=\{ s_7=s_2=0 \}$ \\[0.95cm] \hline

$(\one,\one)_{(1,0)}$ & $\begin{array}{c} (2[K_B^{-1}]-\cS_7) \\ (3[K_B^{-1}]-\cS_9-\cS_7) \end{array}$ & \rule{0pt}{1.1cm}\parbox[c]{2.6cm}{\includegraphics[width=2.6cm]{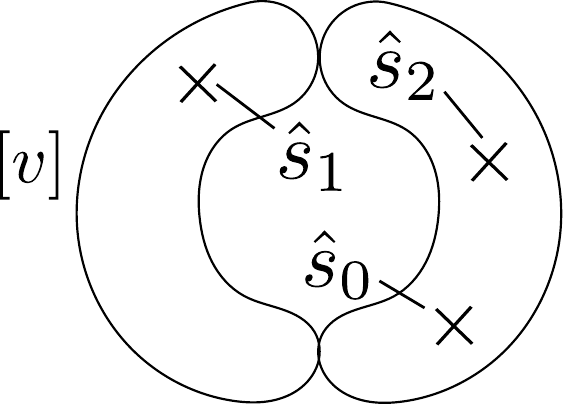}} & $V(I_{(4)}):=\{ s_5=s_1=0 \}$ \\[0.7cm] \hline

$(\one,\one)_{(0,1)}$ & $\begin{array}{c} (2[K_B^{-1}]-\cS_9) \\ (3[K_B^{-1}]-\cS_9-\cS_7) \end{array}$ & \rule{0pt}{1.1cm}\parbox[c]{2.6cm}{\includegraphics[width=2.6cm]{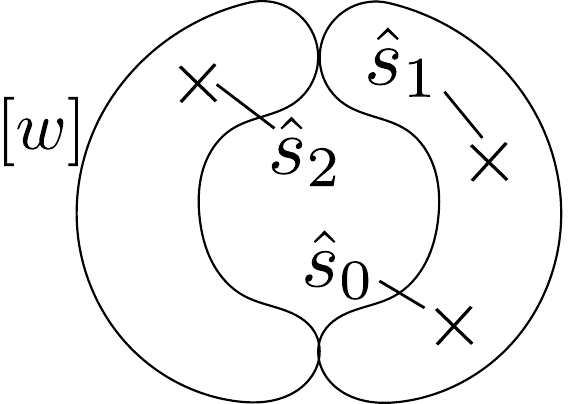}} & $V(I_{(5)}):=\{ s_2=s_1=0 \}$ \\[0.7cm] \hline

$(\one,\two)_{(0,-1/2)}$ & $\cS_9\cdot(6[K_B^{-1}]-2\cS_9-\cS_7)$ & \rule{0pt}{1.35cm}\parbox[c]{2.6cm}{\includegraphics[width=2.6cm]{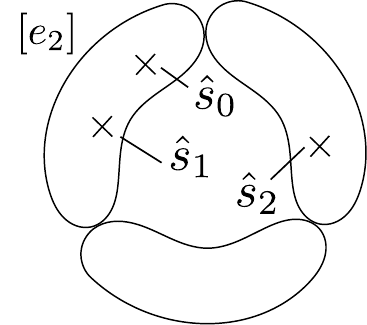}} & $\begin{array}{c} V(I_{(6)}):=\{ s_9=0 \\ s_5s_2^2-s_6s_2s_1+s_7s_1^2=0 \} \end{array}$ \\[0.95cm] \hline

$(\two,\one)_{(-1/2,0)}$ & $\cS_7\cdot(6[K_B^{-1}]-\cS_9-2\cS_7)$ & \rule{0pt}{1.35cm}\parbox[c]{2.6cm}{\includegraphics[width=2.6cm]{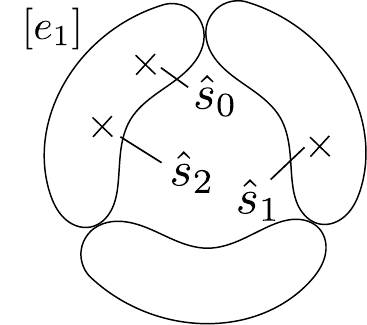}} & $\begin{array}{c} V(I_{(7)}):=\{ s_7=0 \\ s_9s_1^2+s_2s_5^2-s_6s_5s_1=0 \} \end{array}$ \\[0.95cm] \hline

$(\one,\one)_{(1,1)}$ & $\begin{array}{c} 2[K_B^{-1}]\cdot(3[K_B^{-1}]-\cS_7) \\ - \cS_9\cdot(2[K_B^{-1}]-\cS_9) \end{array}$ & \rule{0pt}{1.1cm}\parbox[c]{2.7cm}{\includegraphics[width=2.7cm]{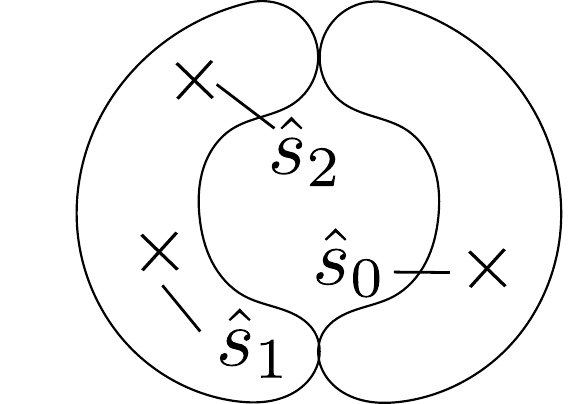}} & $\begin{array}{c} V(I_{(8)}):=\{ s_9s_2-s_7s_5=0 \\ s_9s_1-s_5s_6=0 \}\backslash V(I_{(2)}) \end{array}$ \\[0.7cm] \hline \hline

$(\three,\one)_{(0,0)}$ & \rule{0pt}{0.55cm}$1+\cS_7\cdot\frac{(\cS_7-[K_B^{-1}])}{2}$ & Figure \ref{fig:poly12_codim1} & $s_7=0$\\[0.05cm] \hline

$(\one,\three)_{(0,0)}$ & \rule{0pt}{0.55cm}$1+\cS_9\cdot\frac{(\cS_9-[K_B^{-1}])}{2}$ & Figure \ref{fig:poly12_codim1} & $s_9=0$\\[0.05cm] \hline

\end{tabular}
\caption{\label{tab:poly12_matter}Charged matter representations under SU$(2)^2\times$U(1)$^2$ and corresponding codimension two fibers of $X_{F_{12}}$. The adjoint matter is included for completeness.}
\end{center}
\end{table}
Using these results, we compute the Shioda map \eqref{eq:ShiodaMap}
of the sections $\hat{s}_m$, $m=1,2$, as
\begin{align} \label{eq:ShiodaF12}
\sigma (\hat{s}_m)&=S_m-S_0 + [K_B]	+\tfrac{1}{2} \delta_{m,1} D^{\SU2_1}_{1} + \tfrac{1}{2} \delta_{m,2} D^{\SU2_2}_{1}\, .
\end{align}
Here $S_0$, $S_m$ denote the divisor classes of the sections
$\hat{s}_0$, $\hat{s}_m$, respectively, and  we used
\begin{align}
 \pi(S_1\cdot S_2)=\pi(S_1\cdot S_0)=\pi(S_2\cdot S_0)=0\,,
\end{align}
which follows directly from \eqref{eq:sectionsF12} as well as
\begin{align}
S_m \cdot C^{\SU2_1}_{-\alpha _i}=\delta_{m,2}\, ,\quad
S_m \cdot C^{\SU2_2}_{-\alpha _i}=\delta_{m,1}\,,
\end{align}
which can be deduced from Figure \ref{fig:poly12_codim1}.
Employing these intersection relations along with 
\eqref{eq:SP^2}, we  obtain the height pairing \eqref{eq:anomalycoeff} 
as
\begin{align}
\label{eq:bmnF12}
b_{mn}=\begin{pmatrix}	2[K_B^{-1}]-\frac12 \cS_7 &	[K_B^{-1}]\\
[K_B^{-1}] &	2[K_B^{-1}]-\frac12 \cS_9	\end{pmatrix}_{mn} \, .
\end{align}

To obtain the charged matter spectrum we analyze the codimension two 
singularities of the WSF of
$X_{F_{12}}$. The corresponding  representation under the gauge group
are determined following the general procedure outlined in Section
\ref{sec:ellipticCurvesWithRP}. As before all non-trivial  
representations of the non-Abelian part of $G_{F_{12}}$ are easily read
off from the discriminant. The charged singlets are obtained by the 
primary decompositions of the two complete intersections 
\eqref{eq:charge1Matter} associated to the sections $\hat{s}_1$ and 
$\hat{s}_2$.
We find ten codimension two singularities, eight of
which lead to the matter representations and the corresponding
codimension two fibers in $X_{F_{12}}$ given in the first and third 
column of Table~\ref{tab:poly12_matter}, respectively.
At the remaining loci, namely $s_7=s_6^2-4s_2s_9=0$ and 
$s_9=s_6^2-4s_5s_7$, we find Type $III$ singularities and thus no 
additional matter, cf.~Section \ref{sec:polyF6} for more details.
We note that the matter locus $V(I_{(3)})$ agrees with the singular
locus of the dependent rational section in \eqref{eq:sectionsF12}.
For completeness, matter in the adjoint representation of $G_{F_{12}}$
is also given in the last two rows of Table~\ref{tab:poly12_matter}.

The number of neutral hyper multiplets completes the matter spectrum 
of $X_{F_{12}}$.
It is computed from \eqref{eq:Hneutral} using
the Euler number \eqref{eq:EulerNumbers} of $X_{F_{12}}$. It reads
\beq \label{eq:HneutF12}
H_{\text{neut}} = 16 + 11 [K_B^{-1}]^2 - 6 [K_B^{-1}] \cS_7 + 2 \cS_7^2 - 6 [K_B^{-1}] \cS_9 + \cS_7 \cS_9 + 2 \cS_9^2 \,.
\eeq
There are a base-dependent number $T$ of tensor and $V=7$
vector multiplets.
Finally, we use 
$S_{\text{SU}(2)_1}^b=\{s_7=0\}$, $S_{\text{SU}(2)_2}^b=\{s_9=0\}$,
\eqref{eq:bmnF12}, the charged spectrum in Table \ref{tab:poly12_matter} 
and \eqref{eq:HneutF12} to confirm
that all 6D anomalies in \eqref{eq:6dAnomalies} are canceled.

We conclude with the list of all codimension three singularities
of the WSF of $X_{F_{12}}$ and the corresponding Yukawa points
in Table \ref{tab:poly12_yukawa}.
\begin{table}[htb!]
\begin{center}
\renewcommand{\arraystretch}{1.2}
\begin{tabular}{|c|c|}\hline
Yukawa & Locus \\ \hline
$(\one,\one)_{(1,0)} \cdot (\one,\one)_{(0,1)} \cdot \overline{(\one,\one)_{(1,1)}}$ & $s_2=s_5=s_1=0$ \\ \hline

$(\one,\two)_{(0,-1/2)} \cdot (\one,\two)_{(-1,-1/2)} \cdot (\one,\one)_{(1,1)}$ & $s_9=s_5=0$, $s_6 s_2-s_7 s_1=0$  \\ \hline

$(\two,\one)_{(-1/2,-1)} \cdot (\two,\one)_{(-1/2,0)} \cdot (\one,\one)_{(1,1)}$ & $s_7=s_2=0$, $s_1 s_9-s_5 s_6=0$  \\ \hline

$(\two,\two)_{(1/2,1/2)} \cdot (\two,\one)_{(-1/2,0)} \cdot (\one,\two)_{(0,-1/2)}$ & $s_9=s_7=0$, $s_2s_5 - s_6 s_1=0$  \\ \hline

$(\two,\two)_{(1/2,1/2)} \cdot (\one,\two)_{(-1,-1/2)} \cdot \overline{(\two,\one)_{(-1/2,0)}}$ & $s_9=s_7=s_5=0$  \\ \hline

$(\two,\two)_{(1/2,1/2)} \cdot \overline{(\one,\two)_{(0,-1/2)}} \cdot (\two,\one)_{(-1/2,-1)}$ & $s_9=s_7=s_2=0$  \\ \hline
\end{tabular}
\caption{Codimension three loci and corresponding Yukawa couplings for $X_{F_{12}}$.}
\label{tab:poly12_yukawa}
\end{center}
\end{table}

%%%%%%%%%%%%%%%%%%%%%%%%%%%%%%%%%%%%%%%%%%%%%%%%%%%%%%%%%%%%%%%%%%%%%%%%%%%%%%%%%%%%%%%%%%%%%%%%%
\subsubsection{Polyhedron $F_{14}$: $G_{F_{14}}=\text{SU(3)}\times\text{SU(2)}^2\times\text{U(1)}$}
\label{sec:polyF14}
%%%%%%%%%%%%%%%%%%%%%%%%%%%%%%%%%%%%%%%%%%%%%%%%%%%%%%%%%%%%%%%%%%%%%%%%%%%%%%%%%%%%%%%%%%%%%%%%%

\begin{figure}[H]
\center
\includegraphics[scale=0.4]{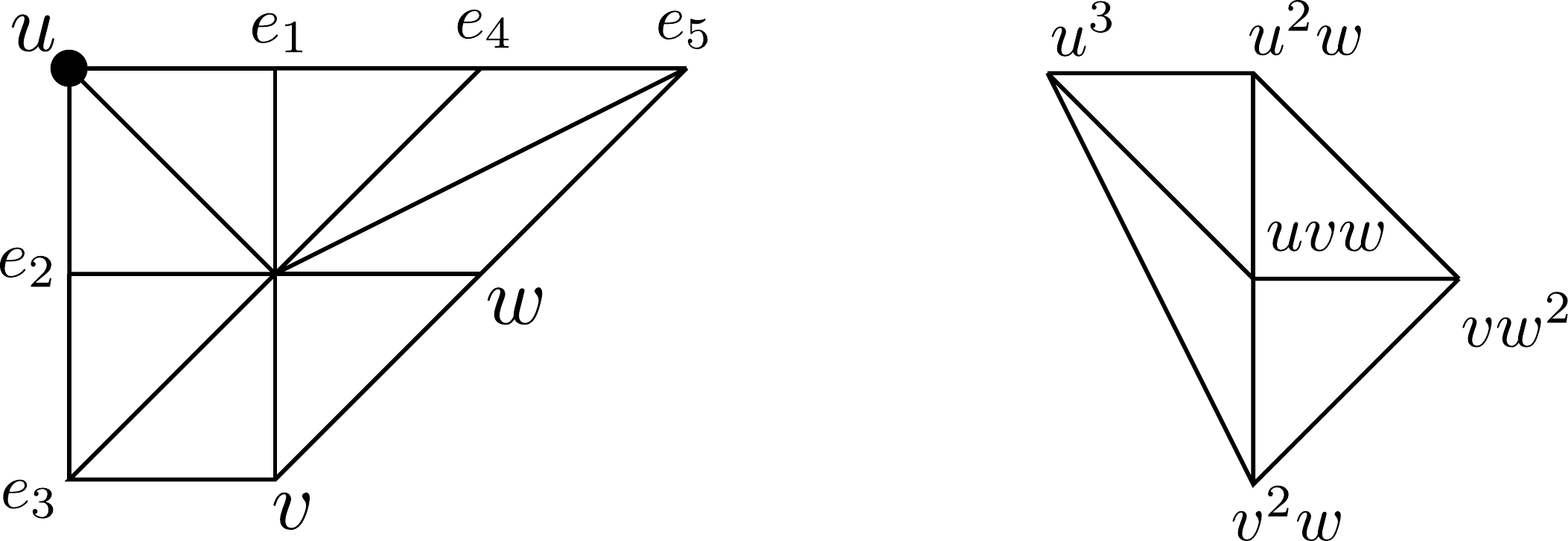}
\caption{\label{fig:poly14_toric}The toric diagram of polyhedron $F_{14}$ and its dual. The zero section is indicated by the dot.}
\end{figure}

Consider the elliptically fibered Calabi-Yau manifold $X_{F_{14}}$ with 
base $B$ and general elliptic fiber  given by the elliptic curve 
$\mathcal{E}$ in $\mathbb{P}_{F_{14}}$.
In Figure \ref{fig:poly14_toric} the toric data of $\mathbb{P}_{F_{14}}$ 
is summarized in terms of its polyhedron $F_{14}$,
a choice of homogeneous coordinates as well as its dual polyhedron 
$F_{3}$ with all monomials (shown in the patch $e_i=1$, $\forall i$) 
corresponding to its integral points.
Note that $\mathbb{P}_{F_{14}}$ is the blow-up of $\mathbb{P}^2$ defined 
by the blow-up map
\begin{align}\label{eq:F14blowup}
u \rightarrow e_1 e_2 e_3 e_4 e_5 u \,,\quad w\rightarrow e_1 e_4^2 e_5^3 w \,, \quad	v\rightarrow e_2 e_3^2 v \,.
% u \rightarrow e_1 u\; &, \; w\rightarrow e_1 w\\
% u \rightarrow e_2 u\; &, \; v\rightarrow e_2 v\\
% e_2 \rightarrow e_3 e_2\; &, \; v\rightarrow e_3 v\\
% e_1 \rightarrow e_4 e_1\; &, \; w\rightarrow e_4 w\\
% e_4 \rightarrow e_5 e_4\; &, \; w\rightarrow e_5 w
\end{align}
The homogeneous coordinates $[u:v:w:e_1:e_2:e_3:e_4:e_5]$ on 
$\mathbb{P}_{F_{14}}$ take values in the line bundles associated to the 
following divisors:
\begin{equation}
\label{tab:poly14_bundles}
\text{
\begin{tabular}{|c|c|}\hline
Section & Divisor class\\ \hline
$u$ & $H-E_1-E_2+\cS_9+[K_B]$ \\ \hline
$v$ & $H-E_2-E_3+\cS_9-\cS_7$\\ \hline
$w$ & $H-E_1-E_4-E_5$\\ \hline
$e_1$ & $E_1-E_4$\\ \hline
$e_2$ & $E_2-E_3$\\ \hline
$e_3$ &  $E_3$\\ \hline
$e_4$ &  $E_4-E_5$\\ \hline
$e_5$ &  $E_5$\\ \hline
\end{tabular}}
\end{equation}
The Stanley-Reisner ideal of $\mathbb{P}_{F_{14}}$ is given by
\begin{align}
\begin{split}
SR&=\{ u e_4, u e_5, uw, uv, ue_3, e_1e_5, e_1 w, e_1v, e_1 e_3, e_1e_2, \\  &\phantom{=} 
e_4w, e_4 v, e_4 e_3, e_4 e_2, e_5v, e_5 e_3, e_5 e_2, w e_3, w e_2, ve_2\} \, .
\end{split}
\end{align}
In order to find the hypersurface equation for $X_{F_{14}}$ we either 
use \eqref{eq:BatyrevFormula} or specialize  \eqref{eq:pF1}
as $s_2=s_3=s_4=s_8=s_{10}=0$ and apply the map \eqref{eq:F14blowup}.
It reads
\begin{align}
% \begin{split}
\label{eq:pF14}
p_{F_{14}}&=s_1 e_1^2 e_2^2 e_3 e_4 u^3+s_5 e_1^2 e_2 e_4^2 e_5^2 u^2 w+s_6 e_1 e_2 e_3 e_4 e_5 u v w+s_7 e_2 e_3^2 v^2 w%\\ &\phantom{=}
+s_9 e_1 e_4^2 e_5^3 v w^2 \, ,
% \end{split}
\end{align}
where the classes of the  sections $s_i$ are given in 
\eqref{eq:cubicsections}.

There are four rational sections of the fibration of $X_{F_{14}}$, two 
of which being linearly independent.
The coordinates of the sections are
\begin{align}
\begin{split} \label{eq:sectionsF14}
\hat{s}_0=X_{F_{14}}\cap\{u=0\}&:\quad [0:1:1:s_7:-s_9:1:1:1] \, ,\\
\hat{s}_1=X_{F_{14}}\cap\{e_5=0\}&:\quad [1:1:s_1:1:1:1:-s_7:0] \, ,\\
X_{F_{14}}\cap\{e_3=0\}&:\quad [1:s_5:1:1:-s_9:0:1:1] \, ,\\
X_{F_{14}}\cap\{v=0\}&:\quad [1:0:s_1:1:1:-s_5:1:1] \,,
\end{split}
\end{align}
where we choose $\hat{s}_0$ as the zero section and $\hat{s}_1$ as the 
generator of the MW-group of $X_{F_{14}}$.

\begin{figure}[h!]
\center
\includegraphics[scale=0.8]{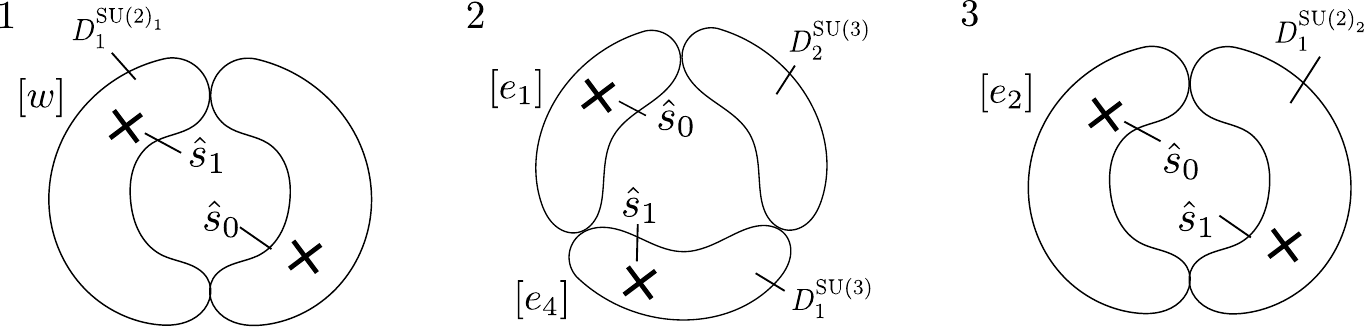}
\caption{\label{fig:poly14_codim1}Codimension one fibers
of $X_{F_{14}}$. The crosses denote the intersections
with the two sections.}
\end{figure}

As a prerequisite for the analysis of the singularities of $X_{F_{14}}$,
we compute its Weierstrass form \eqref{eq:WSF}. This is obtained
by applying Nagell's algorithm to \eqref{eq:pF14}.
The WS-coordinates of the section $\hat{s}_1$ are given by \eqref{eq:WScoordsSecF3} after setting $s_2=s_3=s_4=s_8=0$.
Similarly, we obtain
the functions $f$ and $g$ using the specialization 
$s_2=s_3=s_4=s_8=s_{10}=0$ from the general expressions
\eqref{eq:fcubic} and \eqref{eq:gcubic}, respectively.
To find all codimension one singularities of the WSF of $X_{F_{14}}$ we 
calculate the discriminant from $f$ and $g$. The discriminant 
$\Delta$ factorizes as follows: We find two $I_2$-singularities over the 
divisors $\mathcal{S}^b_{\text{SU}(2)_1}=\{s_1=0\}\cap B$ and 
$\mathcal{S}^b_{\text{SU}(2)_2}=\{s_9=0\}\cap B$ in $B$ and one 
$I_3$-singularity over the divisor 
$\mathcal{S}^b_{\text{SU}(3)}=\{s_7=0\}\cap B$ in $B$. The constraint 
\eqref{eq:pF14} factorizes along these divisors as
\begin{align}
\begin{split}\label{eq:SU2F14}
\SU2_1&:\quad p_{F_{14}}|_{s_1=0}= w \cdot q_2 \, ,\\
\SU3&:\quad p_{F_{14}}|_{s_7=0}= e_1 e_4 \cdot q_3 \, ,\\
\SU2_2&:\quad p_{F_{14}}|_{s_9=0}= e_2 \cdot q_3' \, ,
\end{split}
\end{align}
where $q_2$, $q_3$ and $q_3'$ are the polynomials that remain after 
factoring out $w$, $e_1 e_4$ and $e_2$. The fibers at these three 
codimension one loci are depicted in Figure \ref{fig:poly14_codim1}.
In summary, the  gauge group of $X_{F_{14}}$  is given by
\begin{equation} \label{eq:GF14}
	G_{F_{14}}=\SU3\times \text{SU}(2)^2\times\text{U}(1)\,.
\end{equation}
The divisor classes of the Cartan divisors are calculated in a similar
way as in Section \ref{sec:polyF6}. They read
\begin{align}\label{eq:CartansF14}
D^{\SU2_1}_{1}=[w] \, ,\quad
D^{\SU3}_{1}=[e_4] \, ,\quad
D^{\SU3}_{2}=[s_7]-[e_1]-[e_4]\, ,\quad
D^{\SU2_2}_{1}=[s_9]-[e_2]\, .
\end{align}
Using these results, we compute the Shioda map \eqref{eq:ShiodaMap} of the section $\hat{s}_1$ as
\begin{align} \label{eq:ShiodaF14}
\sigma (\hat{s}_1)=S_1-S_0 +  [K_B] + \tfrac{1}{2}D^{\text{SU}(2)_1}_{1}+\tfrac{1}{3} \big( 2D^{\text{SU}(3)}_{1}+D^{\text{SU}(3)}_{2} \big) +\tfrac{1}{2} D^{\text{SU}(2)_2}_{1}\, .
\end{align}
Here $S_0$, $S_1$ denote the divisor classes of the sections
$\hat{s}_0$, $\hat{s}_1$, respectively, and  we have employed
\beq
	\pi(S_1\cdot S_0)=0\,,
\eeq
which follows from \eqref{eq:sectionsF14} as well as
\begin{align}
S_1 \cdot C^{SU(2)_1}_{-\alpha _1}=1 \, ,\quad
S_1 \cdot C^{SU(3)}_{-\alpha _i}=\left ( \begin{matrix} 1\\0 \end{matrix} \right ) \, ,\quad
S_1 \cdot C^{SU(2)_2}_{-\alpha _1}=1\, ,
\end{align}
which can be deduced from Figure \ref{fig:poly14_codim1}.
Employing \eqref{eq:ShiodaF14}, we  compute the height pairing 
\eqref{eq:anomalycoeff}, using these results and the intersection 
\eqref{eq:SP^2},  as
\begin{align}\label{eq:bmnF14}
 b_{11}=-\frac{1}{2}[K_B]-\frac{1}{6}\cS_7\,.
\end{align}

Next,  we determine the spectrum of charged matter by investigating the 
codimension two singularities of the WSF of $X_{F_{14}}$.
As before, all matter representations under the gauge group $G_{F_{14}}$ are obtained
by application of the techniques discussed in Section~\ref{sec:ellipticCurvesWithRP}. Again, the non-trivial representations
under the non-Abelian part of $G_{F_{14}}$ easily follow from the 
discriminant $\Delta$, while the charged singlets require the primary
decomposition of the locus \eqref{eq:charge1Matter} for the section 
$\hat{s}_1$. We find nine codimension two singularities in $X_{F_{14}}$. 
Six of these lead to the matter representations and the corresponding
codimension two fibers in $X_{F_{14}}$ given in the first and third
column of Table~\ref{tab:poly14_matter}, respectively.
The remaining loci, namely $s_1=s_6^2-4s_5s_7=0$,  
$s_9=s_6^2-4s_5s_7=0$ and $s_6=s_7=0$, support two type $III$ and one 
type $IV$ fiber, respectively, and thus do not support further 
representations. The adjoint representations of $G_{F_{14}}$
are shown in the last three rows of Table~\ref{tab:poly14_matter} for
completeness.

We note that the fiber corresponding to the representation 
$(\two,\one,\two)_0$ is non-split, cf.~Section~\ref{sec:polyF8} 
for a more detailed discussion. We have indicated the fibers that
are exchanged by codimension three monodromies by dashed lines
in Table~\ref{tab:poly14_matter}.

The multiplicities of the charged hyper multiplets are presented in the 
second column of Table \ref{tab:poly14_matter}. They are computed 
directly from all complete intersections $V(I_{(k)})$, $k=1,\ldots, 5$.

The matter spectrum of $X_{F_{14}}$ is completed by the number of neutral
hyper multiplets, which can be computed from \eqref{eq:Hneutral} using
the Euler number \eqref{eq:EulerNumbers} of $X_{F_{14}}$. It is given by
\beq \label{eq:HneutF14}
H_{\text{neut}} = 17 + 11 [K_B^{-1}]^2 - 9 [K_B^{-1}] \cS_7 + 3 \cS_7^2 - 6 [K_B^{-1}] \cS_9 + 2 \cS_7 \cS_9 + 2 \cS_9^2 \,.
\eeq
\begin{table}[H]
\begin{center}
\renewcommand{\arraystretch}{1.2}
\begin{tabular}{|c|c|c|c|}\hline
Representation & Multiplicity & Fiber & Locus \\ \hline
$(\two,\one,\one)_{1/2}$ & $\begin{array}{c} (3[K_B^{-1}] -\cS_7-\cS_9)\\ \times (2 [K_B^{-1}]-\cS_7) \end{array}$ & \rule{0pt}{1.5cm}\parbox[c]{2.9cm}{\includegraphics[width=2.9cm]{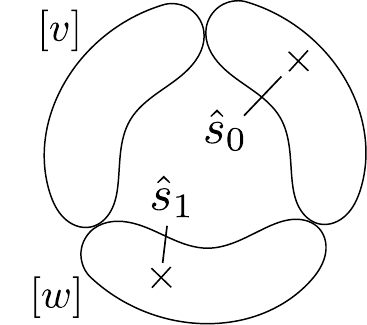}} & $\begin{array}{c} V(I_{(1)}):= \\ \{ s_1=s_5=0 \} \end{array}$  \\[1.1cm] \hline
$(\one,\three,\one)_{-1/3}$ & $\cS_7 (3 [K_B^{-1}]-\cS_7)$ & \rule{0pt}{1.6cm}\parbox[c]{2.9cm}{\includegraphics[width=2.9cm]{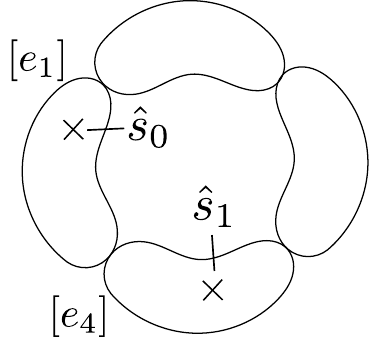}} & $\begin{array}{c} V(I_{(2)}):=\{ s_7=0 \\ s_5s_6-s_1s_9=0 \} \end{array}$  \\[1.2cm] \hline
$(\one,\one,\two)_{1/2}$ & $\cS_9 (2 [K_B^{-1}]-\cS_7)$ & \rule{0pt}{1.5cm}\parbox[c]{2.9cm}{\includegraphics[width=2.9cm]{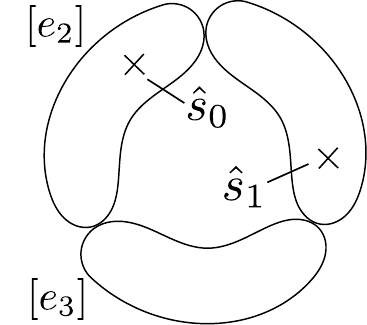}} & $\begin{array}{c} V(I_{(3)}):= \\ \{ s_5=s_9=0 \} \end{array}$ \\[1.1cm] \hline
$(\two,\three,\one)_{1/6}$ & $\cS_7(3[K_B^{-1}]-\cS_7-\cS_9)$ & \rule{0pt}{1.6cm}\parbox[c]{2.9cm}{\includegraphics[width=2.9cm]{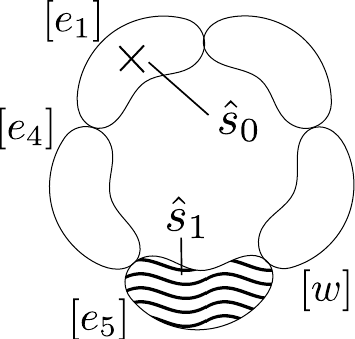}} & $\begin{array}{c} V(I_{(4)}):= \\ \{ s_1=s_7=0 \} \end{array}$  \\[1.3cm] \hline
$(\two,\one,\two)_{0}$ & $\cS_9 (3[K_B^{-1}]-\cS_7-\cS_9)$ & \rule{0pt}{1.6cm}\parbox[c]{2.9cm}{\includegraphics[width=2.9cm]{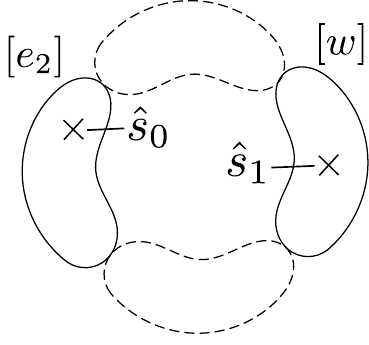}} & $\begin{array}{c} V(I_{(5)}):= \\ \{ s_1=s_9=0 \} \end{array}$   \\[1.2cm] \hline
$(\one,\three, \two)_{1/6}$ & $\cS_7 \cS_9$ & \rule{0pt}{1.6cm}\parbox[c]{2.9cm}{\includegraphics[width=2.9cm]{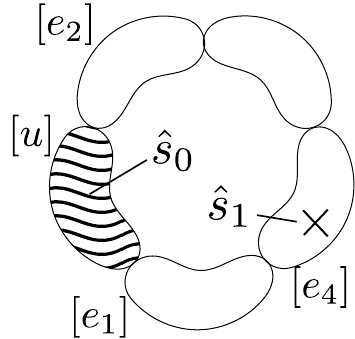}} & $\begin{array}{c} V(I_{(6)}):= \\ \{ s_7=s_9=0 \} \end{array}$  \\[1.3cm] \hline \hline
$(\three,\one,\one)_{0} $ & \rule{0pt}{0.55cm}$\begin{array}{c} 1 + \tfrac12 (2 [K_B^{-1}] - \cS_7 - \cS_9) \\ \times (3 [K_B^{-1}] - \cS_7 - \cS_9) \end{array}$ & Figure \ref{fig:poly14_codim1} & $s_1=0$ \\[0.05cm] \hline
$(\one,8,\one)_{0}$ & \rule{0pt}{0.55cm}$1 + \cS_7 \frac{(\cS_7 - [K_B^{-1}])}{2}$ & Figure \ref{fig:poly14_codim1} & $s_7=0$ \\[0.05cm] \hline
$(\one,\one,\three)_{0}$ & \rule{0pt}{0.55cm}$1 + \cS_9 \frac{(\cS_9 - [K_B^{-1}])}{2}$ & Figure \ref{fig:poly14_codim1} & $s_9=0$ \\[0.05cm] \hline
\end{tabular}
\caption{\label{tab:poly14_matter}Charged matter representations under SU$(3)\times \SU2^2\times$U(1) and corresponding codimension two fibers of $X_{F_{14}}$. The adjoint matter is included for completeness.}
\end{center}
\end{table}
We note that there are a base-dependent number $T$ of tensor  
and $V=15$ vector multiplets.
As a consistency check we confirm cancelation of all 6D anomalies 
following Appendix \ref{app:Anomalies}, employing the divisors 
$\mathcal{S}^b_{\text{SU}(2)_1}$, 
$\mathcal{S}^b_{\text{SU}(2)_2}$,
$\mathcal{S}^b_{\text{SU}(3)}$, 
\eqref{eq:bmnF14}, the spectrum in Table \ref{tab:poly14_matter} 
and \eqref{eq:HneutF14}.

Finally, we list all codimension three singularities of the WSF of 
$X_{F_{14}}$ and the corresponding Yukawa points of an F-theory
compactification to 4D  in Table \ref{tab:poly14_yukawa}.
\begin{table}[htb!]
\begin{center}
\renewcommand{\arraystretch}{1.2}
\begin{tabular}{|c|c|}\hline
Yukawa & Locus \\ \hline
$(\two,\one,\one)_{1/2} \cdot (\one,\three,\one)_{-1/3} \cdot \overline{(\two,\three,\one)_{1/6}}$ & $s_1=s_5=s_7=0$  \\ \hline

$(\two,\one,\one)_{1/2} \cdot \overline{(\one,\one,\two)_{1/2}} \cdot (\two,\one,\two)_{0}$ & $s_1=s_5=s_9=0$ \\ \hline

$(\one,\three,\one)_{-1/3} \cdot (\one,\one,\two)_{1/2} \cdot \overline{(\one,\three,\two)_{1/6}}$ & $s_5=s_7=s_9=0$  \\ \hline

$(\two,\three,\one)_{1/6} \cdot (\two,\one,\two)_{0} \cdot \overline{(\one,\three,\two)_{1/6}}$ & $s_1=s_7=s_9=0$  \\ \hline
\end{tabular}
\caption{\label{tab:poly14_yukawa}Codimension three loci and corresponding Yukawa couplings for $X_{F_{14}}$.}
\end{center}
\end{table}

%%%%%%%%%%%%%%%%%%%%%%%%%%%%%%%%%%%%%%%%%%%%%%%%%%%%%%%%%%%%%%%%%%%%
%%%%%%%%%%%%%%%%%%%%%%%%%%%%%%%%%%%%%%%%%%%%%%%%%%%%%%%%%%%%%%%%%%%%

%%%%%%%%%%%%%%%%%%%%%%%%%%%%%%%%%%%%%%%%%%%%%%%%%%%%%%%%%%%%%%%%%%%%%%%%%%%%%%%%%%%%%%%%%%%%%%%%%%
\subsection{Fibrations with gauge groups of rank 5 and 6 and MW-torsion}
\label{sec:TorsionPolys}
%%%%%%%%%%%%%%%%%%%%%%%%%%%%%%%%%%%%%%%%%%%%%%%%%%%%%%%%%%%%%%%%%%%%%%%%%%%%%%%%%%%%%%%%%%%%%%%%%%

In this section we study the toric hypersurface fibrations constructed
from the fiber polyhedra $F_{13}$, $F_{15}$ and $F_{16}$.
These are the three toric hypersurface fibrations that have
non-trivial Mordell-Weil torsion and give rise to non-simply connected 
gauge groups in F-theory. 

The Calabi-Yau manifold $X_{F_{13}}$
has Mordell-Weil group $\mathbb{Z}_2$, $X_{F_{15}}$ has Mordell-Weil 
group $\mathbb{Z}\oplus\mathbb{Z}_2$ and the fibration $X_{F_{16}}$ has 
Mordell-Weil group $\mathbb{Z}_3$ 
\cite{Braun:2013nqa,Mayrhofer:2014opa}.
We confirm these findings by explicitly working out the WSF
of these toric hypersurface fibrations, which are shown to precisely
take the standard form of WSF's 
with these MW-torsion groups, cf.~\cite{Aspinwall:1998xj}. 

The influence of the MW-torsion on the spectrum of F-theory was 
discussed recently in 
\cite{Mayrhofer:2014opa}. There, the models considered in this section
were also studied, but under the assumption of a holomorphic zero 
section. Here, we relax this condition which results in additional gauge 
groups and matter representations. This has interesting 
consequences for the phenomenology of these models, because we find
that the gauge groups and matter representations of $X_{F_{13}}$ and
$X_{F_{15}}$ are completed precisely into the ones of the 
\emph{Pati-Salam}
and \emph{trinification} model, respectively.

%%%%%%%%%%%%%%%%%%%%%%%%%%%%%%%%%%%%%%%%%%%%%%%%%%%%%%%%%%%%%%%%%%%%%%%%%%%%%%%%%%%%%%%%%%%%%%%%%
\subsubsection{Polyhedron $F_{13}$: $G_{F_{13}}=(\text{SU(4)}\times\text{SU(2)}^2)/\mathbb{Z}_2$}
\label{sec:polyF13}
%%%%%%%%%%%%%%%%%%%%%%%%%%%%%%%%%%%%%%%%%%%%%%%%%%%%%%%%%%%%%%%%%%%%%%%%%%%%%%%%%%%%%%%%%%%%%%%%%

\begin{figure}[ht!]
\center
\includegraphics[scale=0.4]{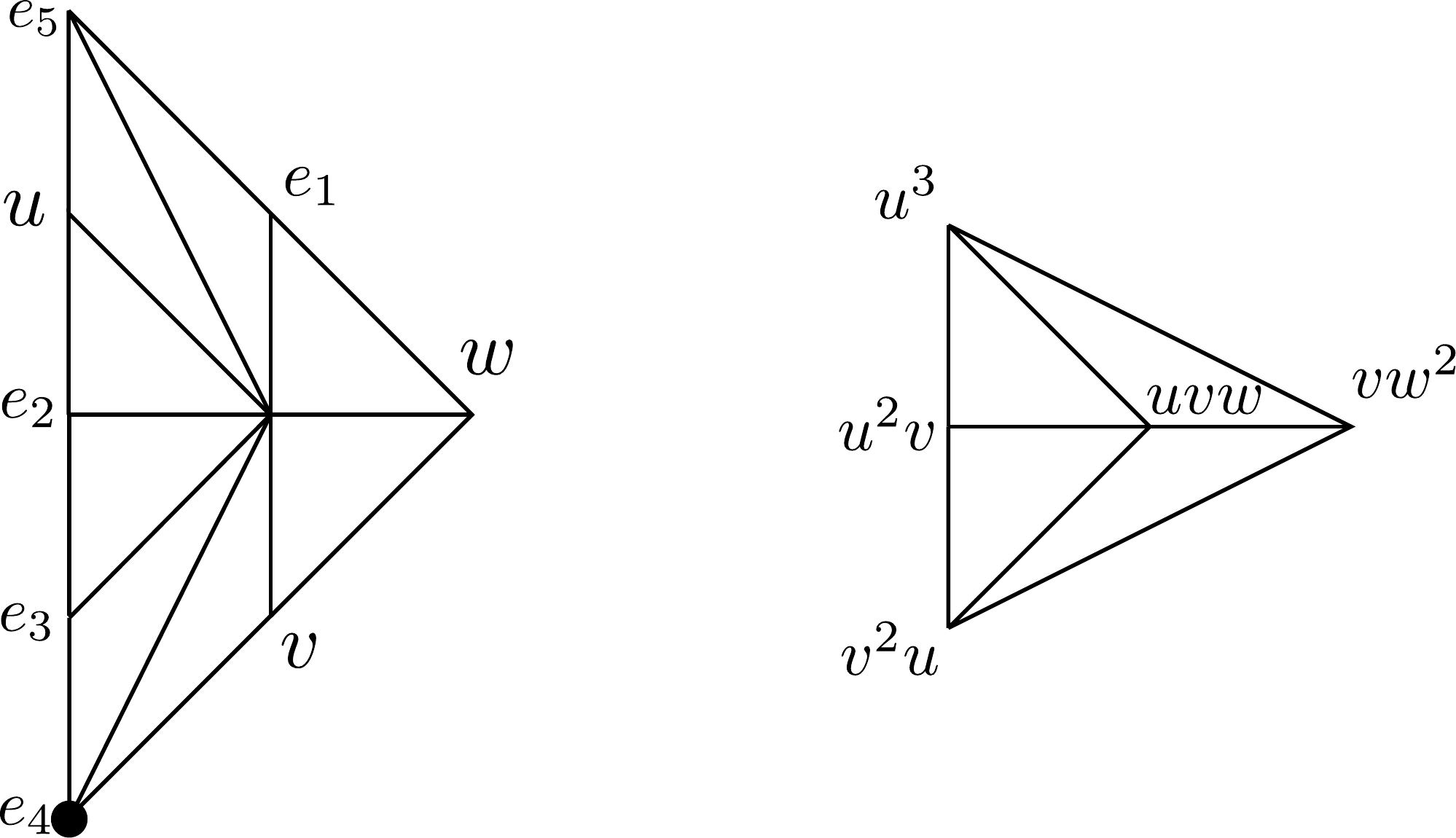}
\caption{\label{fig:poly13_toric}The toric diagram of polyhedron $F_{13}$ and its dual. The zero section is indicated by the dot.}
\end{figure}

Consider the elliptically fibered Calabi-Yau manifold $X_{F_{13}}$ with 
base $B$ and general fiber  given by the elliptic curve $\mathcal{E}$ in 
$\mathbb{P}_{F_{13}}$. The toric diagram of the fiber polyhedron 
$F_{13}$ as well as a choice of homogeneous coordinates and its dual 
polyhedron are depicted in Figure~\ref{fig:poly13_toric}, where we have 
set $e_i=1$, $\forall i$, in the monomials that are associated to the 
integral points of $F_{4}$ by \eqref{eq:BatyrevFormula}.

We note that $\mathbb{P}_{F_{13}}$ is the blow-up of $\mathbb{P}^2$
at five non-generic points, that is defined by
\begin{align}\label{eq:F13blowup}
u \rightarrow e_1 e_2 e_3 e_4 e_5^2 u\; &, \; w\rightarrow e_1 e_5 w \,, \quad v\rightarrow e_2 e_3^2 e_4^3 v \, .
% u \rightarrow e_1 u\; &, \; w\rightarrow e_1 w\\
% u \rightarrow e_2 u\; &, \; v\rightarrow e_2 v\\
% e_2 \rightarrow e_3 e_2\; &, \; v\rightarrow e_3 v\\
% e_3 \rightarrow e_4 e_3\; &, \; v\rightarrow e_4 v\\
% u \rightarrow e_5 u\; &, \; e_1\rightarrow e_5 e_1
\end{align}
The homogeneous coordinates on the fiber after this blow-up are 
$[u:v:w:e_1:e_2:e_3:e_4:e_5]$ and take values in the line bundles 
associated to the divisor classes given by:
\begin{align}
\label{tab:poly13_bundles}
\text{
\begin{tabular}{|c|c|}\hline
Section & Divisor class\\ \hline
$u$ & $H-E_1-E_2-E_5+\cS_9+[K_B]$ \\ \hline
$v$ & $H-E_2-E_3-E_4+\cS_9-\cS_7$\\ \hline
$w$ & $H-E_1$\\ \hline
$e_1$ & $E_1-E_5$\\ \hline
$e_2$ & $E_2-E_3$\\ \hline
$e_3$ &  $E_3-E_4$\\ \hline
$e_4$ &  $E_4$\\ \hline
$e_5$ &  $E_5$\\ \hline
\end{tabular}}
\end{align}
The Stanley-Reisner ideal of $\mathbb{P}_{F_{13}}$ follows from
Figure~\ref{fig:poly15_toric} as
\begin{align}
\begin{split}
SR&=\{ u e_1, uw, uv, u e_4, u e_3, e_5 w, e_5 v, e_5 e_4, e_5 e_3, e_5 e_2, e_1 v, e_1 e_4, e_1 e_3, e_1 e_2, \\ 
&\phantom{=} w e_4, w e_3, we_2, ve_3, ve_2, e_4 e_2\} \, .
\end{split}
\end{align}
We find the hypersurface equation for $X_{F_{13}}$ in the total
space of the fibration \eqref{eq:PFfibration} with $F_i\equiv F_{13}$  
using \eqref{eq:BatyrevFormula} or directly from \eqref{eq:pF1} by 
setting $s_4=s_5=s_7=s_8=s_{10}=0$ and by applying the map 
\eqref{eq:F13blowup}.  We obtain
\begin{align}
\begin{split} \label{eq:pF13}
p_{F_{13}}&= s_1 e_1^2 e_2^2e_3 e_5^4 u^3 + s_2 e_1 e_2^2 e_3^2 e_4^2 e_5^2 u^2 v + s_3 e_2^2 e_3^3 e_4^4 u v^2 + s_6 e_1 e_2 e_3 e_4 e_5 u v w %\\ &\phantom{=}
+s_9 e_1 v w^2\, ,
\end{split}
\end{align}
where the divisor classes of the $s_i$ are given in 
\eqref{eq:cubicsections}.

There are two seemingly rational sections of the fibration of 
$X_{F_{13}}$. However, there is one torsional relation between them, 
which reveals that the MW-group is pure torsion, namely $\mathbb{Z}_2$ 
\cite{Braun:2013nqa}. The coordinates of our choice for 
the zero section $\hat{s}_0$ and the section of order two are
\begin{align}
\begin{split}\label{eq:sectionsF13}
\hat{s}_0=X_{F_{13}}\cap\{e_4=0\}&:\quad [1:s_1:1:1:1:-s_9:0:1]\, ,\\
X_{F_{13}}\cap\{e_5=0\}&:\quad [s_9:1:1:-s_3:1:1:1:0]\, .
\end{split}
\end{align}

The presence of MW-torsion restricts the matter spectrum 
realized in F-theory \cite{Aspinwall:1998xj}. Indeed,
the torsion acts on the gauge group, turning it into a non-simply 
connected group, which reduces its weight lattice, i.e.~the realized 
representations. In \cite{Mayrhofer:2014opa} this has recently been
understood in terms of a geometric $k$-fractional refinement of the 
coweight lattice. In particular, it has been argued that the MW-torsion 
$\mathbb{Z}_2$ of $X_{F_{13}}$ forbids the presence of  fundamental 
matter in this model. We will confirm these findings in the following 
explicit analysis.

We begin by computing the Weierstrass form \eqref{eq:WSF} of 
\eqref{eq:pF13}. As an intermediate step we use the birational map of 
$X_{F_5}$ to the Tate form given in \cite{Cvetic:2013nia} in the
limit $s_5=s_7=s_8=0$. We obtain the local Tate coefficients 
\eqref{eq:TateF13sing} from which we readily compute the functions $f$ 
and $g$, that are given in \eqref{eq:WSeqX13}. We note that the same
WSF arises from the global Tate model  given  \eqref{eq:TateFormZ2MW},
which precisely agrees with the Tate form of a model with $\mathbb{Z}_2$
MW-torsion as argued in \cite{Aspinwall:1998xj}, confirming the 
presence of $\mathbb{Z}_2$ MW-torsion in $X_{F_{13}}$.

\begin{figure}[h!]
\center
\includegraphics[scale=0.8]{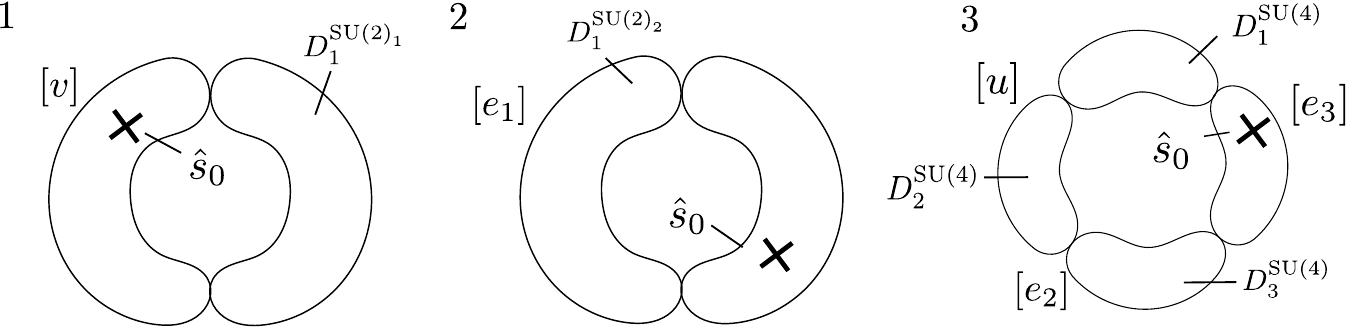}
\caption{\label{fig:poly13_codim1}Codimension one fibers
of $X_{F_{13}}$. The crosses denote the intersections with the zero section.}
\end{figure}

Using these results, we readily compute the discriminant $\Delta$, which 
allows us to  find all codimension one singularities
of the WSF of $X_{F_{13}}$. We find two $I_2$-singularities over the 
divisors $\mathcal{S}^b_{\text{SU}(2)_1}=\{s_1=0\}\cap B$ and 
$\mathcal{S}^b_{\text{SU}(2)_2}=\{s_3=0\}\cap B$ in $B$ as well as an 
$I_4$-singularity over the divisor 
$\mathcal{S}^b_{\text{SU}(4)}=\{s_9=0\}\cap B$ in $B$. Along these
divisors the constraint \eqref{eq:pF13} factorizes as
\begin{align}
\begin{split}\label{eq:SU2F13}
\SU2_1&:\quad p_{F_{13}}|_{s_1=0}= v \cdot q_2 \, ,\\
\SU2_2&:\quad p_{F_{13}}|_{s_3=0}= e_1 \cdot q_3\, ,\\
\SU4&:\quad p_{F_{13}}|_{s_9=0}= u e_2 e_3\cdot q_2' \, ,
\end{split}
\end{align}
where $q_2$, $q_3$ and $q_2'$ are the polynomials that remain after 
factoring out $v$, $e_1$ and $u e_2 e_3$. The corresponding fibers are 
depicted in Figure~\ref{fig:poly13_codim1} and give rise to two SU(2)  
and one SU(4) gauge groups. There is another potential codimension one 
singularity of the WSF of $X_{F_{13}}$, where the fiber of $X_{F_{13}}$ 
splits into two $\mathbb{P}^1$. However, as it is shown in
\cite{Mayrhofer:2014opa}, the torsional MW-group identifies these two 
$\mathbb{P}^1$'s, so that the fiber in the quotient space is a single 
singular $\mathbb{P}^1$.
Thus, there is no additional gauge symmetry and the gauge group of 
$X_{F_{13}}$ is
\begin{equation}\label{eq:GF13}
	G_{F_{13}}=(\SU4 \times \text{SU}(2)^2)/\mathbb{Z}_2\,.
\end{equation}
We note that this is precisely the Pati-Salam group.
The action of the MW-torsion on the gauge group $G_{F_{13}}$
is worked out in Appendix~\ref{app:AppPoly}. To this end,  we show
that the WS-coordinates of the generator of the $\mathbb{Z}_2$ 
MW-torsion, given in \eqref{eq:WScoordsSecF13}, pass through the
WS-coordinates of the singularities in the fiber  at all codimension one 
loci $s_1=0$, $s_3=0$ and $s_9=0$ in \eqref{eq:SU2F13}.
\begin{table}[H]
\begin{center}
\renewcommand{\arraystretch}{1.2}
\begin{tabular}{|c|c|c|c|}\hline
Representation & Multiplicity & Fiber & Locus \\ \hline

$(\two, \two, \one)$ & $\begin{array}{c} (3[K_B^{-1}]-\cS_7-\cS_9) \\ \times ([K_B^{-1}]+\cS_7-\cS_9) \end{array}$ & \rule{0pt}{1.6cm}\parbox[c]{2.9cm}{\includegraphics[width=2.9cm]{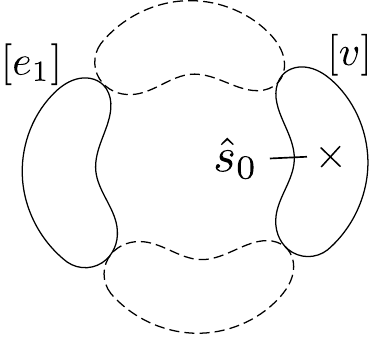}} & $\begin{array}{c} V(I_{(1)}):= \\ \{ s_1=s_3=0 \} \end{array}$  \\[1.1cm] \hline

$(\two, \one, \mathbf{4})$ & $(3[K_B^{-1}]-\cS_7-\cS_9)\cS_9$ & \rule{0pt}{1.9cm}\parbox[c]{2.9cm}{\includegraphics[width=2.9cm]{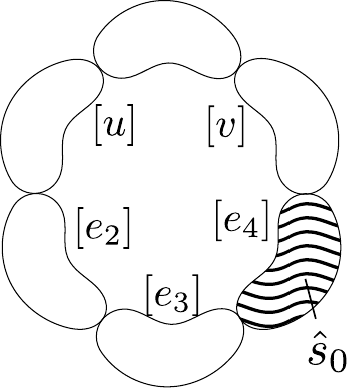}} & $\begin{array}{c} V(I_{(2)}):= \\ \{ s_1=s_9=0 \} \end{array}$  \\[1.4cm] \hline

$(\one, \two, \mathbf{4})$ & $([K_B^{-1}]+\cS_7-\cS_9)\cS_9$ & \rule{0pt}{1.9cm}\parbox[c]{2.9cm}{\includegraphics[width=2.9cm]{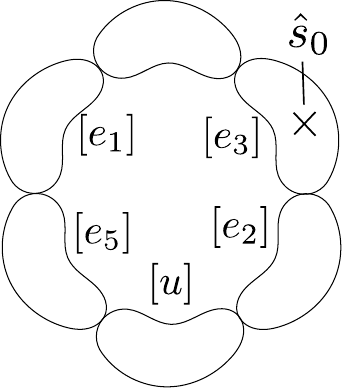}} & $\begin{array}{c} V(I_{(3)}):= \\ \{ s_3=s_9=0 \} \end{array}$  \\[1.4cm] \hline

$(\one, \one, \mathbf{6})$ & $\cS_9 [K_B^{-1}]$ & \rule{0pt}{1.3cm}\parbox[c]{2.9cm}{\includegraphics[width=2.9cm]{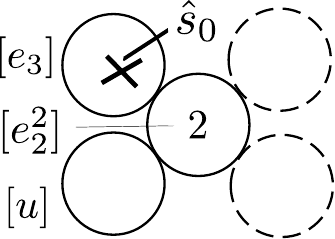}} & $\begin{array}{c} V(I_{(4)}):= \\ \{ s_6=s_9=0 \} \end{array}$  \\[0.9cm] \hline \hline

$(\three,\one,\one)$ & \rule{0pt}{0.85cm}$\begin{array}{c} 1 + \frac{((2 [K_B^{-1}] - \cS_7 - \cS_9))}{2} \\ \times (3 [K_B^{-1}] - \cS_7 - \cS_9) \end{array}$ & Figure~\ref{fig:poly13_codim1} & $s_1=0$ \\ \hline

$(\one,\three,\one)$ & \rule{0pt}{0.85cm}$\begin{array}{c} 1 + \frac{(\cS_7-\cS_9)}{2} \\ \times ([K_B^{-1}]+\cS_7-\cS_9) \end{array}$ & Figure~\ref{fig:poly13_codim1} & $s_3=0$ \\ \hline

$(\one,\one,\mathbf{15})$ & \rule{0pt}{0.6cm}$1 + \cS_9 \frac{(\cS_9 - [K_B^{-1}])}{2}$ & Figure~\ref{fig:poly13_codim1} & $s_9=0$ \\ \hline

\end{tabular}
\caption{\label{tab:poly13_matter}Charged matter representations under $(\SU4\times\SU2^2)/\mathbb{Z}_2$ and corresponding codimension two fibers of $X_{F_{13}}$. The adjoint matter is included for completeness.}
\end{center}
\end{table}
As before, cf.~Section~\ref{sec:polyF6}, we calculate the classes of the 
Cartan divisors of $X_{F_{13}}$ as
\begin{align}
\begin{split}\label{eq:CartansF13}
D^{\SU2_1}_{1}=[s_1]-[v] \, &,\quad
D^{\SU2_2}_{1}=[e_1] \, ,\\
D^{\SU4}_{1}=[s_9]-[u]-[e_2]-[e_3]\, &, \quad D^{\SU4}_{2}=[u] \, ,\quad
D^{\SU4}_{3}=[e_2] \, .
\end{split}
\end{align}

Next, we calculate the charged matter spectrum of $X_{F_{13}}$,
which requires the analysis of all its codimension two singularities. We 
directly read off from the discriminant of $X_{F_{13}}$ the loci of
six codimension two singularities.
Four of these lead to the matter representations in the first column of 
Table~\ref{tab:poly13_matter} that are determined, using the techniques 
discussed in Section \ref{sec:ellipticCurvesWithRP}, from the 
corresponding codimension two fibers in $X_{F_{13}}$ given in the third 
column of the same table. 
The remaining loci $s_1=s_6^2-4s_2s_9=0$ and 
$s_3=s_6^2-4s_2s_9=0$ are both of type $III$ , that we first encountered 
in Section~\ref{sec:polyF6}, and, thus, do not support additional 
matter representations. The three adjoint representations in
the last three rows of Table~\ref{tab:poly13_matter} are shown for 
completeness.

We find three singularities which support the bi-fundamental 
representations and one singularity leading to an anti-symmetric 
representation of $SU(4)$, but no fundamental representation. 
This has been anticipated before, due to the action of MW-torsion
on the gauge group $G_{F_{13}}$  given in \eqref{eq:GF13}.
In addition, we find that the fibers at the loci $V(I_{(1)})$ and 
$V(I_{(4)})$, that correspond to the $(\two,\two,\one)$ and the 
$(\one,\one,\mathbf{6})$ representation, are non-split, 
cf.~Section~\ref{sec:polyF8} for more details. The $\mathbb{P}^1$'s
drawn with a dashed line are interchanged by codimension three
monodromies.

The spectrum of matter of $X_{F_{13}}$ is completed
by the number of neutral hyper multiplets, which is computed from 
\eqref{eq:Hneutral} using the Euler number given in 
\eqref{eq:EulerNumbers} of $X_{F_{13}}$. It reads
\beq \label{eq:HneutF13}
H_{\text{neutral}} = 17 + 11 [K_B^{-1}]^2 - 4 [K_B^{-1}] \cS_7 + 2 \cS_7^2 - 10 [K_B^{-1}] \cS_9 + 4 \cS_9^2 \,.
\eeq
The number $T$ of tensor multiplets is given by the base-dependent 
expression \eqref{eq:Tformula} and we have $V=27$ vector multiplets.
In order to check that all 6D anomalies are canceled we use the 
divisors $S_{\text{SU}(2)_1}^b$, $S_{\text{SU}(2)_2}^b$ and 
$S_{\text{SU}(4)}^b$ as well as the charged spectrum in 
Table~\ref{tab:poly13_matter} and \eqref{eq:HneutF13}. Indeed, we find 
that all 6D anomalies in \eqref{eq:6dAnomalies} are canceled.

We conclude this section with the list of all geometrically allowed 
Yukawa couplings of the charged matter spectrum of $X_{F_{13}}$, that is 
given in Table~\ref{tab:poly13_yukawa}.
\begin{table}[H]
\begin{center}
\renewcommand{\arraystretch}{1.2}
\begin{tabular}{|c|c|}\hline
Yukawa & Locus \\ \hline
$(\two,\two,\one) \cdot \overline{(\two,\one,\mathbf{4})} \cdot (\one,\two,\mathbf{4})$ & $s_1=s_3=s_9=0$ \\ \hline
\end{tabular}
\caption{\label{tab:poly13_yukawa}Codimension three loci and corresponding Yukawa coupling for $X_{F_{13}}$.}
\end{center}
\end{table}

%%%%%%%%%%%%%%%%%%%%%%%%%%%%%%%%%%%%%%%%%%%%%%%%%%%%%%%%%%%%%%%%%%%%%%%%%%%%%%%%%%%%%%%%%%%%%%%%%
\subsubsection{Polyhedron $F_{15}$: $G_{F_{15}}=\text{SU(2)}^4/\mathbb{Z}_2\times\text{U(1)}$}
\label{sec:polyF15}
%%%%%%%%%%%%%%%%%%%%%%%%%%%%%%%%%%%%%%%%%%%%%%%%%%%%%%%%%%%%%%%%%%%%%%%%%%%%%%%%%%%%%%%%%%%%%%%%%

\begin{figure}[H]
\center
\includegraphics[scale=0.4]{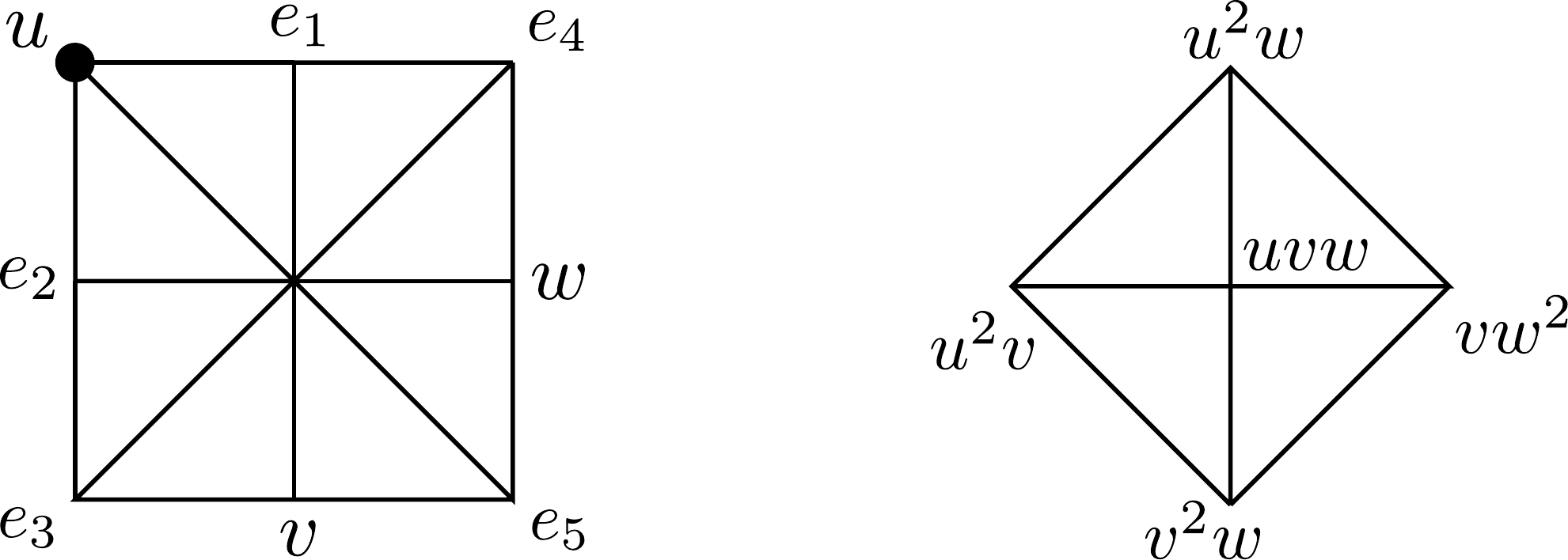}
\caption{\label{fig:poly15_toric}The toric diagram of polyhedron $F_{15}$ and its dual. The zero section is indicated by the dot.}
\end{figure}

We construct an elliptically fibered Calabi-Yau manifold $X_{F_{15}}$ 
over a base $B$ and with general fiber  given by the elliptic curve 
$\mathcal{E}$ in $\mathbb{P}_{F_{15}}$.
The toric data of $\mathbb{P}_{F_{15}}$ is encoded in 
Figure~\ref{fig:poly15_toric}, that shows the polyhedron $F_{15}$,
our convention for projective coordinates as well as the dual 
polyhedron $F_{2}$, with the monomials (in the patch $e_i=1$, $\forall i$) associated to its integral points.
We note that $\mathbb{P}_{F_{15}}$ is obtained from $\mathbb{P}^2$ by the five non-generic blow-ups defined by
\begin{align}\label{eq:F15blowup}
u \rightarrow e_1 e_2 e_3 e_4 u\, , \quad w\rightarrow e_1 e_4^2 e_5 w \,,\quad v\rightarrow e_2 e_3^2 e_5 v \,.
%u \rightarrow e_1 u\; &, \; w\rightarrow e_1 w\\
%u \rightarrow e_2 u\; &, \; v\rightarrow e_2 v\\
%e_2 \rightarrow e_3 e_2\; &, \; v\rightarrow e_3 v\\
%e_1 \rightarrow e_4 e_1\; &, \; w\rightarrow e_4 w\\
%v \rightarrow e_5 v\; &, \; w\rightarrow e_5 w\; .
\end{align}
The homogeneous coordinates on the fiber after this blow-up are  $[u:v:w:e_1:e_2:e_3:e_4:e_5]$. Their divisor classes are given by:
\begin{equation}
\label{tab:poly15_bundles}
\text{
\begin{tabular}{|c|c|}\hline
Section & Divisor class \\ \hline
$u$ & $H-E_1-E_2+\cS_9+[K_B]$ \\ \hline
$v$ & $H-E_2-E_3-E_5+\cS_9-\cS_7$\\ \hline
$w$ & $H-E_1-E_4-E_5$\\ \hline
$e_1$ & $E_1-E_4$\\ \hline
$e_2$ & $E_2-E_3$\\ \hline
$e_3$ & $E_3$\\ \hline
$e_4$ & $E_4$\\ \hline
$e_5$ & $E_5$\\ \hline
\end{tabular}}
\end{equation}
The Stanley-Reisner ideal of $\mathbb{P}_{F_{15}}$ is can be read off
from Figure~\ref{fig:poly15_toric} as
\begin{align}
\begin{split}
SR&=\{ u e_4, uw, u e_5, uv, ue_3, e_1 w, e_1e_5, e_1v, e_1 e_3, e_1e_2, e_4 e_5, e_4 v, e_4 e_3, e_4 e_2, \\ 
&\phantom{=}\phantom{...} wv, w e_3, w e_2, e_5 e_3, e_5 e_2, ve_2 \} \, .
\end{split}
\end{align}
We use \eqref{eq:BatyrevFormula} to find the hypersurface equation for 
$X_{F_{15}}$ in the ambient space given by the fibration 
\eqref{eq:PFfibration} with $\mathbb{P}_{F_i}\equiv 
\mathbb{P}_{F_{15}}$. Alternatively, we can set 
$s_1=s_3=s_4=s_8=s_{10}=0$ in \eqref{eq:pF1} and use the map 
\eqref{eq:F15blowup}. Either ways, we obtain
\begin{align}
\begin{split}\label{eq:pF15}
p_{F_{15}}&=s_2 e_1 e_2^2 e_3^2 u^2 v + s_5 e_1^2 e_2 e_4^2 u^2 w +
 s_6 e_1 e_2 e_3 e_4 e_5 u v w %\\ &\phantom{=}
 + s_7 e_2 e_3^2 e_5^2 v^2 w +  s_9 e_1 e_4^2 e_5^2 v w^2 \,,
\end{split}
\end{align}
where the sections $s_i$ assume values in the line bundles associated
to the divisor classes in \eqref{eq:cubicsections}.

The fibration $X_{F_{15}}$ has four seemingly rational points with one 
linear and one torsional relation between, showing that the MW-group
is $\mathbb{Z}\oplus\mathbb{Z}_2$ \cite{Braun:2013nqa}. 
Their coordinates are
\begin{align}
\begin{split}\label{eq:sectionsF15}
\hat{s}_0=X_{F_{15}}\cap\{u=0\}&:\quad [0:1:1:s_7:-s_9:1:1:1]\,,\\
\hat{s}_1=X_{F_{15}}\cap\{e_4=0\}&:\quad [1:1:s_2:-s_7:1:1:0:1]\,,\\
X_{F_{15}}\cap\{e_5=0\}&:\quad [1:s_5:-s_2:1:1:1:1:0]\,,\\
X_{F_{15}}\cap\{e_3=0\}&:\quad [1:s_5:1:1:-s_9:0:1:1]\,,
\end{split}
\end{align}
where we choose $\hat{s}_0$ as the zero section and $\hat{s}_1$ as
the generator of the free part of the MW-group.

Next, we compute the Weierstrass form \eqref{eq:WSF} of \eqref{eq:pF15}.
Again, we use the birational map from $X_{F_5}$ in \cite{Cvetic:2013nia}
to first obtain the local Tate coefficients \eqref{eq:TateF15sing}, 
which determine the WSF \eqref{eq:WSFF15}. The equivalent global Tate
model in \eqref{eq:TorsionF15} is precisely of the form of
an elliptic fibration with $\mathbb{Z}\oplus\mathbb{Z}_2$ MW-group, that 
has been studied in \cite{Aspinwall:1998xj}.
The WS-coordinates of the section $\hat{s}_1$ are given by \eqref{eq:WScoordsSecF3} after setting $s_1=s_3=s_4=s_8=0$ and 
the torsion point is given in \eqref{eq:WScoordsSecF15}.

\begin{figure}[h!]
\center
\includegraphics[scale=0.8]{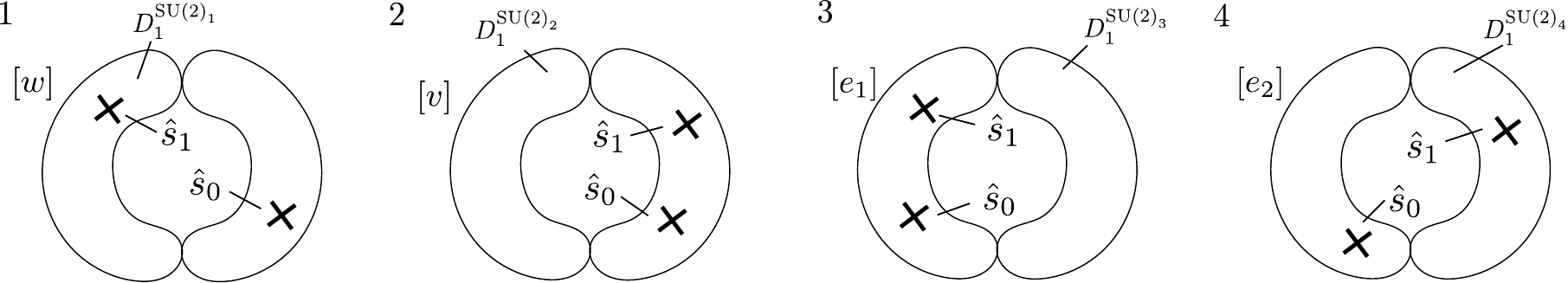}
\caption{\label{fig:poly15_codim1}Codimension one fibers
of $X_{F_{15}}$. The crosses denote the intersections
with the two sections.}
\end{figure}

These results allow us to compute the discriminant $\Delta$ of $X_{F_{15}}$. We find four $I_2$-singularities over the divisors
$\mathcal{S}^b_{\text{SU}(2)_1}=\{s_2=0\}\cap B$,
$\mathcal{S}^b_{\text{SU}(2)_2}=\{s_5=0\}\cap B$, 
$\mathcal{S}^b_{\text{SU}(2)_3}=\{s_7=0\}\cap B$ and 
$\mathcal{S}^b_{\text{SU}(2)_4}=\{s_9=0\}\cap B$ in $B$. Along these
divisors the constraint \eqref{eq:pF15} factorizes as
\begin{align}
\begin{split}\label{eq:SU2F15}
\SU2_1&: \quad p_{F_{15}}|_{s_2=0}= w \cdot q_2\, ,\\
\SU2_2&: \quad p_{F_{15}}|_{s_5=0}= v \cdot q_2'\, ,\\
\SU2_3&: \quad p_{F_{15}}|_{s_7=0}= e_1 \cdot q_3\, ,\\
\SU2_4&: \quad p_{F_{15}}|_{s_9=0}= e_2 \cdot q_3' \, ,
\end{split}
\end{align}
where $q_2$, $q_2'$, $q_3$ and $q_3'$ are the polynomials that remain 
after factoring out $w$, $v$, $e_1$ and $e_2$, respectively. The 
corresponding codimension one fibers in $X_{F_{15}}$ are shown in 
Figure~\ref{fig:poly15_codim1}. In summary, the  gauge group of 
$X_{F_{15}}$ is
\beq \label{eq:GF15}
	G_{F_{15}}=(\text{SU}(2)^4)/\mathbb{Z}_2\times \text{U}(1) \,.
\eeq
As before we confirm the action of the MW-torsion on the non-Abelian 
factors in Appendix~\ref{app:AppPoly} by explicitly working
out the WS-coordinates  \eqref{eq:WScoordsSecF15} of the generator of 
the $\mathbb{Z}_2$ MW-torsion.

In order to calculate the Cartan divisors of $X_{F_{15}}$
we use a similar logic as in Section~\ref{sec:polyF6}. We obtain
the following divisor classes:
\begin{align}\label{eq:CartansF15}
D^{\SU2_1}_{1}=[w] \, ,\quad
D^{\SU2_2}_{1}=[v]\, ,\quad
D^{\SU2_3}_{1}=[s_7]-[e_1]\, ,\quad
D^{\SU2_4}_{1}=[s_9]-[e_2]\, .
\end{align}
With these results at hand, we compute the Shioda map 
\eqref{eq:ShiodaMap} of the section $\hat{s}_1$ as
\begin{align} \label{eq:ShiodaF15}
\sigma (\hat{s}_1)=S_1-S_0 +[K_B]	 +\tfrac{1}{2} D^{\SU2_1}_{1}+\tfrac{1}{2} D^{\SU2_4}_{1}\,.
\end{align}
Here $S_0$, $S_1$ denote the divisor classes of the sections
$\hat{s}_0$, $\hat{s}_1$, respectively, and  we used
\begin{align} \label{eq:S1S0F15}
	\pi(S_1\cdot S_0)=0\,,
\end{align}
which directly follows from \eqref{eq:sectionsF15} as well as
\begin{align}
S_1 \cdot C^{SU(2)_1}_{-\alpha _1}=1 \, ,\quad
S_1 \cdot C^{SU(2)_2}_{-\alpha _1}=0 \, ,\quad
S_1 \cdot C^{SU(2)_3}_{-\alpha _1}=0 \, ,\quad
S_1 \cdot C^{SU(2)_4}_{-\alpha _1}=1\,,
\end{align}
which can be read off from Figure~\ref{fig:poly15_codim1}.
Using \eqref{eq:ShiodaF15}, we  compute the height pairing \eqref{eq:anomalycoeff},
\begin{align}\label{eq:bmnF15}
 b_{11}=-[K_B]\,,
\end{align}
\begin{table}[H]
\begin{center}
\small
\renewcommand{\arraystretch}{1.2}
\begin{tabular}{|c|c|c|c|}\hline
Representation & Multiplicity & Fiber & Locus \\ \hline

$(\two,\two,\one,\one)_{1/2}$ & $\begin{array}{c} (2 [K_B^{-1}]-\cS_7) \\ \times (2[K_B^{-1}]-\cS_9) \end{array}$ & \rule{0pt}{1.5cm}\parbox[c]{2.7cm}{\includegraphics[width=2.7cm]{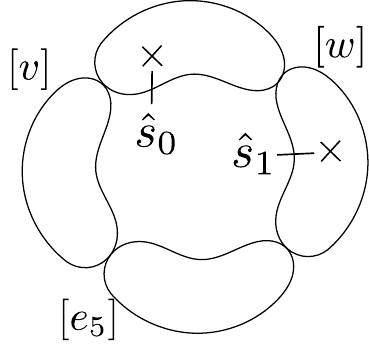}} & $\begin{array}{c} V(I_{(1)}):= \\ \{ s_2=s_5=0 \} \end{array}$ \\[1cm] \hline

$(\two,\one, \two, \one)_{1/2}$ & $(2 [K_B^{-1}]-\cS_9)\cS_7$ & \rule{0pt}{1.5cm}\parbox[c]{2.7cm}{\includegraphics[width=2.7cm]{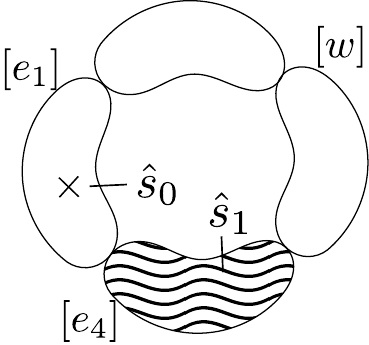}} & $\begin{array}{c} V(I_{(2)}):= \\ \{ s_2=s_7=0 \} \end{array}$ \\[1cm] \hline

$(\two,\one,\one,\two)_{0}$ & $\cS_9(2[K_B^{-1}]-\cS_9)$ & \rule{0pt}{1.5cm}\parbox[c]{2.7cm}{\includegraphics[width=2.7cm]{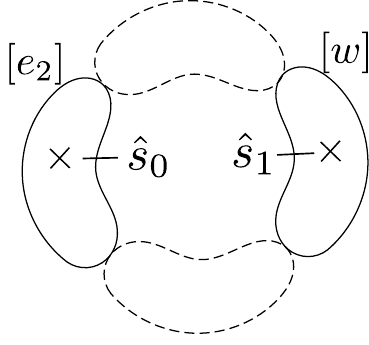}} & $\begin{array}{c} V(I_{(3)}):= \\ \{ s_2=s_9=0 \} \end{array}$ \\[1cm] \hline

$(\one, \two,\two,\one)_{0}$ & $(2 [K_B^{-1}]-\cS_7)\cS_7$ & \rule{0pt}{1.5cm}\parbox[c]{2.7cm}{\includegraphics[width=2.7cm]{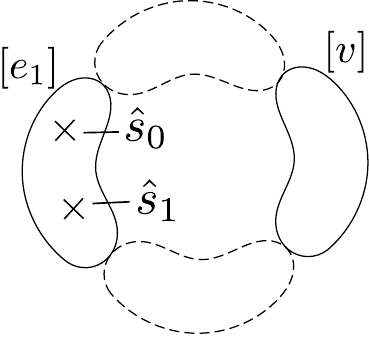}} & $\begin{array}{c} V(I_{(4)}):= \\ \{ s_5=s_7=0 \} \end{array}$ \\[1cm] \hline

$(\one,\one,\two,\two)_{1/2}$ & $\cS_7\cS_9$ & \rule{0pt}{1.5cm}\parbox[c]{2.7cm}{\includegraphics[width=2.7cm]{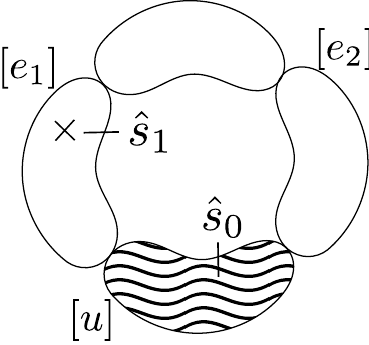}} & $\begin{array}{c} V(I_{(5)}):= \\ \{ s_7=s_9=0 \} \end{array}$ \\[1cm] \hline

$(\one,\two,\one,\two)_{1/2}$ & $(2 [K_B^{-1}]-\cS_7)\cS_9$ & \rule{0pt}{1.45cm}\parbox[c]{2.7cm}{\includegraphics[width=2.7cm]{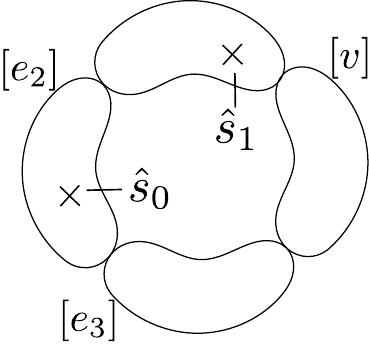}} & $\begin{array}{c} V(I_{(6)}):= \\ \{ s_5=s_9=0 \} \end{array}$ \\[1cm] \hline

$(\one,\one,\one,\one)_{1}$ & $2 [K_B^{-1}]^2$ & \rule{0pt}{1.15cm}\parbox[c]{2.7cm}{\includegraphics[width=2.7cm]{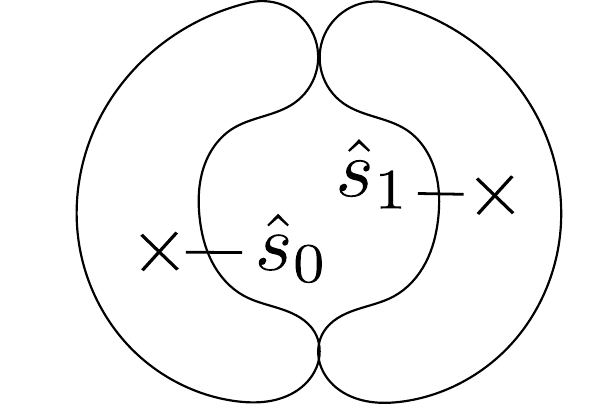}} & $\begin{array}{c} V(I_{(7)}):=\{ s_6=0 \\ s_5s_7-s_2s_9=0 \} \end{array}$ \\[0.7cm] \hline \hline

$(\three,\one,\one,\one)_{0}$ & \rule{0pt}{0.55cm}$1+ \frac{([K_B^{-1}]-\cS_9)}{2} (2 [K_B^{-1}]-\cS_9) $ & Figure \ref{fig:poly15_codim1} & $s_2=0$ \\ \hline

$(\one,\three,\one,\one)_{0}$ & \rule{0pt}{0.55cm}$1+\frac{([K_B^{-1}]-\cS_7)}{2} (2 [K_B^{-1}]-\cS_7) $ & Figure \ref{fig:poly15_codim1} & $s_5=0$ \\ \hline

$(\one,\one,\three,\one)_{0}$ & \rule{0pt}{0.55cm}$1+\cS_7 \frac{(\cS_7-[K_B^{-1}])}{2}$ & Figure \ref{fig:poly15_codim1} & $s_7=0$ \\ \hline

$(\one,\one,\one,\three)_{0}$ & \rule{0pt}{0.55cm}$1+\cS_9 \frac{(\cS_9-[K_B^{-1}])}{2}$ & Figure \ref{fig:poly15_codim1} & $s_9=0$ \\ \hline

\end{tabular}
\caption{\label{tab:poly15_matter}Charged matter representations under $\SU2^4\times$U(1)$/\mathbb{Z}_2$ and corresponding codimension two fibers of $X_{F_{15}}$. The adjoint matter is included for completeness.}
\end{center}
\end{table}
\noindent where we used \eqref{eq:SP^2} and \eqref{eq:S1S0F15}.

Next, we turn to the analysis of the codimension two singularities of 
the WSF of $X_{F_{15}}$ and the determination of the charged matter 
spectrum.
As before, all  representations
are determined from the codimension two fibers in $X_{F_{15}}$
following the procedure presented in
Section~\ref{sec:ellipticCurvesWithRP}. All codimension two singularities are
easily seen from the discriminant $\Delta$.
We find nine codimension two singularities. Seven of
these lead to the matter representations and the corresponding
codimension two fibers in $X_{F_{15}}$ given in the first and third 
column of Table~\ref{tab:poly15_matter}, respectively. At the remaining 
two loci $s_2=s_6^2-4s_5s_7=0$ and $s_9=s_6^2-4s_5s_7=0$, the fiber is 
of Type $III$ , cf.~Section~\ref{sec:polyF6}, which means that there are 
no additional matter representations. Again, we observe the absence
of fundamental matter which is consistent with the action of the 
MW-torsion on the gauge group in \eqref{eq:GF15}.
The spectrum of charged matter is completed by the matter in the
adjoint representations $(\three,\one,\one,\one)_{0}$, $(\one,\three,\one,\one)_{0}$, $(\one,\one,\three,\one)_{0}$ and $(\one,\one,\one,\three)_{0}$ given in
the last four rows of Table~\ref{tab:poly15_matter}.

We emphasize that the representations $(\two,\one,\one,\two)_0$ and 
$(\one,\two,\two,\one)_0$ at the loci $V(I_{(3)})$ and $V(I_{(4)})$, 
respectively, arise from non-split codimension two fibers. The dashed
nodes in Table~\ref{tab:poly15_matter} are interchanged by
a codimension three monodromy.

The total matter spectrum of $X_{F_{15}}$ is completed by the number of 
neutral hyper multiplets, which is computed from \eqref{eq:Hneutral} 
using the Euler number $\chi(X_{F_{15}})$ given in 
\eqref{eq:EulerNumbers}. It reads
\beq \label{eq:HneutF15}
H_{\text{neut}} = 17 + 7 [K_B^{-1}]^2 - 4 [K_B^{-1}] \cS_7 + 2 \cS_7^2 - 4 [K_B^{-1}] \cS_9 + 2 \cS_9^2 \,.
\eeq
The number $T$ of tensor multiplets is base-dependent, 
cf.~\eqref{eq:Tformula}, and we have $V=24$ vector multiplets.
Finally, we use this together with the divisors $S_{\text{SU}(2)_I}^b$, 
$I=1,\ldots,4$, \eqref{eq:bmnF15}, the charged spectrum in 
Table~\ref{tab:poly15_matter}  and \eqref{eq:HneutF15} to check that all 
6D anomalies in \eqref{eq:6dAnomalies} are  canceled.

Finally, we present our analysis of codimension three singularities
of the WSF of $X_{F_{15}}$ and the corresponding Yukawa points in a
compactification to 4D in Table~\ref{tab:poly15_yukawa}.
\begin{table}[htb!]
\begin{center}
\renewcommand{\arraystretch}{1.2}
\begin{tabular}{|c|c|}\hline
Yukawa & Locus \\ \hline
$\overline{(\two,\two,\one,\one)_{1/2}} \cdot (\two,\one,\two,\one)_{1/2} \cdot (\one,\two,\two,\one)_{0}$ & $s_2=s_5=s_7=0$ \\ \hline
$\overline{(\two,\two,\one,\one)_{1/2}} \cdot (\two,\one,\one,\two)_{0} \cdot (\one,\two,\one,\two)_{1/2}$ & $s_2=s_5=s_9=0$ \\ \hline
$\overline{(\two,\one,\two,\one)_{1/2}} \cdot (\two,\one,\one,\two)_{0} \cdot (\one,\one,\two,\two)_{1/2}$ & $s_2=s_7=s_9=0$ \\ \hline
$(\one,\two,\two,\one)_{0} \cdot \overline{(\one,\one,\two,\two)_{1/2}} \cdot (\one,\two,\one,\two)_{1/2}$ & $s_5=s_7=s_9=0$ \\ \hline
\end{tabular}
\caption{\label{tab:poly15_yukawa}Codimension three loci and corresponding Yukawa points for polyhedron $F_{15}$. }
\end{center}
\end{table}
%
%
%
%
%
%
%
%
%
%
%
%
%
%
%
%%%%%%%%%%%%%%%%%%%%%%%%%%%%%%%%%%%%%%%%%%%%%%%%%%%%%%%%%%%%%%%%%%%%%%%%%%%%%%%%%%%%%%%%%%%%%%%%%
\subsubsection{Polyhedron $F_{16}$: $G_{F_{16}}=\text{SU(3)}^3/\mathbb{Z}_3$}
\label{sec:polyF16}
%%%%%%%%%%%%%%%%%%%%%%%%%%%%%%%%%%%%%%%%%%%%%%%%%%%%%%%%%%%%%%%%%%%%%%%%%%%%%%%%%%%%%%%%%%%%%%%%%

\begin{figure}[H]
\center
\includegraphics[scale=0.4]{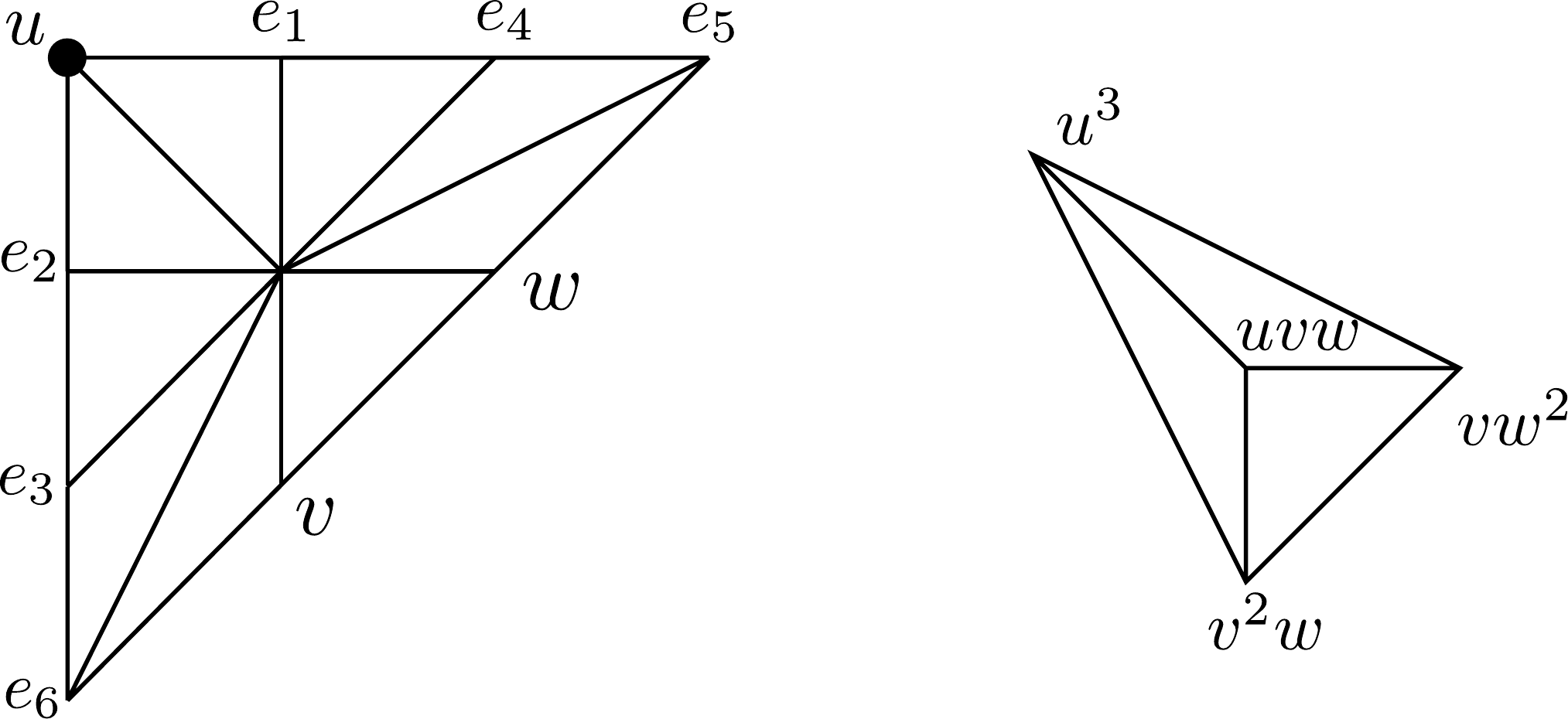}
\caption{\label{fig:poly16_toric}The toric diagram of polyhedron $F_{16}$ and its dual. The zero section is indicated by the dot.}
\end{figure}

Consider the elliptically fibered Calabi-Yau manifold $X_{F_{16}}$ with 
base $B$ and general fiber  given by the elliptic curve $\mathcal{E}$ in 
$\mathbb{P}_{F_{16}}$.
The toric data of $\mathbb{P}_{F_{16}}$ is summarized in
Figure~\ref{fig:poly16_toric}, where the polyhedron $F_{16}$,
a choice of projective coordinates as well as its dual polyhedron $F_1$
are depicted. The monomials associated to the integral points of 
$F_{1}$ are presented in the patch 
$e_i=1$, $\forall i$.
The toric variety $\mathbb{P}_{F_{16}}$ is the six-fold blow-up of 
$\mathbb{P}^2$ at non-generic points, that is defined as
\begin{align}\label{eq:F16blowup}
u \rightarrow e_1 e_2 e_3 e_4 e_5 e_6 u\,, \quad w\rightarrow e_1 e_4^2 e_5^3 w \,,\quad v\rightarrow e_2 e_3^2 e_6^3 v \,.
%u \rightarrow e_1 u\; &, \; w\rightarrow e_1 w\\
%u \rightarrow e_2 u\; &, \; v\rightarrow e_2 v\\
%e_2 \rightarrow e_3 e_2\; &, \; v\rightarrow e_3 v\\
%e_1 \rightarrow e_4 e_1\; &, \; w\rightarrow e_4 w\\
%e_4 \rightarrow e_5 e_4\; &, \; w\rightarrow e_5 w\\
%e_3 \rightarrow e_6 e_3\; &, \; v\rightarrow e_6 v
\end{align}
After this blow-up the projective coordinates on the fiber are 
$[u:v:w:e_1:e_2:e_3:e_4:e_5:e_6]$ and take values in the line bundles 
associated to the following divisor classes:
\begin{equation}
\label{tab:poly16_bundles}
\text{
\begin{tabular}{|c|c|}\hline
Section & Divisor class\\ \hline
$u$ & $H-E_1-E_2+\cS_9+[K_B]$ \\ \hline
$v$ & $H-E_2-E_3-E_6+\cS_9-\cS_7$\\ \hline
$w$ & $H-E_1-E_4-E_5$\\ \hline
$e_1$ & $E_1-E_4$\\ \hline
$e_2$ & $E_2-E_3$\\ \hline
$e_3$ &  $E_3-E_6$\\ \hline
$e_4$ &  $E_4-E_5$\\ \hline
$e_5$ &  $E_5$\\ \hline
$e_6$ &  $E_6$\\ \hline
\end{tabular}}
\end{equation}
The Stanley-Reisner ideal of $\mathbb{P}_{F_{16}}$ can be seen from 
Figure~\ref{fig:poly16_toric}  to be given by
\begin{align}
\begin{split}
SR&=\{ u e_4, u e_5, uw, uv, ue_6, ue_3, e_1e_5, e_1 w, e_1v, e_1e_6, e_1 e_3, e_1e_2, e_4w, e_4 v, e_4e_6, \\
&\phantom{={}} e_4 e_3, e_4 e_2, e_5v, e_5e_6, e_5 e_3, e_5 e_2, we_6, w e_3, w e_2, ve_3, ve_2, e_6e_2\} \, .
\end{split}
\end{align}
We obtain the hypersurface equation  for $X_{F_{16}}$ either employing 
\eqref{eq:BatyrevFormula} or  by setting $s_2=s_3=s_4=s_5=s_8=s_{10}=0$ 
in \eqref{eq:pF1} and applying the map \eqref{eq:F16blowup}. It reads
\begin{align}\label{eq:pF16}
p_{F_{16}}=s_1 e_1^2 e_2^2 e_3 e_4 u^3+ s_6 e_1 e_2 e_3 e_4 e_5 e_6 u v w + s_7 e_2 e_3^2 e_6^3 v^2 w + s_9 e_1 e_4^2 e_5^3 v w^2 \, .
\end{align}
where the divisor classes of the sections $s_i$  are given
in \eqref{eq:cubicsections}. 

There are three rational sections of the fibration of $X_{F_{16}}$ with two 
torsional relations between them which shows that the MW-group is 
$\mathbb{Z}_3$ \cite{Braun:2013nqa}. The coordinates of the sections are
\begin{align}
\begin{split}\label{eq:sectionsF16}
\hat{s}_0=X_{F_{16}}\cap\{u=0\}&:\quad [0:1:1:s_7:-s_9:1:1:1:1]\,,\\
X_{F_{16}}\cap\{e_5=0\}&:\quad [1:1:s_1:1:1:1:-s_7:0:1]\,,\\
X_{F_{16}}\cap\{e_6=0\}&:\quad [1:s_1:1:1:1:-s_9:1:1:0]\,,
\end{split}
\end{align}
where we choose $\hat{s}_0$ as the zero section.

In order to compute the WSF \eqref{eq:WSF} of \eqref{eq:pF16}, we first
compute the Tate form using the birational map from $X_{F_5}$
\cite{Cvetic:2013nia} in the limit $s_2=s_3=s_5=s_8=0$.
The global Tate coefficients are given in \eqref{eq:TateFormZ3MW}, which
is precisely of the form of an elliptic fibration with MW-group
$\mathbb{Z}_3$ \cite{Aspinwall:1998xj}. The WSF is given in 
\eqref{eq:WSFF16} and the WS-coordinates of the torsional section are 
given in \eqref{eq:WScoordsSecF16}. 

\begin{figure}[h!]
\center
\includegraphics[scale=0.8]{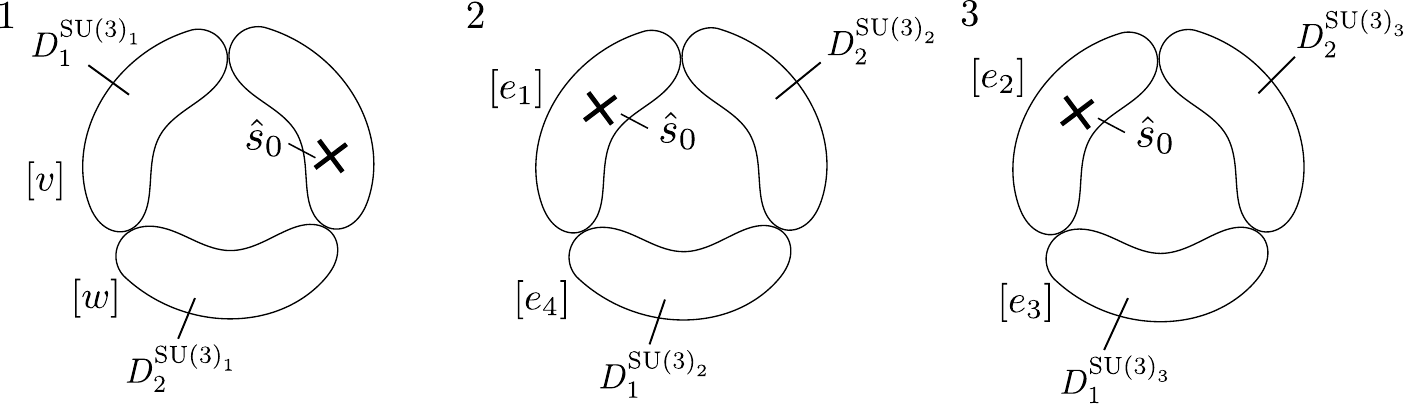}
\caption{\label{fig:poly16_codim1}Codimension one fibers
of $X_{F_{16}}$.}
\end{figure}

We readily compute the discriminant $\Delta$, which allows us to find 
all codimension one singularities of the WSF of $X_{F_{16}}$. We find 
three $I_3$-singularities over the divisors
$\mathcal{S}^b_{\text{SU}(3)_1}=\{s_1=0\}\cap B$,
$\mathcal{S}^b_{\text{SU}(3)_2}=\{s_7=0\}\cap B$ and
$\mathcal{S}^b_{\text{SU}(3)_3}=\{s_9=0\}\cap B$ in $B$. The 
hypersurface constraint \eqref{eq:pF16} factorizes along these divisors
as
\begin{align}
\begin{split}\label{eq:SU2F16}
\SU3_1&:\quad p_{F_{16}}|_{s_1=0}=v w \cdot q_1\,,\\
\SU3_2&:\quad p_{F_{16}}|_{s_7=0}= e_1 e_4 \cdot q_3\,,\\
\SU3_3&:\quad p_{F_{16}}|_{s_9=0}= e_2 e_3\cdot q'_3 \, ,
\end{split}
\end{align}
where $q_1$, $q_3$ and $q_3'$ are homogeneous polynomials in $[u:v:w]$
that remain after factoring out $vw$, $e_1 e_4$ and $e_2 e_3$. The 
corresponding fibers are depicted in Figure~\ref{fig:poly16_codim1}.
In summary, the  gauge group of $X_{F_{16}}$ is
\beq \label{eq:GF16}
	G_{F_{16}}=\text{SU}(3)^3/\mathbb{Z}_3\,.
\eeq
We note that this is precisely the gauge group of the trinification
model.
Here we confirmed the action of the MW-torsion on the non-Abelian
factors in Appendix~\ref{app:AppPoly} by explicitly working
out the Weierstrass coordinates \eqref{eq:WScoordsSecF16} 
of the generator of the $\mathbb{Z}_3$-torsion.

We calculate the classes of the Cartan divisors in the same fashion as 
in Section~\ref{sec:polyF6}. Using \eqref{eq:SU2F16} we obtain
the classes
\begin{align}
\begin{split}\label{eq:CartansF16}
D^{\SU3_1}_{1}=[v]\, &, \quad
D^{\SU3_1}_{2}=[w]\,,\quad
D^{\SU3_2}_{1}=[e_4]\,,\\
D^{\SU3_2}_{2}=[s_7]-[e_1]-[e_4]\, &,\quad
D^{\SU3_3}_{1}=[e_3]\,,\quad
D^{\SU3_3}_{2}=[s_9]-[e_2]-[e_3]\, .
\end{split}
\end{align}

The determination of the charged matter spectrum requires the knowledge
of the codimension two singularities of the WSF of
$X_{F_{16}}$. Again, we then extract the corresponding  representation 
data by application of the general recipe outlined in 
Section~\ref{sec:ellipticCurvesWithRP}. By investigation of the
discriminant of $X_{F_{16}}$ we readily find three codimension
two singularities that lead to bi-fundamental representations.
These and their corresponding
codimension two fibers in $X_{F_{16}}$ are listed in the first and third 
column of Table~\ref{tab:poly16_matter}, respectively.
\begin{table}[h!]
\begin{center}
\renewcommand{\arraystretch}{1.2}
\begin{tabular}{|c|c|c|c|}\hline
Representation & Multiplicity & Fiber & Locus \\ \hline

$(\three,\bar{\three},\one)$ & $\cS_7(3[K_B^{-1}]-\cS_7-\cS_9)$ & \rule{0pt}{1.85cm}\parbox[c]{2.9cm}{\includegraphics[width=2.9cm]{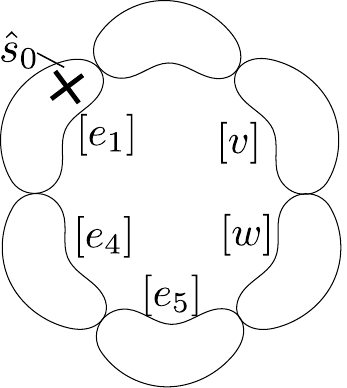}} & $\begin{array}{c} V(I_{(1)}):= \\ \{ s_1=s_7=0 \} \end{array}$  \\[1.45cm] \hline

$(\three,\one,\bar{\three})$ & $\cS_9 (3[K_B^{-1}]-\cS_7-\cS_9)$ & \rule{0pt}{1.85cm}\parbox[c]{2.9cm}{\includegraphics[width=2.9cm]{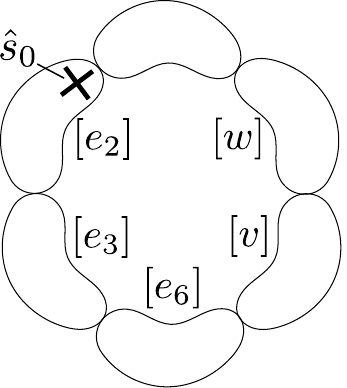}} & $\begin{array}{c} V(I_{(2)}):= \\ \{ s_1=s_9=0 \} \end{array}$   \\[1.45cm] \hline

$(\one,\three, \bar{\three})$ & $\cS_7 \cS_9$ & \rule{0pt}{1.85cm}\parbox[c]{2.9cm}{\includegraphics[width=2.9cm]{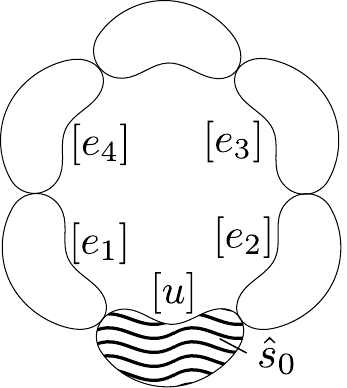}} & $\begin{array}{c} V(I_{(3)}):= \\ \{ s_7=s_9=0 \} \end{array}$  \\[1.45cm] \hline \hline

$(\mathbf{8},\one,\one)$ & \rule{0pt}{0.85cm}$\begin{array}{c} 1 + \frac{((2 [K_B^{-1}] - \cS_7 - \cS_9))}{2} \\ \times (3 [K_B^{-1}] - \cS_7 - \cS_9) \end{array}$ & Figure \ref{fig:poly16_codim1} & $s_1=0$ \\ \hline

$(\one,\mathbf{8},\one)$ & \rule{0pt}{0.6cm}$1 + \cS_7 \frac{(\cS_7 - [K_B^{-1}])}{2}$ & Figure \ref{fig:poly16_codim1} & $s_7=0$ \\ \hline

$(\one,\one,\mathbf{8})$ & \rule{0pt}{0.6cm}$1 + \cS_9 \frac{(\cS_9 - [K_B^{-1}])}{2}$ & Figure \ref{fig:poly16_codim1} & $s_9=0$ \\ \hline
\end{tabular}
\caption{\label{tab:poly16_matter}Charged matter representations under $\SU3^3/\mathbb{Z}_3$ and corresponding codimension two fibers of $X_{F_{16}}$. The adjoint matter is included for completeness.}
\end{center}
\end{table}
%. 
There are three additional codimension two singularities at $s_6=s_9=0$, 
$s_6=s_7=0$ and $s_1=s_6=0$, that, however, yield fibers of Type $IV$ in 
$X_{F_{16}}$, that do not support any additional matter.

Again,  we do not find fundamental matter, confirming the restrictions
imposed on the spectrum of $X_{F_{16}}$ by the MW-torsion.
The spectrum of charged matter is completed by the matter in the
adjoint representations $(\mathbf{8},\one,\one)$, $(\one,\mathbf{8},\one)$ and $(\one,\one,\mathbf{8})$ given in
the last three rows of Table~\ref{tab:poly16_matter}.
We recall that they  do not originate from codimension two fibers of
$X_{F_{16}}$, but are present if the divisors $S_{G}^b$ are
higher genus curves in $B$. The multiplicity of charged hyper multiplets
in the adjoint is given by \eqref{eq:genusformula}.

We complete the matter spectrum of $X_{F_{16}}$ by the number of neutral
hyper multiplets, which is computed employing the Euler number 
\eqref{eq:EulerNumbers} of $X_{F_{16}}$
from \eqref{eq:Hneutral}. It reads
\beq \label{eq:HneutF16}
H_{\text{neut}} = 18 + 11 [K_B^{-1}]^2 - 9 [K_B^{-1}] \cS_7 + 3 \cS_7^2 - 9 [K_B^{-1}] \cS_9 + 3 \cS_7 \cS_9 + 3 \cS_9^2 \,.
\eeq
The base-dependent number $T$ of tensor multiplets is given by 
\eqref{eq:Tformula} and we have $V=24$ vector multiplets.
Finally, we use this together with the divisors $S_{\text{SU}(3)_I}^b$ 
for $I=1,2,3$, the charged spectrum in Table~\ref{tab:poly16_matter} and
\eqref{eq:HneutF16} to confirm cancelation of all 6D anomalies 
\eqref{eq:6dAnomalies}.

By analyzing the codimension three singularities of the WSF of 
$X_{F_{16}}$, we finally calculate all Yukawa
couplings of the charged matter spectrum of $X_{F_{16}}$, that are given
in Table~\ref{tab:poly16_yukawa}.
\begin{table}[htb!]
\begin{center}
\renewcommand{\arraystretch}{1.2}
\begin{tabular}{|c|c|}\hline
Yukawa & Locus \\ \hline
$(\three,\bar{\three},\one) \cdot \overline{(\three,\one,\bar{\three})} \cdot (\one,\three,\bar{\three})$ & $s_1=s_7=s_9=0$ \\ \hline
\end{tabular}
\caption{\label{tab:poly16_yukawa}Codimension three loci and corresponding Yukawa points for $F_{16}$. }
\end{center}
\end{table}

\section{The  Toric Higgs Branch of F-Theory}
\label{sec:Higgsings}

In Section~\ref{sec:AnalysisOfPolytopes} we discussed
in great detail the geometric derivations
of the gauge groups and matter spectra of all genus-one fibrations 
$X_{F_i}$ based on the 16 polyhedra.
Here we show that the effective SUGRA theories obtained from these 
fibrations are not isolated from each other but connected by means of 
the Higgs mechanism.
This section is devoted to the study of those 
transitions and the resulting network of theories summarized in 
Figure~\ref{fig:network}. As we have noted earlier,
this network is nothing but the field theoretic realization 
of the network of extremal transitions relating the $X_{F_i}$ that are 
induced by blowing up/down in the toric ambient varieties 
$\mathbb{P}_{F_i}$ of their genus-one fibers.
This network, to which
we refer to as the toric {\it Higgsing diagram}, is a powerful 
consistency check for the results of 
Section~\ref{sec:AnalysisOfPolytopes} and exhibits some remarkable 
features:
\begin{itemize}
\item The Higgsing diagram is symmetric with respect to the horizontal 
axis corresponding to the self dual polyhedra, all of which have a gauge 
group rank equal to three. 
\item The rank of the gauge groups of a polyhedron and its dual sum up 
to six. 
\item Every toric hypersurface fibration can be reached upon a chain of 
Higgsings starting from one of the three manifolds $X_{F_{13}}$, 
$X_{F_{15}}$ and $X_{F_{16}}$ which exhibit non trivial MW-torsion. 
\item Both the analysis on the geometrical side (see 
Sections~\ref{sec:F1_poly}, \ref{sec:F2_poly} and \ref{sec:F4_poly}) and 
the Higgsings (see Section \ref{S:discretesymm}) lead to the conclusion 
that the MW-torsion in the $X_{F_i}$ with fibers in $F_{13}$, $F_{15}$ 
and $F_{16}$, manifests itself as discrete gauge symmetries in the $X_{F_i}$ 
with fibers in their respective dual polyhedra $F_1$, $F_2$ and $F_4$.  
\end{itemize}

\begin{figure}[t!]
\center
\includegraphics[scale=0.4]{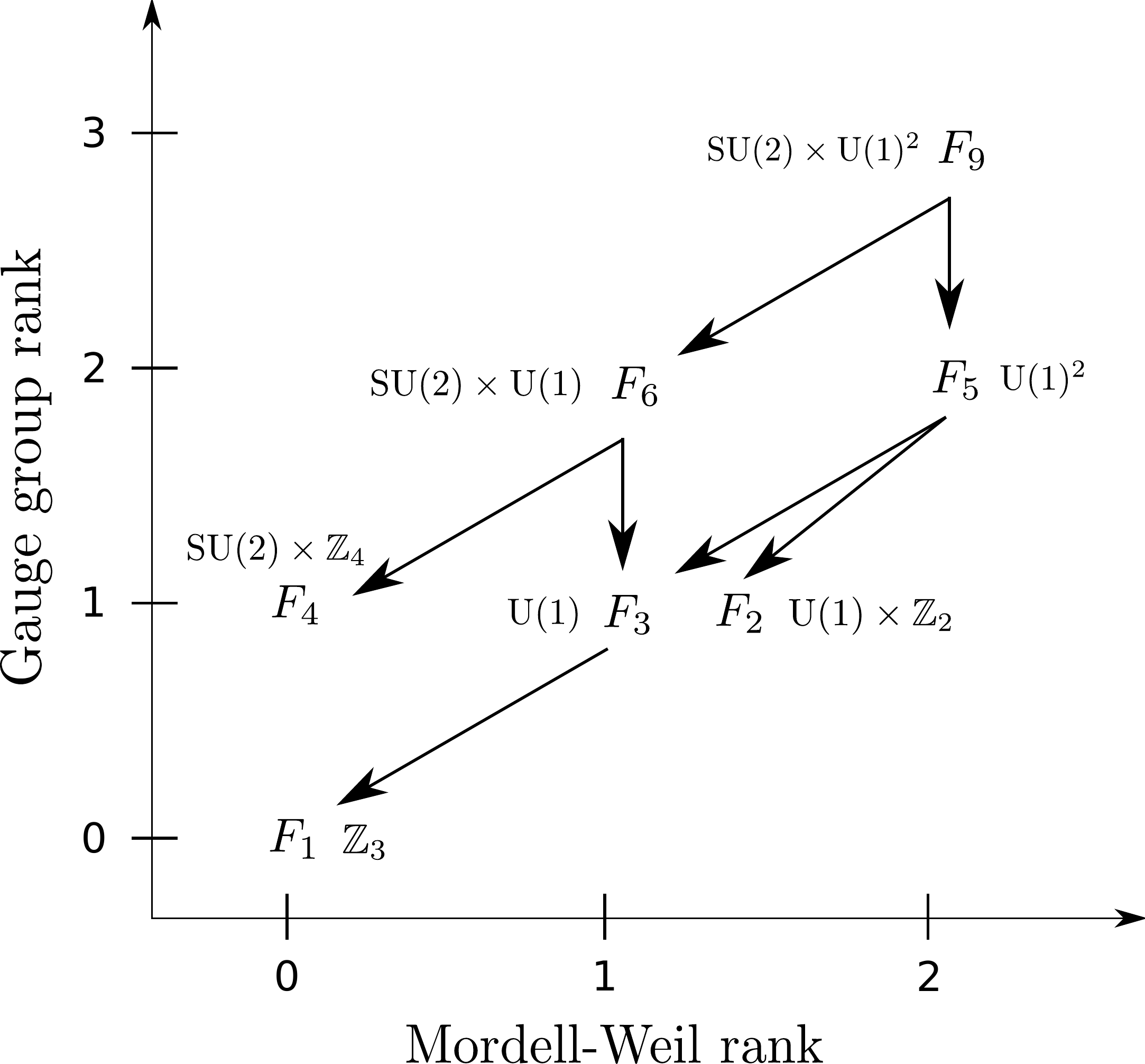}
\caption{\label{fig:network2} A subbranch of the Higgsing chain which we 
use to illustrate certain features relevant for all Higgsings. We use 
the Higgsing from $X_{F_9}\rightarrow X_{F_5}$ to demonstrate the type 
of bundle and charge redefinitions which are needed in order to match a 
Higgsed model with the geometrical computations. The Higgsings to 
$X_{F_1}$, $X_{F_2}$ and $X_{F_4}$ are used to confirm the presence of 
discrete gauge symmetries.}
\end{figure}

In the following we discuss the above features and the Higgsing diagram
in more detail. In order to illustrate the 
relevant features of the Higgsing we focus on a particular sub-branch of 
the Higgsing diagram which we depict in Figure~\ref{fig:network2}. This 
includes the transition of the effective
theory derived from F-theory on 
$X_{F_9}$ to that on $X_{F_5}$. This transition is convenient in order 
to discuss certain (unphysical) redefinitions of the divisor classes 
that are sometimes needed in order to match the field theoretic results 
with the geometrical computation.
In this example we also describe the matching between the gauge group generators 
before and after the Higgsing, and how this information can be inferred 
from the toric diagram (see Section~\ref{S:ToricHiggsings}).
After these
redefinitions we obtain a perfect match of the massless 
spectrum of the effective theory after Higgsing with that obtained from 
the geometrical computation. 
While our results apply for any generic two-dimensional base $B$, 
we also discuss the Higgsing for the specific case of a $\mathbb{P}^2$ 
base. Here we comment on specific boundary strata of the moduli 
space where a specific Higgsing might not be possible but different 
equivalent Higgsings are.

In section \ref{S:discretesymm} we focus on the theories corresponding 
to the genus-one fibrations $X_{F_1}$, $X_{F_2}$ and $X_{F_4}$. The  
field theoretical Higgsings imply the presence of discrete gauge 
symmetries $\mathbb{Z}_3$, $\mathbb{Z}_2$ and $\mathbb{Z}_4$
which confirms the results of the 
geometrical computations. In Section~\ref{S:fullspec} we discuss the 
full chain of Higgs transitions. There we summarize the relevant 
redefinitions of the gauge group generators and divisor classes needed 
to match the spectrum obtained in the geometrical computations of Section~\ref{sec:AnalysisOfPolytopes}. 

\subsection{Toric Higgsing: an example}
\label{S:ToricHiggsings}

We are interested in Higgs transitions 
relating two supersymmetric vacua in a 6D $\mathcal{N}=1$ 
SUGRA theory.
This requires that the 
vacuum expectation value (VEV) of the Higgs field triggering this 
transition
must preserve supersymmetry. 
This is guaranteed by imposing
flatness of the D-term potential along the direction of the VEV. As it has been observed in 
\cite{Honecker:2006qz}, for the case of 
a Higgs field that is not in
the adjoint representation,  
at least two hyper multiplets with identical quantum numbers have
to acquire a VEV in order to achieve a D-flat direction.
 
Furthermore, the Higgs mechanism of  interest here has to relate two 
toric hypersurface fibrations. Transitions of this type have a clear 
geometric interpretation in terms of the toric diagrams. This can be 
seen as follows. First recall that the coefficients $s_i$ which 
appear in the hypersurface constraint of $X_{F_k}$ 
correspond to integral points in the dual polyhedron $F_k^*$, see 
\eqref{eq:BatyrevFormula}. As we have seen in Section \ref{sec:AnalysisOfPolytopes},
many $X_{F_k}$ exhibit hyper multiplets  
at codimension two loci of the form $\{s_i=s_j=0\}$, where $s_i$ and 
$s_j$ are neighboring vertices in the dual polyhedron $F_k^*$ connected 
by an edge. If the Higgs fields are of this type, they lead to a 
\emph{toric Higgsing}, i.e.~the resulting theory is associated to a new 
toric hypersurface fibration $X_{F_{k'}}$, $k'\neq k$.\footnote{Non-toric Higgsings on the other hand are transitions for which the 
resulting fibration can not anymore be described by one of the 16
polyhedra.} 
Here the polyhedron $F_{k'}$ is obtained from $F_{k}$ by blowing up the 
dual polyhedron $F_k^*$ precisely at the edge connecting the vertices
corresponding to $s_i$, $s_j$, respectively, and taking its dual.
In $F_k$, this corresponds to the blow-down associated to removing the 
corner that is dual to the aforementioned edge in $F_{k}^*$. Note that, since the Higgs fields in 
the toric breaking are never in the adjoint representation, the toric 
Higgsing is not rank preserving.  \\
\begin{figure}[t!]
\center
\includegraphics[scale=0.4]{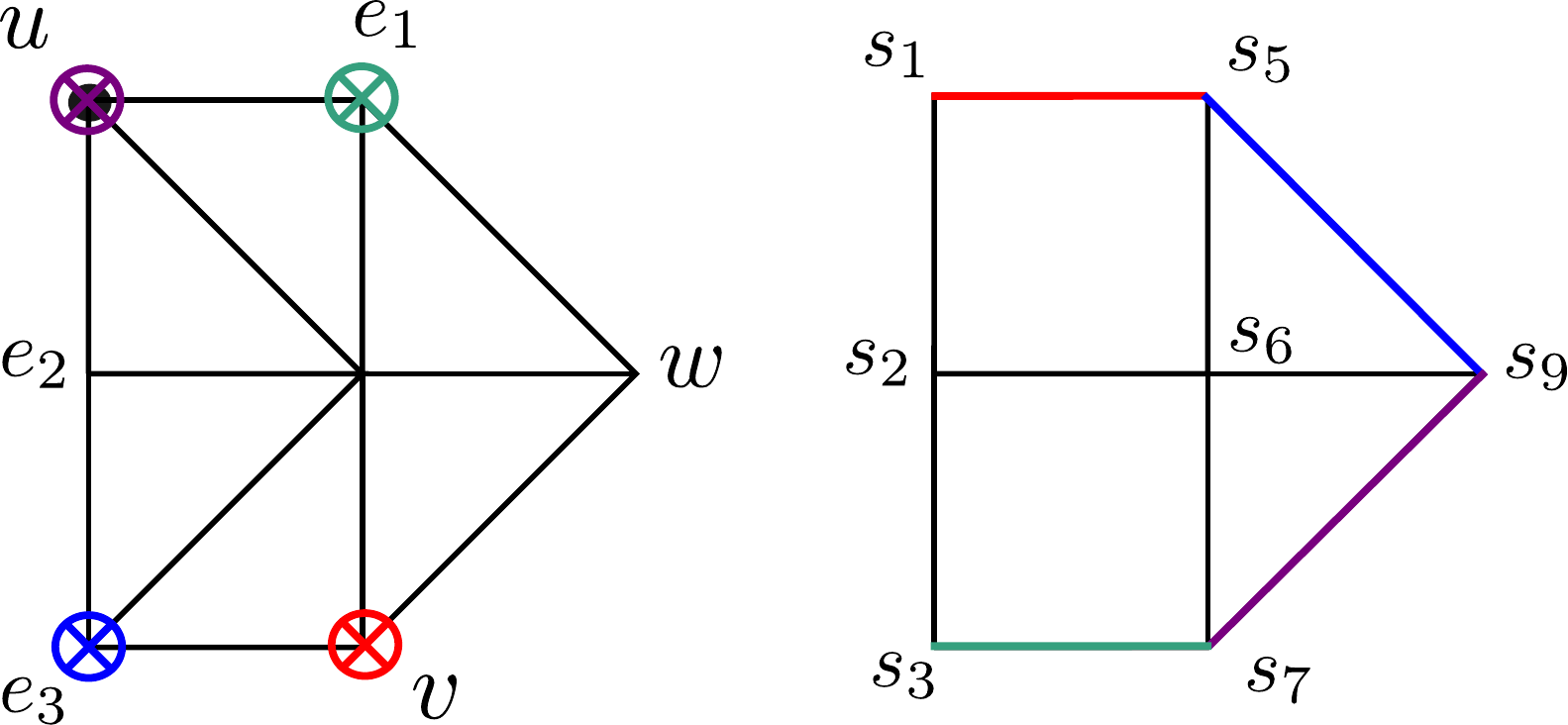}
\centerline{(a)}
\includegraphics[scale=0.4]{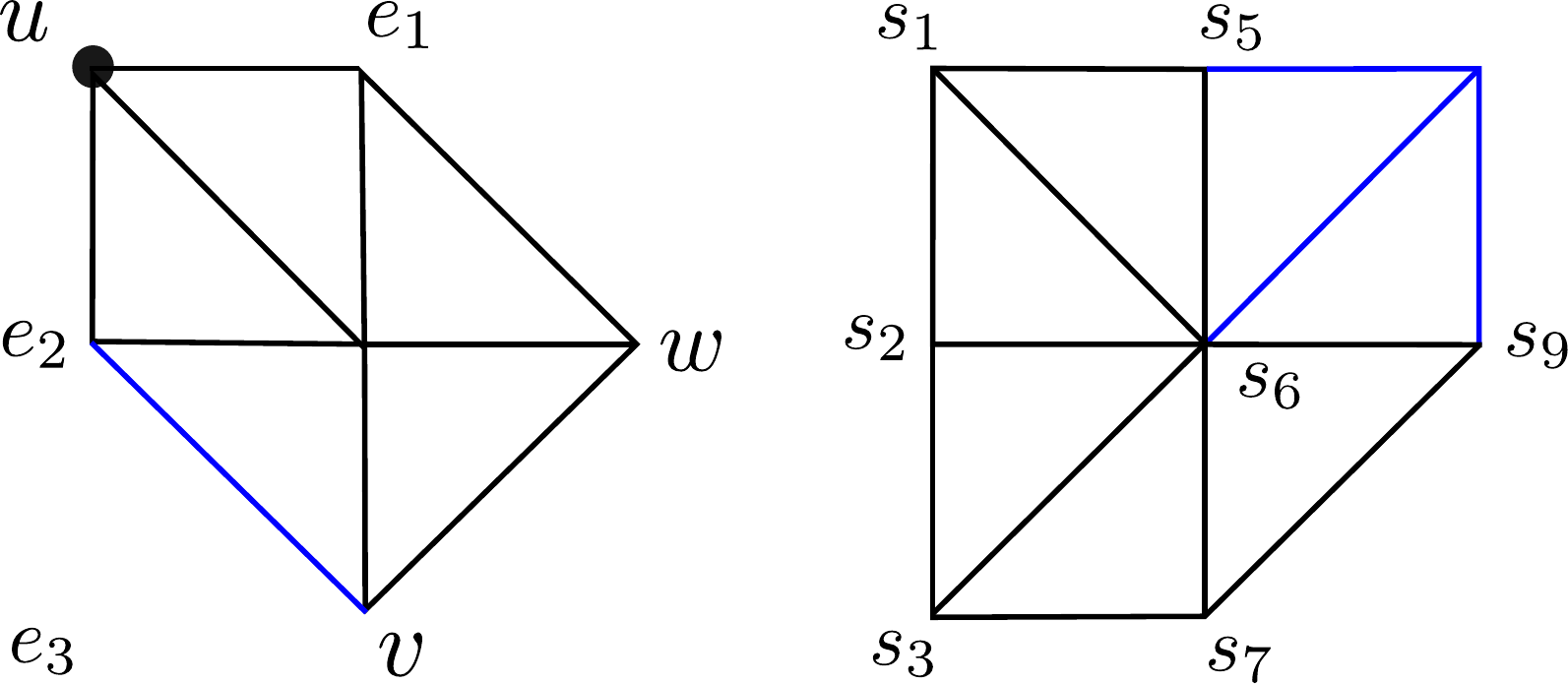}\hspace{1cm}\includegraphics[scale=0.4]{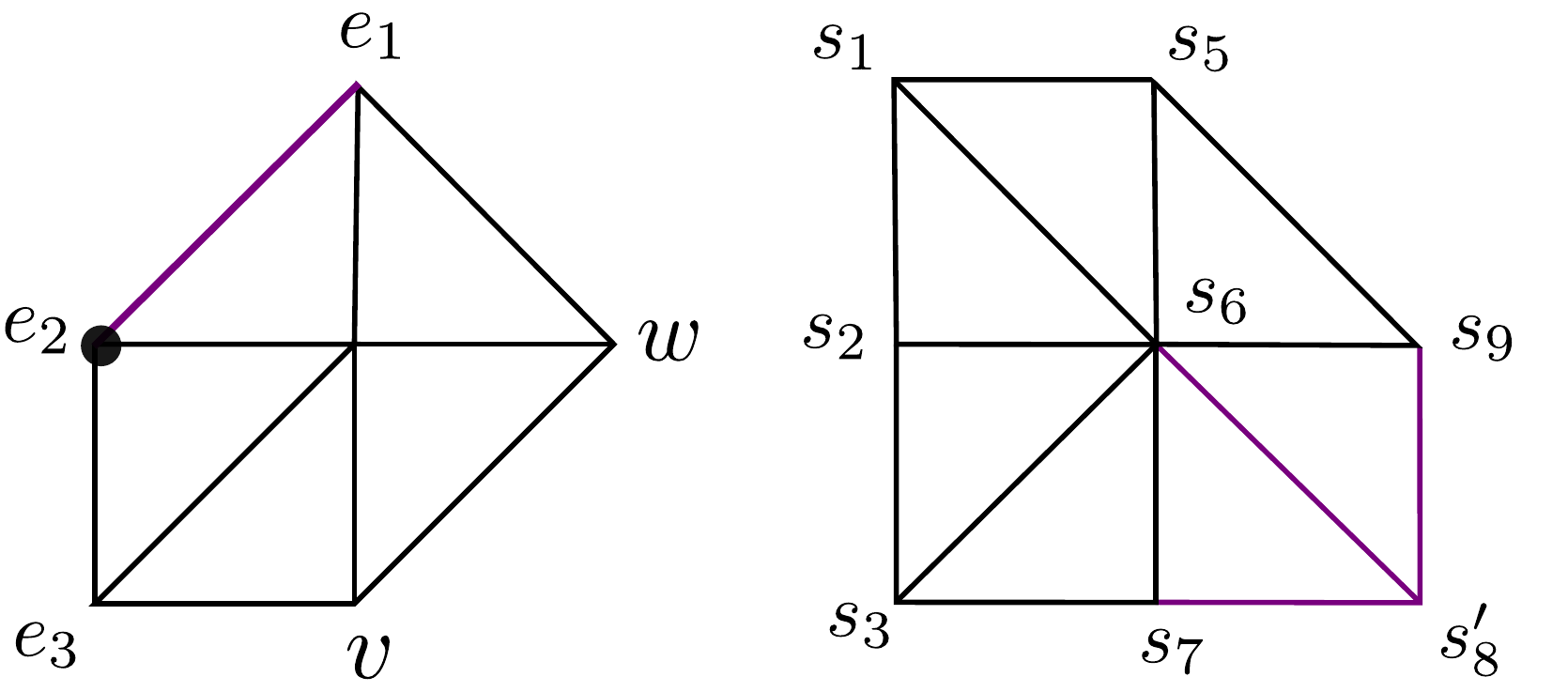}
\centerline{(b)\hspace{5cm}(c)}
\caption{(a) Possible toric Higgsings in $X_{F_9}$. The transitions to $X_{F_5}$ are achieved by Higgs fields in the representations $\two_{(-1,-1/2)}$ (b) or $\two_{(1,3/2)}$ (c).}
\label{fig:HiggsingF9}
\end{figure}

To exemplify this, consider the Calabi-Yau manifold $X_{F_{9}}$. There 
the possible toric Higgs fields are, cf.~Section \ref{sec:polyF9},
\bea
 & \,\,\,\,\,\,\,\,\,\,\one_{(1,2)}\,: \,\, \{s_3=s_7=0\}\,, \qquad  \,\,\,\,\one_{(1,0)}\,: \,\, \{s_1=s_5=0\}\,,&\label{E:F_6}\\
  &\two_{(-1,-1/2)}\,: \,\, \{s_5=s_9=0\}\,, \qquad \two_{(1,3/2)}\,: \,\, \{s_7=s_9=0\}\,.&\label{E:F_5}
\eea
Considering the group theoretical breaking that could be induced by 
these fields, we see that taking the Higgs fields in any of the 
singlet representations in \eqref{E:F_6}, for instance $\one_{(1,2)}$ 
leads to $\SU2 \times \U1$, which coincides with the gauge group of the 
fibration $X_{F_6}$,
while a VEV in the $\two_{(-1,-1/2)}$ or $\two_{(1,3/2)}$ (see 
\eqref{E:F_5}) leads to $\U1^2$, i.e. the gauge group expected for 
$X_{F_5}$. In Figure~\ref{fig:HiggsingF9} (a) we depict the polyhedron
which is dual to $F_9$ on the right and highlight  in different colors 
the edges  corresponding to the fields in \eqref{E:F_6} and 
\eqref{E:F_5}. They have to be blown up, i.e.~subdivided by a new ray,
for each of the possible toric Higgsings. In the actual polyhedron of 
$F_9$, we indicate the vertices that are dual to these edges and get 
cut off in the toric Higgsings. In the following we consider the 
Higgsing from $X_{F_9}$ to $X_{F_5}$ in more detail.
 
In the case of Higgs fields in the representation $\two_{(-1,-1/2)}$, 
the Higgsing
corresponds to a removal of the lower left corner
in the polyhedron $F_9$ and to a blow-up at its dual edge, which
is the edge between $s_5$ and $s_9$, in its dual polyhedron (see 
Figure~\ref{fig:HiggsingF9} (b)). After that, we obtain the toric 
diagram of $F_5$ as given in Section~\ref{sec:polyF5}. Similarly, if we 
pick VEVs in the $\two_{(1,3/2)}$ representation, we observe that after 
the Higgsing, the resulting polyhedron and its dual 
both have to be reflected along the horizontal axis 
(see Figure~\ref{fig:HiggsingF9} (c)) 
in order to recover Figure~\ref{fig:HiggsingF9} (b). 
Thus, the obtained effective theories after these two Higgsings
are physically equivalent.

However, we note that the geometrical computations leading to the 
spectrum of $X_{F_5}$ have been made for the polyhedron given in 
Figure~\ref{fig:HiggsingF9} (b). Thus, whenever we perform a Higgsing 
with the Higgs in the representation $\two_{(1,3/2)}$ (leading to
Figure~\ref{fig:HiggsingF9} (c))  we have to transform certain divisor 
classes in order to match the  
multiplicities of hyper multiplets
resulting from the Higgsing with those found geometrically. More 
general, the ``Higgsed'' polyhedron of $F_9$ can be brought to the 
\emph{canonical} form (i.e. the one used for the computation of the 
matter spectrum and multiplicities) by means of an $SL(2,\mathbb{Z})$ 
transformation, which acts simultaneously on the 
polyhedron and its dual. This transformation determines 
how to transform the divisor classes in order to recover precisely
the effective theory obtained by the geometric computation on $X_{F_5}$,
as we demonstrate next.

\subsubsection{Matching the charged spectrum}
In order to match the charged spectrum in the Higgsed theory arising 
from F-theory on $X_{F_9}$ with that on $X_{F_5}$, we first have to 
relate the generators of the gauge groups before and after the Higgsing. 
In $X_{F_9}$, the $\U1$ generators are (see~\eqref{eq:sectionsF9})
\beq \label{eq:F9Shiodas}
\sigma(\hat{s}_1)=[e_3]-[u]-[K_B^{-1}]\,,\qquad
\sigma(\hat{s}_2)=[e_1]-[u]-[K_B^{-1}]-\cS_7+\tfrac{1}{2}D_1\,,
\eeq
with $D_1$ being the class of the $\SU2$ Cartan divisor given in 
\eqref{eq:CartansF9} as $D_1=\mathcal{S}_9-[e_2]$. The Shioda maps in $X_{F_5}$ are given by (see~\eqref{eq:ShiodaF5})
\beq
\label{eq:F5shiodas}
\sigma(\hat{s}_1^{\prime})=[e_1]-[e_2]-[K_B^{-1}]\,,\qquad
\sigma(\hat{s}_2^{\prime})=[u]-[e_2]-[K_B^{-1}]-\cS_9\,.
\eeq

Let us consider first the canonical Higgsing induced by VEVs in the 
$\two_{(-1,-1/2)}$. As shown in Figure~\ref{fig:HiggsingF9} this 
corresponds to blowing down the divisor $e_3=0$. We see that after 
setting its divisor class $[e_3]=0$, the following relations hold
\beq \label{eq:relationsF9F5Shioda}
\sigma(\hat{s}_1^{\prime})=\sigma(\hat{s}_2)-\sigma(\hat{s}_1)+\tfrac{1}{2}D_1-[K_B^{-1}]+\cS_7-\cS_9\,,\qquad
\sigma(\hat{s}_2^{\prime})=-\sigma(\hat{s}_1)+D_1-2([K_B^{-1}]+\cS_9)\,.
\eeq
Since the vertical divisors $[K_B^{-1}]$, $\cS_7$ and $\cS_9$ do not 
contribute to the $\U1$ charges, these equations allow us to make
contact with the charges in $X_{F_5}$ (which we denote by $Q_1^\prime$ 
and $Q_2^\prime$). Indeed, recalling that Dynkin labels
and U(1)-charges
are computed according to \eqref{eq:DynkinLabel} and 
\eqref{eq:U1charge}, respectively, we translate 
\eqref{eq:relationsF9F5Shioda} into the charge relation\footnote{Note here the importance of the fact that the choices
of zero sections in $X_{F_5}$, cf.~Figure~\ref{fig:poly5_toric},
and $X_{F_9}$ after Higgsing, cf.~Figure~\ref{fig:HiggsingF9} (b),
are different. This corresponds to an $\text{SL}(2,\mathbb{Z})$-transformation of
the U(1)-charges.}
\begin{align} \label{eq:chargesF9F5}
 Q_1^\prime=Q_2-Q_1+T^3\,,\quad Q_2^\prime=-Q_1+2T^3\,,
\end{align}
with $Q_1$, $Q_2$ being the $\U1$-charges  and 
$T^3=\tfrac12 D_1$ the Cartan generator of the $\SU2$ in $X_{F_9}$.

For the Higgsing with fields in the $\two_{(1,3/2)}$ representation we 
can proceed in a similar manner. In this case, according to 
Figure~\ref{fig:HiggsingF9}, one has to set the divisor 
class $[u]=0$. In addition, one must take into account 
that the toric diagram is reflected with respect to the canonical one 
(compare Figure~\ref{fig:HiggsingF9} (c) with 
Figure~\ref{fig:poly5_toric}, including the location of the zero 
section). This implies that the classes of $[e_3]$, $[v]$ and $[e_1]$ in 
$X_{F_9}$ get mapped to $[u]$, $[e_1]$ and $[v]$ in $X_{F_5}$, 
respectively. Hence, the Shioda maps \eqref{eq:F9Shiodas}
for $X_{F_9}$, written in terms of divisor classes on $X_{F_5}$, read
\beq
\label{eq:F9shiodas2}
\sigma(\hat{s}_1)=[u]-[K_B^{-1}]\,,\qquad
\sigma(\hat{s}_2)=[v]-3[K_B^{-1}]+\cS_7+\tfrac{1}{2}D_1\,.
\eeq
Writing both \eqref{eq:F5shiodas} and \eqref{eq:F9shiodas2}, in terms of 
the exceptional divisors $E_1$, $E_2$ as well as the hyperplane class 
$H$ using \eqref{eq:cubicsections}, we find the following relations 
among them 
\beq
\sigma(\hat{s}_1^{\prime})=\sigma(\hat{s}_2)-\sigma(\hat{s}_1)+\tfrac{1}{2}D_1-\cS_9\,,\qquad
\sigma(\hat{s}_2^{\prime})=\sigma(\hat{s}_1)+D_1-2\cS_9\,,
\eeq
from which it follows that the $\U1$-charge redefinition in this case is 
given by
\begin{align}
 Q_1^\prime=Q_2-Q_1+T^3\,,\quad Q_2^\prime=Q_1+2T^3\,.
\end{align}

The charge formulas in both cases agree with the field theory 
expectations i.e.~there is a complete gauge singlet in the decomposition 
of the Higgs field into representations of the residual gauge 
symmetry. The decomposition of the states for both cases is given in 
Table~\ref{tab:F9TOF5}. We observe that, indeed, all charged states in 
$X_{F_5}$ have been reproduced.
\begin{table}[t!]
\begin{center}
\renewcommand{\arraystretch}{1.2}
\begin{tabular}{|c|c|c|}
\cline{2-3}
\multicolumn{1}{c|}{} & VEV: $\two_{(-1,-1/2)}$ &  VEV: $\two_{(1,3/2)}$ \\
\multicolumn{1}{c|}{} & $Q_1^\prime=(Q_2-Q_1+T^3)$&  $Q_1^\prime=(Q_2-Q_1+T^3)$\\
\multicolumn{1}{c|}{} & $Q_2^\prime=(-Q_1+2T^3)$ &  $Q_2^\prime=(Q_1+2T^3)$\\
\hline
$\one_{(1,2)}$ & $\one_{(1,-1)}$ & $\one_{(1,1)}$ \\
$\one_{(1,0)}$ & $\one_{(-1,-1)}$ & $\one_{(-1,1)}$ \\
$\one_{(0,1)}$ & $\one_{(1,0)}$ & $\one_{(1,0)}$ \\
$\one_{(1,1)}$ & $\one_{(0,-1)}$ & $\one_{(0,1)}$ \\
$\two_{(-1,-1/2)}$ & $\one_{(0,0)}+\one_{(1,2)}$ &  $\one_{(1,0)}+\one_{(0,-2)}$\\
$\two_{(1,3/2)}$ & $\one_{(0,-2)}+\one_{(1,0)}$ & $\one_{(1,2)}+\one_{(0,0)}$\\
$\two_{(0,-1/2)}$ & $\one_{(-1,-1)}+\one_{(0,1)}$ & $\one_{(0,1)}+\one_{(-1,-1)}$ \\
$\three_{(0,0)}$ & $\one_{(-1,-2)}+\one_{(1,2)}+\one_{(0,0)}$ & $\one_{(1,2)}+\one_{(-1,-2)}+\one_{(0,0)}$\\
\hline
\end{tabular}
\caption{Possible state decompositions from $X_{F_9}$ to those of $X_{F_5}$ for different Higgses.}
\label{tab:F9TOF5}
\end{center}
\end{table}

In order to match their multiplicities as well, we first recall the
basic fact that a hyper multiplet consists of two half-hypers 
transforming in representations conjugate to each other. Thus, in order
to compute the resulting multiplicities after Higgsing, one has to take 
into account all states transforming under the representation of 
interest together with their complex conjugates. For example, note that 
in the branching induced by $\two_{(-1,-1/2)}$, the states 
$\two_{(0,-1/2)}$ and $\one_{(1,1)}$ decompose as
\begin{align}
\two_{(0,-1/2)} \rightarrow \one_{(-1,-1)}+\one_{(0,1)}\,,\quad 
\one_{(1,1)} \rightarrow \one_{(0,-1)}\,,
\end{align}
thus, the multiplicity of hyper multiplets in the representation 
$\one_{(0,1)}$  after the Higgsing must be computed as the sum of the 
multiplicities of $\two_{(0,-1/2)}$ and $\one_{(1,1)}$ in $X_{F_9}$. 
Similarly, note that every Higgs doublet decomposes into a neutral and a 
charged singlet. In computing the multiplicity of such charged singlets 
after Higgsing, one has to take into account that two of these are 
absorbed as longitudinal components of the massive $W$ bosons from the 
broken SU(2).

In the case of the Higgsing induced by a VEV in the $\two_{(-1,-1/2)}$ 
representation of $X_{F_9}$, we can directly compare the resulting 
multiplicities with the geometric result of Section~\ref{sec:polyF5}. 
However, in the case where we turn on VEVs for the fields
in the representation $\two_{(1,3/2)}$, the multiplicities 
only match after performing a redefinition of divisor classes.
Indeed, we note that on the dual polyhedron, the 
reflection\footnote{Another possibility to 
bring the polyhedron $F_9$ ``Higgsed'' by $\two_{(1,3/2)}$ back to the 
canonical form of $F_5$, is to rotate it by 90 degrees clockwise. In 
this case, the bundle redefinitions are $\cS_7\rightarrow2[K_B^{-1}]-\cS_9$, $\cS_9\rightarrow\cS_7\,.$} 
relating Figure~\ref{fig:HiggsingF9} (c) with 
Figure~\ref{fig:poly5_toric} enforces an exchange of the sections $s_5$ 
and $s_1$ in $X_{F_9}$ with 
$s_7$ and $s_3$ in $X_{F_5}$, respectively. From 
\eqref{eq:cubicsections} we see that this effectively amounts to a shift 
in the bundles $\cS_7$ and $\cS_9$ from $X_{F_9}$ to $X_{F_5}$, which is 
given by
\beq
\label{eq:BundleRedef}
 \cS_7 \rightarrow 2[K_B^{-1}]-\cS_7\,,\qquad
 \cS_9 \rightarrow  \cS_9\,.
\eeq 

Using the shift \eqref{eq:BundleRedef} for the second Higgsing, we find 
that in both cases (either VEVs in $\two_{(-1,-1/2)}$ or 
$\two_{(1,3/2)}$), there is a perfect agreement with the spectrum in 
Table~\eqref{tab:poly5_matter} of the toric hypersurface fibration 
$X_{F_5}$.

\subsubsection{Matching of the neutral spectrum: Higgsing \& Euler 
numbers}

So far we have matched only the charged spectrum of the Higgsed theory
of $X_{F_{9}}$ with that computed geometrically on $X_{F_5}$. In this 
section we work out the counting of complex structure moduli, 
that determine the number of neutral hyper multiplets by 
\eqref{eq:Hneutral}, before and after Higgsing. We show 
that the mismatch of the Hodge numbers 
$h^{(2,1)}(X_{F_5})-h^{(2,1)}(X_{F_9})\geq 0$  precisely agrees
with the amount of massless neutral singlets contributed from the Higgs 
multiplets.

First, let us discuss the geometric side of the matching. For a given 
Calabi-Yau manifold $X$, the amount of complex structure moduli can be inferred from its  Euler number $\chi(X)$ as 
\begin{align}
 h^{(2,1)}(X)= h^{(1,1)}(X)-\frac{\chi(X)}{2}\,,
\end{align}
with $h^{(1,1)}(X)$ given by 
\begin{align}
 h^{(1,1)}(X)= 1+\text{rk}(G_X)+h^{(1,1)}(B)\,,
\end{align}
where $\text{rk}(G_X)$ is the rank of the total gauge group $G_X$ of 
$X$. Thus, in an extremal transition from a toric hypersurface fibration 
$X_{F_i}$ to $X_{F_j}$, with the same base $B$, the change in $h^{(2,1)}$ reads
\begin{align}
 h^{(2,1)}(X_{F_j})-h^{(2,1)}(X_{F_i})=\text{rk}(G_{F_j})-
 \text{rk}(G_{F_i})+\frac{\chi(X_{F_i})-\chi(X_{F_j})}{2}\,.
\end{align}
For the specific Higgsing $X_{F_9}\rightarrow X_{F_5}$, 
their Euler 
numbers are given according to \eqref{eq:EulerNumbers} as 
\begin{align}
\begin{split}
 \chi(X_{F_9})&=-24[K_B^{-1}]^2+4[K_B^{-1}](2\cS_7+3\cS_9)-4(\cS_7^2+\cS_9^2)\,, \\
  \chi(X_{F_5})&=-24[K_B^{-1}]^2+8[K_B^{-1}](\cS_7+\cS_9)-2(2\cS_7^2+2\cS_9^2-\cS_7\cS_9)\,,
\end{split}
\end{align}
so that the  difference in their Hodge numbers $h^{(2,1)}$ is given by 
\begin{align}
\label{eq:miss}
 h^{(2,1)}(X_{F_5})-h^{(2,1)}(X_{F_9})=\cS_9(2[K_B^{-1}]-\cS_7)-1\,.
\end{align}
Here the $-1$ is the contribution from the change in the rank of the 
gauge group.

Next, in the corresponding field theories we consider the canonical 
Higgsing, that is induced by a Higgs in the 
representation $\two_{(-1,-1/2)}$. The multiplicity of 
hyper multiplets in the  representation $\two_{(-1,-1/2)}$
is $\cS_9(2[K_B^{-1}]-\cS_7)$, cf.~Table~\ref{tab:poly9_matter}. 
This is also the number of new uncharged singlets produced
in the Higgsing $X_{F_9}\rightarrow X_{F_5}$. However, out of those 
neutral singlets, exactly one gets massive, as can be seen from the D-
term potential.\footnote{In the case of an adjoint Higgs, all neutral 
hyper multiplets remain massless as the D-term is exactly zero.} 
This is also closely related to Goldstone's theorem: As three vectors 
have been lifted (an entire SU(2) is broken), 
three hyper multiplets must be removed from the 
massless spectrum. Two of these hyper multiplets are charged, as we 
discussed in the previous section, while the third one must be neutral, 
since it provides the longitudinal component of a massive 
$\U1$. In fact, only the simultaneous removal of three vectors and three 
hyper multiplets makes it possible for the purely gravitational anomaly 
in \eqref{eq:6dAnomalies} to cancel after the Higgs mechanism. 

Thus, there 
are precisely $\cS_9(2[K_B^{-1}]-\cS_7)-1$ massless singlets after the 
Higgsing, which precisely agrees with \eqref{eq:miss}.
In other words, we observe that the neutral massless hyper multiplets 
resulting from the Higgs mechanism become the complex structure moduli 
that were gained in the transition $X_{F_9}\rightarrow X_{F_5}$. In this 
work, we  explicitly confirm the matching of the complex structure 
moduli for all toric Higgs transitions between two toric hypersurface fibrations .
\subsubsection{Allowed regions for base $\mathbb{P}^2$}
\label{sec:allowedRegionP2}

While we considered in the previous sections the Higgsings for 
fibrations over an arbitrary two dimensional base, we regard it 
appropriate to devote some time on a concrete example with base 
$\mathbb{P}^2$ in order to discuss some subtleties that might arise.

Choosing the base fixes the range of allowed divisor classes for
$\cS_7$ and $\cS_9$ \cite{Cvetic:2013nia}. Expanding these 
divisors and the canonical class $K_{\mathbb{P}^2}$ in terms of 
the hyperplane class $H_B$ on $\mathbb{P}^2$, we have
\begin{align}
 \cS_7= n_7 H_B \, ,\qquad \cS_9= n_9 H_B \,,\qquad 
 K_{\mathbb{P}^2}=-3H_B\, ,
\end{align}
with $n_7$ and $n_9$ being positive integral coefficients. 

\begin{figure}[h!]
  \centering
  \setlength{\unitlength}{0.1\textwidth}
  \begin{picture}(8.7,4.7)
    \put(0.30,0.4){\includegraphics[width=0.83\textwidth]{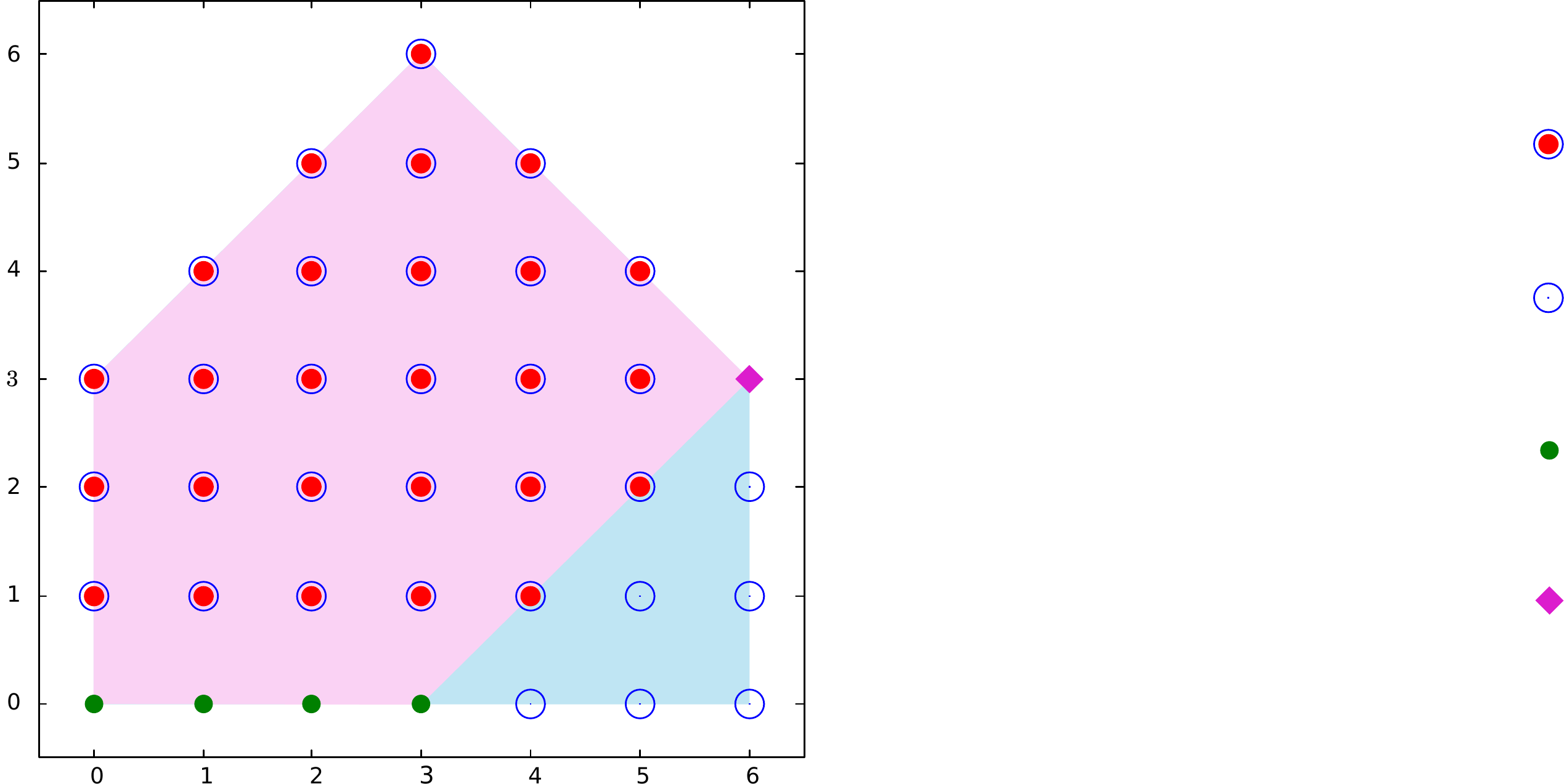}}
    \put(2.51,0.1){\small $n_7$}
    \put(5.19,3.72){\footnotesize Toric Higgsing $X_{F_9}\rightarrow X_{F_5}$ allowed}
    \put(4.85,2.98){\footnotesize Allowed strata for $X_{F_9}$ where the toric}
    \put(4.79,2,78){\footnotesize Higgsing with $\two_{(-1,-1/2)}$ is not possible}
    \put(7.37,2,105){\footnotesize $X_{F_9}\equiv X_{F_5}$ }
    \put(4.99,1,308){\footnotesize $X_{F_5}\equiv X_{F_7}$, adjoint Higgsing allowed }
     \put(0,2.51){\small $n_9$}
  \end{picture}
\caption{\label{fig:allowed_f9} Allowed regions of $\cS_7$ and $\cS_9$ 
for $X_{F_9}$ (blue and purple regions) and $X_{F_5}$ (purple region) 
with base $B=\mathbb{P}^2$.}
\end{figure}
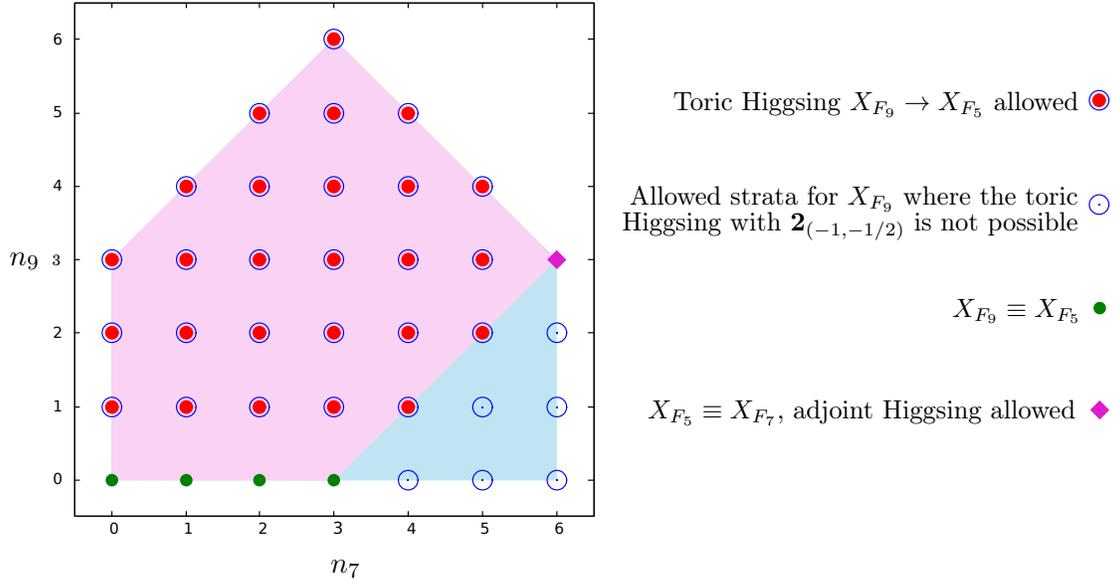

The effectiveness condition on all divisor classes $[s_i]$ in 
\eqref{eq:cubicsections} that occur for a given manifold $X_{F_i}$ 
imposes constraints on the allowed values for $n_7$ and $n_9$. These 
allowed values depend on the choice of the fiber. For the case of 
$X_{F_9}$ and $X_{F_5}$ the allowed regions are depicted in Figure~\ref{fig:allowed_f9}.  A choice of a point in this diagram constitutes a 
consistent fibration. Note that the allowed region for $X_{F_5}$ is 
fully contained inside that of $X_{F_9}$. Thus, there are 
compactifications $X_{F_9}$ inside the blue region in Figure~\ref{fig:allowed_f9}, for which the 
transition to $X_{F_5}$ is not possible because the effectiveness 
condition for some coefficient $s_i$ in $X_{F_5}$ would be violated. 

Indeed, this can also be seen from the field theory perspective, as for 
the points outside the allowed region for $X_{F_5}$, the 
multiplicity of Higgses in the representation $\two_{(-1,-\frac12)}$, 
that reads 
\begin{align}
 \cS_9 \left(2[K_B^{-1}]-\cS_7\right)=n_9 \left(6-n_7 \right)\, ,
\end{align}
is smaller than two. Recall that in order to have a D-flat potential we 
need at least two Higgs fields in the same representation to acquire a 
VEV.\footnote{Note that in this region, a different Higgsing with 
$\two_{(1,\frac32)}$ is possible.}
We also observe that the Higgs mechanism is possible for all points 
$(n_7,n_9)$ in the interior of the allowed region of $X_{F_5}$. However, 
for certain points on the 
boundary of the allowed region for $X_{F_5}$ we see that the amount of 
doublets does not suffice for a supersymmetry preserving Higgsing. 
These points are
\begin{itemize}
\item $0\leq n_7\leq3$, $n_9=0$: Here we see that $s_9$ belongs to the 
trivial bundle. Since in $X_{F_9}$ the locus of the $\SU2$ singularity 
is precisely $\{s_9=0\}$, cf.~\eqref{eq:SU2F9}, it is removed and the 
gauge group of $X_{F_9}$ at these points equals that of $X_{F_5}$, 
namely $\U1\times\U1$. In addition, we see that the spectra of
$X_{F_5}$ and $X_{F_9}$  in Tables~\ref{tab:poly5_matter} and 
\ref{tab:poly9_matter}, respectively, match perfectly. Hence we are 
at points where the strata of the moduli spaces of the two theories 
overlap and a transition among them is trivial.

 \item $n_7=6$, $n_9=3$: At this point there are no states in $X_{F_9}$ 
which transform in the $\two_{(-1,-\frac12)}$, so that the toric 
Higgsing is again not possible. Note also that since $s_9$ does not 
belong to the trivial bundle, the $\SU2$ factor is part of 
the gauge symmetry of the effective theory. However, in the hypersurface 
constraint for $X_{F_5}$, the sections $s_1$, $s_5$ and $s_8$ 
transform in the trivial bundle. Hence, at this particular point one can shift the toric 
coordinates $u$, $v$ and $w$ in order to globally set the section 
$s_1=0$ \cite{Cvetic:2013uta}, resulting in a non-toric U(1). 
This shows that the effective theory of $X_{F_5}$ coincides with that of 
$X_{F_7}$ precisely at the point $n_7=6$, $n_9=3$. Since the rank of
the gauge groups of $X_{F_9}$ and $X_{F_7}$ coincide, no toric Higgsing is possible. 
However, on $X_{F_9}$, one sees that there is 
one hyper multiplet in the adjoint of $\SU2$. In fact one can use this 
field to induce an adjoint (non-toric) Higgsing which leads precisely to 
the effective theory of $X_{F_5}$ at $n_7=6$, $n_9=3$. Indeed, we have 
explicitly computed that at this particular stratum in moduli space the 
numbers of complex structure moduli in $X_{F_7}$ and $X_{F_9}$ coincide. 
Similarly, we have confirmed that the entire matter spectrum in 
$X_{F_7}$ is reproduced (with the correct multiplicities) after the 
adjoint breaking from $X_{F_9}$. We omit all the details and just state
the corresponding $\U1$ redefinitions, in terms of the generators in 
$X_{F_9}$:
\begin{equation}
 Q_1^\prime=-3T_3-2Q_1+Q_2\,,\quad Q_2^\prime=-T_3-2Q_1+Q_2\,,\quad Q_3^\prime=-2T_3-Q_1\,.
\end{equation}
Looking at the Calabi-Yau constraint for $X_{F_9}$ we also observe 
that, by a shift in the toric coordinates, we can set $s_1=0$. Hence, the 
hypersurface constraint for $X_{F_7}$ only contains the  additional 
monomial $s_8 w^2u$ which is absent in the one of $X_{F_9}$. 
At $n_7=6$, $n_9=3$ the coefficient $s_8$ is just a constant, 
cf~\eqref{eq:cubicsections}, i.e.~one degree of freedom. On the field 
theory side, this degree of freedom corresponds precisely to the single 
adjoint Higgs on $X_{F_9}$. 
\end{itemize}
\subsection{Higgsings to theories with discrete gauge symmetries}
\label{S:discretesymm}

From the analysis carried out in Section~\ref{sec:AnalysisOfPolytopes}, 
we observe that the presence of discrete gauge symmetries is exclusive 
to the polyhedra $F_1$, $F_2$ and $F_4$. On the field theory side, we 
can use the Higgs mechanism to track the discrete symmetries as well, since 
these correspond to surviving remnants of broken $\U1$ symmetries. In 
fact, one can use the Higgsing diagram to show that the only Higgs 
mechanism for which the $\U1$s are broken to a discrete subgroup, are 
those leading precisely to $F_1$, $F_2$ and $F_4$, as expected 
geometrically. In this section we want to discuss in some detail those 
Higgsings leading to the toric hypersurface fibrations with discrete 
symmetries. To this end, we focus on the possible Higgs branches of 
$X_{F_6}$, $X_{F_5}$ and $X_{F_3}$.

In many of the transitions considered here, the charge of the Higgs 
fields does not allow us to directly infer whether or not there is a 
non-trivial discrete symmetry. Consider for example the Higgsings from 
$X_{F_9}\rightarrow X_{F_6}$, with the toric Higgses given in 
\eqref{E:F_6}. One possibility is to have VEV fields in the 
representation $\one_{(1,2)}$. In principle one might think that, given 
the charge of the Higgses, there is a discrete remnant of the second 
$\U1$. However, this leftover symmetry is trivial, since there is an 
$SL(2,\mathbb{Z})$ transformation which maps the charge of the Higgses 
from $(1,2)$ to $(1,0)$. 
As discussed in Section~\ref{S:ToricHiggsings}, the field $\one_{(1,0)}$ 
allows for a geometrically equivalent breaking, where it becomes clear 
that no discrete symmetries are present in $X_{F_6}$ as there the U(1) charge is minimal. 

More general, if we 
break a $\U1^n$ gauge symmetry by the VEV of a field $\phi$, there is no 
discrete symmetry left provided the existence of an $SL(n,\mathbb{Z})$ 
which makes its charge minimal. In other words, after the 
$SL(n,\mathbb{Z})$ transformation of the VEV field, the charge 
$(q_1,q_2,\dots,q_n)$ takes the form $(1,0,\dots,0)$. In this new basis 
it is obvious that no discrete gauge group is left after the breaking. In other cases a seeming 
discrete group can be embedded into an unbroken U(1) or the center of a non-Abelian gauge 
group,  showing again the absence of a discrete group.
\subsubsection*{$X_{F_6}$ Higgs branches}

From the spectrum in $X_{F_6}$ given in Table~\ref{tab:poly6_matter}
we see that there are two possible toric Higgsings, depending on whether 
the Higgs fields are taken in the representation $\two_{-3/2}$ or 
$\one_2$. In the first case the $\SU2\times\U1$ symmetry in $X_{F_6}$ is 
broken to a single $\U1$, so that this Higgs branch leads to $X_{F_3}$. 
In the second case the $\SU2$ symmetry remains unbroken, as expected 
from a Higgsing to $X_{F_4}$. The splitting of the states in either 
cases proceeds according to Table~\ref{tab:F6branchings}.
\begin{table}[H]
\begin{center}
\renewcommand{\arraystretch}{1.2}
\begin{tabular}{|c|c||c|}
\cline{2-3}
\multicolumn{1}{c|}{} & VEV: $\two_{-\frac32}$ & VEV: $\one_{2}$  \\
\multicolumn{1}{c|}{} & $Q^\prime=(3T^3-Q)$ & $Q^\prime= Q$ mod $2$ \\
\hline
$\two_{-\frac32}$ & $\one_{0} + \one_{-3}$ & $\two_{\frac12}$ \\
$\two_{\frac12}$ & $\one_{1}$ + $\one_{-2}$ & $\two_{\frac12}$ \\
$\one_{2}$ & $\one_{-2}$ & $\one_{0}$ \\
$\one_{1}$ & $\one_{-1}$ & $\one_{1}$ \\
$\three_{0}$ & $\one_{0}+ \one_3 + \one_{-3}$ & $\three_{0}$   \\
\hline
\end{tabular}
\caption{Branching of representations under the possible Higgsings of $X_{F_6}$ to $X_{F_3}$ and $X_{F_4}$.}
\label{tab:F6branchings}
\end{center}
\end{table}
Using these branchings of the representations on $X_{F_6}$ 
into representations of $G_{F_3}$,
we can compute the multiplicities of the multiplets after Higgsing using
Table~\ref{tab:poly6_matter}. They read 
\begin{align}
\label{eq:F3}
\begin{split}
\one_3 :& \qquad (\cS_9 - \cS_7) \left( [K_B^{-1}]-\cS_7 + \cS_9    \right) \, \\
\one_2 :& \qquad  ([K_B^{-1}]- \cS_9 + \cS_7) \left(6 [K_B^{-1}]- 2 \cS_9 + \cS_7    \right)    \, \\
\one_1 :& \qquad  12[K_B^{-1}]^2 +[K_B^{-1}]\left( 8\cS_7 -\cS_9    \right) -4\cS_7^2+ \cS_7 \cS_9 - \cS_9^2\,.
\end{split}
\end{align}
The above multiplicities agree with our geometrical result for the 
spectrum of $X_{F_3}$, see Table~\ref{tab:poly3_matter}. Note also that naively, due to the 
non-primitive U(1)-charge of the Higgs field $\two_{-\frac32}$, we 
expect a surviving discrete $\mathbb{Z}_3$ symmetry. However, this 
symmetry is contained in the surviving $\U1$ symmetry in $X_{F_3}$, 
i.e.~there is no discrete gauge group on $X_{F_3}$ as expected
geometrically.

In contrast, we see that the U(1)-charge of the VEV $\one_{2}$ 
triggering the transition $X_{F_6}\rightarrow X_{F_4}$ 
is \emph{non-minimal}. Thus, we expect a discrete gauge 
symmetry to be left unbroken, in addition to the $\SU2$ gauge factor. The 
decomposition of representations for this Higgsing is given in 
Table~\ref{tab:F6branchings}. There we see that the doublet after the 
Higgsing carries a half integral charge with respect to the discrete 
gauge factor. Hence, one has to rescale all charges by a factor of two, 
so that all charges become integral. Thus, we see that the resulting 
discrete symmetry is in fact $\mathbb{Z}_4$. Note that in this case, the 
discrete factor is of physical relevance, since we can not embed it into 
the local gauge group of $X_{F_4}$. This implies that in a four 
dimensional theory, there will be gauge invariant couplings which are 
absent due to selection rules imposed by the $\mathbb{Z}_4$-symmetry. 

The multiplicities of charged states following from 
Tables~\ref{tab:F6branchings} and \ref{tab:poly6_matter} are given by  
\begin{align}
\begin{split}
\two_1 :& \qquad  \left( [K_B^{-1}]-\cS_7 + \cS_9    \right)\left( 6[K_B^{-1}]-2\cS_9 + 2\cS_7    \right) \, , \\
%\one_0 :& \qquad  \cS_7(- \cS_9 + 2\cS_7)   \\
\one_2 :& \qquad  6[K_B^{-1}]^2 +[K_B^{-1}]\left( 13\cS_7 -5\cS_9    \right) -3\cS_7^2-2\cS_7 \cS_9 +\cS_9^2 \, , \\
\three_0 :& \qquad  1+([K_B^{-1}]- \cS_7 + \cS_9)\frac{(-\cS_7 + \cS_9)}{2}    \,.
\end{split}
\end{align}
This precisely agrees with the geometrically obtained spectrum of 
$X_{F_4}$ in Table~\ref{tab:poly4_matter}.
We emphasize that charges of the matter states w.r.t~to the 
$\mathbb{Z}_4$, that we have obtained by Higgsing $X_{F_6}$, precisely 
coincide with those computed by intersections with the four-section in
\eqref{eq:ShiodaFourSecF4}. 
\subsubsection*{$X_{F_5}$ Higgs branches}

In $X_{F_5}$ there are two possible toric Higgsing to $X_{F_3}$. The 
Higgs fields in that case are in the representations 
$\one_{(-1,-2)}$ or $\one_{(1,-1)}$, cf.~Table \ref{tab:poly5_matter}. 
The branching of the representations of $X_{F_5}$ into representations 
of $G_{F_3}$ are shown in  
Table~\ref{tab:F5branchings}. The resulting spectrum matches that in 
\eqref{eq:F3}, cf.~Table~\ref{tab:poly3_matter}, up to 
redefinitions of $\cS_7$ and $\cS_9$, that correspond to the 
transformations needed in order to bring the 
resulting polyhedron to its canonical form
in Figure~\ref{fig:F3}.
\begin{table}[H]
\begin{center}
\renewcommand{\arraystretch}{1.2}
\begin{tabular}{|c|c|c||c|}
\cline{2-4}
\multicolumn{1}{c|}{} & VEV: $\one_{(-1,-2)}$ & VEV: $\one_{(-1,1)}$  & VEV: $\one_{(0,2)}$  \\
\multicolumn{1}{c|}{} & $Q_1^\prime=(2Q_1-Q_2)$& $Q_1^\prime=(Q_1+Q_2)$ & $Q_1^\prime=(Q_1)$, $Q_{\mathbb{Z}_2}=Q_2$ mod $2$ \\
\hline
$\one_{(1,-1)}$ & $\one_{3}$& $\one_{0}$& $\one_{(1,-)}$ \\
$\one_{(1,0)}$ & $\one_{2}$& $\one_{1}$& $\one_{(1,+)}$ \\
$\one_{(-1,-2)}$ & $\one_{0}$& $\one_{3}$& $\one_{(1,+)}$ \\
$\one_{(-1,-1)}$ & $\one_{1}$& $\one_{2}$& $\one_{(1,-)}$ \\
$\one_{(0,2)}$ & $\one_{2}$& $\one_{2}$& $\one_{(0,+)}$ \\
$\one_{(0,1)}$ & $\one_{1}$& $\one_{1}$& $\one_{(0,-)}$ \\
\hline
\end{tabular}
\caption{Branching of representations under the possible Higgsings of $X_{F_5}$ to $X_{F_3}$ (first two columns) and $X_{F_2}$ (third column).}
\label{tab:F5branchings}
\end{center}
\end{table}

Similarly, the polyhedron allows for an additional toric Higgsing 
from $X_{F_5}$ to $X_{F_2}$ which is triggered by VEVs in the 
representation $\one_{(0,-2)}$. This leaves the first $\U1$ unbroken, 
together with a remnant discrete $\mathbb{Z}_2$ symmetry from the second 
$\U1$. The multiplicities of the charged matter in $X_{F_2}$
that are deduced by Higgsing from $X_{F_5}$ are obtained from the 
group-theoretical branchings shown in Table~\ref{tab:F5branchings} and 
the spectrum of $X_{F_5}$ in Table~\ref{tab:poly5_matter}. They read
\begin{align}
\begin{split}
\one_{(1,-)} : &\,\qquad 6[K_B^{-1}]^2 +4[K_B^{-1}]\left( \mathcal{S}_9- \mathcal{S}_7    \right)+ 2\cS_7^2 - 2\cS_9^2\,, \\
\one_{(1,+)} :&\,\qquad 6[K_B^{-1}]^2+ 4[K_B^{-1}]\left( \mathcal{S}_7 - \mathcal{S}_9    \right) - 2\mathcal{S}_7^2 + 2\mathcal{S}_9^2\,, \\
\one_{(0,-)} :& \,\qquad 6[K_B^{-1}]^2+ 4[K_B^{-1}]\left(\mathcal{S}_7 + \mathcal{S}_9    \right)-2\mathcal{S}_7^2 - 2\mathcal{S}_9^2\,,
%\one_{(0,+)} :& \,\qquad \mathcal{S}_7\mathcal{S}_9
\end{split}
\end{align}
which precisely matches the geometrical result in Table~\ref{tab:poly2_matter}.
\subsubsection{$X_{F_{3}}$ Higgs branches}

In the spectrum of $X_{F_3}$, the singlet $\one_3$ allows for a toric 
Higgsing. In this transition, the $\U1$ symmetry gets broken to a 
$\mathbb{Z}_3$ subgroup. The decomposition of representations of 
$X_{F_3}$ for this Higgsing is shown in Table~\ref{tab:F3toF1}.
\begin{table}[htb!]
\begin{center}
\renewcommand{\arraystretch}{1.2}
\begin{tabular}{|c|c|}
\cline{2-2}
\multicolumn{1}{c|}{} & VEV: $\one_{3}$  \\
\multicolumn{1}{c|}{} & $Q_{\mathbb{Z}_3}^\prime= q$ mod $3$ \\
\hline
$\one_{3}$ & $\one_{0}$ \\
$\one_{2}$ & $\one_{2}$ \\
$\one_{1}$ & $\one_{1}$ \\
\hline
\end{tabular}
\caption{Branching of representations under the toric Higgsing of $X_{F_3}$ to $X_{F_1}$.}
\label{tab:F3toF1}
\end{center}
\end{table}

As mentioned before, 6D hyper multiplets in the representation $\one_2$ 
under the discrete group $\mathbb{Z}_3$ are equivalent to hyper 
multiplets in the representation  $\one_1$. Hence, there is only one 
type of charged hyper multiplet in $X_{F_1}$. This is in agreement with 
the geometrical computation (see Section~\ref{sec:F1_poly}), where one 
sees a single codimension two locus supporting an $I_2$-fiber. From the Higgsing we can read of the multiplicity of this charged state as
\begin{align}
\one_{1} : &\,\qquad 3\left( 6 [K_B^{-1}]^2 - \cS_7^2 + \cS_7 \cS_9 - \cS_9^2 + [K_B^{-1}](\cS_7 + \cS_9) \right) \, .
\end{align}
\subsection{The complete Higgsing chain}
\label{S:fullspec}

Having summarized the relevant features of the toric Higgsing procedure,  
we devote this section to a complete account on all possible toric 
Higgsings, that are summarized in Appendix~\ref{app:higgschain},  
Tables~\ref{tab:HiggsChain1}-\ref{tab:HiggsChain3}. In these tables
we indicate in the first column, which toric hypersurface fibrations 
$X_{F_i}\rightarrow X_{F_{j}}$, $i\neq j$, are to be related. Then, in 
the second column, we state the possible toric 
Higgsings and which fields are to be identified as the Higgs fields that 
acquire a VEV in the transition.  
We note that all toric Higgsings between the same two toric hypersurface 
fibrations are physically equivalent.  
In the third column the  U(1)-generators on $X_{F_j}$ are expressed in
terms of the U(1)-generators and Cartan generators on $X_{F_i}$.
We have checked explicitly in all Higgsings that all matter charges of 
the fibration $X_{F_j}$ determined in  
Section~\ref{sec:AnalysisOfPolytopes} are obtained. In addition,
in some Higgsings, the unbroken 
non-Abelian gauge group factors are interchanged in the Higgsing 
process. In these cases, the change of the order of non-Abelian
factors is indicated in the third column of the tables in Appendix~\ref{app:higgschain}. 
In addition in most of the cases, an $SL(2,\mathbb{Z})$-transformation 
on the ``Higgsed'' polyhedron is necessary 
in order to bring it into the canonical form used for the geometric
computations in Section~\ref{sec:AnalysisOfPolytopes}. These 
transformations determine a unique redefinition of the divisor
classes, similar as in  \eqref{eq:BundleRedef}, that is necessary
in order to compare the matter multiplicities of the representations
obtained after the Higgsing with the ones obtained by inspecting the 
geometry of $X_{F_j}$. The relevant redefinitions are shown in the last 
column of the tables in Appendix~\ref{app:higgschain}. 

The decompositions of the representations on $X_{F_i}$ under the group 
$G_{F_i}$ into representations of the unbroken gauge group $G_{F_j}$ 
after Higgsing can be found in  Appendix~\ref{app:decomposition} for all  
canonical toric Higgsings. 
We have checked in all cases that the matter spectra obtained by 
Higgsing the theory on $X_{F_i}$ to the one corresponding to $X_{F_j}$  
agree with  those of Section~\ref{sec:AnalysisOfPolytopes}, which 
provides another non-trivial check of the geometric analysis presented 
there.

We conclude this section with one final observation.
As highlighted before, we observe that for every transition between two 
toric hypersurface fibrations $X_{F_{j}} \rightarrow X_{F_{k}}$ there 
exists a dual transition
between $X_{F_{k}^*} \rightarrow X_{F_{j}^*}$. This symmetry of the 
Higgs  diagram in Figure~\ref{fig:network} can be directly understood 
from the interpretation of the toric Higgsing on the level of the fiber 
polyhedron and its dual as we will explain in the following. 

As pointed out in 
Section~\ref{S:ToricHiggsings}, a toric Higgsing acts exactly as a 
blow-down in the original polyhedron $F_j$ and a blow-up $F_j^*$ in its 
dual polyhedron:
\begin{align}
\label{eq:HiggsTransition1}
(F_j \, ,\, F_{j}^*) \stackrel{\text{Higgs}}{\longrightarrow} (F_j  \stackrel{\text{blow-down}}{\longrightarrow} F_k \, , \, F_{j}^* \stackrel{\text{blow-up}}{\longrightarrow}  F_{k}^*) \, ,
\end{align}
with $j>k$.
Next we consider the inverse of the above process: We take $F_{k}^*$ 
as the starting polyhedron with $F_k$ as its dual. From the diagram 
\eqref{eq:HiggsTransition1}, we know that there exists a blow-up map 
from $F_{j}^*$ to $F_{k}^*$. However, now we take its inverse map as the 
blow-down from $F_{k}^*$ to $F_{j}^*$. The same can analogously be 
done for the dual polyhedron $F_k$ whose blow-up map is obtained from 
the inverse of the blow-down map in \eqref{eq:HiggsTransition1}. 
Consequently, we arrive at the following map for the dual Higgs 
transition
\begin{align}
(F_{k}^* \, , \, F_{k}) \stackrel{\text{Dual Higgs}}{\longrightarrow} (F_{k}^* \stackrel{\text{blow-down}}{\longrightarrow} F_{j}^* \, , \, F_{k}  \stackrel{\text{blow-up}}{\longrightarrow} F_{j} ) \, .
\end{align}

The above relation holds for every toric Higgsing and hence shows, that 
every Higgs transition has indeed a dual counterpart.
However in general we observe more equivalent transitions between higher 
polyhedra than in their dual counterparts. In the geometry this reflects 
the fact, that in polyhedra with a larger area, there are more ways to 
embed subpolyhedra. On the field theory side this corresponds to less 
representations, that can be used for the Higgsing. An example is the 
transition $X_{F_{15}} \rightarrow X_{F_{12}}$ which can be equally 
realized by cutting any of the four vertices in the square of $F_{15}$. 
However the dual transition $F_{5} \rightarrow F_{2}$ can only be 
achieved by deleting the unique node and hence there is only one 
Higgsing possible.

\clearpage
\section{Conclusions}
\label{sec:conclusion}

In this work we have analyzed F-theory compactifications on all 
toric hypersurface fibrations $X_{F_i}$. In these 
manifolds the genus-one fibers are
given as a hypersurface in any of the 2D toric varieties 
associated to the reflexive polyhedra $F_i$ depicted in 
Figure \ref{fig:16polytopes}. For each 
of these 16 Calabi-Yau manifolds 
we have computed the full MW-group 
(cf.~with the results of \cite{Braun:2013nqa} for the toric MW-group) 
and determined all codimension one, two and three singularities and the 
corresponding reducible fibers in the crepant resolutions $X_{F_i}$. 
Our work presents the first complete analysis of all these aspects.
In the course of our study, we have encountered 
some codimension two fibers which are non-split and others whose 
complete splitting is visible only by computing their associated prime 
ideals. In addition, we have computed the Hodge numbers of the $X_{F_i}$ 
in the case of Calabi-Yau threefolds.
All these geometric results determine the gauge groups, matter 
representations and Yukawa couplings of the effective SUGRA theories of 
F-theory on these manifolds. We have shown that these 
effective theories are anomaly-free in 6D, which proves in turn the 
completeness of our analysis of codimension one and two singularities of 
all these models \cite{Grassi:2011hq}. 

The gauge groups we have found range from rank zero to six with up
to three \U1-factors corresponding to a rank 
three MW-group of rational sections. The Calabi-Yau manifolds $X_{F_1}$, 
$X_{F_2}$ and $X_{F_4}$,  that are constructed as 
fibrations of the cubic in $\mathbb{P}^2$, the biquadric in 
$\mathbb{P}^1\times \mathbb{P}^1$ and the quartic\footnote{As remarked 
already, this case has been subject of recent attention in 
\cite{Braun:2014oya,Morrison:2014era,Anderson:2014yva}.} in 
$\mathbb{P}^{2}(1,1,2)$, respectively, do not have a section and lead to 
F-theory models with discrete gauge groups $\mathbb{Z}_3$,  
$\mathbb{Z}_2$ and $\mathbb{Z}_4$. We have 
established a relationship between the order of the multi-section in 
these Calabi-Yau manifolds and the order of the discrete gauge group. Furthermore, we have 
shown the existence (and computed the multiplicity) of 
$I_2$-singularities that support matter  charged 
only under these discrete groups. We have also 
a proposal for a ``Shioda-map'' of multi-sections, that allowed 
us to consistently compute the charges of all matter fields under these 
discrete gauge groups. In addition, by an explicit computation of the 
respective generators of their rank one MW-groups, we have 
demonstrated that $X_{F_2}$ and $X_{F_{3}}$ (which is constructed 
as a fibration of the elliptic curve in $dP_1$) both yield effective 
theories with one U(1)-gauge field. Most notably, we have found the 
first F-theory realization of charged singlets with U(1)-charge $q=3$ in $X_{F_3}$. 
Furthermore, the non-trivial consistency conditions  imposed on the 
U(1)-charge lattice of a model in quantum gravity \cite{Banks:2010zn} have been checked.

We emphasize that the non-toric nature of the generator of the MW-group 
of the fibration $X_{F_3}$ was key to obtaining this U(1)-charge. We 
expect that the presence of non-toric sections can lead to more exotic 
U(1)-charge assignments of matter than those that occur in toric cases. 
Such a situation can be desirable as these exotic charge assignments 
could serve to control the phenomenology of particle physics models 
constructed in F-theory. 

Besides these geometrical advances, we have shown that those 
extremal transitions between all toric hypersurface fibrations 
$X_{F_i}$, which are induced by toric blow-downs in the toric varieties 
$\mathbb{P}_{F_i}$, can be described by a corresponding Higgs mechanism 
in the effective theories of F-theory on the $X_{F_i}$. Although this 
correspondence between geometry and physics is expected to hold in 
general, 
also in chiral F-theory compactifications to 4D\footnote{As 4D chirality 
is induced by $G_4$-flux in F-theory, this requires the understanding of 
the behavior of $G_4$-flux during extremal transitions in Calabi-Yau 
fourfolds, which is  discussed in \cite{Intriligator:2012ue}. The 
results of \cite{Bizet:2014uua} will be crucial for carrying out any 
quantitative analysis in this context.}, we considered here the 6D case. 

We have explicitly worked out the full network of all toric Higgs 
transitions, shown in Figure \ref{fig:network}, in the 6D SUGRA theories 
of these F-theory models. Cutting off a vertex in a polyhedron $F_i$ 
corresponds to a blow down in $\mathbb{P}_{F_i}$, which implies the 
removal of a corresponding divisor in $X_{F_i}$. After this blow-down, 
a new monomial can be added to the hypersurface constraint of $X_{F_i}$, 
resulting in a new Calabi-Yau manifold $X_{F_{i'}}$. In the effective 
theory of F-theory on $X_{F_i}$, this transition corresponds to giving a 
VEV to a particular matter field along a D-flat direction and  a 
consequent breakdown of the gauge symmetry of the theory.  
For all extremal transitions, we have identified the relevant matter 
field that has to acquire a VEV and matched the effective theory  
after Higgsing with the one obtained geometrically from F-theory on 
$X_{F_{i'}}$. As an explicit example, for a specific choice of the base 
$B=\mathbb{P}^2$, we have described the transition $X_{F_9}\rightarrow X_{F_5}$ for all strata in moduli
space, that are labeled by $(n_7,n_9)$, 
cf.~Figure \ref{fig:allowed_f9}. 
There we have found that a D-flat Higgsing
in the effective theory is only possible for those points $(n_7,n_9)$, 
that are allowed for both $X_{F_9}$ and $X_{F_5}$. In this 
context, we have also commented on the different gauge groups that 
appear on the boundary of the allowed region for $(n_7,n_9)$ in $X_{F_5}$.

We have found that the full toric Higgs network in Figure~\ref{fig:network}
 is beautifully mirror symmetric under the exchange of 
a polyhedron $F_i$ and its dual $F_i^*$.\footnote{We note that a similar 
observation has been made for elliptically fibered toric $K3$ surfaces 
in \cite{Berglund:1998ej}.} This gave rise to a number of interesting 
observations: The toric hypersurface fibrations $X_{F_i}$ and 
$X_{F_i^*}$ always have the same amount of \U1-symmetries and the 
ranks of their gauge groups, $G_{F_{i}}$ and $G_{F_i^*}$,  
always sum up to six. Indeed, this sum rule of the rank of the gauge group is a 
direct consequence of the sum rule for the volumes of $F_i$ and its dual 
$F_i^*$.\footnote{We thank Albrecht Klemm for explaining this fact to 
us.}
The duality between $X_{F_i}$ and $X_{F_i^*}$ is realized also
on the level of the Higgs transitions, i.e.~for every toric Higgs 
transition $X_{F_i}\rightarrow X_{F_{i'}}$, there is a dual Higgs 
transition $X_{F^*_{i'}}\rightarrow X_{F_i^*}$.
In addition, we have observed that this duality maps 
discrete symmetries in F-theory to Mordell-Weil torsion. 

Finally, let us highlight some features of toric hypersurface 
fibrations, which make them attractive for particle physics 
applications. First, recall that the 
presence of discrete symmetries is a desirable 
feature in field theoretic constructions beyond the standard 
model.\footnote{ The reader is referred to 
\cite{Farrar:1978xj,Dimopoulos:1981dw,Ibanez:1991pr,Dreiner:2005rd,Lee:2010gv} 
for a selection of discrete symmetries which have been invoked in order 
to forbid certain unappealing operators in supersymmetric models.} So 
far, all efforts towards embedding the standard model in F-theory have 
been based on compactifications with a zero section, which are typically 
free of discrete symmetries.\footnote{Instead of discrete symmetries, the 
phenomenology of these models is kept under control by virtue of 
additional $\U1$ symmetries. Discrete symmetries 
can arise by a manual breakdown of these additional $\U1$s. This 
possibility has been studied in e.g \cite{Dudas:2009hu,Krippendorf:2014xba}.} 
Since discrete symmetries arise naturally in genus one-fibrations with 
multi-sections, this type of compactifications constitutes a 
promising new arena for engineering semi-realistic particle physics 
models.

In addition, we have found concrete toric hypersurface fibrations, that 
directly realize the 
gauge group and representations of the Standard Model ($X_{F_{11}}$),
the Pati-Salam model ($X_{F_{13}}$) and the trinification model
($X_{F_{16}}$). Even more interestingly, we have found that the Standard 
Model can be obtained via toric Higgsings from the Pati-Salam or the 
trinification model, both of these models being at the same time the 
two theories with the maximal non-Abelian gauge groups among all toric 
hypersurface fibrations.

\subsection*{Outlook}

For future directions it would be interesting to use the 
effective theories we have obtained for particle physics applications.
Since the Pati-Salam and trinification model are the two 
maximal toric enhancements of the Standard Model, as we have seen, 
they are natural candidates for toric unified model building in F-theory.
It would be fascinating to work out the phenomenological implications of 
this observation.

Clearly, the results of this paper most directly apply to six 
dimensional or to non-chiral four dimensional  
compactifications of F-theory. Thus, a natural 
and most straightforward extension of this work, that would also be 
crucial for phenomenological applications, is the inclusion and 
construction of $G_4$-flux as well as the computation of the chiral 
indices of all the matter representations found in all toric 
hypersurface fibrations $X_{F_i}$, following the recipe and techniques 
described in \cite{Cvetic:2013uta}.

The beautiful realization of mirror symmetry in the Higgs network 
and related observations like the sum rule for the gauge groups and the connection
between Mordell-Weil torsion and discrete symmetries are  topics of further investigation.
It would be exciting to understand all these observations by 
unveiling a common structure underlying all toric hypersurface 
fibrations, for example a master gauge group into which all gauge groups 
$G_{F_i}$ could be embedded. This might require a departure from the 
toric framework.

There has been a lot of recent progress in obtaining matter 
representations in F-theory using deformations or, in physical 
terms, the Higgs mechanism \cite{Grassi:2013kha,Grassi:2014sda}. 
It would be very interesting to see how the results about the Higgs network 
of the toric hypersurface fibrations obtained here can be worked out 
using the deformation techniques employed there. 

Finally, it would be interesting to use the toric hypersurface 
fibrations studied here also for compactifications of M-theory to  
engineer 3D $\mathcal{N}=2$ gauge theories and to
study their Coulomb-branches and phase structures, see 
\cite{Hayashi:2013lra,Esole:2014bka,Esole:2014hya,Braun:2014kla} 
(and also the seminal works 
\cite{Esole:2011sm,Marsano:2011hv}) 
for recent detailed analyses of the phase structure of 3D SU$(N)$-gauge 
theories for all $N\leq 5$. 

\subsection*{Acknowledgments}

We would like to thank Lara Anderson, Ron Donagi, Thomas Grimm, 
Hans Jockers, Jan Keitel, Albrecht Klemm, Craig Lawrie, Dave Morrison, 
Sakura Sch\"afer-Nameki, Wati Taylor, Timo Weigand and 
especially Mirjam Cveti\v{c}  for valuable discussions and correspondence. 
D.K.~thanks the Bethe Center for Theoretical Physics Bonn and the Theory Division of CERN 
for hospitality during completion of the project. This work is supported 
by the  DOE grant DE-SC0007901 (D.K., H.P.), the NSF String Vacuum Project 
Grant No. NSF PHY05-51164 (H.P.) and the Dean's Funds for Faculty Working 
Group (D.K.). The work of D.M., P.O.~and J.R.~is partially supported by a 
scholarship of the Bonn-Cologne Graduate School BCGS, the SFB-Transregio 
TR33 The Dark Universe (Deutsche Forschungsgemeinschaft) and the European 
Union 7th network program Unification in the LHC era 
(PITN-GA-2009-237920). 

\clearpage
%========================================================================
%
%
\appendix

%%%%%%%%%%%%%%%%%%%%%%%%%%%%%%%%%%%%%%%%%%%%%%%%%%%%%%%%%%%%%%%%%%%%%%%%%%%%%%%%%%%%%%%%%%%%%%%%%%%%%%
%%%%%%%%%%%%%%%%%%%%%%%%%%%%%%%%%%%%%%%%%%%%%%%%%%%%%%%%%%%%%%%%%%%%%%%%%%%%%%%%%%%%%%%%%%%%%%%%%%%%%%
\section{Anomaly Cancellation Conditions in 6D}
\label{app:Anomalies}
%%%%%%%%%%%%%%%%%%%%%%%%%%%%%%%%%%%%%%%%%%%%%%%%%%%%%%%%%%%%%%%%%%%%%%%%%%%%%%%%%%%%%%%%%%%%%%%%%%%%%%
%%%%%%%%%%%%%%%%%%%%%%%%%%%%%%%%%%%%%%%%%%%%%%%%%%%%%%%%%%%%%%%%%%%%%%%%%%%%%%%%%%%%%%%%%%%%%%%%%%%%%%

In this appendix we summarize the consistency relations that have to be 
obeyed by an anomaly-free 6D SUGRA theory.
We follow the conventions and notations of 
\cite{Taylor:2011wt,Park:2011wv,Cvetic:2013nia}, to which we also refer 
for further details.

There are three qualitatively different types of anomalies, the pure 
gravitational anomalies, the mixed gauge-gravitational anomalies and
the pure gauge anomalies. Depending on the number of gauge group 
factors, mixed anomalies between different gauge group factors are
present. A theory is referred to as anomaly-free if all one-loop 
anomalies are canceled by the contributions from the anomalous 
variations of Green-Schwarz (GS) counter-terms.

For an effective SUGRA theory in 6D, the anomaly cancellation 
conditions read:
\bea
\label{eq:6dAnomalies}
 \text{tr}R^4&:&H-V+ 29T=273\,,\quad  (\text{tr}R^2)^2:\,\,9-T=a\cdot a \,\,\,\,\,\,\,\,(\text{Pure gravitational})\nn\\
  \text{tr}F_\kappa^2\text{tr}R^2&:& \textstyle{ -\frac{1}{6}}\left( A_{adj_\kappa}-\sum_{\mathbf{R}} x_{\mathbf{R}} A_{\mathbf{ R}}\right)=a \cdotp \left( \frac{b_{\kappa}}{\lambda_\kappa}\right) \qquad\qquad\,\,\,\,\,\, (\text{Non-Abelian-gravitational})\nn \\
   F_mF_n\text{tr}R^2&:&-\textstyle{\frac{1}{6}}\sum_{\underline{q}} x_{q_m, q_n} q_m q_n=a\cdotp b_{mn} \qquad\qquad\qquad\qquad\,\,\,\,\,\,\,\,\, (\text{Abelian-gravitational})\nn\\
 \text{tr}F_\kappa^4&:& B_{adj_\kappa} - \sum_{\mathbf{R}} x_{\mathbf{R}} B_{\mathbf{R}} = 0\,,\qquad\qquad\qquad\qquad\qquad\,\,\,\,\,\,\,\,\,\,  (\text{Pure non-Abelian})\nn \\
\text{tr}F_\kappa^2\text{tr}F_\kappa^2&:& \textstyle{\frac{1}{3}} \left( \sum_{\mathbf{R}} x_{\mathbf{R}} C_{\mathbf{R}}-  C_{adj_\kappa}  \right) = \left( \frac{b_\kappa}{\lambda_\kappa}\right)^2\nn \\
  	F_m F_nF_kF_l&:&\sum_{\underline{q}} x_{q_m,q_n,q_k,q_l} q_m q_n q_k q_l=b_{(mn} \cdotp b_{kl)}\qquad\qquad\qquad\quad\,\,\,\,\,\,\,\,\, (\text{Pure Abelian})\nn \\
   F_{m}F_{n}\text{tr}F_\kappa^2&:&\sum_{\mathbf{R},q_m,q_n} x_{\mathbf{R},q_m,q_n} q_m q_n A_{\mathbf{R}} = \left( \frac{b_\kappa}{\lambda_{\kappa}} \right) \cdotp b_{mn}\,\,\qquad\qquad\quad(\text{Non-Abelian-Abelian} )\nn\\
F_{m}\text{tr}F_\kappa^3&:&\sum_{\mathbf{R},q_m} x_{\mathbf{R},q_m} q_i E_{\mathbf{R}} = 0\,.
\eea
Here, we have given the terms in the 6D anomaly 
polynomial, whose coefficients are the respective anomalies.
The Ricci tensor is denoted by $R$ and the field strengths
of the non-Abelian and Abelian gauge field for the gauge group
factor $G_\kappa$ and the $m^{\text{th}}$ U(1) are denoted by
$F_\kappa$ and $F_m$, respectively. The overall number of hyper, 
vector and tensor multiplets is 
denoted by $H$, $V$ and $T$, respectively and the variables 
`$x_{\cdot}$'  denote the multiplicities of certain charged
hyper multiplets: $x_{\mathbf{R}}$,  $x_{\mathbf{R},q_m}$ and 
$x_{\mathbf{R},q_m,q_n}$  are the number of hyper multiplets in 
the  representation $\mathbf{R}$, in the 
representation $\mathbf{R}$ with charge $q_m$ under U(1)$_m$ and 
in the  representation $\mathbf{R}$ with 
charges $\{q_m,q_n\}$ under U(1)$_m
\times$U(1)$_n$, respectively; $x_{q_m,q_n}$ and  
$x_{q_m,q_n,q_k,q_l}$ denote the
number of matter hyper multiplets with charges $(q_m,q_n)$ and 
$(q_m,q_n,q_k,q_l)$ under $\text{U}(1)_m\times \text{U}(1)_n$ and 
$\text{U}(1)_m\times \text{U}(1)_n\times\text{U}(1)_k\times \text{U}
(1)_l$, respectively.  

In the contributions from the GS counter-terms, $a$, $b_{\kappa}$ and $b_{mn}$ are the anomaly 
coefficients. These transform as vectors of ${\rm SO}(1,T)$, and are 
determined by the underlying microscopic theory.
In our F-theory compactification these coefficients can be readily interpreted in terms of geometrical objects. We have 
\beq \label{eq:bmnSU5}
a=[K_B]\,,\qquad b_\kappa=\mathcal{S}^b_{G_\kappa}\,,\qquad b_{mn}= 
-\pi({\sigma}({\hat s}_n)\cdot {\sigma}({\hat s}_m))\, ,
\eeq
where $K_B$ is the canonical divisor of $B$, 
$\mathcal{S}_{G_\kappa}^b$ is the divisor on $B$ 
defined in \eqref{eq:defSb} supporting the non-Abelian group $G_\kappa$ 
and $\pi({\sigma}({\hat s}_n)\cdot {\sigma}({\hat s}_m))$ is 
the N\'eron-Tate height pairing defined in\eqref{eq:anomalycoeff}.
Under these identifications, the inner product in \eqref{eq:6dAnomalies} 
is replaced by the intersection pairing on the base $B$.

In addition, in the anomalies \eqref{eq:6dAnomalies}, we have made use 
of the following group theoretical relations between traces in different 
representations $\mathbf{R}$:
\beq \label{eq:groupfactors}
\text{tr}_{\mathbf{R}} F_\kappa ^2=A_{\mathbf{R}}\text{tr} F_\kappa ^2\,,\qquad \text{tr}_{\mathbf{R}} F_\kappa ^3=E_{\mathbf{R}}\text{tr} F_\kappa ^3\,,\qquad\text{tr}_{\mathbf{R}} F_\kappa ^4=B_{\mathbf{R}}\text{tr} F_\kappa ^4+C_{\mathbf{R}}(\text{tr} F_\kappa ^2)^2\,.
\eeq
Here `tr' denotes the trace with respect to the fundamental 
representation, while $\text{tr}_{\mathbf{R}}$ is the trace for a given  
representation ${\mathbf{R}}$. 
For $\kappa=\text{SU}(N)$ with $N>3$, the group theory factors in 
\eqref{eq:groupfactors} assume the following
values:\footnote{See \cite{Taylor:2011wt} for further details.}
\beq
\text{
\begin{tabular}{|c|c|c|c|c|c|} \hline
 Representation & Dimension & $A_{\mathbf{R}}$ & $B_{\mathbf{R}}$ & $C_{\mathbf{R}}$ & $E_{\mathbf{R}}$  \\ \hline
 Fundamental & $N$ &1 & 1& 0& 1 \rule{0pt}{1Em}\\ \hline
 Adjoint & $N^2-1$ &$2N$ & $2N$  & 6 & $N+4$%$2N$\rule{0pt}{1Em} 
 \\ \hline
 Antisymmetric& $N(N-1)/2$ & $N-2$ & $N-8$ & 3 & $N-4$%This I took from Erler
 \rule{0pt}{1Em}\\ \hline
\end{tabular}
}
\eeq
For the specific case of $\SU2$ and $\SU3$, the coefficients 
$A_{\mathbf{R}}$ coincide with those given in the table. In contrast to 
that, the coefficients $B_{\mathbf{R}}$ and $E_{\mathbf{R}}$ are equal 
to zero in both cases. The actual coefficient $C_{\mathbf{R}}$ can be 
computed using the values for $B_{\mathbf{R}}$ and  $C_{\mathbf{R}}$ in 
the above table, as $C_{\mathbf{R}}+\tfrac12 B_{\mathbf{R}}$, for 
$N=2,3$. Finally, the coefficient $\lambda_\kappa$ in 
(\ref{eq:6dAnomalies}) corresponds to the group normalization constant 
defined by $\lambda_\kappa=2c_{\kappa}/E_{\mathbf{adj}_\kappa}$, where 
$c_{\kappa}$ is the dual Coxeter number for the group $G_\kappa$ and 
$E_{\mathbf{adj}_\kappa}$ is $E_{\mathbf{R}}$ for the adjoint 
representation.  For $G_\kappa=\text{SU}(N)$, we have 
$\lambda_{\kappa}=1$.

%%%%%%%%%%%%%%%%%%%%%%%%%%%%%%%%%%%%%%%%%%%%%%%%%%%%%%%%%%%%%%%%%%%%%%%%%%%%%%%%%%%%%%%%%%%%%%%%%%%%%%
%%%%%%%%%%%%%%%%%%%%%%%%%%%%%%%%%%%%%%%%%%%%%%%%%%%%%%%%%%%%%%%%%%%%%%%%%%%%%%%%%%%%%%%%%%%%%%%%%%%%%%
\section{Additional Data on Toric Hypersurface Fibrations}
\label{app:AppPoly}
%%%%%%%%%%%%%%%%%%%%%%%%%%%%%%%%%%%%%%%%%%%%%%%%%%%%%%%%%%%%%%%%%%%%%%%%%%%%%%%%%%%%%%%%%%%%%%%%%%%%%%
%%%%%%%%%%%%%%%%%%%%%%%%%%%%%%%%%%%%%%%%%%%%%%%%%%%%%%%%%%%%%%%%%%%%%%%%%%%%%%%%%%%%%%%%%%%%%%%%%%%%%%

In this appendix we provide the explicit expressions for $f$ and $g$ of 
the WSF of the Jacobian fibrations of $X_{F_1}$, $X_{F_2}$ and 
$X_{F_4}$. Additionally, we present the explicit WS-coordinates of the 
rational sections of $X_{F_3}$, $X_{F_5}$ and $X_{F_7}$. The  
functions $f$, $g$ as well as the WS-coordinates of the rational 
sections of all other toric hypersurface fibrations $X_{F_i}$ can be 
obtained by specializing the ones presented here. 
Finally, we derive the Tate form, the WSF and the WS-coordinates of 
the generators of the MW-torsion of the toric hypersurface fibrations 
$X_{F_{13}}$, $X_{F_{15}}$ and $X_{F_{16}}$.

\subsection*{WSF of $J(X_{F_1})$}

Here we explicitly write out the polynomials $f$ and $g$ of the WSF of 
the Jacobian fibration $J(X_{F_1})$. The discriminant $\Delta$ is 
calculated straightforwardly from these quantities but is  omitted here 
due to the length of its explicit form. The functions $f$, $g$ in 
the WSF of \eqref{eq:pF1} read
\footnotesize
\begin{align}
\begin{split}
\label{eq:fcubic}
f&=\frac{1}{48} (-(s_6^2 - 4 (s_5 s_7 + s_3 s_8 + s_2 s_9))^2 + 
    24 (-s_6 (s_{10} s_2 s_3 - 9 s_1 s_{10} s_4 + s_4 s_5 s_8 \\%\nonumber\\
&\phantom{=}+ s_2 s_7 s_8 + s_3 s_5 s_9 +
          s_1 s_7 s_9) + 
      2 (s_{10} s_3^2 s_5 + s_1 s_7^2 s_8 + s_2 s_3 s_8 s_9 + s_1 s_3 s_9^2 \\
&\phantom{=}+  s_7 (s_{10} s_2^2 - 3 s_1 s_{10} s_3 + s_3 s_5 s_8 + s_2 s_5 s_9) + 
         s_4 (-3 s_{10} s_2 s_5 + s_2 s_8^2 + (s_5^2 - 3 s_1 s_8) s_9)))) %\nonumber
\end{split}
\end{align}
\begin{align}
\label{eq:gcubic}
\begin{split}
g&=\frac{1}{864} ((s_6^2 - 4 (s_5 s_7 + s_3 s_8 + s_2 s_9))^3 - 
   36 (s_6^2 - 4 (s_5 s_7 + s_3 s_8 + s_2 s_9)) \\%\times \nonumber\\
&\phantom{=}\times (-s_6 (s_{10} s_2 s_3 - 9 s_1 s_{10} s_4 + 
         s_4 s_5 s_8 + s_2 s_7 s_8 + s_3 s_5 s_9 + s_1 s_7 s_9) \\%+ \nonumber\\
&\phantom{=}+  2 (s_{10} s_3^2 s_5 + s_1 s_7^2 s_8 + s_2 s_3 s_8 s_9 + s_1 s_3 s_9^2 + 
         s_7 (s_{10} s_2^2 - 3 s_1 s_{10} s_3 + s_3 s_5 s_8 + s_2 s_5 s_9) \\%+\nonumber\\
&\phantom{=}+  s_4 (-3 s_{10} s_2 s_5 + s_2 s_8^2 + (s_5^2 - 3 s_1 s_8) s_9))) + 
   216 ((s_{10} s_2 s_3 - 9 s_1 s_{10} s_4 + s_4 s_5 s_8 \\%+\nonumber\\
&\phantom{=}+ s_2 s_7 s_8 + s_3 s_5 s_9 + 
        s_1 s_7 s_9)^2 + 4 (-s_1 s_{10}^2 s_3^3 - s_1^2 s_{10} s_7^3 - 
         s_4^2 (27 s_1^2 s_{10}^2 + s_{10} s_5^3 \\%+\nonumber\\
&\phantom{=}+ s_1 (-9 s_{10} s_5 s_8 + s_8^3)) + 
         s_{10} s_3^2 (-s_2 s_5 + s_1 s_6) s_9 - s_1 s_3^2 s_8 s_9^2 \\%- \nonumber\\
&\phantom{=}-  s_7^2 (s_{10} (s_2^2 s_5 - 2 s_1 s_3 s_5 - s_1 s_2 s_6) + 
            s_1 s_8 (s_3 s_8 + s_2 s_9)) \\%- \\
&\phantom{=}-  s_3 s_7 (s_{10} (-s_2 s_5 s_6 + s_1 s_6^2 + s_2^2 s_8 + 
               s_3 (s_5^2 - 2 s_1 s_8) + s_1 s_2 s_9) \\%+ \nonumber\\
&\phantom{=}+  s_9 (s_2 s_5 s_8 - s_1 s_6 s_8 + s_1 s_5 s_9)) + 
         s_4 (-s_{10}^2 (s_2^3 - 9 s_1 s_2 s_3) \\%+ \nonumber\\
&\phantom{=}+ s_{10} (s_6 (-s_2 s_5 s_6 + s_1 s_6^2 + s_2^2 s_8) + 
               s_3 (s_5^2 s_6 - s_2 s_5 s_8 - 3 s_1 s_6 s_8)) \\%+\nonumber\\
&\phantom{=}+ (s_{10} (2 s_2^2 s_5 + 3 s_1 s_3 s_5 - 
                  3 s_1 s_2 s_6) + 
               s_8 (-s_3 s_5^2 + s_2 s_5 s_6 - s_1 s_6^2 - s_2^2 s_8 + 
                  2 s_1 s_3 s_8)) s_9 \\%+\nonumber\\
&\phantom{=}+ (-s_2 s_5^2 + s_1 s_5 s_6 + 
               2 s_1 s_2 s_8) s_9^2 - s_1^2 s_9^3 + 
            s_7 (s_{10} (2 s_2 s_5^2 - 3 s_1 s_5 s_6 + 3 s_1 s_2 s_8 + 
                  9 s_1^2 s_9) \\%- \nonumber\\
&\phantom{=}- s_8 (s_2 s_5 s_8 - s_1 s_6 s_8 + s_1 s_5 s_9))))))%\nonumber
\end{split}
\end{align}
\normalsize
\subsection*{WSF of $J(X_{F_2})$ and the cubic form of the biquadric}

First, we present the explicit expressions for $f$ and $g$ in the WSF of 
the Jacobian fibration $J(X_{F_2})$, where we omit the expression of 
the discriminant $\Delta$. The functions $f$, $g$ in 
the WSF of \eqref{eq:biquadric-F2} read
\footnotesize
\begin{align} 
\begin{split}\label{eq:f-F2}
f&= \frac{1}{48} \left[ -(-4 b_1 b_{10} + b_6^2 - 4 (b_5 b_7 + b_3 b_8 + b_2 b_9))^2 + 
   24 (-b_6 (b_{10} b_2 b_5 + b_2 b_7 b_8 + b_3 b_5 b_9 + b_1 b_7 b_9) + 2 (b_{10} \right. \\%+ \nonumber \\ 
&\phantom{=} \left .\times (b_1 b_5 b_7 + b_2^2 b_8 + b_3 (b_5^2 - 4 b_1 b_8) + b_1 b_2 b_9) + 
         b_7 (b_1 b_7 b_8 + b_2 b_5 b_9) + 
         b_3 (b_5 b_7 b_8 + b_2 b_8 b_9 + b_1 b_9^2))) \right ]\,,
\end{split}
\end{align}
\normalsize
and
\footnotesize
\begin{align} 
\begin{split}\label{eq:g-F2}
g&= \frac{1}{864} \left[ (-4 b_1 b_{10} + b_6^2 - 4 (b_5 b_7 + b_3 b_8 + b_2 b_9))^3 - 36 (-4 b_1 b_{10} + b_6^2 - 
      4 (b_5 b_7 + b_3 b_8 + b_2 b_9)) \right . \\%\times \nonumber \\ 
  &\phantom{=} \times (-b_6 (b_{10} b_2 b_5 + b_2 b_7 b_8 + b_3 b_5 b_9 + b_1 b_7 b_9) + 
      2 (b_{10} (b_1 b_5 b_7 + b_2^2 b_8 + b_3 (b_5^2 - 4 b_1 b_8) + b_1 b_2 b_9) \\%+ \nonumber \\ 
  &\phantom{=} + b_7 (b_1 b_7 b_8 + b_2 b_5 b_9) + b_3 (b_5 b_7 b_8 + b_2 b_8 b_9 + b_1 b_9^2))) + 
   216 ((b_{10} b_2 b_5 + b_2 b_7 b_8 + b_3 b_5 b_9 + b_1 b_7 b_9)^2 \\%- \nonumber \\ 
  &\phantom{=} - 4 (b_2 b_3 b_5 b_7 b_8 b_9 + b_1^2 b_{10} (-4 b_{10} b_3 b_8 + b_7^2 b_8 + b_3 b_9^2) + 
         b_{10} (b_3^2 b_5^2 b_8 + b_2^2 b_5 b_7 b_8 \\%+ \nonumber \\ 
  &\phantom{=} + b_2 b_3 (-b_5 b_6 b_8 + b_2 b_8^2 + b_5^2 b_9)) + b_1 (b_{10}^2 (b_3 b_5^2 + b_2^2 b_8) + b_2 b_7^2 b_8 b_9 + b_3^2 b_8 b_9^2 + b_3 b_7 (b_7 b_8^2 - b_6 b_8 b_9 + b_5 b_9^2) \\%+ \nonumber \\ 
  &\phantom{=} \left . + b_{10} (-4 b_3^2 b_8^2 + b_3 b_6 (b_6 b_8 - b_5 b_9) + b_2 b_7 (-b_6 b_8 + b_5 b_9))))) \right ]\, .
\end{split}
\end{align}
\normalsize

Second, the explicit expressions of the $\tilde{s}_i$ in 
\eqref{eq:F2-as-cubic} obtained by mapping  $X_{F_2}$ to $X_{F_5}$ read
\footnotesize
\begin{align} 
\begin{split}\label{eq:trans-F5F2}
\tilde{s}_1 &= b_1\,, \\% \nonumber \\ 
\tilde{s}_2 &= \frac{1}{b_8}\left(b_2 b_8 - b_1 b_9 - b_1 \sqrt{-4 b_{10} b_8 + b_9^2}\right)\,,\\% \nonumber \\ 
\tilde{s}_3 &= \frac{1}{b_8^2} \left(-2 b_1 b_{10} b_8 + 2 b_3 b_8^2 - b_2 b_8 b_9 + b_1 b_9^2 - 
 b_2 b_8 \sqrt{-4 b_{10} b_8 + b_9^2} + b_1 b_9 \sqrt{-4 b_{10} b_8 + b_9^2}\right)\,,\\% \nonumber \\ 
\tilde{s}_5 &= b_5\,,\\%  \nonumber \\ 
\tilde{s}_6 &= \frac{1}{b_8}\left(b_6 b_8 - b_5 b_9 - b_5 \sqrt{-4 b_{10} b_8 + b_9^2}\right)\,,\\%  \nonumber \\ 
\tilde{s}_7 &= \frac{1}{b_8^2}\left(-2 b_{10} b_5 b_8 + 2 b_7 b_8^2 - b_6 b_8 b_9 + b_5 b_9^2 - 
 b_6 b_8 \sqrt{-4 b_{10} b_8 + b_9^2} + b_5 b_9 \sqrt{-4 b_{10} b_8 + b_9^2} \right)\,,\\%  \nonumber \\ 
\tilde{s}_8 &= b_8\,,\\%  \nonumber \\ 
\tilde{s}_9 &= -\sqrt{-4 b_{10} b_8 + b_9^2} \,.
\end{split}
\end{align}
\normalsize

\subsection*{WSF of $J(X_{F_4})$}

The explicit expressions for $f$ and $g$ in the WSF of 
the Jacobian fibration $J(X_{F_4})$ associated to $X_{F_4}$ with 
hypersurface equation \eqref{eq:quartic_hypersurface} read
\footnotesize
\begin{align}
\begin{split}
f_4&=\tfrac{1}{48} [-24 d_9 (-2 d_5 d_6^2 + d_4 d_6 d_7 - 2 d_3 d_6 d_8 + d_2 d_7 d_8 - 
      2 d_1 d_8^2 - 2 d_2 d_4 d_9 + 8 d_1 d_5 d_9) %\\%-\nonumber\\&\phantom{=}
      - (d_7^2 - 4 (d_6 d_8 + d_3 d_9))^2]\,,
\end{split}
\end{align}
\normalsize
and
\footnotesize
\begin{align}
\begin{split}
g_4&=\tfrac{1}{864} [36 d_9 (-2 d_5 d_6^2 + d_4 d_6 d_7 - 2 d_3 d_6 d_8 + d_2 d_7 d_8 - 
      2 d_1 d_8^2 - 2 d_2 d_4 d_9 + 8 d_1 d_5 d_9) %\\ \times \nonumber \\&\phantom{=}\times
      (d_7^2 - 4 (d_6 d_8 + d_3 d_9)) \\
&\phantom{=} + (d_7^2 - 4 (d_6 d_8 + d_3 d_9))^3 + 
   216 d_9^2 [4 d_2 d_5 d_6 d_7 - 4 d_1 d_5 d_7^2 %\nonumber
 + d_2^2 d_8^2 + d_4 (-2 d_2 d_6 d_8 + 4 d_1 d_7 d_8) \\
 &\phantom{=} - 4 d_2^2 d_5 d_9 + 
      d_4^2 (d_6^2 - 4 d_1 d_9) - 4 d_3 (d_5 d_6^2 + d_1 d_8^2 - 4 d_1 d_5 d_9)]]\,.
\end{split}
\end{align}
\normalsize

\subsection*{WS-coordinates of the non-toric section of $X_{F_3}$}
\label{app:F3}

As we have shown in Section~\ref{sec:F3_poly}, there is one additional 
rational section, besides the toric section $\hat{s}_0$, of the 
fibration of $X_{F_3}$. The section $\hat{s}_1$ has coordinates 
$[x_{1}:y_{1}:z_{1}]$ in the WSF that are given by
\footnotesize
\begin{align}
\begin{split}\label{eq:WScoordsSecF3}
x_1 &= \frac{1}{12} (12 s_1^2 s_9^6+4 (2 s_2 (s_5^2-3 s_1 s_8)-3 s_1 s_5 s_6) s_9^5+((s_6^2-4 s_5
   s_7) s_5^2+12 (s_2^2+2 s_1 s_3) s_8^2-4 (4 s_3 s_5^2 \\%+\nn\\&
  &\phantom{=} + s_2 s_6 s_5-3 s_1 (s_6^2+2 s_5 s_7)) s_8) s_9^4-2 s_8 (-4 (s_6 s_7+3 s_4 s_8) s_5^2+(s_6^3-10 s_3 s_8 s_6+4 s_2 s_7
   s_8) s_5 \\%\nn\\&
  &\phantom{=} + 2 s_8 (9 s_1 s_6 s_7+6 s_1 s_4 s_8+s_2 (s_6^2+6 s_3 s_8))) s_9^3+s_8^2 (s_6^4-2 s_5 s_7 s_6^2-8 s_5^2 s_7^2+12 (s_3^2+2 s_2 s_4) s_8^2 \\%\nn\\&
  &\phantom{=} - 4 (9 s_4 s_5 s_6-s_7 (5 s_2 s_6+6 s_1 s_7)+s_3 (s_6^2+2 s_5 s_7)) s_8) s_9^2-2 s_8^3 (12 s_3 s_4 s_8^2+2 (s_7 (s_3 s_6+4 s_2 s_7) \\%\nn\\&
  &\phantom{=} - 3 s_4 (s_6^2+2 s_5 s_7)) s_8+s_6 s_7 (s_6^2-4 s_5 s_7)) s_9+s_8^4 ((s_6^2-4 s_5 s_7) s_7^2+4 (2 s_3 s_7-3 s_4 s_6) s_8 s_7+12 s_4^2 s_8^2))\,, \\% \nn\\
y_1 &=\frac{1}{2} (2 s_1^3 s_9^9+s_1 (2 s_2 (s_5^2-3 s_1 s_8)-3 s_1 s_5 s_6) s_9^8+((s_3 s_5^2-s_2 s_6 s_5+s_1 (s_6^2-s_5 s_7)) s_5^2+6 s_1 (s_2^2  \\%\nn\\&
  &\phantom{=} + s_1 s_3) s_8^2 + (-2 s_2^2 s_5^2+2 s_1 s_2 s_6 s_5+s_1 (3 s_1 (s_6^2+2 s_5 s_7)-4 s_3 s_5^2)) s_8) s_9^7-s_8 (2 (s_2^3+6 s_1 s_3 s_2  \\%\nn\\&
  &\phantom{=} + 3 s_1^2 s_4) s_8^2 - (s_5 s_6 s_2^2+(6 s_3 s_5^2-4 s_1 (s_6^2+2 s_5 s_7)) s_2+s_1 (6 s_4 s_5^2+2 s_3 s_6 s_5-9 s_1 s_6 s_7)) s_8+s_5 (3 s_4 s_5^3 \\%\nn\\&
  &\phantom{=} + 2 s_3 s_6 s_5^2-3 s_2 s_7 s_5^2-2 s_2 s_6^2 s_5+s_1 s_6 s_7 s_5+2 s_1 s_6^3)) s_9^6+s_8^2 (s_1 s_6^4-s_2 s_5 s_6^3+s_3 s_5^2 s_6^2+7 s_1 s_5 s_7 s_6^2 \\%\nn\\&
  &\phantom{=} + 9 s_4 s_5^3 s_6-8 s_2 s_5^2 s_7 s_6+s_1 s_5^2 s_7^2+6 (s_3 (s_2^2+s_1 s_3)+2 s_1 s_2 s_4) s_8^2-s_3 s_5^3 s_7+(-4 s_3^2 s_5^2-8 s_2 s_4 s_5^2 \\%\nn\\&
  &\phantom{=} - 6 s_1 s_4 s_6 s_5+s_2^2 s_6^2+6 s_1^2 s_7^2+2 s_2 (s_2 s_5+7 s_1 s_6) s_7+s_3 (2 s_1 (s_6^2+2 s_5 s_7)-6 s_2 s_5 s_6)) s_8) s_9^5 \\%\nn\\&
  &\phantom{=} - s_8^3 (s_8 (6 s_2 s_8-5 s_5 s_6) s_3^2-5 s_6 s_7 (s_5^2-2 s_1 s_8) s_3+5 s_7 (s_6 s_8 s_2^2-s_5 (s_6^2+s_5 s_7) s_2+2 s_1 s_7 s_8 s_2 \\%\nn\\&
  &\phantom{=} + s_1 s_6 (s_6^2+2 s_5 s_7))+s_4 (5 (2 s_6^2+s_5 s_7) s_5^2-10 (s_3 s_5+s_2 s_6) s_8 s_5+6 (s_2^2+2 s_1 s_3) s_8^2)) s_9^4 \\%\nn\\&
  &\phantom{=} + s_8^4 (2 (s_3^3+6 s_2 s_4 s_3+3 s_1 s_4^2) s_8^2-(6 s_4^2 s_5^2+s_3^2 s_6^2-4 (s_2^2+2 s_1 s_3) s_7^2+2 s_3 (s_3 s_5-3 s_2 s_6) s_7 \\%\nn\\&
   &\phantom{=} + 2 s_4 (s_2 s_6^2+7 s_3 s_5 s_6-3 s_1 s_7 s_6+2 s_2 s_5 s_7)) s_8+5 (s_4 s_5 s_6 (s_6^2+2 s_5 s_7)+s_7 (s_7 (2 s_1 s_6^2-s_2 s_5 s_6+s_1 s_5 s_7) \\%\nn\\&
  &\phantom{=} - s_3 s_5 (s_6^2+s_5 s_7)))) s_9^3-s_8^5 (3 s_8 (2 s_2 s_8-3 s_5 s_6) s_4^2+(s_6^4+(7 s_5 s_7-4 s_3 s_8) s_6^2+2 s_2 s_7 s_8 s_6+s_5^2 s_7^2-8 s_3 s_5 s_7 s_8 \\%\nn\\&
  &\phantom{=} + 6 s_8 (s_8 s_3^2+s_1 s_7^2)) s_4+s_7 (s_6 s_8 s_3^2-(s_6^3+8 s_5 s_7 s_6-6 s_2 s_7 s_8) s_3+s_7 (9 s_1 s_6 s_7+s_2 (s_6^2-s_5 s_7)))) s_9^2 \\%\nn\\&
  &\phantom{=} + s_8^6 (3 s_8 (-s_6^2-2 s_5 s_7+2 s_3 s_8) s_4^2+s_7 (2 s_6^3+s_5 s_7 s_6-2 s_3 s_8 s_6+4 s_2 s_7 s_8) s_4+s_7^2 (2 s_8 s_3^2-2 s_6^2 s_3 \\%-\nn\\&
  &\phantom{=} - 3 s_5 s_7 s_3+3 s_1 s_7^2+2 s_2 s_6 s_7)) s_9+s_8^7 (-2 s_8^2 s_4^3+3 s_6 s_7 s_8 s_4^2+s_7^2 (-s_6^2+s_5 s_7-2 s_3 s_8) s_4+s_7^3 (s_3 s_6-s_2 s_7)))\,, \\%\nn \\
z_1 &= s_7 s_8^2+s_9 (s_5 s_9-s_6 s_8)\, .
\end{split}
\end{align}
\normalsize
\subsection*{WS-coordinates of the two rational sections of $X_{F_5}$}

In addition to $\hat{s}_0$, there are two rational sections of the 
fibration of $X_{F_5}$. The WS-coordinates have been worked out first in 
\cite{Borchmann:2013jwa,Cvetic:2013nia}. We reproduce these results here
for convenience.

The section $\hat{s}_1$ has coordinates $[x_1:y_1:z_1]$ in the WSF given 
by
\begin{align}
\begin{split}\label{eq:secsF51}
x_1 &=  \tfrac{1}{12} (s_6^2 - 4 s_5 s_7 + 8 s_3 s_8 - 4 s_2 s_9)\,, \\ 
y_1 &= \tfrac{1}{2} (s_3 s_6 s_8 - s_2 s_7 s_8 - s_3 s_5 s_9 + s_1 s_7 s_9)\,,  \\ 
z_1 &=  1 \,.
\end{split}
\end{align}
The section $\hat{s}_2$ has coordinates $[x_2:y_2:z_2]$ in the WSF given by
\begin{align}
\begin{split}\label{eq:secsF52}
  x_2 &=  \tfrac{1}{12} (12 s_7^2 s_8^2 + s_9^2 (s_6^2 + 8 s_3 s_8 - 4 s_2 s_9) + 
    4 s_7 s_9 (-3 s_6 s_8 + 2 s_5 s_9))\,, \\ 
  y_2&= \tfrac{1}{2} (2 s_7^3 s_8^3 + s_3 s_9^3 (-s_6 s_8 + s_5 s_9) + 
    s_7^2 s_8 s_9 (-3 s_6 s_8 + 2 s_5 s_9)  \\ 
  &\phantom{=}+s_7 s_9^2 (s_6^2 s_8 + 2 s_3 s_8^2 - s_5 s_6 s_9 - s_2 s_8 s_9 + s_1 s_9^2)\,, \\ 
  z_2&= s_9  \,.
\end{split}
\end{align}
\subsection*{WS-coordinates of the three rational sections of $X_{F_7}$}

There are three rational sections of the fibration of $X_{F_7}$ besides 
$\hat{s}_0$ \cite{Braun:2013nqa,Cvetic:2013qsa}. The following
results have been obtained using the birational map in 
\cite{Cvetic:2013nia} in the special case $s_1=0$. 

The coordinates of 
$\hat{s}_1$ in the WSF, denoted by $[x_1:y_1:z_1]$, are
\begin{align}
\begin{split}\label{eq:secsF71}
x_1 &=  \frac{1}{12}  (s_6^2 s_7^2 - 4 s_5 s_7^3 + 8 s_3 s_7^2 s_8 - 12 s_3 s_6 s_7 s_9 + 
   8 s_2 s_7^2 s_9 + 12 s_3^2 s_9^2)\,, \\ 
y_1 &= \frac{1}{2} (-2 s_3^3 s_9^3 + s_2 s_7^3 (-s_7 s_8 + s_6 s_9) + 
  s_3^2 s_7 s_9 (-2 s_7 s_8 + 3 s_6 s_9) \\
&\phantom{=} + s_3 s_7^2 (s_6 s_7 s_8 - s_6^2 s_9 + s_9 (s_5 s_7 - 2 s_2 s_9)))\,,  \\ 
z_1 &=  s_7 \,.
\end{split}
\end{align}
Similarly, the coordinates of $\hat{s}_2$ in WSF, denoted by 
$[x_2:y_2:z_2]$, read
\begin{align}
\begin{split}\label{eq:secsF72}
x_2 &=  \frac{1}{12}(s_6^2 - 4 s_5 s_7 - 4 s_3 s_8 + 8 s_2 s_9)\,, \\ 
y_2 &= \frac{1}{2}(s_2 s_7 s_8 + s_3 s_5 s_9 - s_2 s_6 s_9)\,,  \\ 
z_2 &=  1 \,.
\end{split}
\end{align}
Finally, the coordinates of $\hat{s}_3$ in the WSF, that we denote by
$[x_3:y_3:z_3]$, are
\begin{align}
\begin{split}\label{eq:secsF73}
  x_3 &=  \frac{1}{12}(12 s_7^2 s_8^2 + s_9^2 (s_6^2 + 8 s_3 s_8 - 4 s_2 s_9) + 
    4 s_7 s_9 (-3 s_6 s_8 + 2 s_5 s_9))\,, \\ 
  y_3&= \frac{1}{2}(-2 s_7^3 s_8^3 + s_7^2 s_8 s_9 (3 s_6 s_8 - 2 s_5 s_9) + 
    s_3 s_9^3 (s_6 s_8 - s_5 s_9) \\
  &\phantom{=} + 
    s_7 s_9^2 (-s_6^2 s_8 + s_5 s_6 s_9 + s_8 (-2 s_3 s_8 + s_2 s_9))) \\ 
  z_3 &= s_9  \,.
\end{split}
\end{align}
% 

%%%%%%%%%%%%%%%%%%%%%%%%%%%%%%%%%%%%%%%%%%%%%%%%%%%%%%%%%%%%%%%%%%%%%%%%%%%%%%%%%%%%%%%%%%%%%%%%%
\subsection*{Tate form, WSF and the MW-torsion of $X_{F_{13}}$}
%%%%%%%%%%%%%%%%%%%%%%%%%%%%%%%%%%%%%%%%%%%%%%%%%%%%%%%%%%%%%%%%%%%%%%%%%%%%%%%%%%%%%%%%%%%%%%%%%

In this appendix we determine two Tate forms and the WSF of
$X_{F_{13}}$. We explicitly derive the WS-coordinates of its
torsional section and use this to show, that the MW-torsion
acts on all codimension one singularities in $X_{F_{13}}$, i.e.~on
all non-Abelian gauge group factors in $G_{F_{13}}$.

We directly employ
the birational map in \cite{Cvetic:2013nia} for $s_5= s_7= s_8=0$
to obtain a Tate form for the hypersurface constraint 
\eqref{eq:pF13} of $X_{F_{13}}$. The Tate coefficients we naively obtain 
read
\beq \label{eq:TateF13sing}
	a_1= \frac{s_6^2 + 2 s_2 s_9}{s_6}\,,\quad  a_2= -\frac{s_2 s_9 (-s_6^2 + s_2 s_9)}{s_6^2}\,,\quad a_3 = s_2 s_6 s_9\,,\quad a_4 = s_1 s_3 s_9^2\,,\quad a_6 = s_1 s_2 s_3 s_9^3\,.
\eeq
Due to the poles at $s_6=0$, this Tate model is clearly globally ill-defined. However, there
exists an equivalent Tate model with WSF identical to the one of
\eqref{eq:TateF13sing}. It reads
\beq \label{eq:TateFormZ2MW}
  y^2+s_6 xy z=x \left(x^2+s_1 s_9^2 s_3 z^4-s_2 s_9 x z^2\right)\,.
\eeq
This is precisely of the form of an elliptic curve with
$\mathbb{Z}_2$ MW-torsion given in
\cite{Aspinwall:1998xj}, after the
shift $y\rightarrow y-\frac12 s_6 x z$. 
Thus, the MW-group is indeed $\mathbb{Z}_2$, in agreement with the 
results of  \cite{Braun:2013nqa,Mayrhofer:2014opa}.

The following analysis is presented in the patch $z=1$ without loss of 
generality. The Weierstrass equation of \eqref{eq:TateFormZ2MW} takes the form
\beq \label{eq:WSeqX13}
y^2=\frac{1}{864} (s_6^2 - 4 s_2 s_9 - 12 x) (s_6^4 - 8 s_2 s_6^2 s_9 + 16 s_2^2 s_9^2 - 
   72 s_1 s_3 s_9^2 - 6 s_6^2 x + 24 s_2 s_9 x - 72 x^2)\,.
\eeq
The coordinates of the section of order two in \eqref{eq:TateFormZ2MW} 
are 
\beq
\label{eq:WScoordsSecF13}
\{x=\tfrac{1}{12}(s_6^2-4s_2 s_9),\,y=0\}\,.
\eeq
At the loci of codimension one singularities, $s_1=0$ and $s_3=0$, the 
WSF takes the form
\beq
y^2=\tfrac{1}{864} (s_6^2 - 4 s_2 s_9 - 12 x)^2 (s_6^2 - 4 s_2 s_9 + 6 x)\,,
\eeq
and on the locus $s_9=0$ it reads
\beq
y^2=\tfrac{1}{864} (s_6^2 - 12 x)^2 (s_6^2 + 6 x)\,,
\eeq
showing that the $A_4$- and both $A_2$-singularities at  codimension one
in the fibration are located exactly at the point of order two. This 
implies that the $\mathbb{Z}_2$ associated to the torsional section acts 
on all non-Abelian gauge group factors,  rendering the gauge group 
\beq
G_{F_{13}} = (\text{SU}(4)\times \text{SU}(2)^2)/\mathbb{Z}_2\,.
\eeq

\subsection*{Tate form, WSF and the MW-torsion of $X_{F_{15}}$}

Here, we determine two Tate forms and the WSF of
$X_{F_{15}}$. The explicit WS-coordinates of its
torsional section allow us to show that the MW-torsion
acts on all codimension one singularities in $X_{F_{15}}$, i.e.~on
all non-Abelian gauge group factors in $G_{F_{15}}$.

We apply the birational map in \cite{Cvetic:2013nia} for 
$s_1\!\!=\!\!s_3\!\!=\!\!s_8\!\!=\!0$ to \eqref{eq:pF15} to obtain the Tate 
coefficients
\beq \label{eq:TateF15sing}
	a_1=\frac{s_6^2 - 2 s_5 s_7 + 2 s_2 s_9}{s_6}\,,\quad a_2=\frac{s_2 s_6^2 s_9 - (s_5 s_7 - s_2 s_9)^2}{s_6^2}\,,\quad  a_3=s_2s_6s_9\,,\quad a_4=a_6=0\,.
\eeq
Clearly this Tate form has poles at $s_6=0$ and is ill-defined. 
Fortunately, there exists an equivalent Tate model that has 
the same Weierstrass equation as \eqref{eq:TateF15sing}. It reads
\beq \label{eq:TorsionF15}
y^2+s_6xyz=x(x^2-(s_7 s_5 + s_2 s_9)x z^2+s_2s_7 s_5  s_9z^4)\,.
\eeq
Now we see that this elliptic curve is precisely of the form of the 
elliptic curve with a $\mathbb{Z}_2\oplus \mathbb{Z}$ MW-group studied
in \cite{Aspinwall:1998xj}. This result agrees with
the findings in  \cite{Braun:2013nqa,Mayrhofer:2014opa}. 
We note that the Tate coefficients \eqref{eq:TateF15sing} can
be parametrized as $a_1=\gamma_1$, $a_2=-(\gamma_2+\delta_2)$, $a_4=\gamma_2\delta_2$ according to \cite{Mayrhofer:2014opa}, where
we obtain $\gamma_1=s_6$, $\gamma_2=s_3 s_8$ and $\delta_2=s_2 s_9$.

Let us work in the patch $z=1$  without loss of 
generality. The WSF of \eqref{eq:TateF15sing}, \eqref{eq:TorsionF15} reads 
\bea 
y^2&=&\tfrac{1}{864} (s_6^2 - 4 s_5 s_7 - 4 s_2 s_9 - 12 x) (s_6^4 - 8 s_5 s_6^2 s_7 + 
   16 s_5^2 s_7^2 - 8 s_2 s_6^2 s_9 \nn\\&& -40 s_2 s_5 s_7 s_9 + 16 s_2^2 s_9^2 - 
   6 s_6^2 x + 24 s_5 s_7 x + 24 s_2 s_9 x - 72 x^2)\,,\label{eq:WSFF15}
\eea
in which the coordinates of the point of order two are
\beq 
\label{eq:WScoordsSecF15}
 \{x=\tfrac{1}{12}(s_6^2 - 4 s_5 s_7 - 4 s_2 s_9 ) ,\, y=0 \}\,.
\eeq
At $s_2=0$ and $s_9=0$, the location of two SU(2) singularities, the WSF 
simplifies to
\beq
y^2= \tfrac{1}{864} (s_6^2 - 4 s_5 s_7 - 12 x)^2 (s_6^2 - 4 s_5 s_7 + 6 x)\,,
\eeq
and at $s_5=0$ and $s_7=0$, the location of the two other SU(2)'s, it reads
\beq
y^2= \tfrac{1}{864} (s_6^2 - 4 s_2 s_9 - 12 x)^2 (s_6^2 - 4 s_2 s_9 + 6 x)\,.
\eeq
Thus,  the section of order two goes through all $A_2$-singularities, 
which implies that the $\mathbb{Z}_2$ acts on all non-Abelian gauge 
group factors, i.e.~
\beq
G_{F_{15}} = (\text{SU}(2)^4)/\mathbb{Z}_2\times \text{U}(1)\,.
\eeq

\subsection*{Tate form, WSF and the MW-torsion of $X_{F_{16}}$}

Here, we compute a Tate form and the WSF of
$X_{F_{16}}$. Using the explicit WS-coordinates of the
$\mathbb{Z}_3$ MW-generator, we show that the MW-torsion
acts on the entire gauge group $G_{F_{16}}$.

The Tate form of the hypersurface equation \eqref{eq:pF16}
is obtained employing the birational map of \cite{Cvetic:2013nia} for 
$s_2= s_3= s_5= s_8= 0$:
\beq \label{eq:TateFormZ3MW}
  y^2+s_6 x y z-s_1 s_7 s_9 z^3=x^3\,.
\eeq
This is the normal form of an elliptic
curve with $\mathbb{Z}_3$ torsion  \cite{Aspinwall:1998xj}, in 
agreement with \cite{Braun:2013nqa,Mayrhofer:2014opa}. 

We work in the patch $z=1$ in the following, without loss of 
generality. The WSF reads
\beq
\label{eq:WSFF16}
y^2-\left(\tfrac{s_1 s_7 s_9}{2}\right)^2=\tfrac{1}{864} (s_6^2 - 12 x) (s_6^4 + 36 s_1 s_6 s_7 s_9 - 6 s_6^2 x - 72 x^2)\,,
\eeq
from which we obtain the WS-coordinates of the order three section as
\beq \label{eq:WScoordsSecF16}
\{x=\tfrac{1}{12}s_6^2,\,y=\tfrac{s_1 s_7 s_9}{2}\}\,.
\eeq 
This section passes through the $A_3$ singularities as we see from 
\eqref{eq:WSFF16} at $s_1=0$, $s_7=0$, $s_9=0$:
\beq
y^2=\tfrac{1}{864} (s_6^2 - 12 x)^2 (s_6^2 + 6 x)\,.
\eeq
Thus, the $\mathbb{Z}_3$ acts on all SU(3)'s and the gauge group is
\beq
G_{F_{16}} = \text{SU}(3)^3/\mathbb{Z}_3\,.
\eeq

%%%%%%%%%%%%%%%%%%%%%%%%%%%%%%%%%%%%%%%%%%%%%%%%%%%%%%%%%%%%%%%%%%%%%%%%%%%%%%%%%%%%%%%%%%%%%%%%%%%%%%
%%%%%%%%%%%%%%%%%%%%%%%%%%%%%%%%%%%%%%%%%%%%%%%%%%%%%%%%%%%%%%%%%%%%%%%%%%%%%%%%%%%%%%%%%%%%%%%%%%%%%%
\section{Euler Numbers of the Calabi-Yau Threefolds $X_{F_i}$}
\label{app:Eulernumbers}
%%%%%%%%%%%%%%%%%%%%%%%%%%%%%%%%%%%%%%%%%%%%%%%%%%%%%%%%%%%%%%%%%%%%%%%%%%%%%%%%%%%%%%%%%%%%%%%%%%%%%%
%%%%%%%%%%%%%%%%%%%%%%%%%%%%%%%%%%%%%%%%%%%%%%%%%%%%%%%%%%%%%%%%%%%%%%%%%%%%%%%%%%%%%%%%%%%%%%%%%%%%%%

In this section we present the explicit expressions for the Euler 
numbers of all Calabi-Yau threefolds $X_{F_i}$ that are constructed as 
toric hypersurface fibrations over an arbitrary two-fold base $B$ with 
their fibrations parametrized by two divisors $\cS_7$ and $\cS_9$, see 
Section \ref{sec:BasicIngredients}. 

The Euler numbers are computed using the presentation of the vertical 
cohomology ring of $X_{F_i}$ as a quotient ring in its divisors, see 
\cite{Mayr:1996sh,Klemm:1996hh}, and the adjunction formula.
For a detailed explanation in an F-theory context and  many 
explicit examples, we refer the reader to \cite{Cvetic:2013uta}.

The following table contains our results for the Euler numbers of 
$X_{F_i}$, where we denote the first Chern class of the base $B$ 
by $c_1$, implicitly invoke Poincar\'{e} duality between divisors and 
forms and suppress the integral over $B$:
\beq \label{eq:EulerNumbers}
\text{
\begin{tabular}{|c|c|} \hline
Manifold & Euler number $\chi(X_{F_i})$\\ \hline
$X_{F_1}$ & $-6 (4 c_1^2 - c_1 \cS_7 + \cS_7^2 - c_1 \cS_9 - \cS_7 \cS_9 + \cS_9^2)$ \\
$X_{F_2}$ & $-4 (6 c_1^2 - 2 c_1 \cS_7 + \cS_7^2 - 2 c_1 \cS_9 + \cS_9^2)$ \\
$X_{F_3}$ & $-2 (12 c_1^2 - 3 c_1 \cS_7 + 3 \cS_7^2 - 4 c_1 \cS_9 - 2 \cS_7 \cS_9 + 2 \cS_9^2)$ \\
$X_{F_4}$ & $-4 (6 c_1^2 - 2 c_1 \cS_7 + 3 \cS_7^2 - 2 c_1 \cS_9 - 2 \cS_7 \cS_9 + \cS_9^2)$ \\
$X_{F_5}$ & $-2 (12 c_1^2 -4 c_1 \cS_7 + 2 \cS_7^2 -4 c_1 \cS_9 - \cS_7 \cS_9 + 2\cS_9^2)$ \\
$X_{F_6}$ & $-2 (12 c_1^2 - 4 c_1 \cS_7 + 4 \cS_7^2 - 4 c_1 \cS_9 - 3 \cS_7 \cS_9 + 2 \cS_9^2)$ \\
$X_{F_7}$ & $-4 (4 c_1^2 + \cS_7^2 - \cS_7 \cS_9 + \cS_9^2 - c_1 (\cS_7 + \cS_9))$ \\
$X_{F_8}$ & $-2 (12 c_1^2 - 5 c_1 \cS_7 + 3 \cS_7^2 - 4 c_1 \cS_9 - 2 \cS_7 \cS_9 + 2 \cS_9^2)$ \\
$X_{F_9}$ & $-4 (6 c_1^2 - 2 c_1 \cS_7 + \cS_7^2 - 3 c_1 \cS_9 + \cS_9^2)$ \\
$X_{F_{10}}$ & $-6 (4 c_1^2 - 2 c_1 \cS_7 + 2 \cS_7^2 - c_1 \cS_9 - 2 \cS_7 \cS_9 + \cS_9^2)$ \\
$X_{F_{11}}$ & $-2 (12 c_1^2 - 4 c_1 \cS_7 + 2 \cS_7^2 - 7 c_1 \cS_9 - \cS_7 \cS_9 + 3 \cS_9^2)$ \\
$X_{F_{12}}$ & $-2 (12 c_1^2 - 6 c_1 \cS_7 + 2 \cS_7^2 - 6 c_1 \cS_9 + \cS_7 \cS_9 + 2 \cS_9^2)$ \\
$X_{F_{13}}$ & $-4 (6 c_1^2 - 2 c_1 \cS_7 + \cS_7^2 - 5 c_1 \cS_9 + 2 \cS_9^2)$ \\
$X_{F_{14}}$ & $-2 (12 c_1^2 - 9 c_1 \cS_7 + 3 \cS_7^2 - 6 c_1 \cS_9 + 2 \cS_7 \cS_9 + 2 \cS_9^2)$ \\
$X_{F_{15}}$ & $-4 (4 c_1^2 - 2 c_1 \cS_7 + \cS_7^2 - 2 c_1 \cS_9 + \cS_9^2)$ \\
$X_{F_{16}}$ & $-6 (4 c_1^2 - 3 c_1 \cS_7 + \cS_7^2 - 3 c_1 \cS_9 + \cS_7 \cS_9 + \cS_9^2)$ \\\hline
\end{tabular}
}
\eeq

%%%%%%%%%%%%%%%%%%%%%%%%%%%%%%%%%%%%%%%%%%%%%%%%%%%%%%%%%%%%%%%%%%%%%%%%%%%%%%%%%%%%%%%%%%%%%%%%%%%%%%
%%%%%%%%%%%%%%%%%%%%%%%%%%%%%%%%%%%%%%%%%%%%%%%%%%%%%%%%%%%%%%%%%%%%%%%%%%%%%%%%%%%%%%%%%%%%%%%%%%%%%%
\section{The Full Higgs Chain of Toric Hypersurface Fibrations}
\label{app:higgschain}
%%%%%%%%%%%%%%%%%%%%%%%%%%%%%%%%%%%%%%%%%%%%%%%%%%%%%%%%%%%%%%%%%%%%%%%%%%%%%%%%%%%%%%%%%%%%%%%%%%%%%%
%%%%%%%%%%%%%%%%%%%%%%%%%%%%%%%%%%%%%%%%%%%%%%%%%%%%%%%%%%%%%%%%%%%%%%%%%%%%%%%%%%%%%%%%%%%%%%%%%%%%%%

The complete Higgs chain of all toric Higgs transition 
$X_{F_i}\rightarrow X_{j}$, the  relevant
hyper multiplets acquiring VEVs, the U(1)-generators and the necessary 
redefinitions of divisors classes are shown in Tables 
\ref{tab:HiggsChain1}, \ref{tab:HiggsChain2} and \ref{tab:HiggsChain3}.
\begin{table}[H]
\begin{center}
\footnotesize
\renewcommand{\arraystretch}{1.2}
\begin{tabular}{|c|c|c|c|}\hline
Higgs transition & VEV & U(1) Generators & Divisor Class Matching \\ \hline
\multirow{5}{*}{$X_{F_{16}} \rightarrow X_{F_{14}}$} & $(\three,\one,\bar{\three})$ &  $Q=-(T_1^8+T_3^8)$ & trivial \\ \cline{2-4}
&  \multirow{2}{*}{$(\one,\three,\bar{\three})$} & \multirow{2}{*}{$Q=-(T_2^8+T_3^8)$} & $\cS_7\rightarrow \cS_9$ \\%\parbox[c]{\widthof{$\cS_9\rightarrow 3[K_B^{-1}]-\cS_7-\cS_9$}}{\center $\cS_7\rightarrow \cS_9$ \\$\cS_9\rightarrow 3[K_B^{-1}]-\cS_7-\cS_9$} 
&&&$\cS_9\rightarrow 3[K_B^{-1}]-\cS_7-\cS_9$\\ \cline{2-4} 
&  \multirow{2}{*}{$(\three,\bar{\three},\one)$} & \multirow{2}{*}{$Q=-(T_1^8+T_2^8)$} & $\cS_7\rightarrow \cS_9$ \\%\parbox[c]{\widthof{$\cS_9\rightarrow \cS_7$}}{\center $\cS_7\rightarrow \cS_9$ \\$\cS_9\rightarrow \cS_7$} 
&&&$\cS_9\rightarrow \cS_7$\\ \hline
\multirow{8}{*}{$X_{F_{15}} \rightarrow X_{F_{12}}$} & \multirow{2}{*}{$(\two,\two,\one,\one)_{(1/2)}$} & $Q_1^\prime=T^3_2+Q$ & \multirow{2}{*}{trivial} \\ 
&&$Q_2^\prime=T^3_1+Q$ &  \\ \cline{2-4}
&\multirow{2}{*}{$(\two,\one,\two,\one)_{(1/2)}$} & $Q_1^\prime=T_3^3+Q$ & $\cS_7\rightarrow 2[K_B^{-1}]-\cS_7$ \\%\parbox[c]{\widthof{$\cS_7\rightarrow 2[K_B^{-1}]-\cS_7$}}{\center $\cS_7\rightarrow 2[K_B^{-1}]-\cS_7$\\$\cS_9\rightarrow \cS_7$} \\
&&$Q_2^\prime=T^3_1+Q$& $\cS_9\rightarrow \cS_7$\\ \cline{2-4}
&\multirow{2}{*}{$(\one,\one,\two,\two)_{(1/2)}$} & $Q_1^\prime=T^3_4+Q$ & $\cS_7\rightarrow 2[K_B^{-1}]-\cS_9$ \\%\parbox[c]{\widthof{$\cS_9\rightarrow 2[K_B^{-1}]-\cS_7$}}{\center $\cS_7\rightarrow 2[K_B^{-1}]-\cS_9$\\$\cS_9\rightarrow 2[K_B^{-1}]-\cS_7$} \\
&&$Q_2^\prime=T_3^3+Q$& $\cS_9\rightarrow 2[K_B^{-1}]-\cS_7$\\ \cline{2-4}
&\multirow{2}{*}{$(\one,\two,\one,\two)_{(1/2)}$} & $Q_1^\prime=T^3_4+Q$ & $\cS_7\rightarrow \cS_9$ \\%\parbox[c]{\widthof{$\cS_9\rightarrow 2[K_B^{-1}]-\cS_7$}}{\center $\cS_7\rightarrow \cS_9$ \\$\cS_9\rightarrow 2[K_B^{-1}]-\cS_7$} \\
&&$Q_2^\prime=T^3_2+Q$& $\cS_9\rightarrow 2[K_B^{-1}]-\cS_7$\\ \hline
% &\multirow{2}{*}{$(\two,\one,\one,\two)_{(0)}$} & to do & no toric interpretation \\
% &&& \\ \cline{2-4}
% &\multirow{2}{*}{$(\one,\two,\two,\one)_{(0)}$} & to do & no toric interpretation \\
% &&& \\ \hline
\multirow{4}{*}{$X_{F_{14}} \rightarrow X_{F_{12}}$} & \multirow{2}{*}{$(\two,\three,\one)_{1/6}$} & $Q_1^\prime =T^8+2Q$ & \multirow{2}{*}{trivial} \\ 
& & $Q_2^\prime=2T^8-T^3+Q$ & \\ \cline{2-4}
& \multirow{2}{*}{$(\one,\three,\two)_{1/6}$} & $Q_1^\prime=2T^8-T^3+Q$ & $\cS_7 \rightarrow \cS_9$\\%\parbox[c]{\widthof{$\cS_9\rightarrow 3[K_B^{-1}]-\cS_7-\cS_9$}}{\center $\cS_7 \rightarrow \cS_9$ \\ $\cS_9\rightarrow 3[K_B^{-1}]-\cS_7-\cS_9$} \\ 
&&$Q_1^\prime =T^8+2Q$& $\cS_9\rightarrow 3[K_B^{-1}]-\cS_7-\cS_9$\\ \hline
\multirow{4}{*}{$X_{F_{14}} \rightarrow X_{F_{11}}$}  & \multirow{2}{*}{$(\one,\one,\two)_{1/2}$} & \multirow{2}{*}{$Q^\prime=Q_1-T^3$} &  $\cS_7\rightarrow \cS_9$ \\
& & & $\cS_9\rightarrow 2[K_B^{-1}]-\cS_7$\\ \cline{2-4}%\parbox[c]{\widthof{$\cS_9\rightarrow 2[K_B^{-1}]-\cS_7$}}{\center $\cS_7\rightarrow \cS_9$ \\ $\cS_9\rightarrow 2[K_B^{-1}]-\cS_7$}\\ \cline{2-4}
& \multirow{2}{*}{$(\two,\one,\one)_{1/2}$} & \multirow{2}{*}{$Q^\prime=Q-T^3$} & $\cS_7\rightarrow \cS_9$\\%\parbox[c]{\widthof{$\cS_9\rightarrow 2[K_B^{-1}]-\cS_7$}}{\center $\cS_7\rightarrow \cS_9$ \\ $\cS_9\rightarrow 2[K_B^{-1}]-\cS_7$} 
&&&$\cS_9\rightarrow 2[K_B^{-1}]-\cS_7$\\ \hline
\multirow{3}{*}{$X_{F_{13}}\rightarrow X_{F_{11}}$} & $(\two,\one,\four)$ & $Q^\prime=T^{15}-T^3$ & trivial \\ \cline{2-4}
& \multirow{2}{*}{$(\one,\two,\four)$} & \multirow{2}{*}{$Q^\prime=T^{15}-T^3$} & $\cS_7\rightarrow 2[K_B^{-1}]-\cS_7$\\%\parbox[c]{\widthof{$\cS_7\rightarrow 2[K_B^{-1}]-\cS_7$}}{\center $\cS_7\rightarrow 2[K_B^{-1}]-\cS_7$ \\ $\cS_9\rightarrow \cS_9$} 
&&&$\cS_9\rightarrow \cS_9$\\ \hline
% 
% & \multirow{2}{*}{ $(\two,\one)_{(-1/2,0)}$ } & $Q_1^\prime=(T_3^1-Q_1)$ & NON TORIC?!? \\ 
% & &  $Q_2^\prime=Q_2$ & ok \\ \cline{2-4}
% 
\multirow{4}{*}{$X_{F_{12}}\rightarrow X_{F_{9}}$} & \multirow{2}{*}{$(\two,\one)_{(-1/2,-1)}$} &  $Q_1^\prime=Q_1-T^3$ & \multirow{2}{*}{trivial} \\
&& $Q_2^\prime=Q_2-2T^3$ & \\ \cline{2-4}
& \multirow{2}{*}{$(\one,\two)_{(-1,-1/2)}$} &  $Q_1^\prime=2T^3-Q_1$ &  $\cS_7\rightarrow \cS_9$\\
&& $Q_2^\prime=T^3-Q_2$ & $\cS_9\rightarrow 2[K_B^{-1}]-\cS_7$\\ \hline% \parbox[c]{\widthof{$\cS_9\rightarrow 2[K_B^{-1}]-\cS_7$}}{\center $\cS_7\rightarrow \cS_9$ \\ $\cS_9\rightarrow 2[K_B^{-1}]-\cS_7$}\\ \hline
\multirow{4}{*}{$X_{F_{12}} \rightarrow X_{F_8}$} & \multirow{2}{*}{$(\one,\one)_{(1,0)}$} &\multirow{2}{*}{$ Q^\prime=Q_2$} & $\cS_7\rightarrow \cS_7$ \\%\parbox[c]{\widthof{$\cS_9\rightarrow [K_B^{-1}]-\cS_7+\cS_9$}}{\center $\cS_7\rightarrow \cS_7$ \\ $\cS_9\rightarrow [K_B^{-1}]-\cS_7+\cS_9$}%$\tilde{\cS}_9 = \cS_9 + \cS_7 -[K_B^{-1}]$ 
&&&$\cS_9\rightarrow [K_B^{-1}]-\cS_7+\cS_9$\\ \cline{2-4}
&  \multirow{2}{*}{$(\one,\one)_{(0,1)}$} & $Q^\prime= Q_1\, , SU(2)^\prime_1=SU(2)^2$ & $\cS_7\rightarrow [K_B^{-1}]-\cS_7+\cS_9$\\%\parbox[c]{\widthof{$\cS_7\rightarrow [K_B^{-1}]-\cS_7+\cS_9$}}{\center $\cS_7\rightarrow [K_B^{-1}]-\cS_7+\cS_9$ \\ $\cS_9\rightarrow \cS_7$}\\
&&$SU(2)^\prime_2=SU(2)^1$ & $\cS_9\rightarrow \cS_7$ \\ \hline

\end{tabular}
\caption{Data of toric Higgs transitions for F-theory compactified on 
$X_{F_{16}}-X_{F_{12}}$.}
\label{tab:HiggsChain1}
\end{center}
\end{table}
\begin{table}[H]
\begin{center}
\renewcommand{\arraystretch}{1.2}
\begin{tabular}{|c|c|c|c|}\hline
Higgs transition & VEV & U(1) Generators & Divisor Class Matching \\ \hline

\multirow{3}{*}{$X_{F_{12}} \rightarrow X_{F_7}$} & \multirow{3}{*}{$(\two,\two)_{(\frac12,\frac12)}$} & $Q_1^\prime=2Q_2 - Q_1 + T^3_1$ & \multirow{3}{*}{\vspace{1cm}\parbox[]{\widthof{$\cS_9\rightarrow 2[K_B^{-1}]-\cS_7$}}{\center $\cS_7\rightarrow \cS_9$ \\ $\cS_9\rightarrow 2[K_B^{-1}]-\cS_7$}}\\%$\tilde{\cS}_7 \rightarrow 2[K_B^{-1}]-\cS_9 $ \\ 
&& $ Q_2^\prime=Q_1 -Q_2-T_1^3-T_2^3 $ & \\%$\tilde{\cS}_9 \rightarrow \cS_7$ \\
&& $Q_3^\prime=Q_2 +2T_1^3+T_2^3$ & \\ \hline
\multirow{2}{*}{$X_{F_{11}}\rightarrow X_{F_{10}}$} & \multirow{2}{*}{$(\one,\one)_{-1}$} & \multirow{2}{*}{-} & $\cS_7 \rightarrow \cS_7$ \\ 
&&& $\cS_9 \rightarrow [K_B^{-1}] -\cS_7+\cS_9$ \\ \hline
\multirow{2}{*}{$X_{F_{11}}\rightarrow X_{F_{9}}$} & \multirow{2}{*}{$(\three,\two)_{-\frac16}$} & $Q_1^\prime = Q-2T_1^8 + T_2^3$ & \multirow{2}{*}{trivial} \\
 && $Q_2^\prime = 2T_2^3 - 3T_1^8$ & \\ \hline
\multirow{2}{*}{$X_{F_{11}}\rightarrow X_{F_{8}}$} & \multirow{2}{*}{$(\three,\one)_{\frac13}$}  & \multirow{2}{*}{$Q^\prime=2T^{8}-Q$} & $\cS_7 \rightarrow \cS_9$  \\
&&& $\cS_9 \rightarrow \cS_7$  \\ \hline
$X_{F_{10}} \rightarrow X_{F_6}$ & $(\three,\two)$ & $Q^\prime=T^8-2T^3$ & trivial \\ \hline
%\multirow{2}{*}{$F_{10} \rightarrow F_6$} & \multirow{2}{*}{$(\three,\two)$} & \multirow{2}{*}{ $Q^\prime=T^8-2T^3$} & $\ell_x\rightarrow [K_B^{-1}]-\cS_7$\\%\parbox[c]{\widthof{$l_y\rightarrow [K_B^{-1}]-\cS_9$}}{\center $\ell_x\rightarrow [K_B^{-1}]-\cS_7$ \\ $\ell_y\rightarrow [K_B^{-1}]-\cS_9$} \\
%&&&$\ell_y\rightarrow [K_B^{-1}]-\cS_9$\\ \hline
\multirow{4}{*}{$X_{F_{9}} \rightarrow X_{F_6}$} & \multirow{2}{*}{$\one_{(1,2)}$} & \multirow{2}{*}{$Q^\prime=2Q_1-Q_2$} &  $\cS_7\rightarrow 2 [K_B^{-1}] - \cS_7$ \\
&&& $\cS_9 \rightarrow [K_B^{-1}] - \cS_7 + \cS_9$ \\ \cline{2-4}
& \multirow{2}{*}{$\one_{(1,0)}$} & \multirow{2}{*}{$Q^\prime=Q_2$} & $\cS_7\rightarrow  \cS_7 $\\ 
&&& $\cS_9 \rightarrow [K_B^{-1}] - \cS_7 + \cS_9$ \\ \hline
\multirow{4}{*}{$X_{F_{9}} \rightarrow X_{F_5}$} & \multirow{2}{*}{$\two_{(-1,-\frac12)}$} & $Q_1^\prime= Q_2-Q_1 + T^3$ & \multirow{2}{*}{trivial} \\
&& $Q_2^\prime=-Q_1+2T^3$ & \\ \cline{2-4}
&\multirow{2}{*}{$\two_{(1,\frac32)}$} & $Q^\prime_1=Q_2-Q_1+T^3$ & $\cS_7\rightarrow 2[K_B^{-1}]-\cS_7$\\%\parbox[c]{\widthof{$\cS_7\rightarrow 2[K_B^{-1}]-\cS_7$}}{\center $\cS_7\rightarrow 2[K_B^{-1}]-\cS_7$ \\ $\cS_9\rightarrow \cS_9$} \\
&& $Q^\prime_2=Q_1 + 2T^3$ & $\cS_9\rightarrow \cS_9$\\ \hline 
$X_{F_{8}} \rightarrow X_{F_6}$ & $(\two,\one)_{1}$ & $Q^\prime=(Q+2T_1^3)$ & trivial \\ \hline
\multirow{2}{*}{$X_{F_{8}} \rightarrow X_{F_5}$} & \multirow{2}{*}{$(\two,\two)_{1/2}$} & $Q_1^\prime=Q-T^1_2$ & \multirow{2}{*}{trivial} \\
& & $Q_2^\prime=2(T_3^1-T^2_3)$ & \\ \hline
\multirow{12}{*}{$X_{F_{7}} \rightarrow X_{F_5}$} & \multirow{2}{*}{$\one_{(0,-1,0)}$} & $Q_1^\prime=-Q_1+Q_3$  & \multirow{2}{*}{trivial} \\
& & $Q_2^\prime=Q_3$ & \\ \cline{2-4}
& \multirow{2}{*}{$\one_{(1,1,0)}$} &  $Q_1^\prime=Q_1-Q_2$ & $\cS_7\rightarrow [K_B^{-1}]-\cS_7+\cS_9$ \\%\parbox[c]{\widthof{$\cS_7\rightarrow [K_B^{-1}]-\cS_7+\cS_9$}}{\center $\cS_7\rightarrow [K_B^{-1}]-\cS_7+\cS_9$ \\ $\cS_9\rightarrow \cS_9$}\\
& &  $Q_2^\prime = Q_3$ & $\cS_9\rightarrow \cS_9$ \\ \cline{2-4}
& \multirow{2}{*}{$\one_{(2,1,1)}$} &  $Q_1^\prime= Q_2-Q_3$ & $\cS_7\rightarrow 2[K_B^{-1}]-\cS_7$ \\%\parbox[c]{\widthof{$\cS_9\rightarrow [K_B^{-1}]-\cS_7+\cS_9$}}{\center $\cS_7\rightarrow 2[K_B^{-1}]-\cS_7$ \\ $\cS_9\rightarrow [K_B^{-1}]-\cS_7+\cS_9$}\\
& &  $Q_2^\prime=Q_1-Q_2-Q_3$ & $\cS_9\rightarrow [K_B^{-1}]-\cS_7+\cS_9$\\ \cline{2-4}
& \multirow{2}{*}{$\one_{(0,1,1)}$} & $Q_1^\prime = -Q_1$ & $\cS_7\rightarrow \cS_9$ \\%\parbox[c]{\widthof{$\cS_9\rightarrow [K_B^{-1}]-\cS_7+\cS_9$}}{\center $\cS_7\rightarrow \cS_9$ \\ $\cS_9\rightarrow [K_B^{-1}]-\cS_7+\cS_9$}\\
& & $Q_2^\prime = Q_2-Q_3$ & $\cS_9\rightarrow [K_B^{-1}]-\cS_7+\cS_9$\\ \cline{2-4}
& \multirow{2}{*}{$\one_{(-2,-1,-2)}$} & $Q_1^\prime = Q_1-Q_3$ & $\cS_7\rightarrow 2[K_B^{-1}]-\cS_9$ \\%\parbox[c]{\widthof{$\cS_9\rightarrow 2[K_B^{-1}]-\cS_9$}}{\center $\cS_7\rightarrow 2[K_B^{-1}]-\cS_7$ \\ $\cS_9\rightarrow 2[K_B^{-1}]-\cS_9$}\\
& &$Q_2^\prime = Q_1-2Q_2$ & $\cS_9\rightarrow 2[K_B^{-1}]-\cS_7$\\ \cline{2-4}
& \multirow{2}{*}{$\one_{(1,1,2)}$} & $Q_1^\prime = Q_1 - Q_2 $ & $\cS_7\rightarrow [K_B^{-1}]-\cS_7+\cS_9$ \\%\parbox[c]{\widthof{$\cS_7\rightarrow [K_B^{-1}]-\cS_7+\cS_9$}}{\center $\cS_7\rightarrow [K_B^{-1}]-\cS_7+\cS_9$ \\ $\cS_9\rightarrow 2[K_B^{-1}]-\cS_7$}\\
& & $Q_2^\prime = Q_1 + Q_2 - 2 Q_3$ & $\cS_9\rightarrow 2[K_B^{-1}]-\cS_7$\\ \hline
\end{tabular}
\caption{Data of toric Higgs transitions for F-theory compactified on 
$X_{F_{12}}-X_{F_{7}}$.}
\label{tab:HiggsChain2}
\end{center}
\end{table}
\begin{table}[H]
\begin{center}
\renewcommand{\arraystretch}{1.2}
\begin{tabular}{|c|c|c|c|}\hline
Higgs transition & VEV & U(1) Generators & Divisor Class Matching \\ \hline
$X_{F_6} \rightarrow X_{F_4}$ & $\one_2$ & $Q^\prime_{\mathbb{Z}_4}=2Q\text{ mod } 4$ & trivial \\ \hline
    %\multirow{2}{*}{$F_6 \rightarrow F_4$} & \multirow{2}{*}{$\one_2$} & \multirow{2}{*}{$Q^\prime_{\mathbb{Z}_4}=2Q\text{ mod } 4$} & $\cS_7\rightarrow [K_B^{-1}]+D_x-D_y$ \\%\parbox[c]{\widthof{$\cS_7\rightarrow [K_B^{-1}]+D_x-D_y$}}{\center $\cS_7\rightarrow [K_B^{-1}]+D_x-D_y$ \\ $\cS_9\rightarrow [K_B^{-1}]+2D_x$}\\ \hline
    %&&&$\cS_9\rightarrow [K_B^{-1}]+2D_x$\\ \hline
$X_{F_{6}} \rightarrow X_{F_{3}}$ & $\two_{-\frac32}$ & $Q^\prime=3T^3-Q$ & trivial \\ \hline
\multirow{3}{*}{{$X_{F_{5}} \rightarrow X_{F_{3}}$}} & \multirow{2}{*}{$\one_{(-1,-2)}$} & \multirow{2}{*}{$Q_1^\prime=2Q_1-Q_2$} & $\cS_7\rightarrow \cS_9$ \\%\parbox[c]{\widthof{$\cS_9\rightarrow \cS_7$}}{\center $\cS_7\rightarrow \cS_9$ \\ $\cS_9\rightarrow \cS_7$} \\
&&& $\cS_9\rightarrow \cS_7$ \\ \cline{2-4} 
& $\one_{(-1,1)}$ & $Q_1^\prime=Q_1+Q_2$ & trivial \\ \hline
\multirow{2}{*}{{$X_{F_{5}} \rightarrow X_{F_{2}}$}} & \multirow{2}{*}{$\one_{(0,-2)}$} & $Q_1^\prime=Q_1$ & \multirow{2}{*}{trivial} \\%\parbox[c]{\widthof{$\cS_9\rightarrow \cS_9$}}{\center $\cS_7\rightarrow \cS_7$ \\ $\cS_9\rightarrow \cS_9$}\\
&& $Q_{\mathbb{Z}_2}=Q_2$ mod $2$ & \\ \hline
$X_{F_{3}} \rightarrow X_{F_{1}}$ & $\one_{3}$ & $Q_{\mathbb{Z}_3}^\prime= Q$ mod $3$ & trivial \\ \hline
\end{tabular}
\caption{Data of toric Higgs transitions for F-theory compactified on $X_{F_{6}}-X_{F_{3}}$.}
\label{tab:HiggsChain3}
\end{center}
\end{table}

%%%%%%%%%%%%%%%%%%%%%%%%%%%%%%%%%%%%%%%%%%%%%%%%%%%%%%%%%%%%%%%%%%%%%%%%%%%%%%%%%%%%%%%%%%%%%%%%%%%%%%
%%%%%%%%%%%%%%%%%%%%%%%%%%%%%%%%%%%%%%%%%%%%%%%%%%%%%%%%%%%%%%%%%%%%%%%%%%%%%%%%%%%%%%%%%%%%%%%%%%%%%%
\section{Group Theoretical Decomposition of Representations}
\label{app:decomposition}
%%%%%%%%%%%%%%%%%%%%%%%%%%%%%%%%%%%%%%%%%%%%%%%%%%%%%%%%%%%%%%%%%%%%%%%%%%%%%%%%%%%%%%%%%%%%%%%%%%%%%%
%%%%%%%%%%%%%%%%%%%%%%%%%%%%%%%%%%%%%%%%%%%%%%%%%%%%%%%%%%%%%%%%%%%%%%%%%%%%%%%%%%%%%%%%%%%%%%%%%%%%%%
In Tables~\ref{tab:GroupSplitting1}, \ref{tab:GroupSplitting2}, 
\ref{tab:GroupSplitting3}, \ref{tab:GroupSplitting4}, 
\ref{tab:GroupSplitting5} and \ref{tab:GroupSplitting6} we show the 
explicit decompositions of representations of the group $G_{F_i}$ 
into representations of the unbroken group $G_{F_j}$ for each 
Higgs transition $X_{F_i}\rightarrow X_{F_j}$.

\begin{table}[htb!]
\begin{center}
\renewcommand{\arraystretch}{1.2}
\begin{tabular}{|c|c|c|}\hline
Breaking & Starting Multiplets & Target Multiplets\\ \hline
\multirow{9}{*}{\parbox[c]{\widthof{VEV: $(\three,\one,\bar{\three})$}}{\center $X_{F_{16}} \rightarrow X_{F_{14}}$ \\ VEV: $(\three,\one,\bar{\three})$}} & $(\three,\bar{\three},\one)$ & $(\two,\bar{\three},\one)_{-1/6}+(\one,\bar{\three},\one)_{1/3}$  \\ 
& \multirow{2}{*}{$(\three,\one,\bar{\three})$} & $(\two,\one,\two)_{0}+(\two,\one,\one)_{1/2}$ \\
&  & $+(\two,\one,\one)_{-1/2}+(\one,\one,\one)_{0}$  \\ 
& $(\one,\three,\bar{\three})$ & $(\one,\three,\two)_{1/6}+(\one,\three,\one)_{-1/3}$   \\ 
& \multirow{2}{*}{$(\eight,\one,\one)$} & $(\three,\one,\one)_{0}+(\two,\one,\one)_{1/2}$ \\
&  & $+(\two,\one,\one)_{-1/2}+(\one,\one,\one)_{0}$   \\ 
& $(\one,\eight,\one)$ & $(\one,\eight,\one)_{0}$  \\ 
& \multirow{2}{*}{$(\one,\one,\eight)$} & $(\one,\one,\three)_{0}+(\one,\one,\two)_{1/2}$ \\
&  & $+(\one,\one,\two)_{-1/2}+(\one,\one,\one)_{0}$   \\ \hline
\end{tabular}
\caption{Group theoretical decompositions of representation in toric 
Higgsings of $X_{F_{16}}$.}
\label{tab:GroupSplitting1}
\end{center}
\end{table}

\begin{table}[htb!]
\begin{center}
\renewcommand{\arraystretch}{1.2}
\begin{tabular}{|c|c|c|}\hline
Breaking & Starting Multiplets & Target Multiplets\\ \hline
\multirow{12}{*}{\parbox[c]{\widthof{ VEV: $(\two,\two,\one,\one)_{(1/2)}$}}{\center $X_{F_{15}} \rightarrow X_{F_{12}}$ \\ VEV: $(\two,\two,\one,\one)_{(1/2)}$}} & \multirow{2}{*}{$(\two,\two,\one,\one)_\frac12$} & $(\one,\one)_{(1,1)}+(\one,\one)_{(1,0)}$ \\
&  & $+(\one,\one)_{(0,1)}+(\one,\one)_{(0,0)}$ \\
& $(\two,\one,\two,\one)_\frac12$  & $(\two,\one)_{(\frac12,1)} + (\two,\one)_{(\frac12,0)}$ \\
& $(\two,\one,\one, \two)_0$ & $(\one,\two)_{(0,\frac12)} + (\one,\two)_{(0,-\frac12)}$ \\
& $(\one,\two,\two, \one)_0$ & $(\two,\one)_{(\frac12,0)} + (\two,\one)_{(-\frac12,0)}$  \\
& $(\one,\one,\two,\two)_\frac12$ & $(\two,\two)_{(\frac12,\frac12)}$ \\
& $(\one,\two,\one, \two)_\frac12$ & $(\one,\two)_{(1,\frac12)} + (\one,\two)_{(,0\frac12)}$  \\
& $(\one,\one,\one, \one)_1$ & $(\one,\one)_{(1,1)}$ \\
& $(\three,\one,\one, \one)_0$ & $(\one,\one)_{(0,1)} + (\one,\one)_{(0,-1)} + (\one,\one)_{(0,0)}$ \\
& $(\one,\three,\one, \one)_0$ & $(\one,\one)_{(1,0)} + (\one,\one)_{(-1,0)} + (\one,\one)_{(0,0)}$ \\
& $(\one,\one,\three, \one)_0$ & $(\three,\one)_{(0,0)}$ \\
& $(\one,\one,\one, \three)_0$ & $(\one,\three)_{(0,0)}$ \\ \hline
\multirow{11}{*}{\parbox[c]{\widthof{ VEV: $(\two,\three,\one)_{1/6}$}}{\center $X_{F_{14}} \rightarrow X_{F_{12}}$ \\ VEV: $(\two,\three,\one)_{1/6}$}} & $(\two,\one,\one)_{1/2}$ & $(\one,\one)_{(1,0)}+(\one,\one)_{(1,1)}$  \\ 
& $(\one,\three,\one)_{1/3}$ & $(\two,\one)_{(-1/2,0)}+(\one,\one)_{(-1,-1)}$  \\ 
& $(\one,\one,\two)_{1/2}$ & $(\one,\two)_{(1,1/2)}$   \\ 
& \multirow{2}{*}{$(\two,\three,\one)_{1/6}$} & $(\two,\one)_{(1/2,0)}+(\two,\one)_{(1/2,1)}$ \\
&  & $+(\one,\one)_{(0,-1)}+(\one,\one)_{(0,0)}$  \\ 
& $(\two,\one,\two)_{0}$ & $(\one,\two)_{(0,-1/2)}+(\one,\two)_{(0,1/2)}$   \\ 
& $(\one,\three, \two)_{1/6}$ & $(\two,\two)_{(1/2,1/2)}+(\one,\two)_{(0,-1/2)}$  \\ 
& $(\three,\one,\one)_{0} $ &  $(\one,\one)_{(0,0)}+(\one,\one)_{(0,1)}+(\one,\one)_{(0,-1)}$\\ 
& \multirow{2}{*}{$(\one,\eight,\one)_{0}$} & $(\three,\one)_{(0,0)}+(\one,\one)_{(0,0)}$ \\
&  & $+(\two,\one)_{(1/2,1)}+(\two,\one)_{(-1/2,-1)}$\\ 
& $(\one,\one,\three)_{0}$ & $(\one,\three)_{(0,0)}$\\ \hline
\multirow{7}{*}{\parbox[c]{\widthof{VEV: $(\one,\one,\two)_{1/2}$}}{\center $X_{F_{14}} \rightarrow X_{F_{11}}$ \\ VEV: $(\one,\one,\two)_{1/2}$}} & $(\two,\one,\one)_{1/2}$ & $(\two,\one)_{1/2}$\\
& $(\one,\three,\one)_{1/3}$ & $(\one,\three)_{1/3}$\\
& $(\one,\one,\two)_{1/2}$ & $(\one,\one)_{0} + (\one,\one)_{1}$\\
& $(\two,\three,\one)_{1/2}$ & $(\two,\three)_{1/6} $\\
& $(\two,\one,\two)_{0}$ & $(\one,\two)_{1/2} + (\one,\two)_{-1/2}$\\
& $(\one,\three,\two)_{1/6}$ & $(\three,\one)_{2/3} + (\three,\one)_{-1/3}$\\
& $(\one,\one,\three)_{0}$ & $(\one,\one)_{1} + (\one,\one)_{-1}$\\ \hline
\end{tabular}
\caption{Group theoretical decompositions of representation in toric 
Higgsings of  $X_{F_{15}}-X_{F_{14}}$.}
\label{tab:GroupSplitting2}
\end{center}
\end{table}

\begin{table}[htb!]
\begin{center}
\renewcommand{\arraystretch}{1.2}
\begin{tabular}{|c|c|c|}\hline
Breaking & Starting Multiplets & Target Multiplets\\ \hline
\multirow{7}{*}{\parbox[c]{\widthof{VEV: $(\two,\one,\four)$}}{\center $X_{F_{13}}\rightarrow X_{F_{11}}$ \\ VEV: $(\two,\one,\four)$}} & $(\two,\two,\one)$ & $(\two,\one)_{-1/2} + (\two,\one)_{1/2} $\\
& $(\two,\one,\four)$ & $(\one,\one)_{-1} + (\one,\one)_{0} + (\one,\three)_{-1/3} + (\one,\three)_{\frac23}$ \\
& $(\one,\one,\six)$ & $(\one,\three)_{-\frac13} + (\one,\overline{\three})_{\frac13}$\\
& $(\two,\two,\four)$ & $(\two,\three)_{1/6} + (\two,\one)_{-1/2} $\\
& $(\one,\one,\fiveteen)$ & $(\one,\eight)_{0} + (\one,\one)_{0} + (\one,\three)_{\frac23} + (\one,\overline{\three})_{-\frac23}$\\
& $(\one,\three,\one)$ & $(\three,\one)_{0} $\\
& $(\three,\one,\one)$ & $(\one,\one)_{0} + (\one,\one)_{-1} + (\one,\one)_{1}$\\ \hline
\multirow{10}{*}{\parbox[c]{\widthof{VEV: $(\two,\one)_{(-1/2,-1)}$}}{\center $X_{F_{12}}\rightarrow X_{F_{9}}$ \\ VEV: $(\two,\one)_{(-1/2,-1)}$}} & $(\two,\two)_{(1/2,1/2)}$ & $\two_{(0,-1/2)}+\two_{(1,3/2)}$  \\
& $(\one,\two)_{(-1,-1/2)}$ & $\two_{(-1,-1/2)}$  \\
& $(\two,\one)_{(-1/2,-1)}$ & $\one_{(-1,-2)}+\one_{(0,0)}$  \\
& $(\one,\one)_{(1,0)}$ & $\one_{(1,0)}$  \\
& $(\one,\one)_{(0,1)}$ & $\one_{(0,1)}$  \\
& $(\one,\two)_{(0,-1/2)}$ & $\two_{(0,-1/2)}$  \\
& $(\two,\one)_{(-1/2,0)}$ & $\one_{(-1,-1)}+\one_{(0,1)}$  \\
& $(\one,\one)_{(1,1)}$ & $\one_{(1,1)}$  \\
& $(\one,\three)_{(0,0)}$ & $\three_{(0,0)}$ \\
& $(\three,\one)_{(0,0)}$ & $\one_{(-1,-2)}+\one_{(1,2)}+\one_{(0,0)}$ \\ \hline
\multirow{10}{*}{\parbox[c]{\widthof{VEV: $(\one,\one)_{(1,0)}$}}{\center $X_{F_{12}} \rightarrow X_{F_8}$ \\ VEV: $(\one,\one)_{(1,0)}$}} & $(\two,\two)_{(1/2,1/2)}$ & $(\two,\two)_{1/2}$ \\
& $(\one,\two)_{(-1,-1/2)}$ & $(\one,\two)_{-1/2}$  \\
& $(\two,\one)_{(-1/2,-1)}$ & $(\two,\one)_{-1}$  \\
& $(\one,\one)_{(1,0)}$ & $(\one,\one)_{0}$  \\
& $(\one,\one)_{(0,1)}$ & $(\one,\one)_{1)}$ \\
& $(\one,\two)_{(0,-1/2)}$ & $(\one,\two)_{-1/2}$ \\
& $(\two,\one)_{(-1/2,0)}$ & $(\two,\one)_{0}$  \\
& $(\one,\one)_{(1,1)}$ & $(\one,\one)_{1}$  \\
& $(\one,\three)_{(0,0)}$ & $(\one,\three)_{0}$ \\
& $(\three,\one)_{(0,0)}$ & $(\three,\one)_{0}$ \\ \hline
\end{tabular}
\caption{Group theoretical decompositions of representation in toric 
Higgsings of  $X_{F_{13}}-X_{F_{12}}$.}
\label{tab:GroupSplitting3}
\end{center}
\end{table}

\begin{table}[htb!]
\begin{center}
\renewcommand{\arraystretch}{1.2}
\begin{tabular}{|c|c|c|}\hline
Breaking & Starting Multiplets & Target Multiplets\\ \hline
\multirow{10}{*}{\parbox[c]{\widthof{ VEV: $(\two,\two)_{(\frac12,\frac12)}$}}{\center $X_{F_{12}} \rightarrow X_{F_7}$ \\ VEV: $(\two,\two)_{(\frac12,\frac12)}$}} & $(\two,\two)_{(1/2,1/2)}$ & \parbox[c]{\widthof{$\one_{(-1,-1,-2)}+\one_{(0,0,0)}$}}{\center $\one_{(-1,0,-1)}+\one_{(0,1,1)}$ \\ $\one_{(-1,-1,-2)}+\one_{(0,0,0)}$}  \\
& $(\one,\two)_{(-1,-1/2)}$ & $\one_{(0,0,1)}$ +$\one_{(0,-1,0)}$  \\
& $(\two,\one)_{(-1/2,-1)}$ & $\one_{(1,0,0)}$+$\one_{(-2,-1,-2)}$  \\
& $(\one,\one)_{(1,0)}$ & $\one_{(1,1,0)}$ \\
& $(\one,\one)_{(0,1)}$ & $\one_{(2,1,1)}$  \\
& $(\one,\two)_{(0,-1/2)}$ & $\one_{(1,1,1)}$ + $\one_{(1,0,0)}$ \\
& $(\two,\one)_{(-1/2,0)}$ & $\one_{(0,0,1)}$ + $\one_{(1,1,1)}$ \\
& $(\one,\one)_{(1,1)}$ &  $\one_{(1,0,1)}$ \\
& $(\one,\three)_{(0,0)}$ &   $\one_{(0,1,1)}$ +  $\one_{(0,-1,-1)}$\\
& $(\three,\one)_{(0,0)}$ &  $\one_{(1,1,2)}$ +  $\one_{(-1,-1,-2)}$ \\ \hline
$X_{F_{11}}\rightarrow X_{F_{10}}$ & \multicolumn{2}{c|}{trivial}\\ \hline
\multirow{7}{*}{\parbox[c]{\widthof{VEV: $(\three,\two)_{-\frac16}$}}{\center $X_{F_{11}}\rightarrow X_{F_{9}}$ \\ VEV: $(\three,\two)_{-\frac16}$}} & $(\three,\two)_{-\frac16}$ & $\one_{(0,0)} + \one_{(1,2)} + \two_{(0,\frac12)} + \two_{(-1,-\frac32)} $\\
& $(\one,\two)_{\frac12}$ & $\one_{(1,1)} + \one_{(0,-1)} $ \\
& $(\three,\one)_{-\frac23}$ & $\one_{(0,1)} + \two_{(-1,-\frac12)}$\\
& $(\three,\one)_{\frac13}$  & $\one_{(1,1)} + \two_{(0,-\frac12)} $\\
& $(\one,\one)_{-1}$ & $\one_{(1,0)}$\\
& $(\eight,\one)_0$ & $\two_{(1,\frac32)} +\two_{(-1,-\frac32)} + \one_{(0,0)} + \three_{(0,0)}   $\\
& $(\one, \three)_0 $ & $\one_{(0,0)} + \one_{(1,2)} + \one_{(-1,-2)}   $\\ \hline
\multirow{7}{*}{\parbox[c]{\widthof{VEV: $(\three,\one)_{\frac13}$}}{\center $X_{F_{11}}\rightarrow X_{F_{8}}$ \\ VEV: $(\three,\one)_{\frac13}$}} & $(\three,\two)_{-\frac16}$ & $(\two,\two)_{1/2} + (\one,\two)_{-1/2} $\\
& $(\one,\two)_{\frac12}$ & $(\one,\two)_{-\frac12} $ \\
& $(\three,\one)_{-\frac23}$ & $(\one,\one)_{-1} + (\two,\one)_{0}$\\
& $(\three,\one)_{\frac13}$  & $(\two,\one)_{0} + (\one,\one)_{0} $\\
& $(\one,\one)_{-1}$ & $(\one,\one)_{1}$\\
& $(\eight,\one)_0$ & $(\three,\one)_{0} + (\two,\one)_{1} + (\two,\one)_{-1} + (\one,\one)_{0} $\\
& $(\one, \three)_0 $ & $(\one, \three)_0$\\ \hline
\multirow{5}{*}{\parbox[c]{\widthof{$X_{F_{10}} \rightarrow X_{F_6}$}}{\center $X_{F_{10}} \rightarrow X_{F_6}$ \\ VEV: $(\three,\two)$}} & $(\three,\two)$ & $\two_{-1/2}+\two_{3/2}+\one_{-2}+\one_0$  \\
& $(\one,\two)$ & $\one_{1}+\one_{-1}$  \\
& $(\three,\one)$ & $\one_{1/2}+\one_{-1}$  \\
& $(\one,\three)$ & $\one_0+\one_2+\one_{-2}$ \\
& $(\eight,\one)$ & $\three_0+\two_{3/2}+\two_{-3/2}+\one_{0}$ \\ \hline
\end{tabular}
\caption{Group theoretical decompositions of representation in toric 
Higgsings of  $X_{F_{12}}-X_{F_{10}}$.}
\label{tab:GroupSplitting4}
\end{center}
\end{table}

\begin{table}[htb!]
\begin{center}
\renewcommand{\arraystretch}{1.2}
\begin{tabular}{|c|c|c|}\hline
Breaking & Starting Multiplets & Target Multiplets\\ \hline
\multirow{8}{*}{\parbox[c]{\widthof{$X_{F_{9}} \rightarrow X_{F_6}$}}{\center $X_{F_{9}} \rightarrow X_{F_6}$ \\ VEV: $\one_{(1,2)}$}} & $\one_{(1,2)}$ & $\one_{0}$ \\
& $\one_{(1,0)}$ & $\one_{2}$ \\
& $\one_{(0,1)}$ & $\one_{-1}$ \\
& $\one_{(1,1)}$ & $\one_{1}$ \\
& $\two_{(-1,-1/2)}$ & $\two_{-3/2}$ \\
& $\two_{(1,-3/2)}$ & $\two_{1/2}$ \\
& $\two_{(0,-1/2)}$ & $\two_{-1/2}$ \\
& $\three_{(0,0)}$ & $\three_{0}$ \\ \hline
\multirow{8}{*}{\parbox[c]{\widthof{$X_{F_{9}} \rightarrow X_{F_5}$}}{\center $X_{F_{9}} \rightarrow X_{F_5}$ \\ VEV: $\two_{(-1,-1/2)}$}} & $\one_{(1,2)}$ & $\one_{0}$ \\
& $\one_{(1,0)}$ & $\one_{(-1,-1)}$ \\
& $\one_{(0,1)}$ & $\one_{(1,0)}$ \\
& $\one_{(1,1)}$ & $\one_{(0,-1)}$ \\
& $\two_{(-1,-1/2)}$ & $\one_{(0,0)} + \one_{(1,2)}$ \\
& $\two_{(1,-3/2)}$ & $\one_{(0,-2)} + \one_{(1,0)}$ \\
& $\two_{(0,-1/2)}$ & $\one_{(-1,-1)} + \one_{(0,1)}$ \\
& $\three_{(0,0)}$ & $\one_{(-1,-2)} + \one_{(1,2)} + \one_{(0,0)}$ \\ \hline
\multirow{7}{*}{\parbox[c]{\widthof{VEV: $(\two,\one)_{1}$}}{\center $X_{F_{8}} \rightarrow X_{F_6}$ \\ VEV: $(\two,\one)_{1}$}} & $(\two,\two)_{1/2}$ & $\two_{\frac12}+\two_{-\frac32}$\\
& $(\one,\two)_{1/2}$ & $\two_{\frac12}$ \\
& $(\two,\one)_{1}$ & $\one_{0}+\one_{2}$\\
& $(\two,\one)_0$  & $\one_{1}+\one_{-1}$\\
& $(\one,\one)_{1}$ & $\one_{1}$\\
& $(\three,\one)_0$ & $\one_{0}+\one_{2}+\one_{2}$\\
& $(\one, \three)_0$ & $(\one, \three)_0$\\ \hline
\multirow{7}{*}{\parbox[c]{\widthof{VEV: $(\two,\two)_{\frac12}$}}{\center $X_{F_{8}} \rightarrow X_{F_5}$ \\ VEV: $(\two,\two)_{\frac12}$}} & $(\two,\two)_{1/2}$ & $\one_{(-1,0)}+\one_{(-1,-2)}+\one_{(0,2)}+\one_{(0,0)}$\\
& $(\one,\two)_{1/2}$ & $\one_{(0,-1)}+\one_{(1,1)}$ \\
& $(\two,\one)_{1}$ & $\one_{(1,1)}+\one_{(1,-1)}$\\
& $(\two,\one)_0$  & $\one_{(0,1)}+\one_{(0,-1)}$\\
& $(\one,\one)_{1}$ & $\one_{(-1,-1)}$\\
& $(\three,\one)_0$ & $\one_{(0,2)}+\one_{(0,-2)}+\one_{(0,0)}$\\
& $(\one, \three)_0$ & $\one_{(1,2)}+\one_{(-1,-2)}+\one_{(0,0)}$\\ \hline
\end{tabular}
\caption{Group theoretical decompositions of representation in toric 
Higgsings of  $X_{F_{9}}-X_{F_{8}}$.}
\label{tab:GroupSplitting5}
\end{center}
\end{table}
\begin{table}[htb!]
\begin{center}
\small
\renewcommand{\arraystretch}{1.2}
\begin{tabular}{|c|c|c|}\hline
Breaking & Starting Multiplets & Target Multiplets\\ \hline
\multirow{10}{*}{\parbox[c]{\widthof{VEV: $\one_{(0,-1,0)}$}}{\center $X_{F_{7}} \rightarrow X_{F_5}$ \\ VEV: $\one_{(0,-1,0)}$}} & $\one_{(1,1,0)}$ & $\one_{(-1,0)}$ \\
& $\one_{(0,-1,0)}$ & $\one_{(0,0)}$ \\
& $\one_{(2,1,1)}$ & $\one_{(-1,1)}$ \\
& $\one_{(0,1,1)}$ & $\one_{(1,1)}$ \\
& $\one_{(-2,-1,-2)}$ & $\one_{(0,-2)}$ \\
& $\one_{(1,1,2)}$ & $\one_{(1,2)}$ \\
& $\one_{(1,0,0)}$ & $\one_{(-1,0)}$ \\
& $\one_{(0,0,1)}$ & $\one_{(1,1)}$ \\
& $\one_{(1,0,1)}$ & $\one_{(0,1)}$ \\
& $\one_{(1,1,1)}$ & $\one_{(0,1)}$ \\ \hline
\multirow{5}{*}{\parbox[c]{\widthof{$X_{F_{6}} \rightarrow X_{F_{4}}$}}{\center $X_{F_{6}} \rightarrow X_{F_{4}}$ \\ VEV: $\one_{2}$}} & $\two_{-\frac32}$ & $\two_{\frac12}$ \\
& $\two_{\frac12}$ & $\two_{\frac12}$ \\
& $\one_{2}$ & $\one_{0}$ \\
& $\one_{1}$ & $\one_{1}$ \\
& $\three_{0}$ & $\three_{0}$   \\ \hline
\multirow{5}{*}{\parbox[c]{\widthof{$X_{F_{6}} \rightarrow X_{F_{3}}$}}{\center $X_{F_{6}} \rightarrow X_{F_{3}}$ \\ VEV: $\two_{-\frac32}$}} & $\two_{-\frac32}$ & $\one_{0} + \one_{-2}$ \\
& $\two_{\frac12}$ & $\one_{1}$ + $\one_{-2}$ \\
& $\one_{2}$ & $\one_{-2}$ \\
& $\one_{1}$ & $\one_{-1}$ \\
& $\three_{0}$ & $\one_{0}+ \one_3 + \one_{-3}$ \\ \hline
\multirow{6}{*}{\parbox[c]{\widthof{VEV: $\one_{(-1,-2)}$}}{\center $X_{F_{5}} \rightarrow X_{F_{3}}$ \\ VEV: $\one_{(-1,-2)}$}} & $\one_{(1,-1)}$ & $\one_{3}$ \\
& $\one_{(1,0)}$ & $\one_{2}$ \\
& $\one_{(-1,-2)}$ & $\one_{0}$ \\
& $\one_{(-1,-1)}$ & $\one_{1}$ \\
& $\one_{(0,2)}$ & $\one_{2}$ \\
& $\one_{(0,1)}$ & $\one_{1}$ \\ \hline
\multirow{6}{*}{\parbox[c]{\widthof{$X_{F_{5}} \rightarrow X_{F_{2}}$}}{\center $X_{F_{5}} \rightarrow X_{F_{2}}$ \\ VEV: $\one_{(0,2)}$}} & $\one_{(1,-1)}$ & $\one_{(1,-)}$ \\
& $\one_{(1,0)}$ & $\one_{(1,+)}$ \\
& $\one_{(-1,-2)}$ & $\one_{(1,+)}$ \\
& $\one_{(-1,-1)}$ & $\one_{(1,-)}$ \\
& $\one_{(0,2)}$ & $\one_{(0,+)}$ \\
& $\one_{(0,1)}$ & $\one_{(0,-)}$ \\ \hline
\multirow{3}{*}{\parbox[c]{\widthof{$X_{F_{3}} \rightarrow X_{F_{1}}$}}{\center $X_{F_{3}} \rightarrow X_{F_{1}}$ \\ VEV: $\one_{3}$}} & $\one_{3}$ & $\one_{0}$ \\
& $\one_{2}$ & $\one_{2}$ \\
& $\one_{1}$ & $\one_{1}$ \\ \hline
\end{tabular}
\caption{Group theoretical decompositions of representation in toric 
Higgsings of  $X_{F_{7}}-X_{F_{3}}$.}
\label{tab:GroupSplitting6}
\end{center}
\end{table}

% #################################
% #        Bibliography           #
% #################################
\clearpage
\bibliographystyle{utphys}	% (uses file "plain.bst")
\bibliography{ref}

\providecommand{\href}[2]{#2}\begingroup\raggedright\begin{thebibliography}{100}

\bibitem{Vafa:1996xn}
C.~Vafa, ``{Evidence for F theory},''
  \href{http://dx.doi.org/10.1016/0550-3213(96)00172-1}{{\em Nucl.Phys.}
  {\bfseries B469} (1996) 403--418},
\href{http://arxiv.org/abs/hep-th/9602022}{{\ttfamily arXiv:hep-th/9602022
  [hep-th]}}.
%%CITATION = HEP-TH/9602022;%%.

\bibitem{Morrison:1996na}
D.~R. Morrison and C.~Vafa, ``{Compactifications of F theory on Calabi-Yau
  threefolds. 1},'' \href{http://dx.doi.org/10.1016/0550-3213(96)00242-8}{{\em
  Nucl.Phys.} {\bfseries B473} (1996) 74--92},
\href{http://arxiv.org/abs/hep-th/9602114}{{\ttfamily arXiv:hep-th/9602114
  [hep-th]}}.
%%CITATION = HEP-TH/9602114;%%.

\bibitem{Morrison:1996pp}
D.~R. Morrison and C.~Vafa, ``{Compactifications of F theory on Calabi-Yau
  threefolds. 2.},'' \href{http://dx.doi.org/10.1016/0550-3213(96)00369-0}{{\em
  Nucl.Phys.} {\bfseries B476} (1996) 437--469},
\href{http://arxiv.org/abs/hep-th/9603161}{{\ttfamily arXiv:hep-th/9603161
  [hep-th]}}.
%%CITATION = HEP-TH/9603161;%%.

\bibitem{Donagi:2008ca}
R.~Donagi and M.~Wijnholt, ``{Model Building with F-Theory},''
  \href{http://arxiv.org/abs/0802.2969}{{\ttfamily arXiv:0802.2969 [hep-th]}}.

\bibitem{Beasley:2008dc}
C.~Beasley, J.~J. Heckman, and C.~Vafa, ``{GUTs and Exceptional Branes in
  F-theory - I},'' \href{http://dx.doi.org/10.1088/1126-6708/2009/01/058}{{\em
  JHEP} {\bfseries 01} (2009) 058},
\href{http://arxiv.org/abs/0802.3391}{{\ttfamily arXiv:0802.3391 [hep-th]}}.
%%CITATION = 0802.3391;%%.

\bibitem{Hayashi:2008ba}
H.~Hayashi, R.~Tatar, Y.~Toda, T.~Watari, and M.~Yamazaki, ``{New Aspects of
  Heterotic--F Theory Duality},''
  \href{http://dx.doi.org/10.1016/j.nuclphysb.2008.07.031}{{\em Nucl.Phys.}
  {\bfseries B806} (2009) 224--299},
\href{http://arxiv.org/abs/0805.1057}{{\ttfamily arXiv:0805.1057 [hep-th]}}.
%%CITATION = ARXIV:0805.1057;%%.

\bibitem{Beasley:2008kw}
C.~Beasley, J.~J. Heckman, and C.~Vafa, ``{GUTs and Exceptional Branes in
  F-theory - II: Experimental Predictions},''
  \href{http://dx.doi.org/10.1088/1126-6708/2009/01/059}{{\em JHEP} {\bfseries
  01} (2009) 059},
\href{http://arxiv.org/abs/0806.0102}{{\ttfamily arXiv:0806.0102 [hep-th]}}.
%%CITATION = 0806.0102;%%.

\bibitem{Hori:2006dk}
K.~Hori and D.~Tong, ``{Aspects of Non-Abelian Gauge Dynamics in
  Two-Dimensional N=(2,2) Theories},''
  \href{http://dx.doi.org/10.1088/1126-6708/2007/05/079}{{\em JHEP} {\bfseries
  0705} (2007) 079},
\href{http://arxiv.org/abs/hep-th/0609032}{{\ttfamily arXiv:hep-th/0609032
  [hep-th]}}.
%%CITATION = HEP-TH/0609032;%%.

\bibitem{Jockers:2012zr}
H.~Jockers, V.~Kumar, J.~M. Lapan, D.~R. Morrison, and M.~Romo, ``{Nonabelian
  2D Gauge Theories for Determinantal Calabi-Yau Varieties},''
  \href{http://dx.doi.org/10.1007/JHEP11(2012)166}{{\em JHEP} {\bfseries 1211}
  (2012) 166},
\href{http://arxiv.org/abs/1205.3192}{{\ttfamily arXiv:1205.3192 [hep-th]}}.
%%CITATION = ARXIV:1205.3192;%%.

\bibitem{Marsano:2009gv}
J.~Marsano, N.~Saulina, and S.~Schafer-Nameki, ``{Monodromies, Fluxes, and
  Compact Three-Generation F-theory GUTs},''
  \href{http://dx.doi.org/10.1088/1126-6708/2009/08/046}{{\em JHEP} {\bfseries
  0908} (2009) 046},
\href{http://arxiv.org/abs/0906.4672}{{\ttfamily arXiv:0906.4672 [hep-th]}}.
%%CITATION = ARXIV:0906.4672;%%.

\bibitem{Blumenhagen:2009yv}
R.~Blumenhagen, T.~W. Grimm, B.~Jurke, and T.~Weigand, ``{Global F-theory
  GUTs},'' \href{http://dx.doi.org/10.1016/j.nuclphysb.2009.12.013}{{\em
  Nucl.Phys.} {\bfseries B829} (2010) 325--369},
\href{http://arxiv.org/abs/0908.1784}{{\ttfamily arXiv:0908.1784 [hep-th]}}.
%%CITATION = ARXIV:0908.1784;%%.

\bibitem{Heckman:2013pva}
J.~J. Heckman, D.~R. Morrison, and C.~Vafa, ``{On the Classification of 6D
  SCFTs and Generalized ADE Orbifolds},''
  \href{http://dx.doi.org/10.1007/JHEP05(2014)028}{{\em JHEP} {\bfseries 1405}
  (2014) 028},
\href{http://arxiv.org/abs/1312.5746}{{\ttfamily arXiv:1312.5746 [hep-th]}}.
%%CITATION = ARXIV:1312.5746;%%.

\bibitem{DelZotto:2014hpa}
M.~Del~Zotto, J.~J. Heckman, A.~Tomasiello, and C.~Vafa, ``{6d Conformal
  Matter},''
\href{http://arxiv.org/abs/1407.6359}{{\ttfamily arXiv:1407.6359 [hep-th]}}.
%%CITATION = ARXIV:1407.6359;%%.

\bibitem{Heckman:2014qba}
J.~J. Heckman, ``{More on the Matter of 6D SCFTs},''
\href{http://arxiv.org/abs/1408.0006}{{\ttfamily arXiv:1408.0006 [hep-th]}}.
%%CITATION = ARXIV:1408.0006;%%.

\bibitem{Douglas:2006xy}
M.~R. Douglas and W.~Taylor, ``{The Landscape of intersecting brane models},''
  \href{http://dx.doi.org/10.1088/1126-6708/2007/01/031}{{\em JHEP} {\bfseries
  0701} (2007) 031},
\href{http://arxiv.org/abs/hep-th/0606109}{{\ttfamily arXiv:hep-th/0606109
  [hep-th]}}.
%%CITATION = HEP-TH/0606109;%%.

\bibitem{Cvetic:2014gia}
M.~Cveti{\v c}, J.~Halverson, D.~Klevers, and P.~Song, ``{On finiteness of Type
  IIB compactifications: Magnetized branes on elliptic Calabi-Yau
  threefolds},'' \href{http://dx.doi.org/10.1007/JHEP06(2014)138}{{\em JHEP}
  {\bfseries 1406} (2014) 138},
\href{http://arxiv.org/abs/1403.4943}{{\ttfamily arXiv:1403.4943 [hep-th]}}.
%%CITATION = ARXIV:1403.4943;%%.

\bibitem{Grassi:2012qw}
A.~Grassi and V.~Perduca, ``{Weierstrass models of elliptic toric K3
  hypersurfaces and symplectic cuts},''
\href{http://arxiv.org/abs/1201.0930}{{\ttfamily arXiv:1201.0930 [math.AG]}}.
%%CITATION = ARXIV:1201.0930;%%.

\bibitem{Braun:2013yya}
A.~P. Braun, Y.~Kimura, and T.~Watari, ``{On the Classification of Elliptic
  Fibrations modulo Isomorphism on K3 Surfaces with large Picard Number},''
\href{http://arxiv.org/abs/1312.4421}{{\ttfamily arXiv:1312.4421 [math.AG]}}.
%%CITATION = ARXIV:1312.4421;%%.

\bibitem{Douglas:2014ywa}
M.~R. Douglas, D.~S. Park, and C.~Schnell, ``{The Cremmer-Scherk Mechanism in
  F-theory Compactifications on K3 Manifolds},''
  \href{http://dx.doi.org/10.1007/JHEP05(2014)135}{{\em JHEP} {\bfseries 1405}
  (2014) 135},
\href{http://arxiv.org/abs/1403.1595}{{\ttfamily arXiv:1403.1595 [hep-th]}}.
%%CITATION = ARXIV:1403.1595;%%.

\bibitem{Kumar:2009ac}
V.~Kumar, D.~R. Morrison, and W.~Taylor, ``{Mapping 6D N = 1 supergravities to
  F-theory},'' \href{http://dx.doi.org/10.1007/JHEP02(2010)099}{{\em JHEP}
  {\bfseries 1002} (2010) 099},
\href{http://arxiv.org/abs/0911.3393}{{\ttfamily arXiv:0911.3393 [hep-th]}}.
%%CITATION = ARXIV:0911.3393;%%.

\bibitem{Kumar:2010ru}
V.~Kumar, D.~R. Morrison, and W.~Taylor, ``{Global aspects of the space of 6D N
  = 1 supergravities},'' \href{http://dx.doi.org/10.1007/JHEP11(2010)118}{{\em
  JHEP} {\bfseries 1011} (2010) 118},
\href{http://arxiv.org/abs/1008.1062}{{\ttfamily arXiv:1008.1062 [hep-th]}}.
%%CITATION = ARXIV:1008.1062;%%.

\bibitem{Grimm:2012yq}
T.~W. Grimm and W.~Taylor, ``{Structure in 6D and 4D N=1 supergravity theories
  from F-theory},'' \href{http://dx.doi.org/10.1007/JHEP10(2012)105}{{\em JHEP}
  {\bfseries 1210} (2012) 105},
\href{http://arxiv.org/abs/1204.3092}{{\ttfamily arXiv:1204.3092 [hep-th]}}.
%%CITATION = ARXIV:1204.3092;%%.

\bibitem{Anderson:2014gla}
L.~B. Anderson and W.~Taylor, ``{Geometric constraints in dual F-theory and
  heterotic string compactifications},''
\href{http://arxiv.org/abs/1405.2074}{{\ttfamily arXiv:1405.2074 [hep-th]}}.
%%CITATION = ARXIV:1405.2074;%%.

\bibitem{Park:2011wv}
D.~S. Park and W.~Taylor, ``{Constraints on 6D Supergravity Theories with
  Abelian Gauge Symmetry},''
  \href{http://dx.doi.org/10.1007/JHEP01(2012)141}{{\em JHEP} {\bfseries 1201}
  (2012) 141},
\href{http://arxiv.org/abs/1110.5916}{{\ttfamily arXiv:1110.5916 [hep-th]}}.
%%CITATION = ARXIV:1110.5916;%%.

\bibitem{Kumar:2010am}
V.~Kumar, D.~S. Park, and W.~Taylor, ``{6D supergravity without tensor
  multiplets},'' \href{http://dx.doi.org/10.1007/JHEP04(2011)080}{{\em JHEP}
  {\bfseries 1104} (2011) 080},
\href{http://arxiv.org/abs/1011.0726}{{\ttfamily arXiv:1011.0726 [hep-th]}}.
%%CITATION = ARXIV:1011.0726;%%.

\bibitem{Bonetti:2013fma}
F.~Bonetti, T.~W. Grimm, and T.~G. Pugh, ``{Non-Supersymmetric F-Theory
  Compactifications on Spin(7) Manifolds},''
  \href{http://dx.doi.org/10.1007/JHEP01(2014)112}{{\em JHEP} {\bfseries 1401}
  (2014) 112},
\href{http://arxiv.org/abs/1307.5858}{{\ttfamily arXiv:1307.5858 [hep-th]}}.
%%CITATION = ARXIV:1307.5858;%%.

\bibitem{Bonetti:2013nka}
F.~Bonetti, T.~W. Grimm, E.~Palti, and T.~G. Pugh, ``{F-Theory on Spin(7)
  Manifolds: Weak-Coupling Limit},''
  \href{http://dx.doi.org/10.1007/JHEP02(2014)076}{{\em JHEP} {\bfseries 1402}
  (2014) 076},
\href{http://arxiv.org/abs/1309.2287}{{\ttfamily arXiv:1309.2287 [hep-th]}}.
%%CITATION = ARXIV:1309.2287;%%.

\bibitem{Morrison:2012np}
D.~R. Morrison and W.~Taylor, ``{Classifying bases for 6D F-theory models},''
  \href{http://dx.doi.org/10.2478/s11534-012-0065-4}{{\em Central Eur.J.Phys.}
  {\bfseries 10} (2012) 1072--1088},
\href{http://arxiv.org/abs/1201.1943}{{\ttfamily arXiv:1201.1943 [hep-th]}}.
%%CITATION = ARXIV:1201.1943;%%.

\bibitem{Morrison:2012js}
D.~R. Morrison and W.~Taylor, ``{Toric bases for 6D F-theory models},''
  \href{http://dx.doi.org/10.1002/prop.201200086}{{\em Fortsch.Phys.}
  {\bfseries 60} (2012) 1187--1216},
\href{http://arxiv.org/abs/1204.0283}{{\ttfamily arXiv:1204.0283 [hep-th]}}.
%%CITATION = ARXIV:1204.0283;%%.

\bibitem{Martini:2014iza}
G.~Martini and W.~Taylor, ``{6D F-theory models and elliptically fibered
  Calabi-Yau threefolds over semi-toric base surfaces},''
\href{http://arxiv.org/abs/1404.6300}{{\ttfamily arXiv:1404.6300 [hep-th]}}.
%%CITATION = ARXIV:1404.6300;%%.

\bibitem{Grimm:2010ez}
T.~W. Grimm and T.~Weigand, ``{On Abelian Gauge Symmetries and Proton Decay in
  Global F-theory GUTs},''
  \href{http://dx.doi.org/10.1103/PhysRevD.82.086009}{{\em Phys.Rev.}
  {\bfseries D82} (2010) 086009},
  \href{http://arxiv.org/abs/1006.0226}{{\ttfamily arXiv:1006.0226 [hep-th]}}.

\bibitem{Krause:2011xj}
S.~Krause, C.~Mayrhofer, and T.~Weigand, ``{$G_4$ flux, chiral matter and
  singularity resolution in F-theory compactifications},''
  \href{http://dx.doi.org/10.1016/j.nuclphysb.2011.12.013}{{\em Nucl.Phys.}
  {\bfseries B858} (2012) 1--47},
\href{http://arxiv.org/abs/1109.3454}{{\ttfamily arXiv:1109.3454 [hep-th]}}.
%%CITATION = ARXIV:1109.3454;%%.

\bibitem{Grimm:2011fx}
T.~W. Grimm and H.~Hayashi, ``{F-theory fluxes, Chirality and Chern-Simons
  theories},'' \href{http://dx.doi.org/10.1007/JHEP03(2012)027}{{\em JHEP}
  {\bfseries 1203} (2012) 027},
\href{http://arxiv.org/abs/1111.1232}{{\ttfamily arXiv:1111.1232 [hep-th]}}.
%%CITATION = ARXIV:1111.1232;%%.

\bibitem{Park:2011ji}
D.~S. Park, ``{Anomaly Equations and Intersection Theory},'' {\em JHEP}
  {\bfseries 1201} (2012) 093,
\href{http://arxiv.org/abs/1111.2351}{{\ttfamily arXiv:1111.2351 [hep-th]}}.
%%CITATION = ARXIV:1111.2351;%%.

\bibitem{Cvetic:2012xn}
M.~Cveti{\v c}, T.~W. Grimm, and D.~Klevers, ``{Anomaly Cancellation And
  Abelian Gauge Symmetries In F-theory},''
  \href{http://dx.doi.org/10.1007/JHEP02(2013)101}{{\em JHEP} {\bfseries 1302}
  (2013) 101},
\href{http://arxiv.org/abs/1210.6034}{{\ttfamily arXiv:1210.6034 [hep-th]}}.
%%CITATION = ARXIV:1210.6034;%%.

\bibitem{Mayrhofer:2012zy}
C.~Mayrhofer, E.~Palti, and T.~Weigand, ``{U(1) symmetries in F-theory GUTs
  with multiple sections},''
\href{http://arxiv.org/abs/1211.6742}{{\ttfamily arXiv:1211.6742 [hep-th]}}.
%%CITATION = ARXIV:1211.6742;%%.

\bibitem{Braun:2013yti}
V.~Braun, T.~W. Grimm, and J.~Keitel, ``{New Global F-theory GUTs with U(1)
  symmetries},''
\href{http://arxiv.org/abs/1302.1854}{{\ttfamily arXiv:1302.1854 [hep-th]}}.
%%CITATION = ARXIV:1302.1854;%%.

\bibitem{Borchmann:2013jwa}
E.~P. J.~Borchmann, C.~Mayrhofer and T.~Weigand, ``{Elliptic fibrations for
  SU(5) x U(1) x U(1) F-theory vacua},''
\href{http://arxiv.org/abs/1303.5054}{{\ttfamily arXiv:1303.5054 [hep-th]}}.
%%CITATION = ARXIV:1303.5054;%%.

\bibitem{Cvetic:2013nia}
M.~Cveti{\v c}, D.~Klevers, and H.~Piragua, ``{F-Theory Compactifications with
  Multiple U(1)-Factors: Constructing Elliptic Fibrations with Rational
  Sections},'' \href{http://dx.doi.org/10.1007/JHEP06(2013)067}{{\em JHEP}
  {\bfseries 1306} (2013) 067},
\href{http://arxiv.org/abs/1303.6970}{{\ttfamily arXiv:1303.6970 [hep-th]}}.
%%CITATION = ARXIV:1303.6970;%%.

\bibitem{Grimm:2013oga}
T.~W. Grimm, A.~Kapfer, and J.~Keitel, ``{Effective action of 6D F-Theory with
  U(1) factors: Rational sections make Chern-Simons terms jump},''
\href{http://arxiv.org/abs/1305.1929}{{\ttfamily arXiv:1305.1929 [hep-th]}}.
%%CITATION = ARXIV:1305.1929;%%.

\bibitem{Braun:2013nqa}
V.~Braun, T.~W. Grimm, and J.~Keitel, ``{Geometric Engineering in Toric
  F-Theory and GUTs with U(1) Gauge Factors},''
\href{http://arxiv.org/abs/1306.0577}{{\ttfamily arXiv:1306.0577 [hep-th]}}.
%%CITATION = ARXIV:1306.0577;%%.

\bibitem{Cvetic:2013uta}
M.~Cveti{\v c}, A.~Grassi, D.~Klevers, and H.~Piragua, ``{Chiral
  Four-Dimensional F-Theory Compactifications With SU(5) and Multiple
  U(1)-Factors},''
\href{http://arxiv.org/abs/1306.3987}{{\ttfamily arXiv:1306.3987 [hep-th]}}.
%%CITATION = ARXIV:1306.3987;%%.

\bibitem{Borchmann:2013hta}
J.~Borchmann, C.~Mayrhofer, E.~Palti, and T.~Weigand, ``{SU(5) Tops with
  Multiple U(1)s in F-theory},''
\href{http://arxiv.org/abs/1307.2902}{{\ttfamily arXiv:1307.2902 [hep-th]}}.
%%CITATION = ARXIV:1307.2902;%%.

\bibitem{Cvetic:2013jta}
M.~Cveti{\v c}, D.~Klevers, and H.~Piragua, ``{F-Theory Compactifications with
  Multiple U(1)-Factors: Addendum},''
  \href{http://dx.doi.org/10.1007/JHEP12(2013)056}{{\em JHEP} {\bfseries 1312}
  (2013) 056},
\href{http://arxiv.org/abs/1307.6425}{{\ttfamily arXiv:1307.6425 [hep-th]}}.
%%CITATION = ARXIV:1307.6425;%%.

\bibitem{Cvetic:2013qsa}
M.~Cveti{\v c}, D.~Klevers, H.~Piragua, and P.~Song, ``{Elliptic fibrations
  with rank three Mordell-Weil group: F-theory with U(1) x U(1) x U(1) gauge
  symmetry},'' \href{http://dx.doi.org/10.1007/JHEP03(2014)021}{{\em JHEP}
  {\bfseries 1403} (2014) 021},
\href{http://arxiv.org/abs/1310.0463}{{\ttfamily arXiv:1310.0463 [hep-th]}}.
%%CITATION = ARXIV:1310.0463;%%.

\bibitem{Mayrhofer:2014opa}
C.~Mayrhofer, D.~R. Morrison, O.~Till, and T.~Weigand, ``{Mordell-Weil Torsion
  and the Global Structure of Gauge Groups in F-theory},''
\href{http://arxiv.org/abs/1405.3656}{{\ttfamily arXiv:1405.3656 [hep-th]}}.
%%CITATION = ARXIV:1405.3656;%%.

\bibitem{Aldazabal:1996du}
G.~Aldazabal, A.~Font, L.~E. Ibanez, and A.~Uranga, ``{New branches of string
  compactifications and their F theory duals},''
  \href{http://dx.doi.org/10.1016/S0550-3213(96)00699-2}{{\em Nucl.Phys.}
  {\bfseries B492} (1997) 119--151},
\href{http://arxiv.org/abs/hep-th/9607121}{{\ttfamily arXiv:hep-th/9607121
  [hep-th]}}.
%%CITATION = HEP-TH/9607121;%%.

\bibitem{Klemm:1996hh}
A.~Klemm, P.~Mayr, and C.~Vafa, ``{BPS states of exceptional noncritical
  strings},''
\href{http://arxiv.org/abs/hep-th/9607139}{{\ttfamily arXiv:hep-th/9607139
  [hep-th]}}.
%%CITATION = HEP-TH/9607139;%%.

\bibitem{Klemm:2004km}
A.~Klemm, M.~Kreuzer, E.~Riegler, and E.~Scheidegger, ``{Topological string
  amplitudes, complete intersection Calabi-Yau spaces and threshold
  corrections},'' \href{http://dx.doi.org/10.1088/1126-6708/2005/05/023}{{\em
  JHEP} {\bfseries 0505} (2005) 023},
\href{http://arxiv.org/abs/hep-th/0410018}{{\ttfamily arXiv:hep-th/0410018
  [hep-th]}}.
%%CITATION = HEP-TH/0410018;%%.

\bibitem{Esole:2011cn}
M.~Esole, J.~Fullwood, and S.-T. Yau, ``{$D_5$ elliptic fibrations: non-Kodaira
  fibers and new orientifold limits of F-theory},''
\href{http://arxiv.org/abs/1110.6177}{{\ttfamily arXiv:1110.6177 [hep-th]}}.
%%CITATION = ARXIV:1110.6177;%%.

\bibitem{Grimm:2011tb}
T.~W. Grimm, M.~Kerstan, E.~Palti, and T.~Weigand, ``{Massive Abelian Gauge
  Symmetries and Fluxes in F-theory},''
  \href{http://dx.doi.org/10.1007/JHEP12(2011)004}{{\em JHEP} {\bfseries 1112}
  (2011) 004},
\href{http://arxiv.org/abs/1107.3842}{{\ttfamily arXiv:1107.3842 [hep-th]}}.
%%CITATION = ARXIV:1107.3842;%%.

\bibitem{Braun:2014nva}
A.~P. Braun, A.~Collinucci, and R.~Valandro, ``{The fate of U(1)'s at strong
  coupling in F-theory},''
  \href{http://dx.doi.org/10.1007/JHEP07(2014)028}{{\em JHEP} {\bfseries 1407}
  (2014) 028},
\href{http://arxiv.org/abs/1402.4054}{{\ttfamily arXiv:1402.4054 [hep-th]}}.
%%CITATION = ARXIV:1402.4054;%%.

\bibitem{Aspinwall:1998xj}
P.~S. Aspinwall and D.~R. Morrison, ``{Nonsimply connected gauge groups and
  rational points on elliptic curves},'' {\em JHEP} {\bfseries 9807} (1998)
  012,
\href{http://arxiv.org/abs/hep-th/9805206}{{\ttfamily arXiv:hep-th/9805206
  [hep-th]}}.
%%CITATION = HEP-TH/9805206;%%.

\bibitem{Braun:2014oya}
V.~Braun and D.~R. Morrison, ``{F-theory on Genus-One Fibrations},''
\href{http://arxiv.org/abs/1401.7844}{{\ttfamily arXiv:1401.7844 [hep-th]}}.
%%CITATION = ARXIV:1401.7844;%%.

\bibitem{Morrison:2014era}
D.~R. Morrison and W.~Taylor, ``{Sections, multisections, and U(1) fields in
  F-theory},''
\href{http://arxiv.org/abs/1404.1527}{{\ttfamily arXiv:1404.1527 [hep-th]}}.
%%CITATION = ARXIV:1404.1527;%%.

\bibitem{Anderson:2014yva}
L.~B. Anderson, I.~Garc{\'\i}a-Etxebarria, T.~W. Grimm, and J.~Keitel,
  ``{Physics of F-theory compactifications without section},''
\href{http://arxiv.org/abs/1406.5180}{{\ttfamily arXiv:1406.5180 [hep-th]}}.
%%CITATION = ARXIV:1406.5180;%%.

\bibitem{Huang:2013yta}
M.-x. Huang, A.~Klemm, and M.~Poretschkin, ``{Refined stable pair invariants
  for E-, M- and [p,q]-strings},''
\href{http://arxiv.org/abs/1308.0619}{{\ttfamily arXiv:1308.0619 [hep-th]}}.
%%CITATION = ARXIV:1308.0619;%%.

\bibitem{Huang:2014nwa}
M.-x. Huang, A.~Klemm, J.~Reuter, and M.~Schiereck, ``{Quantum geometry of del
  Pezzo surfaces in the Nekrasov-Shatashvili limit},''
\href{http://arxiv.org/abs/1401.4723}{{\ttfamily arXiv:1401.4723 [hep-th]}}.
%%CITATION = ARXIV:1401.4723;%%.

\bibitem{kodaira1963compact}
K.~Kodaira, ``On compact analytic surfaces: Ii,'' {\em The Annals of
  Mathematics} {\bfseries 77} no.~3, (1963) 563--626.

\bibitem{tate1975algorithm}
J.~Tate, ``Algorithm for determining the type of a singular fiber in an
  elliptic pencil,'' {\em Modular functions of one variable IV} (1975) 33--52.

\bibitem{Bershadsky:1996nh}
M.~Bershadsky, K.~A. Intriligator, S.~Kachru, D.~R. Morrison, V.~Sadov, {\em
  et~al.}, ``{Geometric singularities and enhanced gauge symmetries},''
  \href{http://dx.doi.org/10.1016/S0550-3213(96)90131-5}{{\em Nucl.Phys.}
  {\bfseries B481} (1996) 215--252},
\href{http://arxiv.org/abs/hep-th/9605200}{{\ttfamily arXiv:hep-th/9605200
  [hep-th]}}.
%%CITATION = HEP-TH/9605200;%%.

\bibitem{Katz:2011qp}
S.~Katz, D.~R. Morrison, S.~Schafer-Nameki, and J.~Sully, ``{Tate's algorithm
  and F-theory},'' \href{http://dx.doi.org/10.1007/JHEP08(2011)094}{{\em JHEP}
  {\bfseries 1108} (2011) 094},
\href{http://arxiv.org/abs/1106.3854}{{\ttfamily arXiv:1106.3854 [hep-th]}}.
%%CITATION = ARXIV:1106.3854;%%.

\bibitem{Lawrie:2012gg}
C.~Lawrie and S.~Schafer-Nameki, ``{The Tate Form on Steroids: Resolution and
  Higher Codimension Fibers},''
\href{http://arxiv.org/abs/1212.2949}{{\ttfamily arXiv:1212.2949 [hep-th]}}.
%%CITATION = ARXIV:1212.2949;%%.

\bibitem{Candelas:1996su}
P.~Candelas and A.~Font, ``{Duality between the webs of heterotic and type II
  vacua},'' \href{http://dx.doi.org/10.1016/S0550-3213(96)00410-5}{{\em
  Nucl.Phys.} {\bfseries B511} (1998) 295--325},
\href{http://arxiv.org/abs/hep-th/9603170}{{\ttfamily arXiv:hep-th/9603170
  [hep-th]}}.
%%CITATION = HEP-TH/9603170;%%.

\bibitem{Bouchard:2003bu}
V.~Bouchard and H.~Skarke, ``{Affine Kac-Moody algebras, CHL strings and the
  classification of tops},'' {\em Adv.Theor.Math.Phys.} {\bfseries 7} (2003)
  205--232,
\href{http://arxiv.org/abs/hep-th/0303218}{{\ttfamily arXiv:hep-th/0303218
  [hep-th]}}.
%%CITATION = HEP-TH/0303218;%%.

\bibitem{Kuntzler:2014ila}
M.~Kuntzler and S.~Schafer-Nameki, ``{Tate Trees for Elliptic Fibrations with
  Rank one Mordell-Weil group},''
\href{http://arxiv.org/abs/1406.5174}{{\ttfamily arXiv:1406.5174 [hep-th]}}.
%%CITATION = ARXIV:1406.5174;%%.

\bibitem{Lin:2014qga}
L.~Lin and T.~Weigand, ``{Towards the Standard Model in F-theory},''
\href{http://arxiv.org/abs/1406.6071}{{\ttfamily arXiv:1406.6071 [hep-th]}}.
%%CITATION = ARXIV:1406.6071;%%.

\bibitem{Krauss:1988zc}
L.~M. Krauss and F.~Wilczek, ``{Discrete Gauge Symmetry in Continuum
  Theories},''
\href{http://dx.doi.org/10.1103/PhysRevLett.62.1221}{{\em Phys.Rev.Lett.}
  {\bfseries 62} (1989) 1221}.
%%CITATION = PRLTA,62,1221;%%.

\bibitem{Banks:1989ag}
T.~Banks, ``{Effective Lagrangian Description of Discrete Gauge Symmetries},''
\href{http://dx.doi.org/10.1016/0550-3213(89)90589-0}{{\em Nucl.Phys.}
  {\bfseries B323} (1989) 90}.
%%CITATION = NUPHA,B323,90;%%.

\bibitem{Ibanez:1991wt}
L.~E. Ibanez and G.~G. Ross,
``{Should discrete symmetries be anomaly free?},''.
%%CITATION = CERN-TH-6000-91 ETC.;%%.

\bibitem{silverman2009arithmetic}
J.~H. Silverman, {\em The arithmetic of elliptic curves}, vol.~106.
\newblock Springer, 2009.

\bibitem{lang1959rational}
S.~Lang and A.~Neron, ``Rational points of abelian varieties over function
  fields,'' {\em American Journal of Mathematics} (1959) 95--118.

\bibitem{mazur1977modular}
B.~Mazur, ``Modular curves and the eisenstein ideal,'' {\em Publications
  Math{\'e}matiques de l'Institut des Hautes {\'E}tudes Scientifiques}
  {\bfseries 47} no.~1, (1977) 33--186.

\bibitem{mazur1978rational}
B.~Mazur and D.~Goldfeld, ``Rational isogenies of prime degree,'' {\em
  Inventiones mathematicae} {\bfseries 44} no.~2, (1978) 129--162.

\bibitem{nakayama1985weierstrass}
N.~Nakayama, ``On weierstrass models,'' {\em Algebraic geometry and commutative
  algebra,} {\bfseries Vol. II} (1988) pp. 405--431. Kinokuniya.

\bibitem{an2001jacobians}
S.~Y. An, S.~Y. Kim, D.~C. Marshall, S.~H. Marshall, W.~G. McCallum, and A.~R.
  Perlis, ``Jacobians of genus one curves,'' {\em Journal of Number Theory}
  {\bfseries 90} no.~2, (2001) 304--315.

\bibitem{Morrison:2012ei}
D.~R. Morrison and D.~S. Park, ``{F-Theory and the Mordell-Weil Group of
  Elliptically-Fibered Calabi-Yau Threefolds},''
  \href{http://dx.doi.org/10.1007/JHEP10(2012)128}{{\em JHEP} {\bfseries 1210}
  (2012) 128},
\href{http://arxiv.org/abs/1208.2695}{{\ttfamily arXiv:1208.2695 [hep-th]}}.
%%CITATION = ARXIV:1208.2695;%%.

\bibitem{Grimm:2010ks}
T.~W. Grimm, ``{The N=1 effective action of F-theory compactifications},''
  \href{http://dx.doi.org/10.1016/j.nuclphysb.2010.11.018}{{\em Nucl.Phys.}
  {\bfseries B845} (2011) 48--92},
\href{http://arxiv.org/abs/1008.4133}{{\ttfamily arXiv:1008.4133 [hep-th]}}.
%%CITATION = ARXIV:1008.4133;%%.

\bibitem{Denef:2008wq}
F.~Denef, ``{Les Houches Lectures on Constructing String Vacua},''
\href{http://arxiv.org/abs/0803.1194}{{\ttfamily arXiv:0803.1194 [hep-th]}}.
%%CITATION = ARXIV:0803.1194;%%.

\bibitem{Taylor:2011wt}
W.~Taylor, ``{TASI Lectures on Supergravity and String Vacua in Various
  Dimensions},''
\href{http://arxiv.org/abs/1104.2051}{{\ttfamily arXiv:1104.2051 [hep-th]}}.
%%CITATION = ARXIV:1104.2051;%%.

\bibitem{Witten:1996qb}
E.~Witten, ``{Phase transitions in M theory and F theory},''
  \href{http://dx.doi.org/10.1016/0550-3213(96)00212-X}{{\em Nucl.Phys.}
  {\bfseries B471} (1996) 195--216},
\href{http://arxiv.org/abs/hep-th/9603150}{{\ttfamily arXiv:hep-th/9603150
  [hep-th]}}.
%%CITATION = HEP-TH/9603150;%%.

\bibitem{Katz:1996ht}
S.~H. Katz, D.~R. Morrison, and M.~R. Plesser, ``{Enhanced gauge symmetry in
  type II string theory},''
  \href{http://dx.doi.org/10.1016/0550-3213(96)00331-8}{{\em Nucl.Phys.}
  {\bfseries B477} (1996) 105--140},
\href{http://arxiv.org/abs/hep-th/9601108}{{\ttfamily arXiv:hep-th/9601108
  [hep-th]}}.
%%CITATION = HEP-TH/9601108;%%.

\bibitem{hartshorne1977algebraic}
R.~Hartshorne, {\em Algebraic geometry}.
\newblock No.~52. Springer, 1977.

\bibitem{fulton1993introduction}
W.~Fulton, {\em Introduction to toric varieties}.
\newblock No.~131. Princeton University Press, 1993.

\bibitem{david2011toric}
D.~A. Cox, J.~B. Little, and H.~K. Schenck, {\em Toric varieties}.
\newblock American Mathematical Soc., 2011.

\bibitem{Cox:1993fz}
D.~A. Cox, ``{The Homogeneous coordinate ring of a toric variety, revised
  version},''
\href{http://arxiv.org/abs/alg-geom/9210008}{{\ttfamily arXiv:alg-geom/9210008
  [alg-geom]}}.
%%CITATION = ALG-GEOM/9210008;%%.

\bibitem{Batyrev:1994hm}
V.~V. Batyrev, ``{Dual polyhedra and mirror symmetry for Calabi-Yau
  hypersurfaces in toric varieties},'' {\em J.Alg.Geom.} {\bfseries 3} (1994)
  493--545,
\href{http://arxiv.org/abs/alg-geom/9310003}{{\ttfamily arXiv:alg-geom/9310003
  [alg-geom]}}.
%%CITATION = ALG-GEOM/9310003;%%.

\bibitem{Singular}
W.~Decker, G.-M. Greuel, G.~Pfister, and H.~Sch\"onemann, ``{\sc Singular}
  {3-1-6} --- {A} computer algebra system for polynomial computations.''
  \url{http://www.singular.uni-kl.de}, 2012.

\bibitem{Banks:2010zn}
T.~Banks and N.~Seiberg, ``{Symmetries and Strings in Field Theory and
  Gravity},'' \href{http://dx.doi.org/10.1103/PhysRevD.83.084019}{{\em
  Phys.Rev.} {\bfseries D83} (2011) 084019},
\href{http://arxiv.org/abs/1011.5120}{{\ttfamily arXiv:1011.5120 [hep-th]}}.
%%CITATION = ARXIV:1011.5120;%%.

\bibitem{Hayashi:2014kca}
H.~Hayashi, C.~Lawrie, D.~R. Morrison, and S.~Schafer-Nameki, ``{Box Graphs and
  Singular Fibers},'' \href{http://dx.doi.org/10.1007/JHEP05(2014)048}{{\em
  JHEP} {\bfseries 1405} (2014) 048},
\href{http://arxiv.org/abs/1402.2653}{{\ttfamily arXiv:1402.2653 [hep-th]}}.
%%CITATION = ARXIV:1402.2653;%%.

\bibitem{Friedman:1997yq}
R.~Friedman, J.~Morgan, and E.~Witten, ``{Vector bundles and F theory},''
  \href{http://dx.doi.org/10.1007/s002200050154}{{\em Commun.Math.Phys.}
  {\bfseries 187} (1997) 679--743},
\href{http://arxiv.org/abs/hep-th/9701162}{{\ttfamily arXiv:hep-th/9701162
  [hep-th]}}.
%%CITATION = HEP-TH/9701162;%%.

\bibitem{Honecker:2006qz}
G.~Honecker and M.~Trapletti, ``{Merging Heterotic Orbifolds and K3
  Compactifications with Line Bundles},''
  \href{http://dx.doi.org/10.1088/1126-6708/2007/01/051}{{\em JHEP} {\bfseries
  0701} (2007) 051},
\href{http://arxiv.org/abs/hep-th/0612030}{{\ttfamily arXiv:hep-th/0612030
  [hep-th]}}.
%%CITATION = HEP-TH/0612030;%%.

\bibitem{Grassi:2011hq}
A.~Grassi and D.~R. Morrison, ``{Anomalies and the Euler characteristic of
  elliptic Calabi-Yau threefolds},''
\href{http://arxiv.org/abs/1109.0042}{{\ttfamily arXiv:1109.0042 [hep-th]}}.
%%CITATION = ARXIV:1109.0042;%%.

\bibitem{Intriligator:2012ue}
K.~Intriligator, H.~Jockers, P.~Mayr, D.~R. Morrison, and M.~R. Plesser,
  ``{Conifold Transitions in M-theory on Calabi-Yau Fourfolds with Background
  Fluxes},'' \href{http://dx.doi.org/10.4310/ATMP.2013.v17.n3.a2}{{\em
  Adv.Theor.Math.Phys.} {\bfseries 17} (2013) 601--699},
\href{http://arxiv.org/abs/1203.6662}{{\ttfamily arXiv:1203.6662 [hep-th]}}.
%%CITATION = ARXIV:1203.6662;%%.

\bibitem{Bizet:2014uua}
N.~C. Bizet, A.~Klemm, and D.~V. Lopes, ``{Landscaping with fluxes and the E8
  Yukawa Point in F-theory},''
\href{http://arxiv.org/abs/1404.7645}{{\ttfamily arXiv:1404.7645 [hep-th]}}.
%%CITATION = ARXIV:1404.7645;%%.

\bibitem{Berglund:1998ej}
P.~Berglund and P.~Mayr, ``{Heterotic string / F theory duality from mirror
  symmetry},'' {\em Adv.Theor.Math.Phys.} {\bfseries 2} (1999) 1307--1372,
\href{http://arxiv.org/abs/hep-th/9811217}{{\ttfamily arXiv:hep-th/9811217
  [hep-th]}}.
%%CITATION = HEP-TH/9811217;%%.

\bibitem{Farrar:1978xj}
G.~R. Farrar and P.~Fayet, ``{Phenomenology of the Production, Decay, and
  Detection of New Hadronic States Associated with Supersymmetry},''
\href{http://dx.doi.org/10.1016/0370-2693(78)90858-4}{{\em Phys.Lett.}
  {\bfseries B76} (1978) 575--579}.
%%CITATION = PHLTA,B76,575;%%.

\bibitem{Dimopoulos:1981dw}
S.~Dimopoulos, S.~Raby, and F.~Wilczek, ``{Proton Decay in Supersymmetric
  Models},''
\href{http://dx.doi.org/10.1016/0370-2693(82)90313-6}{{\em Phys.Lett.}
  {\bfseries B112} (1982) 133}.
%%CITATION = PHLTA,B112,133;%%.

\bibitem{Ibanez:1991pr}
L.~E. Ibanez and G.~G. Ross, ``{Discrete gauge symmetries and the origin of
  baryon and lepton number conservation in supersymmetric versions of the
  standard model},''
\href{http://dx.doi.org/10.1016/0550-3213(92)90195-H}{{\em Nucl.Phys.}
  {\bfseries B368} (1992) 3--37}.
%%CITATION = NUPHA,B368,3;%%.

\bibitem{Dreiner:2005rd}
H.~K. Dreiner, C.~Luhn, and M.~Thormeier, ``{What is the discrete gauge
  symmetry of the MSSM?},''
  \href{http://dx.doi.org/10.1103/PhysRevD.73.075007}{{\em Phys.Rev.}
  {\bfseries D73} (2006) 075007},
\href{http://arxiv.org/abs/hep-ph/0512163}{{\ttfamily arXiv:hep-ph/0512163
  [hep-ph]}}.
%%CITATION = HEP-PH/0512163;%%.

\bibitem{Lee:2010gv}
H.~M. Lee, S.~Raby, M.~Ratz, G.~G. Ross, R.~Schieren, {\em et~al.}, ``{A unique
  $Z_4^R$ symmetry for the MSSM},''
  \href{http://dx.doi.org/10.1016/j.physletb.2010.10.038}{{\em Phys.Lett.}
  {\bfseries B694} (2011) 491--495},
\href{http://arxiv.org/abs/1009.0905}{{\ttfamily arXiv:1009.0905 [hep-ph]}}.
%%CITATION = ARXIV:1009.0905;%%.

\bibitem{Dudas:2009hu}
E.~Dudas and E.~Palti, ``{Froggatt-Nielsen models from E(8) in F-theory
  GUTs},'' \href{http://dx.doi.org/10.1007/JHEP01(2010)127}{{\em JHEP}
  {\bfseries 1001} (2010) 127},
\href{http://arxiv.org/abs/0912.0853}{{\ttfamily arXiv:0912.0853 [hep-th]}}.
%%CITATION = ARXIV:0912.0853;%%.

\bibitem{Krippendorf:2014xba}
S.~Krippendorf, D.~K. Mayorga~Pena, P.-K. Oehlmann, and F.~Ruehle, ``{Rational
  F-Theory GUTs without exotics},''
  \href{http://dx.doi.org/10.1007/JHEP07(2014)013}{{\em JHEP} {\bfseries 1407}
  (2014) 013},
\href{http://arxiv.org/abs/1401.5084}{{\ttfamily arXiv:1401.5084 [hep-th]}}.
%%CITATION = ARXIV:1401.5084;%%.

\bibitem{Grassi:2013kha}
A.~Grassi, J.~Halverson, and J.~L. Shaneson, ``{Matter From Geometry Without
  Resolution},''
\href{http://arxiv.org/abs/1306.1832}{{\ttfamily arXiv:1306.1832 [hep-th]}}.
%%CITATION = ARXIV:1306.1832;%%.

\bibitem{Grassi:2014sda}
A.~Grassi, J.~Halverson, and J.~L. Shaneson, ``{Non-Abelian Gauge Symmetry and
  the Higgs Mechanism in F-theory},''
\href{http://arxiv.org/abs/1402.5962}{{\ttfamily arXiv:1402.5962 [hep-th]}}.
%%CITATION = ARXIV:1402.5962;%%.

\bibitem{Hayashi:2013lra}
H.~Hayashi, C.~Lawrie, and S.~Schafer-Nameki, ``{Phases, Flops and F-theory:
  SU(5) Gauge Theories},''
  \href{http://dx.doi.org/10.1007/JHEP10(2013)046}{{\em JHEP} {\bfseries 1310}
  (2013) 046},
\href{http://arxiv.org/abs/1304.1678}{{\ttfamily arXiv:1304.1678 [hep-th]}}.
%%CITATION = ARXIV:1304.1678;%%.

\bibitem{Esole:2014bka}
M.~Esole, S.-H. Shao, and S.-T. Yau, ``{Singularities and Gauge Theory
  Phases},''
\href{http://arxiv.org/abs/1402.6331}{{\ttfamily arXiv:1402.6331 [hep-th]}}.
%%CITATION = ARXIV:1402.6331;%%.

\bibitem{Esole:2014hya}
M.~Esole, S.-H. Shao, and S.-T. Yau, ``{Singularities and Gauge Theory Phases
  II},''
\href{http://arxiv.org/abs/1407.1867}{{\ttfamily arXiv:1407.1867 [hep-th]}}.
%%CITATION = ARXIV:1407.1867;%%.

\bibitem{Braun:2014kla}
A.~P. Braun and S.~Schafer-Nameki, ``{Box Graphs and Resolutions I},''
\href{http://arxiv.org/abs/1407.3520}{{\ttfamily arXiv:1407.3520 [math.AG]}}.
%%CITATION = ARXIV:1407.3520;%%.

\bibitem{Esole:2011sm}
M.~Esole and S.-T. Yau, ``{Small resolutions of SU(5)-models in F-theory},''
\href{http://arxiv.org/abs/1107.0733}{{\ttfamily arXiv:1107.0733 [hep-th]}}.
%%CITATION = ARXIV:1107.0733;%%.

\bibitem{Marsano:2011hv}
J.~Marsano and S.~Schafer-Nameki, ``{Yukawas, G-flux, and Spectral Covers from
  Resolved Calabi-Yau's},''
  \href{http://dx.doi.org/10.1007/JHEP11(2011)098}{{\em JHEP} {\bfseries 1111}
  (2011) 098},
\href{http://arxiv.org/abs/1108.1794}{{\ttfamily arXiv:1108.1794 [hep-th]}}.
%%CITATION = ARXIV:1108.1794;%%.

\bibitem{Mayr:1996sh}
P.~Mayr, ``{Mirror symmetry, N=1 superpotentials and tensionless strings on
  Calabi-Yau four folds},''
  \href{http://dx.doi.org/10.1016/S0550-3213(97)00196-X}{{\em Nucl.Phys.}
  {\bfseries B494} (1997) 489--545},
\href{http://arxiv.org/abs/hep-th/9610162}{{\ttfamily arXiv:hep-th/9610162
  [hep-th]}}.
%%CITATION = HEP-TH/9610162;%%.

\end{thebibliography}\endgroup

\end{document}